\documentclass[11pt,a4paper]{article}
\usepackage{jheppub}
\usepackage{graphicx} 
\usepackage{amsmath}
\usepackage{amssymb} 
\usepackage{dsfont}
\usepackage{xcolor}
\usepackage{hyperref}
\usepackage[normalem]{ulem} 
\usepackage{amsthm} 

\usepackage{braket}
\usepackage{cleveref}
\usepackage{verbatim}
\preprint{CERN-TH-2025-040}
\title{Operator K-complexity in DSSYK: Krylov complexity equals bulk length}

\author[a]{M. Ambrosini,}

\author[b]{E. Rabinovici,} 

\author[c]{A. S\'{a}nchez-Garrido,}

\author[d]{R. Shir}

\author[a]{and J. Sonner}

\affiliation[a]{Department of Theoretical Physics, University of Geneva, 24 quai Ernest-Ansermet, 1214 Gen\`eve 4, Switzerland} 

\affiliation[b]{Racah Institute of Physics, The Hebrew University, Jerusalem 9190401, Israel}
\affiliation[b]{CERN, Theoretical Physics Department, CH-1211 Geneva 23, Switzerland}

\affiliation[c]{School of Mathematical Sciences, University of Southampton, SO17 1BJ, U.K.}

\affiliation[d]{Department of Physics and Materials Science, University of Luxembourg, L-1511 Luxembourg}

\emailAdd{Marco.Ambrosini@unige.ch}
\emailAdd{eliezer@mail.huji.ac.il}
\emailAdd{A.Sanchez-Garrido@soton.ac.uk}
\emailAdd{ruth.shir@uni.lu}
\emailAdd{Julian.Sonner@unige.ch}

\abstract{In this paper we study the notion of complexity under time evolution in chaotic quantum systems with holographic duals. Continuing on from our previous work, we turn our attention to the issue of Krylov complexity upon the insertion of a class of single-particle operators in the double-scaled SYK model. Such an operator is described by a matter-chord insertion, which splits the theory into left/right sectors, allowing us, via chord-diagram technology, to compute two different notions of complexity associated to the operator insertion: first a Krylov operator complexity, and second the Krylov complexity of a state obtained by an operator acting on the thermofield double state.
We will provide both an analytic proof and detailed numerical evidence, that both Krylov complexities arise from a recursively defined basis of states characterized by a constant total chord number. As a consequence, in all cases we are able to establish that Krylov complexity is given by the expectation value of a length operator acting on the Hilbert space of the theory, expressed in terms of basis states, organized by left and right chord number.
We find analytic expressions for the semiclassical limit of K-complexity, and study how the size of the operator encodes the scrambling dynamics upon the matter insertion in Krylov language. We furthermore determine the effective Hamiltonian governing the evolution of K-complexity, showing that evolution on the Krylov chain can equivalently be understood as a particle moving in a Morse potential. A particular type of triple scaling limit allows to access the gravitational sector of the theory, in which the geometrical nature of K-complexity is assured by virtue of being a total chord length, in an analogous fashion to what was found in \cite{Rabinovici:2023yex} for the K-complexity of the thermofield double state.}

\begin{document}

\maketitle

\section{Introduction and overview}
In this paper we continue our study of properties of a  type of complexity named Krylov Complexity or simply K-Complexity. In particular we will investigate the K-complexity of a class of operators inserted in double-scaled SYK, and show that they can be understood in geometric terms suggesting an interpretation of geometric length in the bulk. 
Quantifying the complexity of quantum evolution is a central topic of interest, both from a practical point of view of exploring the limits of quantum information processing and quantum computing, but also from an a-priori much less expected perspective, namely that of quantum gravity \cite{Susskind:2014rva, Stanford:2014jda, Brown:2015bva}. In both of these research directions, the notion of Krylov complexity \cite{Parker:2018yvk} is emerging as a reliable quantifier of both operator \cite{Barbon:2019wsy,Jian:2020qpp,Rabinovici:2020ryf,Rabinovici:2022beu} and state complexity \cite{Balasubramanian:2022tpr,Rabinovici:2023yex}. In this approach one associates the notion of quantum complexity with the spread of the system's wavefunction in a certain optimally chosen basis, the so-called Krylov basis \cite{Parker:2018yvk,Barbon:2019wsy,Dymarsky:2019elm,Rabinovici:2020ryf,Dymarsky:2021bjq,Rabinovici:2022beu,Balasubramanian:2022tpr,Takahashi:2024hex,Erdmenger:2023wjg,Nandy:2024htc}. Such a K-complexity can be associated with quantum operators, where operator growth under Heisenberg evolution is measured with respect to a certain iteratively constructed basis of the Hilbert space of operators, starting from a seed operator ${\cal O}_{\rm seed}$, complemented by its commutators $\left[  H, \left[H,\ldots {\cal O}_{\rm seed}\right]\right]$. Krylov complexity can also be defined for states, where Schrödinger evolution is formulated with respect to a certain iteratively constructed basis of states living in the original Hilbert space, based on a reference state $\left|\Psi\right\rangle$, complemented by states obtained by successive applications of the Hamiltonian $H\left|\Psi\right\rangle, H^2\left|\Psi\right\rangle,\ldots $.

It has been proposed \cite{Susskind:2014rva} that quantum complexity should play a pivotal role in the holographic dictionary between AdS quantum gravity and its boundary field-theory dual. A number of scenarios have been put forward in this direction \cite{Susskind:2014rva, Stanford:2014jda, Brown:2015bva}, which all associate a geometric construction in the bulk to an abstract notion of complexity on the boundary. In fact, a number of geometric diffeomorphism-invariant quantities can be defined, which all exhibit the expected  phenomenology of holographic complexity (before saturation) \cite{Susskind:2014rva, Belin:2021bga}.  It is therefore of great importance to find circumstances in which a concretely defined notion of complexity in the boundary theory can be mapped to a geometric quantity in the bulk. Progress in this direction can be made in double-scaled SYK (DSSYK), a certain quantum-mechanical system, defined as a limit of the class of $p-$body Sachdev-Ye-Kitaev models, and its AdS$_2$ dual description. The $p-$body SYK model possesses a highly interesting double scaling limit where both the cluster size $p$ and the total number of Fermions, $N$, go to infinity, leaving $\lambda = \frac{2p^2}{N}$ fixed  \cite{Berkooz:2018jqr, Berkooz_chords, Garcia-Garcia:2017pzl, Cotler:2016fpe}. In this case, it can be shown that the Krylov basis plays an important and direct role in holographic bulk reconstruction \cite{Lin:2022rbf, Rabinovici:2023yex},  leading to the identification of bulk wormhole length with boundary Krylov complexity of the thermofield double state in the semiclassical limit \cite{Rabinovici:2023yex}. In summary then, Krylov complexity of the thermofield double state in DSSYK can be directly shown to map to wormhole length under the holographic bulk-boundary map, resulting in a precise entry for a notion of complexity in the AdS/CFT dictionary. It is natural to ask whether a similar holographic correspondence can be established for Krylov operator complexity. The results of this paper indeed establish that Krylov complexity directly measures chord number in the bulk Hilbert space, and thus has a natural interpretation of bulk length in the semiclassical limit.

\subsection{Overview}\label{subsect:Overview}
Previous work, \cite{Rabinovici:2023yex}, established a direct geometric manifestation of Krylov complexity by making use of the bulk reconstruction map of \cite{Lin:2022rbf}. The precise quantity that was considered in that work was the Krylov state complexity of the thermofield double state $|TFD\rangle = |0\rangle$, identified with the zero chord state in the chordial treatment of DSSYK \cite{Berkooz:2018jqr,Berkooz:2022mfk,Berkooz:2024lgq}. The chord-diagram techniques relevant for such a description will be reviewed in Section \ref{sec.ToolsTrade} below, including the generalization to matter chords \cite{Berkooz:2018jqr,Berkooz:2022mfk,Berkooz:2024lgq, Lin:2022rbf, Lin:2023trc}, which allows us to study operator Krylov complexity in this work. As anticipated in \cite{Lin:2022rbf}, the Krylov basis in DSSYK is exactly what one obtains by the Gram-Schmidt orthonormalization procedure on the successive states $\left\{|0 \rangle, H|0\rangle, H^2 | 0\rangle, \ldots  \right\}$, giving rise to the Krylov basis $\left\{|0 \rangle, |1\rangle, | 2\rangle, \ldots  \right\}$ that is shown to coincide with the chord number base spanning the bulk Hilbert space under the reconstruction map of \cite{Lin:2022rbf}.

Given that Krylov operator complexity is equally of interest as its state complexity cousin, it is natural to ask whether a similiar story holds true also for operator complexity. We address this question in the present paper. Interestingly, there are two different notions of Krylov complexity associated to a given seed operator $\mathcal{O}_{\rm seed}$. 
\begin{enumerate}
\item Krylov operator complexity: this is the original complexity measure of operator growth, proposed in \cite{Parker:2018yvk}. Note that this complexity will become trivial for an operator which is a symmetry of $H$, and, in particular, once a limit is taken that takes $\mathcal{O}_{\rm seed}$ to the identity operator. We will show that operator complexity can be studied in an effective two-sided Hilbert space, spanned by chord states of the form $|n_L, n_R \rangle$, under evolution of a Hamiltonian $H_R - H_L$.
\item Krylov state complexity of a reference state, perturbed by the application of the operator $\mathcal{O}_{\rm seed}$: this notion of complexity will approach the Krylov state complexity of \cite{Rabinovici:2023yex} when the operator approaches the identity, and stay non-trivial in the limit. We will show that this complexity can be studied in the effective two-sided Hilbert space evolving by $H_R + H_L$.
\end{enumerate}
In order to study the first notion of complexity, we establish a mapping between the Krylov basis built from the Gram-Schmidt procedure on the set of states\footnote{Defined on the Hilbert space of operators acting on the original Hilbert space of states in DSSYK.}, $ \left| \mathcal{O}_{\rm seed} \right)$, $\left| [H,\mathcal{O}_{\rm seed}] \right), \ldots,$ $\left|\left[  H, \left[H,\ldots {\cal O}_{\rm seed}\right]\right]\right)$, and states $| n_L, n_R\rangle$ of \textit{fixed chord number} $n=n_L + n_R$ in the language of \cite{Lin:2023trc}. In this way one finds that $\left\{|0), |1), |2), \ldots   \right\}$, that is the Krylov basis of DSSYK in the presence of a matter operator, again coincides with the fixed chord number basis, and is therefore related to bulk length.

This result can be obtained by several different approaches, on which we will elaborate.
We establish the analytic form of the Lanczos coefficients and Krylov elements of matter operators in DSSYK by sewing together analytic small$-n$ and asymptotic large$-n$ results giving the Lanczos coefficients, as well as Krylov elements, the so-called ``binomial states" $|\psi_n\rangle  $, whose explicit form will be given in Eq. \eqref{Binomial_Ansatz}. This is achieved both analytically in various parameter limits, and by giving strong numerical evidence that the binomial form of the Krylov elements is correct in the for generic $n$, at small $\lambda = \frac{2p^2}{N}$.

This paper is organised as follows: we begin with an introduction to what is by now a large and powerful set of mathematical tools that allows a very explicit description of the physics of DSSYK as well as a very concrete bulk-boundary mapping. This chord-diagram based language is reviewed and extended to suit the present study in \cref{sec.ToolsTrade}. In \cref{Sec:Operator_KC} \& \ref{sect:OTFD_KC} we then use the machinery to obtain analytical and numerical results on both operator K-complexity of a seed operator ${\cal O}_{\rm seed}$ (\cref{Sec:Operator_KC}) and state K-complexity of the ${\cal O}_{\rm seed}$ deformed TFD state (\cref{sect:OTFD_KC}).  The paper ends with a discussion and outlook section, followed by a number of appendices which provide more detail regarding analytical and numerical aspects of \cref{Sec:Operator_KC} \& \ref{sect:OTFD_KC}. We will now turn to a technical introduction of the tools of the DSSYK trade.

\section{The tools of the trade}\label{sec.ToolsTrade}
In this paper, we study the complexity of operators in the DSSYK model, which necessitates the introduction of a certain amount of technology. 
We start by first reviewing the sector of DSSYK that contains no matter, its effective Hamiltonian, chord states and operators. We then move on to mostly review the one-particle sector containing a single operator insertion with its attendant chord states and operators. 
This prepares the ground for calculating its boundary operator K-complexity.  We also discuss the triple-scaling limit and the equivalence between the DSSYK model in this limit and JT gravity.  We recall the identification of K-complexity of the time evolving thermofield double with the length of a particular wormhole.

The point of departure in defining the DSSYK model, is the SYK model itself, which is a many-body system of all-with-all interacting fermions with $p$-range interaction.
The SYK model is an ensemble averaged model of $N$ interacting Majorana fermions with Hamiltonian:
\begin{eqnarray} \label{H_SYK}
    H = i^{p/2} \sum_{1\leq i_1 < \dots < i_p \leq N} J_{i_1 \dots i_p}\,\psi_{i_1}\dots \psi_{i_p},
\end{eqnarray}
where $\left\{\psi_i,\psi_j\right\}=2\delta_{ij}$, the coefficients $J_{i_1 \dots i_p}$ are random and taken from a distribution of zero mean and variance given by $\langle J_{i_1\dots i_{\tilde{p}}} J_{j_1\dots j_{\tilde{p}}}\rangle = \frac{J^2}{\lambda}\binom{N}{p}^{-1} \delta_{i_1 j_1}\delta_{i_2 j_2} \dots \delta_{i_p j_{\tilde{p}}}$\footnote{for simplicity in the numerical analysis we will instead take, as done in \cite{Berkooz:2018jqr}, the variance $\langle J_{i_1\dots i_{\tilde{p}}} J_{j_1\dots j_{\tilde{p}}}\rangle = J^2\binom{N}{p}^{-1}\delta_{i_1 j_1}\delta_{i_2 j_2} \dots \delta_{i_p j_{\tilde{p}}}$. We will go back to the former normalization when we discuss the continuum approximation of complexities or their connection to gravity, and to do so it will be sufficient to multiply the Lanczos coefficients by a factor $J/\sqrt{\lambda}$.}, and $\lambda$ is defined below.\\
DSSYK is the limit of SYK where both the length of the Majorana monomials, $p$, and the number of Majoranas, $N$, are taken to infinity while the ratio $\lambda \equiv 2p^2/N$ is kept fixed, see \cite{Berkooz_chords, Berkooz:2018jqr}. In the limit $N\to \infty$, $p\to \infty$ and $ \lambda $ fixed, the \textit{ensemble averaged moments} are given by: 
\begin{equation} \label{H_moments}
    M_{2k} \equiv \langle \, \textrm{Tr}(H^{2k}) \,\rangle = \frac{J^{2k}}{\lambda^k} \sum_{\substack{\text{chord diagrams}\\ \text{ with $k$ chords}}} q^{\text{\# intersections}}. 
\end{equation}
where $q=e^{-\lambda}$. Here, \textit{chord diagrams with $k$ chords} are circles (which represent the Trace operation) with $2k$ marked points on the circumference (representing Hamiltonian insertions) and chords connecting pairs of points. Every intersection of two chords adds a multiplicative factor of $q$ to the value of the chord diagram. 
For example, 
\begin{eqnarray}
    M_4 = J^4/\lambda^2 \Big[\raisebox{-10pt}{\includegraphics[scale=0.2]{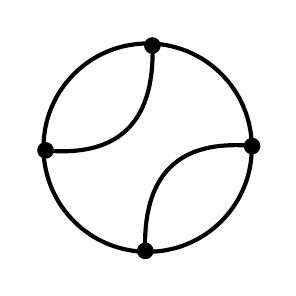} }+\raisebox{-10pt}{\includegraphics[scale=0.2]{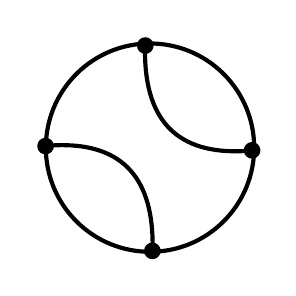} }+\raisebox{-10pt}{\includegraphics[scale=0.2]{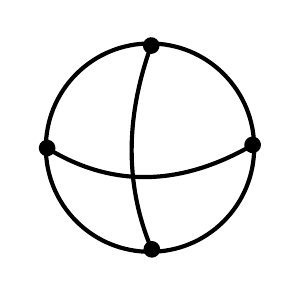} }\Big]  = (J^4/\lambda^2)( 2 + q).
\end{eqnarray}

\subsection{Chord states without matter}
The moments described in \eqref{H_moments} give rise to an effective description of the DSSYK Hamiltonian \cite{Berkooz_chords, Berkooz:2018jqr}. In DSSYK, the ensemble-averaged effective Hamiltonian acquires a concise form in terms of creation and annihilation operators:
\begin{eqnarray} \label{H_a_ad}
    H = \frac{J}{\sqrt{\lambda}} (a + a^\dagger).
\end{eqnarray}
The operators $a^\dagger$ and $a$ act as creation and annihilation operators (in a manner described below) over a particular basis known as the \textit{chord} basis, $\{\ket{n}\}_{n=0}^\infty$. Chord basis elements are conceived by cutting chord diagrams in two, leaving $n$ loose ends called `open' chords. An $n$-chord state can be represented diagrammatically by
\begin{eqnarray}
    \ket{n} = \raisebox{-9pt}{\includegraphics[scale=0.4]{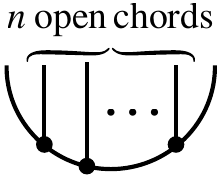} }
\end{eqnarray}

The action of the creation and annihilation operators on the chord basis elements is non-trivial; while creating a chord amounts to adding an open chord to the set of existing $n$ chords,
\begin{eqnarray}
    a^\dagger \ket{n}= \ket{n+1},
\end{eqnarray}
annihilating a chord involves `closing' it, which may involve crossing other chords in the process. The closing of a chord can be done in several ways: either crossing one existing chord, two existing chords, and up to $n-1$ existing chords, contributing a $1+q+q^2+\dots +q^{n-1}=(1-q^n)/(1-q)$ factor to the annihilation operation:
\begin{eqnarray} \label{a_ad_matterless}
     a\ket{n} = [n]_q\ket{n-1}, \quad a\ket{0}=0
\end{eqnarray}
where the \textit{q-number}, $[n]_q$, is defined by\footnote{See e.g. \hyperlink{https://en.wikipedia.org/wiki/Q-analog}{https://en.wikipedia.org/wiki/Q-analog}} 
\begin{eqnarray}
\label{qnumber_definition}
    [n]_q \equiv \frac{1-q^n}{1-q},
\end{eqnarray}
we will sometimes use the notation $[n]$ for simplicity of notation.

The inner product among chord basis elements is non-trivial and given by
\begin{eqnarray} \label{InProd_matterless}
    \langle m | n \rangle = \delta_{mn} [n]_q!  ~,
\end{eqnarray}
where $[n]_q!\equiv [1]_q[2]_q\dots [n]_q$ is the \textit{q-factorial}, and $\braket{0|0}=1$.
This inner-product can be understood as all ways to connect a bra chord state with a ket chord state (of the same number of chords) with each other. For example
$\braket{3|3} = \raisebox{-10pt}{\includegraphics[scale=0.2]{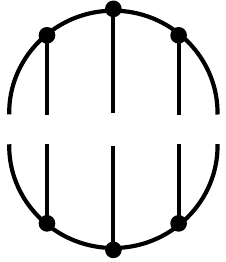} }=\raisebox{-8pt}{\includegraphics[scale=0.2]{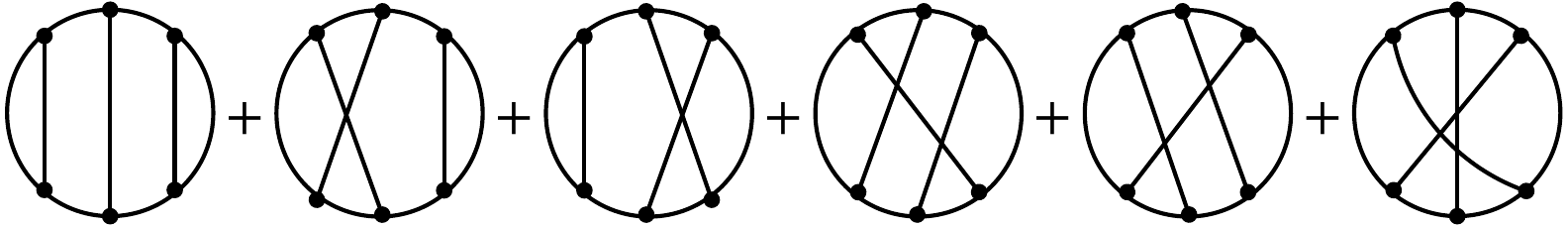} } = 1+ 2q +2q^2+q^3=[3]_q!$.

Finally, we mention that in the (un-normalized) chord basis, the Hamiltonian \eqref{H_a_ad} takes the form
{\small\begin{equation} \label{H_matterless_UN}
        H   \overset{*}{=}  \frac{J}{\sqrt{\lambda(1-q)}} \begin{pmatrix}
            0 & 1-q &  &  &  & \\
            1& 0 & 1-q^2 &  &  & \\
             & 1 & 0 & 1-q^3 &  & \\
             &  & 1 & 0 & \ddots & \\
             &  &  & \ddots & \ddots &  
\end{pmatrix}~.
\end{equation}

The chord state $\ket{0}$ in DSSYK is identified with the normalized infinite-temperature thermofield-double (TFD) state, $|\textrm{TFD} \rangle = \frac{1}{\sqrt{\mathcal{N}}}\sum_E \ket{E}$ in the non-averaged theory (where $\mathcal{N}\to +\infty$ is used to impose normalization of the state to $1$), and, as argued in \cite{Lin:2022rbf}, its evolution is given by the Hamiltonian \eqref{H_matterless_UN}:
\begin{eqnarray} 
    e^{-iH_{\textsc{syk}}t}\ket{\textrm{TFD}}  \to e^{-iHt}\ket{0} \label{eq:Mapping}
\end{eqnarray}
where on the left-hand side $H_{\textsc{syk}}$ stands for a particular realization of \eqref{H_SYK} and on the right-hand side $H$ is given by the effective Hamiltonian \eqref{H_matterless_UN}. By \eqref{eq:Mapping} we mean that disorder-average of expectation values computed with respect to the state on the LHS are reproduced by expectation values with respect to the state on the RHS.

In the next section we will review \cite{Rabinovici:2023yex}, where it was shown that the \textit{normalized} chord states are the Krylov basis elements for the Krylov problem involving the effective DSSYK Hamiltonian with initial state given by the infinite-temperature TFD state which becomes the zero-chord state, $\ket{0}$, in the effective theory. Indeed, the DSSYK Hamiltonian \eqref{H_matterless_UN} becomes symmetric with entries identified with the Lanczos coefficients in the normalized chord basis, see \eqref{H_matterless}. 

\subsection{K-complexity and its holographic dual in DSSYK without matter}
\label{sec:recap_dssyk_nomatter}
The Krylov basis $\{\ket{\psi_n}\}_{n=0}^\infty$ for the chord state $\ket{0}$ evolving under the effective Hamiltonian \eqref{H_a_ad} is constructed using the Lanczos algorithm \cite{viswanath1994recursion}. With the effective Hamiltonian, \eqref{H_a_ad}, the definitions of $a$ and $a^\dagger$, \eqref{a_ad_matterless}, and the chord inner product \eqref{InProd_matterless}, the Krylov basis is found to be
\begin{eqnarray}
    \ket{\psi_0} &=& \ket{0},\\
    \ket{\psi_n} &=& \frac{\ket{n}}{\sqrt{\braket{n|n}}} 
\end{eqnarray}
and 
\begin{equation}\label{bn_DSSYK_noMatter}
    b_{n}= \frac{J}{\sqrt{\lambda }} \sqrt{\frac{1-q^n}{1-q}} , \quad n=1,2,\dots,
\end{equation}
are identified with the Lanczos coefficients. The Krylov basis is an orthonormal basis, $\braket{\psi_m|\psi_n}=\delta_{mn}$, here identified with the normalized chord number states. 
We note that an equivalent way to obtain these Lanczos coefficients is to use the relationship between moments and Lanczos coefficients. 
In the effective theory, the moments of the Hamiltonian, $M_{2k}$, are equivalent to moments of the ensemble-averaged survival probability $\braket{\mathrm{TFD}|e^{-iH_{\textsc{syk}}t}|\mathrm{TFD}} \to \braket{0|e^{-iHt}|0} $.
The Lanczos coefficients read-off from these moments are given by \eqref{bn_DSSYK_noMatter}.

The \textit{Krylov basis} states are then given by \textit{normalized chord number states}, and, in this basis, the \textit{ensemble averaged} effective Hamiltonian is given by {\small\begin{equation} \label{H_matterless}
        H   \overset{*}{=}  \frac{J}{\sqrt{\lambda(1-q)}} \begin{pmatrix}
            0 & \sqrt{1-q} &  &  &  & \\
            \sqrt{1-q}& 0 & \sqrt{1-q^2} &  &  & \\
             & \sqrt{1-q^2} & 0 & \sqrt{1-q^3} &  & \\
             &  & \sqrt{1-q^3} & 0 & \ddots & \\
             &  &  & \ddots & \ddots &  
\end{pmatrix}~.
\end{equation}
We will now perform a number of steps, as first done in \cite{Lin:2022rbf}, which will allow us to connect this Hamiltonian with the Hamiltonian of JT gravity \cite{Harlow:2018tqv}. Defining a canonical conjugate momentum, $\hat{p}$, to the \textit{chord number} operator, $\hat{n}$, such that $[\hat{n},\hat{p}]=i \mathds{1}$\footnote{In general there are some subtleties due to $n$ being discrete, but, in the continuum limit we will be taking, the usual canonical commutation holds, so that $p=-i\partial_n$ and $[\hat{n},\,p]=i\mathds{1}$ (cf. \cite{ferrari_discrete_lattice1}, \cite{ferrari_discrete_lattice2}).}, the effective Hamiltonian \eqref{H_matterless} takes the form
\begin{equation}
    H = \frac{J}{\sqrt{\lambda (1-q)}}\left( e^{i\hat{p}} \sqrt{1-e^{-\lambda\hat{n}}} + \sqrt{1-e^{-\lambda\hat{n}}} e^{-i\hat{p}}\right)\,.
\end{equation}
Now, a \textit{length} observable, $\hat{l} = \lambda \hat{n}$\footnote{We could have also defined the length as $l=\lambda l_f n$, in terms of an arbitrary length scale $l_f$. As in this case, we usually suppress this reference scale, but it may be useful to restore it, with the variable change $l\to l/l_f$, when performing the holographic matching, in order to identify it with the AdS length $l_{AdS}$.}, can be defined, resulting in the Hamiltonian
\begin{equation} \label{H_DSSYK_k_l}
    H = \frac{J}{\sqrt{\lambda (1-q)}}\left( e^{i\lambda \hat{k}} \sqrt{1-e^{-\hat{l}}} + \sqrt{1-e^{-\hat{l}}} e^{-i\lambda \hat{k}}\right)\,,
\end{equation}
where $\hat{k} = \hat{p}/\lambda$. By definition,
Krylov complexity is given by the time-evolving expectation value of $\hat{n}$ \cite{Rabinovici:2023yex}:
\begin{eqnarray} \label{KC_DSSYK_noMatter}
     \lambda C_K(t) = \lambda \langle \hat{n}(t) \rangle =\langle \hat{l}(t) \rangle =  \langle 0| e^{iHt} \, \hat{l} \, e^{-iHt} |0\rangle .
\end{eqnarray}
Later, we will also be interested in the limit of small $\lambda$ which connects with the usual SYK, and is one of the conditions to probe JT gravity. The limit $\lambda \to 0$ with $ n\to \infty$, such that $l=\lambda n$ is kept fixed, renders Krylov space continuous and allows us to treat the Krylov space dynamics semi-classically.  A summary of the results for the Lanczos coefficients and Krylov complexity in DSSYK without matter at small $\lambda$ are shown in Figure \ref{fig:DSSYK_noMatter}. In particular, K-complexity for DSSYK at small $\lambda$ shows two distinct time regimes: a quadratic early-time behavior and a linear late-time behavior. When analyzing the case of operator K-complexity at small $\lambda$ we will see that, depending on the `size' of the DSSYK operator, a new time regime will emerge, reflecting the operator's `scrambling'.
\begin{figure}
    \centering
    \includegraphics[width=0.5\linewidth]{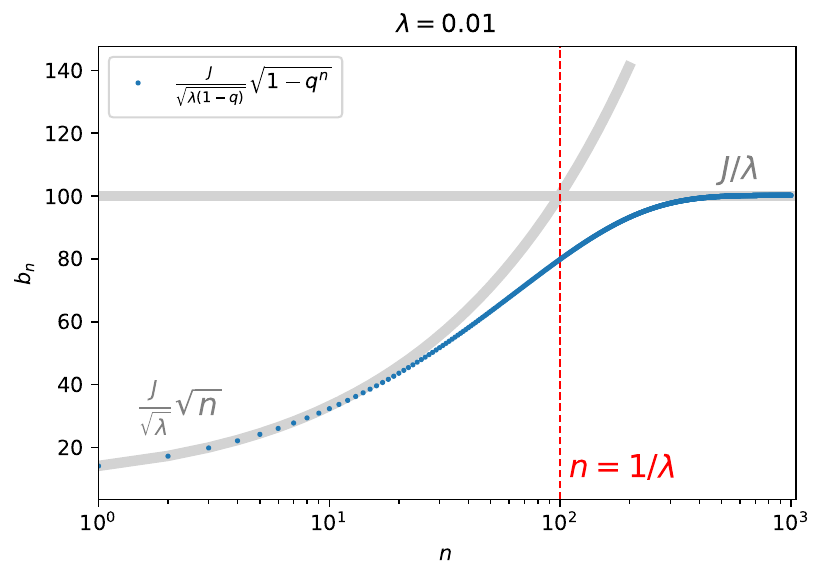}\includegraphics[width=0.5\linewidth]{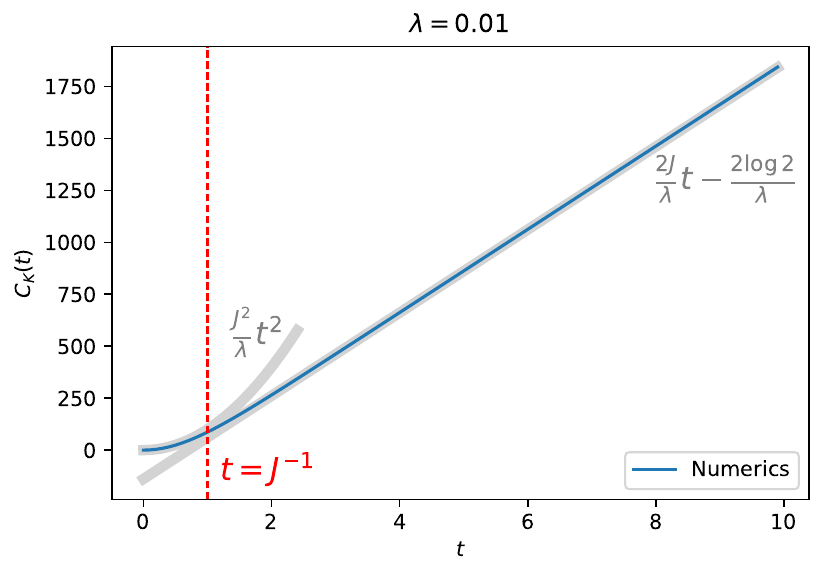}
    \caption{Summary of results for DSSYK without matter for $\lambda=0.01$ and $J=1$. \textbf{Left:} the Lanczos coefficients as given by \eqref{bn_DSSYK_noMatter}. Their behavior switches from $\sim \sqrt{n}$ behavior to constant behavior at $n=1/\lambda$. \textbf{Right:} Krylov complexity switches from $\sim t^2$ behavior to linear behavior at a time $t_*=1/J$. The blue line shows the numerical solution gotten from solving the Krylov problem for the Hamiltonian \eqref{H_matterless}.}
    \label{fig:DSSYK_noMatter}
\end{figure}

In \cite{Lin:2022rbf} it was shown that the Hilbert space of DSSYK in the \textit{triple scaling} limit is the same as the Hilbert space of 2d JT gravity. The triple scaling limit consists of taking $\lambda \to 0$ and  $l \to \infty$ while keeping the ratio $e^{-l}/(2\lambda)^2\equiv e^{-\tilde{l}}$ fixed. Here, $\tilde{l}$ is known as the \textit{renormalized} length $\tilde{l} = l - 2 \log \left( \frac{1}{2\lambda}\right)$. The triple scaling limit is equivalent to taking $\lambda \to 0$ and zooming into the low energy regime of DSSYK where the density-of-states is given by $ \rho(E) \propto \sinh(2\pi \sqrt{E})$.
Taking the triple scaled limit in \eqref{H_DSSYK_k_l}, the \textit{triple scaled} Hamiltonian \cite{Lin:2022rbf} is given by 
\begin{equation}\label{H_triple_scaled}
    H = E_0 +2 \lambda J \left(\frac{\hat{\tilde{k}}^2}{2}+2 e^{-\widehat{\tilde{l}}} \right) + \mathit{O}(\lambda^2),
\end{equation} 
where $E_0 = - \frac{2J}{\lambda} + \mathit{O}(\lambda^0)$.
The leading-order piece of the above Hamiltonian takes the form of \textit{Liouville}
quantum mechanics, 
connecting the low-energy regime of DSSYK with JT gravity \cite{Lin:2022rbf, Lin:2023trc}. In a similar manner to \eqref{KC_DSSYK_noMatter}, K-complexity in this limit is given by the expectation value of the renormalized length, $\lambda \widetilde{C}_K(t) =  \langle \widehat{\tilde{l}}(t)\rangle$. At small $\lambda$, the wavefunction is peaked around its expectation value and it is possible to solve for $\tilde{l}(t)$ classically. Solving the equations of motion obtained from \eqref{H_triple_scaled} with initial conditions $\tilde{l}(0)=x_0$ and $\dot{\tilde{l}}(0)=0$\footnote{The second initial condition is motivated by the Lanczos coefficients \eqref{bn_DSSYK_noMatter}, which become proportional to the velocity of a particle moving on the Krylov chain in the semiclassical limit.} one finds:\\
\begin{equation}
    \tilde{l}(t)/l_f=\lambda \widetilde{C}_K(t) =  2 \log \Big[ \cosh \left( \sqrt{\lambda J E }\, t \right)\Big] -\log \left(\frac{E}{4\lambda J}\right)
\end{equation}
where $E = 4 \lambda J e^{-x_0}$ is the energy of the solution. Here, $l_f$ is an arbitrary reference length in DSSYK that we usually omit and we reinstate here via $\Tilde{l}\to \Tilde{l}/l_f$ in order to provide its entry in the holographic dictionary. This result can be compared with the renormalized wormhole length in JT gravity \cite{Harlow:2018tqv},
\begin{equation}
   \tilde{l}(t)/l_{AdS} = 2 \log\Big[\cosh\left(\sqrt{\frac{E}{2l_{AdS}\phi_b}}t \right) \Big]  - \log \left(\frac{l_{AdS}\phi_b E}{2} \right)
\end{equation}
where $\phi_b$ is the value of the dilaton field on the boundary of AdS$_2$ and $\Phi_h$ is the value of the dilaton on the (observer dependent) horizon; $E$ is the ADM energy on the boundary, $E=\frac{2\Phi_h^2}{\phi_b}$, and $t$ here is the boundary time. 

In the table below we summarize the exact boundary-bulk correspondence between triple-scaled SYK and JT gravity. The duality requires the following parameter identifications in the holographic dictionary:
\begin{equation}\label{eq:hol_dict_nomatt}
    \frac{1}{l_{AdS}\phi_b}=2\lambda J\quad \text{and}\quad l_f=l_{AdS} ~.
\end{equation}

\begin{center}
\begin{tabular}{|p{7cm}|p{7cm}|}
\hline
    \textbf{Boundary} & \textbf{Bulk}  \\ \hline
    triple-scaled SYK & JT gravity  \\ \hline
    $H = - \frac{2J}{\lambda} +2 \lambda J \left(\frac{l_f^2 k^2}{2}+2 e^{-\tilde{l}/l_f} \right)$ & 
    $H = \frac{1}{l_{AdS}\phi_b} \left( \frac{l_{AdS}^2P^2}{2} + 2 e^{-\tilde{l}/l_{AdS}} \right)$ \\ \hline
    Krylov basis are $|\tilde{l}\rangle$ states & Hilbert space consists of states with well defined wormhole length $|\tilde{l}\rangle$ \\ \hline
    K-complexity in the semiclassical limit & Normalized wormhole length in JT gravity \\ \parbox{5cm}{\begin{align*}
    \lambda \widetilde{C}_K(t) /l_f=&  2 \log \Big[\sqrt{\frac{4\lambda J}{E} }\cosh \left( \sqrt{\lambda J E }\, t \right)\Big]
    \end{align*}} & \parbox{5cm}{\begin{align*}
   \frac{\tilde{l}(t)}{l_{AdS}} =& 2 \log\Big[\sqrt{\frac{2}{l_{AdS}\phi_b E}}\cosh\left(\sqrt{\frac{E}{2l_{AdS}\phi_b}}t \right) \Big]
    \end{align*}}\\
\hline
\end{tabular}
\end{center}

We can also use a different, but equivalent, parametrization, when discussing the holographic dictionary. Let us consider the case where $x_0=0$ and $\Phi_h=1$, if we reinsert the value of the boundary energy $E$ on both sides we obtain:
\begin{equation}
    \lambda \widetilde{C}_K(t) =  2 \log \left[ \cosh \left( 2\lambda J t \right)\right]\quad\longleftrightarrow \quad \tilde{l}/l_{AdS} = 2 \log\left[\cosh\left(\frac{r_s}{l_{AdS}^2} t \right) \right],
\end{equation}
where we used that $\Phi_h/\phi_b=r_s/l_{AdS}$. This matching identifies an entry in the holographic dictionary equivalent to \eqref{eq:hol_dict_nomatt}, but stated in terms of the Schwarzschild radius $r_s$:
\begin{equation}\label{eq:hol_dict_nomatt_rs}
    2\lambda J =\frac{r_s}{l_{AdS}^2} ~.
\end{equation}

\subsection{Operators in DSSYK}
In \cite{Rabinovici:2023yex} we focused on state Krylov complexity, here we will focus on operator Krylov complexity for a random operator, as motivated in \cite{Berkooz_chords}, of the form:
\begin{equation}\label{random_operator}
    \mathcal{O} = i^{\tilde{p}/2}\sum_{1\leq i_1< \dots < i_{\tilde{p}} \leq N} O_{i_1\dots i_{\tilde{p}}} \psi_{i_1} \psi_{i_2}\dots \psi_{i_{\tilde{p}}}
\end{equation}
where $O_{i_1\dots i_{\tilde{p}}}$ are random, taken from a distribution with zero mean and variance given by
\begin{equation}
    \langle O_{i_1\dots i_{\tilde{p}}} O_{j_1\dots j_{\tilde{p}}}\rangle = \binom{N}{\tilde{p}}^{-1} \delta_{i_1 j_1}\delta_{i_2 j_2} \dots \delta_{i_p j_{\tilde{p}}}
\end{equation}
taken also to be independent of the random couplings of the Hamiltonian.

In the first part of this work, we will focus on computing operator Krylov complexity for a random operator \eqref{random_operator} in DSSYK.  In \cite{Berkooz_chords, Berkooz:2018jqr} it was shown that the auto-correlation function, 
\begin{equation}\label{autocorrelation_function}
    C(t)=(\mathcal{O}|\mathcal{O}(t)) = \frac{1}{\mathrm{Tr}(\mathds{1})} \mathrm{Tr} \big[ \mathcal{O} e^{iHt} \mathcal{O} e^{-iHt}\big],
\end{equation}
with $H$ given by \eqref{H_SYK}, can be given a closed form in DSSYK, using chord diagram technology. 
In the following, we set $\mathrm{Tr}(\mathds{1})=1$.
Using the Baker-Campbell-Hausdorff formula this expression can be expanded in terms of nested commutators:
\begin{equation}
    C(t) = \sum_{n=0}^\infty \frac{(it)^{2n}}{(2n)!} \mathrm{Tr}\big( \mathcal{O} [H,[H,\dots,[H,\mathcal{O}]\dots]] \big) .
\end{equation}
We will be be interested in the ensemble average of the auto-correlation function which amounts to the computation of the \textit{ensemble averaged} moments, defined as
\begin{equation} \label{moments_tr}
    \mu_{2n} \equiv \langle \mathrm{Tr}\big( \mathcal{O} [H,[H,\dots,[H,\mathcal{O}]\dots]] \big) \rangle,
\end{equation}
with $2n$ nested commutators.
The expression (\ref{moments_tr}) can be reduced to a sum of terms of the form $\mathrm{Tr}(\mathcal{O}H^{k_1}\mathcal{O}H^{k_2})$, in which $k_1+k_2 = 2n$. For example
\begin{equation}\label{mu2}
    \mu_2 =  2\, \langle \mathrm{Tr}(\mathcal{O} H^2 \mathcal{O})\rangle -2 \, \langle \mathrm{Tr}(\mathcal{O} H \mathcal{O} H)\rangle
\end{equation}
\begin{equation}\label{mu4}
    \mu_4 =   2\, \langle \mathrm{Tr}(\mathcal{O} H^4 \mathcal{O})\rangle -8 \, \langle \mathrm{Tr}(\mathcal{O} H^3 \mathcal{O} H)\rangle  + 6 \langle \mathrm{Tr}(\mathcal{O} H^2 \mathcal{O} H^2)\rangle.
\end{equation}
A general expression for the ensemble averaged moments is given by
\begin{equation} \label{AC_moments}
    \mu_{2n} = \sum_{k=0}^{2n} \binom{2n}{k} (-1)^k \langle \mathrm{Tr}(\mathcal{O} H^{2n-k} \mathcal{O} H^k)\rangle ~.
\end{equation}
In \cite{Berkooz:2018jqr, Berkooz_chords} it was shown that expressions of the form $\langle\mathrm{Tr}(\mathcal{O} H^{k_1} \mathcal{O} H^{k_2})\rangle$ can be evaluated using chord diagrams as we discuss below.

\subsection{Moments from chord diagrams}\label{Sec:Moments_from_chord_diagrams} 
We showed that the expression for the moments $\mu_{2n}$ given in (\ref{moments_tr}) can be reduced to a sum of terms of the form $\mathrm{Tr}(\mathcal{O}H^{k_1}\mathcal{O}H^{k_2})$, with $k_1+k_2=2n$, which in turn can be computed via \textit{marked} chord diagrams.  Such marked chord diagrams have one marked chord which represents an $\mathcal{O}$-$\mathcal{O}$ contraction, and $n$ chords representing $H$-$H$ contractions. On one side of the marked chord there are $k_1$ insertions of $H$ and on the other side of the marked chord there are $k_2$ insertions of $H$. We then connect the $H$ chords in all possible ways and count the number of intersections between $H$ chords with themselves and $H$ chords with the $\mathcal{O}$ chord.  Each $H$-$H$ intersection gives a factor of $q=e^{-\lambda}$ (recall that $\lambda = 2p^2/N$) , while a $H$-$\mathcal{O}$ intersection contributes a $\tilde{q} = e^{-\tilde{\lambda}}$ factor, where $\tilde{\lambda}= 2 p \tilde{p}/N$.  This procedure provides an expression for $\mathrm{Tr}(\mathcal{O}H^{k_1}\mathcal{O}H^{k_2})$, as follows \cite{Berkooz_chords, Berkooz:2018jqr}:
\begin{equation} \label{chord_diag_sum}
    \langle\mathrm{Tr}(\mathcal{O} H^{k_1} \mathcal{O} H^{k_2})\rangle = \left(\frac{J}{\sqrt{\lambda}} \right)^{k_1+k_2} \sum_{\pi \in \text{marked chord diagrams}} q^{k_{HH}} \tilde{q}^{k_{H\mathcal{O}}}
\end{equation}
where $k_{HH}$ counts the number of times $H$ chords intersect each other, $k_{H\mathcal{O}}$ counts the number of $H$ chords intersecting an $\mathcal{O}$ chord, and $\pi$ denotes generically a tuple $(k_{HH},k_{H\mathcal{O}})$. 

Below we provide a few examples of the use of marked chord diagrams to compute moments of the autocorrelation function \eqref{autocorrelation_function}. For $\mathcal{O}=\mathds{1}$ the moments are expected to be zero since the autocorrelation function is constant. This is equivalent to setting $\tilde{q}=1$ which means that there is no cost for an operator chord crossing a Hamiltonian chord in the chord diagram; in other words, the identity operator is an operator of length $\tilde{p}=0$ which implies $\tilde{\lambda}=0$ and thus $\tilde{q}=1$. We may check that this is the case for the first few moments below. 

\begin{itemize}
    \item Contributions to the second moment, from \eqref{mu2}:
\begin{equation} \label{mu2_result}
    \mu_2 = 2\, \langle \mathrm{Tr}(\mathcal{O} H^2 \mathcal{O})\rangle -2 \, \langle \mathrm{Tr}(\mathcal{O} H \mathcal{O} H)\rangle = 2\frac{J^2}{\lambda}(1-\tilde{q}).
\end{equation}
Indeed, $\tilde{q}\to 1$ takes $\mu_2$ to zero. In anticipation of the Krylov discussion later, we mention here that $\mu_2=b_1^2$ where $b_1$ is the first Lanczos coefficient. This implies a one-dimensional Krylov space which is expected for the identity operator.\footnote{Note that, for a Krylov dimension $K$, the Lanczos algorithm halts by definition when $b_K=0$, so in this case when $\tilde{q}=1$ we have $K=1$ and there is only one Krylov element, $\mathcal{O}_0=\mathcal{O} = \mathds{1}$.}

    \item Chord diagram contributions to $\mu_4$:  
\begin{equation}
    \langle\mathrm{Tr}(\mathcal{O} H^{4} \mathcal{O} )\rangle = \raisebox{-1em}{\includegraphics[scale=0.2]{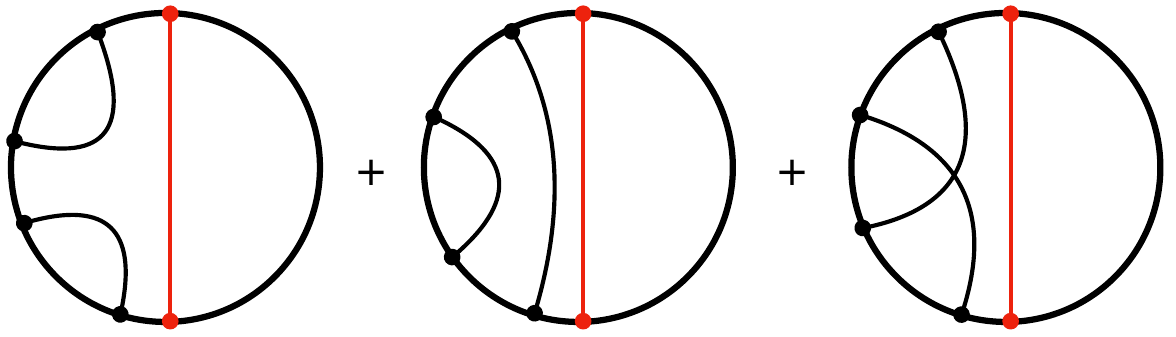}} =\frac{J^4}{\lambda^2}( 2 + q)
\end{equation}
\begin{equation}
    \langle\mathrm{Tr}(\mathcal{O} H^{3} \mathcal{O} H)\rangle = \raisebox{-1em}{\includegraphics[scale=0.2]{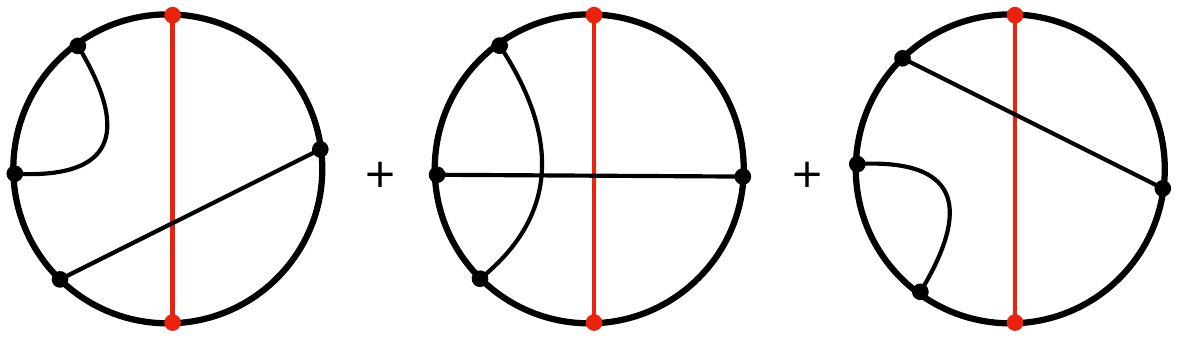}} = \frac{J^4}{\lambda^2}\,\tilde{q}(2+q)
\end{equation}
\begin{equation}
    \langle\mathrm{Tr}(\mathcal{O} H^{2} \mathcal{O} H^2)\rangle = \raisebox{-1em}{\includegraphics[scale=0.2]{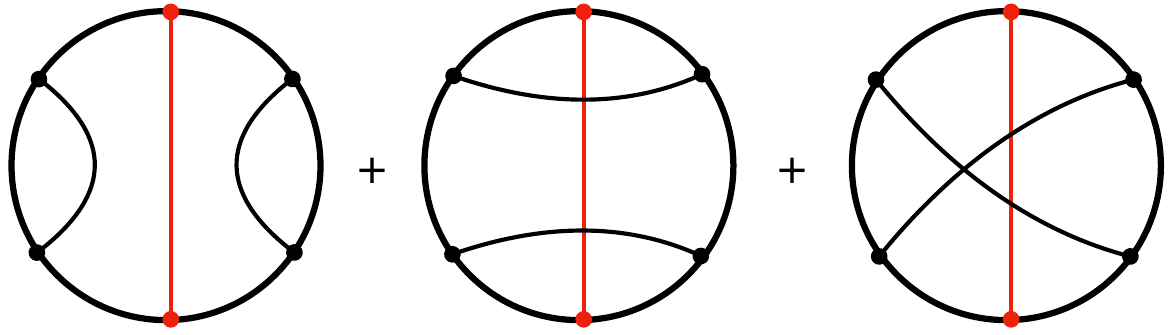}} =\frac{J^4}{\lambda^2}\,[ 1 +\tilde{q}^2 (1 + q)]
\end{equation}
When $\tilde{q}= 1$, there is no cost for the operator line to cross the $H$ lines and therefore we get back for each of these expressions the result for $\langle \mathrm{Tr}(H^4)\rangle$.
In total, from \eqref{mu4}, we find that
\begin{equation} \label{mu4_result}
    \mu_4 = 2\frac{J^4}{\lambda^2}\, (-1 + \tilde{q}) (-5 - q + 3 \tilde{q} (1 + q)) .
\end{equation}
Also here, when $\tilde{q}= 1$, we have $\mu_4=0$ as expected for a unity operator. 

\item In a similar manner, $\mu_6$ can be computed:
\begin{equation} \label{mu6_result}
    \mu_6 = 2 \frac{J^6}{\lambda^3}\,(1 - \tilde{q}) [35 + 10 \tilde{q}^2 (1 + q) (1 + q + q^2) + q (21 + q (3 + q)) - 
   5 \tilde{q} (1 + q) (7 + q (4 + q))]
\end{equation}
which again is zero for $\tilde{q} \to 1$. 
\end{itemize}

Finally, we may note for reference that the moments computed above are compatible, via the recursion method (cf. equation 3.12 in \cite{Rabinovici:2023yex}), with the first three exact operator Lanczos coefficients computed in section \ref{subsect:KrylovBasis}, cf. expressions \eqref{b1_Exact}-\eqref{b3_Exact}.

\subsection{Chord states with operator insertion}\label{subsect:Right_And_Left_Hamiltonians}
In \cite{Lin:2022rbf}, it was proposed that the insertion of an operator on the TFD state gives rise to states which can be described by the number of open chords on the left of an open operator chord, $n_L$, and the number of open chords on the right of the operator chord, $n_R$. This constructs  the one-particle sector, $\mathcal{H}_{1p}$, of the DSSYK Hilbert space in which our discussion takes place. The states in the one-particle sector are denoted by $|n_L,n_R\rangle$ and can be described diagrammatically by
\begin{eqnarray}
    \ket{n_L,n_R}=\raisebox{-8pt}{\includegraphics[scale=0.35]{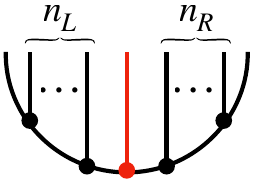} }.
\end{eqnarray}
These states are a useful basis in the construction of the time evolved state
\begin{equation} \label{operator_to_HR_HL}
    e^{i t H_\textsc{syk}} \mathcal{O} e^{-itH_\textsc{syk}} |\mathrm{TFD}\rangle \to e^{it H_L} e^{-it H_R }|n_L=0,n_R=0\rangle ,
\end{equation}
where on the left-hand side $H_\textsc{syk}$ stands for \eqref{H_SYK} and on the right-hand side $H_L$ and $H_R$ provide an effective description for the time evolution of the operator in the ensemble-averaged theory as we discuss below.

To write down the left and right Hamiltonians we need to introduce creation and annihilation operators which act on the $\ket{n_L,n_R}$ states. The creation operators $a_{L/R}^\dagger$ create a Hamiltonian chord to the left/right of all existing chords:
\begin{align}
    a_L^\dagger |n_L, n_R\rangle &= |n_L+1, n_R\rangle \label{aLdagger}\\
     a_R^\dagger |n_L, n_R\rangle &= |n_L, n_R+1\rangle ~.\label{aRdagger}
\end{align}
This can be represented diagrammatically, for example, for $a_L$,  as
\begin{eqnarray}
    a_L^\dagger \raisebox{-9pt}{\includegraphics[scale=0.35]{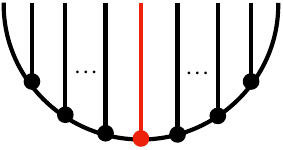} } &=&\raisebox{-9pt}{\includegraphics[scale=0.35]{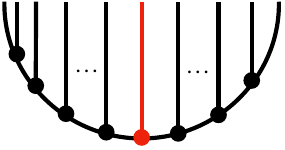} }.
\end{eqnarray}
It is useful to introduce left and right annihilation operators:
\begin{align}
    \alpha_L |n_L, n_R\rangle &= |n_L-1, n_R\rangle \label{alphaL}\\
    \alpha_R |n_L, n_R\rangle &= |n_L, n_R-1\rangle ~. \label{alphaR}
\end{align}
\textit{Attention:} $a$ and $\alpha^\dagger$ are not Hermitian conjugates of each other with respect to the one-particle sector inner product, reviewed in \ref{subsect:inner_product}. They have different notation for this reason.
Following \cite{Lin:2023trc}, the left and right Hamiltonians, $H_L$ and $H_R$, can be written as follows:
\begin{align}
     H_L & = \frac{J}{\sqrt{\lambda}} \left( a_L^\dagger +\alpha_L \,\frac{1-q^{n_L}}{1-q} +\alpha_R \, \tilde{q} \, q^{n_L} \frac{1-q^{n_R}}{1-q} \right)\label{HL}\\
     H_R & = \frac{J}{\sqrt{\lambda}} \left( a_R^\dagger +\alpha_R \,\frac{1-q^{n_R}}{1-q} +\alpha_L \, \tilde{q} \, q^{n_R} \frac{1-q^{n_L}}{1-q} \right) \label{HR}
\end{align}

The expressions for the left and right Hamiltonians can be made simpler if one introduces the left and right chord annihilation operators:
\begin{align}
    a_L &= \alpha_L \,\frac{1-q^{n_L}}{1-q} +\alpha_R \, \tilde{q} \, q^{n_L} \frac{1-q^{n_R}}{1-q} \label{aL_long} \\
    a_R &= \alpha_R \,\frac{1-q^{n_R}}{1-q} +\alpha_L \, \tilde{q} \, q^{n_R} \frac{1-q^{n_L}}{1-q} \label{aR_long}~.
\end{align}
Note that $n_{L/R}$ should be understood as an operator, measuring the number of $n_{L/R}$ in the chord state it acts on, and thus the order of operations among $\alpha_{L/R}$ and $n_{L/R}$ is important. 

The operation $a_{L/R}$ can be understood diagrammatically as closing a chord by taking it all the way to the left/right, for example:
\begin{eqnarray*}
    a_L \raisebox{-8pt}{\includegraphics[scale=0.3]{chord_aL0.pdf} } &=&\raisebox{-8pt}{\includegraphics[scale=0.3]{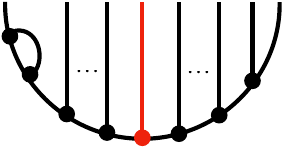} }+\raisebox{-8pt}{\includegraphics[scale=0.3]{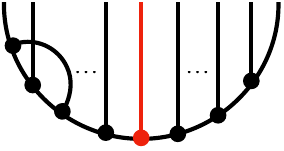} } +\dots +\raisebox{-8pt}{\includegraphics[scale=0.3]{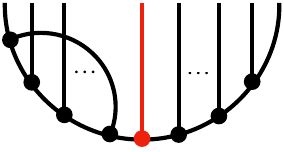} }+\raisebox{-8pt}{\includegraphics[scale=0.3]{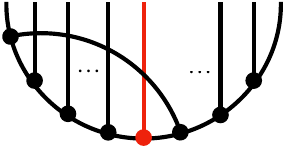} }+\dots+ \raisebox{-8pt}{\includegraphics[scale=0.3]{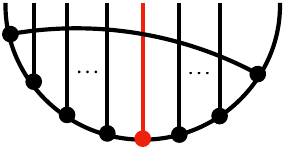} } 
\end{eqnarray*}
which represents
\begin{eqnarray*}
    a_L \ket{n_L,n_R}
    &=& \ket{n_L-1,n_R}+q\ket{n_L-1,n_R}+\dots + q^{n_L-1}\ket{n_L-1,n_R}\nonumber\\
    &&+\tilde{q}\ket{n_L,n_R-1}+\dots+ \tilde{q}q^{n_R-1}\ket{n_L,n_R-1},
\end{eqnarray*}
and can be re-summed as \eqref{aL_long}. 

Introducing $\Delta = \tilde{p}/p$, such that $\tilde{q} = q^\Delta = e^{-\frac{2 p \tilde{p}}{N}}$, and using the $q$-number notation, $a_L$ and $a_R$ can be written more simply as:
\begin{align} 
    a_L &= \alpha_L \,[n_L]_q +\alpha_R \,  q^{n_L+\Delta} [n_R]_q \label{aL} \\
    a_R &= \alpha_R \,[n_R]_q +\alpha_L \,  q^{n_R+\Delta} [n_L]_q ~.  \label{aR}
\end{align} 

In \cite{Lin:2023trc} it is argued that indeed $a_{L/R}$ is the adjoint of $a_{L/R}^\dagger$ with respect to the one-particle sector inner product (cf. section \ref{subsect:inner_product}). We reproduce the proof diagrammatically in the next section where the chord inner-product is introduced.

These definitions make the expressions for the left and right Hamiltonians simple:
\begin{align}
    H_L &=\frac{J}{\sqrt{\lambda}} \left( a_L^\dagger +a_L \right)\label{HL_v2}\\
    H_R &= \frac{J}{\sqrt{\lambda}} \left( a_R^\dagger +a_R \right)~.  \label{HR_v2}
\end{align}
Using \eqref{aL} and \eqref{aR} it can be shown that \cite{Lin:2023trc}:
\begin{align}
    [a_L,a_R] = [a_L^\dagger, a_R^\dagger] & = 0\\
    [a_L, a_R^\dagger] =  [a_R, a_L^\dagger] & = q^{n_L+n_R+\Delta} ~,
\end{align}
which imply that $H_L$ and $H_R$ commute: 
\begin{align}
    [H_L,H_R] &\propto [a_L^\dagger +a_L, a_R^\dagger +a_R] = [a_L^\dagger, a_R^\dagger] +[a_L^\dagger, a_R]+ [a_L, a_R^\dagger ] +[a_L, a_R] \nonumber\\
    &= 0-q^{n_L+n_R+\Delta}+q^{n_L+n_R+\Delta}+0 = 0~.
\end{align}

Since $[H_L,H_R]=0$, the time-evolution of an operator inserted on the TFD state of DSSYK, as given in \eqref{operator_to_HR_HL}, is described by the time evolution of the state $|n_L=0, n_R=0\rangle$ under the effective Hamiltonian $H_R-H_L$:
\begin{equation} \label{operator_evolution}
    e^{i t H_{\textsc{syk}}} \mathcal{O} e^{-itH_{\textsc{syk}}} |\mathrm{TFD}\rangle \to e^{it H_L} e^{-it H_R }|n_L=0,n_R=0\rangle = e^{-it(H_R-H_L)}|0,0\rangle ~.
\end{equation}
Furthermore, for the infinite-temperature TFD state, the ensemble-averaged moments of the auto-correlation function $\langle \mathrm{TFD}| \mathcal{O} e^{i t H_{\textsc{syk}}} \mathcal{O} e^{-itH_{\textsc{syk}}} |\mathrm{TFD}\rangle$ given by \eqref{moments_tr} are equivalent to the moments of the survival probability $\langle 0,0|e^{-it(H_R-H_L)}|0,0\rangle$:
\begin{equation}
\label{averaged_moments_00state}
    \mu_{2n}=\langle \mathrm{Tr}\big( \mathcal{O} \underset{\text{$2n$ nested commutators}}{[H,[H,\dots,[H,\mathcal{O}]\dots]]} \big) \rangle = \langle 0,0|(H_R-H_L)^{2n}|0,0\rangle ~.
\end{equation}

In short, this construction maps the time evolution of an operator to the Schrödinger evolution of a state, in the effective or averaged theory, belonging to the so-called \textit{one-particle sector} of the Hilbert space \cite{Lin:2023trc}, where the map is correct in the sense that the averaged auto-correlation function of such an operator at infinite temperature coincides with the fidelity of the Schrödinger-evolving state.
An alternative (yet equivalent) approach for the computation of the averaged operator moments \eqref{averaged_moments_00state}, presented in \cite{Berkooz:2018jqr}, consists on mapping the operator $\mathcal{O}$ to an effective operator in the averaged theory that acts on the \textit{zero-particle sector} and which evolves in the Heisenberg picture. In appendix \ref{Appx:Lanczos_W_T} we review this construction and discuss certain subtleties of the application of the Lanczos algorithm to such an effective operator.

\subsection{Inner-product of chord states with operator insertion} \label{subsect:inner_product}

Finally, we discuss the inner product between operator chord states. Given two basis elements $|n_L^\prime,n_R^\prime\rangle$ and $|n_L,n_R\rangle$ in the one-particle basis, the inner-product $\langle n_L', n_R' | n_L,n_R\rangle$ is defined in \cite{Lin:2022rbf} in a way that makes it consistent with the evaluation of correlation functions via chord diagrams and, more specifically, with the definition of bras and kets given a consistent slicing of chord diagrams. In a similar manner to the zero-particle state inner product, the one-particle state inner product consists of all ways to connect chords from the bra with chords from the ket, connecting the operator chord from the bra with that of the ket and the Hamiltonian chords of the bra with those of the ket. We shall not review such a construction here, but it will suffice to say that the inner product is fully defined via the recursion relation
\begin{align}
    &\langle n_L', n_R' | n_L,n_R\rangle  \label{inner_rec_start}\\
    &= [n_L]_q \langle n_L'-1, n_R' | n_L-1,n_R\rangle +q^{\Delta+n_L}[n_R]_q \langle n_L'-1, n_R' | n_L,n_R-1\rangle \label{RRL}\\
    &= [n_R]_q \langle n_L', n_R'-1 | n_L,n_R-1\rangle +q^{\Delta+n_R}[n_L]_q \langle n_L', n_R'-1 | n_L-1,n_R\rangle \label{RRR}~.
\end{align}
Iterating this recursion up to the point where e.g. the ket state is $|n_L,n_R\rangle=|0,0\rangle$ allows one to show that
\cite{Lin:2022rbf, Lin:2023trc} the inner product $\langle n_L', n_R' | n_L,n_R\rangle$ is non-zero if and only if $n_L+n_R=n_L'+n_R'$.
This feature can be explained as follows:
 Since \eqref{RRL} lowers $n_L'$ by one and \eqref{RRR} lowers $n_R'$ by one, to lower both the $n_L', n_R'$ indices down to zero, one needs to perform in total $n_L'+n_R'$ steps. The key observation is that every time a recursion step of either \eqref{RRL} or \eqref{RRR} is used, the \textit{total} $n_L'+n_R'\equiv n'$ is lowered by one \textit{and} the \textit{total} $n_L+n_R \equiv n$ is lowered by one for \textit{every} term at that recursion step. Invoking the symmetry of the inner product it can be assumed, without loss of generality, that $n<n'$. In this case, after $n$ recurrence step, we have $n=0$ which means that, for each term, either: one of the $n_L, n_R$ is negative or $n_L=n_R=0$.  If one of them is negative the recursion gives zero (it is equivalent to acting with $a_{L}|0,n_R\rangle=0 $ or $a_{R}|n_L,0\rangle=0 $). If $n_L, n_R$ are both are zero then any further recursion step of either type $\eqref{RRL}$ or \eqref{RRR} will give zero because of the $[n_L=0]=[n_R=0]=0$ coefficients.

Additionally, considering the original construction leading to this definition for the inner-product in the one-particle sector, it is possible \cite{Lin:2022rbf} to supplement the recursion \eqref{inner_rec_start} by some boundary conditions that relate it to the inner product in the matterless sector: 
\begin{align}
\label{Inner_product_recursion_boundarycondition}
    &\langle 0 , n_R^\prime | n_L,n_R\rangle = \widetilde{q}^{~n_L}\langle n_R^\prime | n_L + n_R\rangle~. 
\end{align}
Now, using symmetry of the inner product, and recalling that inner product in the zero-particle sector is given by \cite{Lin:2022rbf}
\begin{equation}
    \label{Inner_product_zero_sector}
    \langle m | n\rangle = \delta_{mn}[n]_q!~,
\end{equation}
where $[n]_q! = \prod_{k=1}^n [k]_q$ is the $q$-factorial (with $[0]_q!=1$), one can combine \eqref{Inner_product_recursion_boundarycondition} and \eqref{inner_rec_start}-\eqref{RRR} to show more directly that states $|n_L^\prime,n_R^\prime\rangle$ and $|n_L,n_R\rangle$ are orthogonal whenever they belong to different total chord number sectors \cite{Lin:2022rbf}, namely
\begin{equation}
    \label{orthog_chord_sectors}
    \langle n_L^\prime,n_R^\prime|n_L,n_R\rangle = 0 \qquad \text{if}\quad n_L^\prime + n_R^\prime \neq n_L+n_R~.
\end{equation}
However, generically states $|k,n-k\rangle$ and $|k^\prime,n-k^\prime\rangle$
with equal total chord number
are not necessarily orthogonal and will generically have a non-zero overlap. See appendix \ref{appx:numerics_inner_product} for some further discussion on the structure and symmetries of this inner-product, beyond the standard inner-product symmetry.

For completeness, let us note that in \cite{Lin:2023trc} the recurrence relations \eqref{RRL} and \eqref{RRR} are solved to find a closed form for the inner product $\langle n_L', n_R' | n_L,n_R\rangle$. As explained above, it is non-zero only if $n_L'+n_R'=n_L+n_R$. Introducing the notation
\begin{equation}
    n_L-n_R = 2y, \quad n_L'-n_R' = 2y', \quad n_L'+n_R'=n=n_R+n_L
\end{equation}
and assuming, without loss of generality that 
\begin{equation}
    y>y', \quad y>0
\end{equation}
where the first condition is a result of the symmetry of the inner product and the second condition is a result of the left-right symmetry (see appendix \ref{appx:numerics_inner_product}), \cite{Lin:2023trc} find the inner product to be
\begin{equation}
\label{inner_product_recursion_solution}
    \langle n_L', n_R' | n_L,n_R\rangle = \sum_{0\leq k \leq n_R} q^{k^2+k(2\Delta + y-y')+\Delta(y-y')} \frac{[n_L]![n_R]![n_L']![n_R']!}{[k]![y-y'+k]![n_L'-k]![n_R-k]!}~.
\end{equation}

The operators $a_{L/R}$ and $a_{L/R}^\dagger$ are Hermitian conjugates of one another under the chord inner product. This can be seen diagrammatically:
\begin{center}
    \includegraphics[scale=0.4]{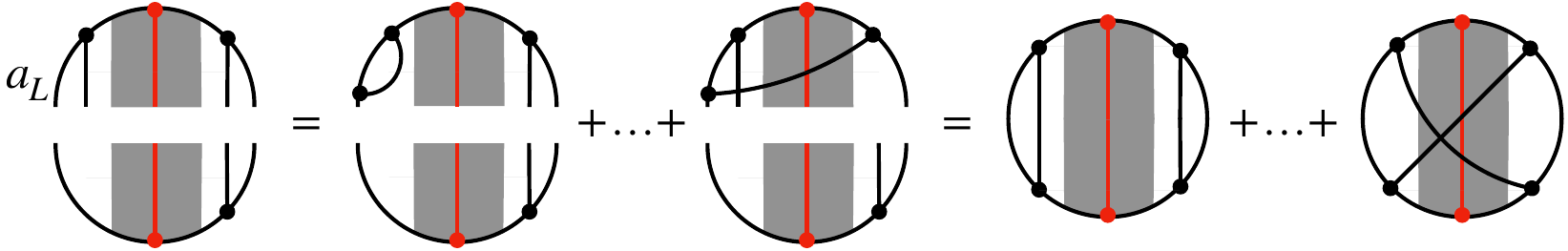}
    \includegraphics[scale=0.4]{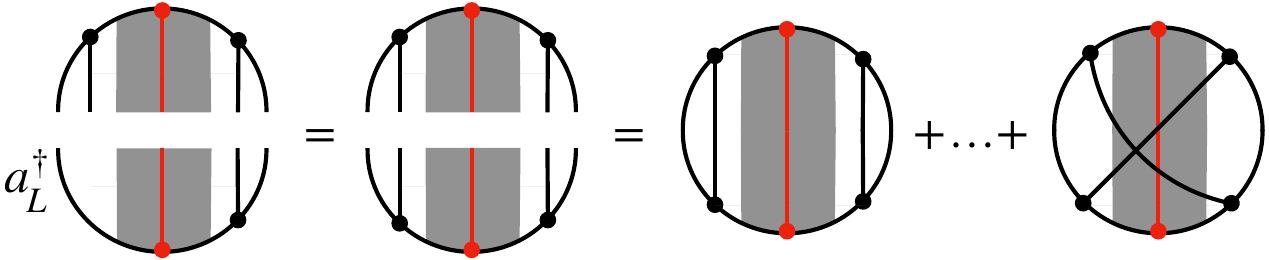}
\end{center}
where the first line shows the action of $a_L$ on the ket state which consists of all ways of taking an open chord and `closing' it to the left of all existing chords; then, the inner product is taken between this linear combination and the bra state. The second line shows the action of $a_L^\dagger$ on the bra state which consists of adding a Hamiltonian chord to the left of all existing chords, then taking the inner product with the ket state. The result of these two procedures is the same. 

Using the tools introduced in this section, we turn to evaluate the operator K-complexity of DSSYK in the next section.

\section{Operator K-complexity} \label{Sec:Operator_KC}
The time-evolution of the operator-inserted state given in \eqref{operator_evolution} defines a Krylov problem with a total Hamiltonian $H_R-H_L$ and initial state $|n_L=0, n_R=0\rangle$, in the sense that having an initial state and a time evolution generator in a Hilbert space are the sufficient ingredients to implement the Lanczos algorithm and to define K-complexity accordingly \cite{Parker:2018yvk}. The solution to this Krylov problem is, however, less immediate as compared to the matterless case studied in \cite{Rabinovici:2023yex}: in the latter case, the chord basis consists of a set of orthogonal states $|n\rangle$ such that $H^m |0\rangle$ is a linear combination of states $|n\rangle$ with $n=0,\dots,m$; where $H$ is the effective Hamiltonian for the averaged theory in the zero-particle sector, cf. \eqref{H_a_ad}. This allows for the identification of the normalized version of the state $|n\rangle$ with the $n$-th Krylov element, up to a global phase that was found to be zero studying the specific action of $H$ on chord number eigenstates\footnote{Equivalently, it can be shown inductively that the normalized states $|n\rangle$ satisfy the Lanczos algorithm with Lanczos coefficients that can be directly read from the expression of $H$ in coordinates over the chord basis \cite{Rabinovici:2023yex}.}. In the present case of the Schr\"{o}dinger evolution generated by $H_R-H_L$ in the one-particle sector, as explained in Section \ref{sec.ToolsTrade}, one considers a basis of states $|n_L,n_R\rangle$, defined through
\begin{equation}
\label{1p_basis_generic_state}
    |n_L, n_R\rangle = (a_L^\dagger)^{n_L} (a_R^\dagger)^{n_R} |0,0\rangle~,
\end{equation}
and which has the property (by construction \cite{Lin:2022rbf, Lin:2023trc}) that each state is a linear combination of states of the form $H_L^{m_L}H_R^{m_R}|0,0\rangle$ with $m_L+m_R\leq n_L + n_R\equiv n$, where $n$ denotes the total chord number of the state. However, this time the total chord number eigenvalue is degenerate. To be explicit, let us note that the one-particle Hilbert space, $\mathcal{H}_{1p}$, is arranged as follows:
\begin{equation}
    \label{Hilbert_space_sum_sectors}
    \mathcal{H}_{1p} = \bigoplus_{n\geq 0} \mathcal{H}_{1p}^{(n)}~,
\end{equation}
where each fixed chord number sector is defined as
\begin{eqnarray}
    \label{Fixed_chord_number_sector_def}
    \mathcal{H}_{1p}^{(n)} = \text{span}\left\{ |k,n-k\rangle ~:~k=0,\dots,n  \right\}~,
\end{eqnarray}
whose dimension is therefore $\dim \mathcal{H}_{1p}^{(n)}=n+1$. Consequently, the operation 
\begin{eqnarray}
    \label{HR-HR_power_action}
    \left(H_R-H_L\right)^m|0,0\rangle
\end{eqnarray}
will result in a sum of certain specific linear combinations of states within each of the fixed chord number sectors in the range $n=0,\dots,m$, and it is not \textit{a priori} clear which orthogonal directions will be selected by the Lanczos algorithm \cite{Lanczos:1950zz,viswanath1994recursion}, stated in section \ref{subsect:KrylovBasis}, when constructing the orthonormal Krylov basis $\left\{ |\psi_n\rangle \right\}$, whose elements are labeled by a single index. Furthermore, the fact that the basis $\left\{ |n_L,n_R\rangle \right\}$ is itself not orthogonal with respect to the chord inner product of the one-particle sector (cf. section \ref{subsect:inner_product}) adds a further technical complication to the problem.

To summarize, observing the form of the time evolution generator $H_R - H_L$ and its action over the initial state $|0,0\rangle$ does not allow for an immediate identification of the orthonormal Krylov basis, and a careful study of the Lanczos algorithm is required in order to construct the Krylov basis elements $|\psi_n\rangle$ and, in particular, in order to assess whether they are exact eigenstates of total chord number. Note that this amounts to a non-trivial statement given the degeneracy of the fixed total chord number sectors\footnote{In this sense, the setup of the problem is reminiscent of the task of building the Krylov basis for an operator in a quantum many-body system with k-local interactions: operator space admits a basis of operators of fixed size, but subspaces of fixed size are highly degenerate (cf. \cite{Parker:2018yvk, Roberts:2014isa, Roberts:2018mnp}, or equation (2.66) in \cite{Sanchez-Garrido:2024pcy}), in a way such that Krylov basis elements need not be exact size eigenstates, and in fact for a maximal operator Krylov dimension in a finite system \cite{Rabinovici:2020ryf} they cannot be so because they need to span all orthogonal directions in Krylov space. However, the analysis in DSSYK will differ from the intuitive description just given because the chord eigenbasis $\left\{|n_L,n_R\rangle\right\}$ and the Krylov basis $\left\{|\psi_n\rangle\right\}$ have the same cardinality, namely $\aleph_0$ (countable infinity), which allows Krylov elements to be total chord number eigenstates. They will indeed be found to be so in the regime where the bulk dual is described by JT gravity.}. 

The Lanczos algorithm will take the states successively generated by \eqref{HR-HR_power_action} and build an orthonormal Krylov basis $\left\{|\psi_n\rangle\right\}_{n\geq 0}$ out of them, which will be a basis adapted for describing the time evolution of the state
\begin{equation}
    \label{Time_evolving_state_HRminusHL_Schr}
    |\psi(t)\rangle = e^{-it(H_R-H_L)}|0,0\rangle~,
\end{equation}
as announced in section \ref{subsect:Right_And_Left_Hamiltonians}. This basis may be used to expand \eqref{Time_evolving_state_HRminusHL_Schr} as
\begin{equation}
    \label{psi_t_in_Krylov_basis}
    |\psi(t)\rangle = \sum_{n\geq 0} \psi_n (t) |\psi_n\rangle~,
\end{equation}
and the Krylov space wave functions $\psi_n(t)$ may be used in order to define the Krylov complexity of $|\psi(t)\rangle$ in the usual manner \cite{Parker:2018yvk, Rabinovici:2020ryf, Balasubramanian:2022tpr, Rabinovici:2023yex}:
\begin{equation}
    \label{KC_oneparticlestate_def}
    C_K(t)=\sum_{n\geq 0} n ~|\psi_n(t)|^2 = \sum_{n\geq 0} n ~ \psi_n(t) \psi_n(-t)~.
\end{equation}
 The second equality holds because the time evolution generator is hermitian \cite{Sanchez-Garrido:2024pcy}. Note that equation \eqref{KC_oneparticlestate_def} may be envisioned as the expectation value of the Krylov complexity operator $\widehat{C_K}$ on the state $|\psi(t)\rangle$ given in \eqref{psi_t_in_Krylov_basis} or in \eqref{Time_evolving_state_HRminusHL_Schr}, where the Krylov complexity operator is
\begin{equation}
\label{KC_operator_def}
\widehat{C_K} := \sum_{n\geq 0} ~n~ |\psi_n\rangle \langle \psi_n|~.
\end{equation}

A considerable fraction of this paper is devoted to explicitly building the Krylov basis elements, $|\psi_n\rangle$. We will test them against the map between the bulk and boundary Hilbert spaces in order to discuss the extent to which they correspond to length operator eigenstates in the bulk theory. The understanding of the bulk dual of the Krylov basis is necessary in order to connect from first principles the K-complexity \eqref{KC_oneparticlestate_def} to the length expectation value as a function of time in a bulk setup including matter.
Technically, we do so by exploiting the fact that in DSSYK the Heisenberg evolution of the operator of interest is mapped via \eqref{operator_evolution} to the Schrödinger evolution of the state $|0,0\rangle\in\mathcal{H}_{1p}$,
which represents in the averaged theory the TFD state perturbed by the operator insertion. 
For an alternative approach \textit{à la Heisenberg}, see appendix \ref{Appx:Lanczos_W_T}, and for a generic proof of the bijective correspondence between the (operator) Krylov basis elements of an operator $\mathcal{O}$, seen as an element of operator space, and the (state) Krylov basis of the $|\textrm{TFD}\rangle$ perturbed by an operator insertion, say $O_L|\textrm{TFD}\rangle$, seen as a state in the doubled Hilbert space of states, we refer to \cite{Sanchez-Garrido:2024pcy}. 

\subsection{Krylov basis and Lanczos coefficients from the operator chord basis}\label{subsect:KrylovBasis}
As explained above, we wish to perform the Lanczos algorithm for the total Hamiltonian $H_R-H_L$ with initial state $|\psi_0\rangle = |0,0\rangle$, as this will produce the Krylov basis $\left\{|\psi_n\rangle\right\}$ adapted to the Schrödinger evolution of the state $|\psi(t)\rangle$ in \eqref{Time_evolving_state_HRminusHL_Schr}. Let us begin by presenting the form of the Lanczos recursion satisfied by the Krylov basis elements\footnote{The most generic form of the Lanczos recursion for states given a hermitian Hamiltonian includes \cite{Lanczos:1950zz,viswanath1994recursion} a second set of Lanczos coefficients $a_n=\langle \psi_n | (H_R-H_L)|\psi_n\rangle$. Such coefficients, however, vanish whenever odd moments are zero, i.e. $\langle 0,0 | (H_R-H_L)^{2n+1} | 0,0\rangle=0$ $\forall n \geq 0$, which is the case due to the Gaussian nature of the random couplings in the disordered Hamiltonian with which the DSSYK model is defined.}:
\begin{equation}
    \label{Lanczos_recursion_1p}
    b_n|\psi_n\rangle = (H_R-H_L)|\psi_{n-1}\rangle - b_{n-1}|\psi_{n-2}\rangle~,
\end{equation}
where $|\psi_0\rangle = |0,0\rangle$, $\langle\psi_m|\psi_n\rangle = \delta_{mn}$, and the boundary conditions may be taken to be $b_{0}=1$, $|\psi_{-1}\rangle = \mathbf{0}$ (the zero vector in the Hilbert space). In an iterative or inductive approach to this recursion, it is often convenient to define the non-normalized Krylov elements $|A_n\rangle$, given by the right-hand side of \eqref{Lanczos_recursion_1p}, so that the \textit{Lanczos coefficients} are given by $b_n=\lVert A_n \rVert$.

Let us now state the explicit form of the total Hamiltonian appearing in \eqref{Lanczos_recursion_1p}, combining equations \eqref{HL_v2} and \eqref{HR_v2} or \eqref{HL} and \eqref{HR}. For the sake of clarity in the present discussion, we shall nevertheless use a normalization of the Hamiltonian in which the overall prefactor in the above-mentioned equations is taken to be $J$ instead of $J/\sqrt{\lambda}$, as done in \cite{Berkooz:2018jqr}. We shall return to the Hamiltonian normalization $J/\sqrt{\lambda}$ later on, in the discussions relevant to the gravitational regime of the theory~:
\begin{align}
    H_R-H_L &= J\left( a_R^\dagger - a_L^\dagger + a_R - a_L \right)\label{Total_Hamiltonian_minus_line1} \\ &= J\left[a_R^\dagger - a_L^\dagger + \alpha_R [n_R]_q \left(1 - \widetilde{q}q^{~n_L}\right) - \alpha_L [n_L]_q \left(1 - \widetilde{q}q^{~n_R}\right)\right] ~. \label{Total_Hamiltonian_minus_line2}
\end{align}
By inspection of \eqref{Total_Hamiltonian_minus_line2} we can see that whenever $H_R-H_L$ acts on some vector  belonging to the $n$-th total chord number sector, $\mathcal{H}_{1p}^{(n)}$, it will produce a linear combination of states belonging to $\mathcal{H}_{1p}^{(n+1)}$ and $\mathcal{H}_{1p}^{(n-1)}$. We can already anticipate that the Krylov vectors constructed by the Lanczos algorithm \eqref{Lanczos_recursion_1p} have the property that
\begin{eqnarray}
    \label{Krylov_vertor_uptpnsector}
    |\psi_n\rangle \in \bigoplus_{m=0}^n \mathcal{H}_{1p}^{(m)}~,
\end{eqnarray}
but as we shall shortly show, generically the projection of $(H_L-H_R)|\psi_{n-1}\rangle$ over $\mathcal{H}_{1p}^{(n-2)}$ need not cancel against the projection over the same subspace of $b_{n-1}|\psi_{n-2}\rangle$, and therefore $|\psi_{n}\rangle$ is not entirely contained in $\mathcal{H}_{1p}^{(n)}$ and, instead, develops a ``tail'' over the sectors of smaller total chord number. In what follows we shall explore, both numerically and analytically, how the Hamiltonian and operator parameters $\lambda$ and $\Delta$ (or equivalently $q$ and $\widetilde{q}$) control such a tail, and under what circumstances the Krylov elements satisfy $|\psi_n\rangle \in \mathcal{H}_{1p}^{(n)}$, thus being simultaneously eigenstates of the total chord number operator and of the Krylov complexity operator \eqref{KC_operator_def}.

Through direct computation one can show (cf.~appendix \ref{appx:Details_Lanczos}) that the states $|\psi_n\rangle$ defined through
\begin{equation}
    \label{Binomial_Ansatz}
    |\psi_n\rangle = \frac{1}{\prod_{k=0}^n (b_k/J)}|\chi_n\rangle~,\qquad |\chi_n\rangle := \sum_{k=0}^n (-1)^k \binom{n}{k}|k,n-k\rangle~, 
\end{equation}
are orthonormal and solve the Lanczos recursion \eqref{Lanczos_recursion_1p} for $n=0,1,2,3$ given the following Lanczos coefficients (recall we use the convention where $b_0\equiv 1$):
\begin{align}
    &b_1^2 = 2J^2(1-\widetilde{q})~, \label{b1_Exact}\\ &b_2^2=J^2[3+q-\widetilde{q}(1+3q)]~, \label{b2_Exact}\\ &b_3^2 = J^2\frac{(1 + q) (10 + (-5 + \widetilde{q}) \widetilde{q} + 
   q + (-14 + \widetilde{q}) \widetilde{q} q + (1 + 5 \widetilde{q} (-1 + 2 \widetilde{q})) q^2)}{b_2^2}~. \label{b3_Exact}
\end{align}
Note that the limit in which the operator becomes the identity is $\Delta\to 0$ for fixed $\lambda$, which is equivalent to $\widetilde{q}\to 1$. Such a limit implies $b_1\to 0$, which is consistent as anticipated in section \ref{Sec:Moments_from_chord_diagrams}: in the absence of operator insertions, the TFD state is stationary under the evolution generated by $H_R-H_L$, therefore having a one-dimensional Krylov space, which is reflected in the fact that the Lanczos algorithm terminates at step $n=1$ by hitting a zero at $b_1=0$ \cite{Rabinovici:2020ryf}. It can also be verified that the moments given in \eqref{mu2_result}, \eqref{mu4_result} and \eqref{mu6_result} give rise to these Lanczos coefficients (modulo using the same Hamiltonian normalization in both cases).

For $n\geq 4$, the Ansatz \eqref{Binomial_Ansatz} is no longer an exact solution of the Lanczos recursion \eqref{Lanczos_recursion_1p} for arbitrary $q$ and $\widetilde{q}$. The generic, exact solution of the Lanczos algorithm for arbitrary values of the parameters $q$ and $\widetilde{q}$ 
does deviates from the binomial Ansatz \eqref{Binomial_Ansatz} for $n\geq 4$ but reduces to it for the initial values of $n$. In order to be able to identify a range of parameters for which the statement \eqref{Binomial_Ansatz} is correct within a controlled approximation, we use an inductive approach. We first will uncover what it would take for \eqref{Binomial_Ansatz} to be exactly correct. 
In this inductive analysis, assuming that the binomial Ansatz \eqref{Binomial_Ansatz} is correct for the Krylov elements $|\psi_m\rangle$ for $m=0,\dots,n$ and for some (by now unspecified) Lanczos coefficients $b_m$ with $m=1,\dots,n$, the next (non-normalized) Krylov element satisfies:
\begin{align}
    |A_{n+1}\rangle &= (H_R-H_L)|\psi_n\rangle - b_n|\psi_{n-1}\rangle \label{Lanczos_Induction_Anplus1_line1}\\
    &= J(a_R^\dagger - a_L^\dagger)|\psi_n\rangle + J(a_R-a_L)|\psi_n\rangle - b_n|\psi_{n-1}\rangle \label{Lanczos_Induction_Anplus1_line2}\\
    &= \frac{J}{(b_1/J)\dots (b_n/J)}\left[(a_R^\dagger - a_L^\dagger)|\chi_n\rangle + (a_R-a_L)|\chi_n\rangle - \frac{b_n^2}{J^2}|\chi_{n-1}\rangle \right] ~.\label{Lanczos_Induction_Anplus1_line3}
\end{align}
It is now possible to show that the first term in \eqref{Lanczos_Induction_Anplus1_line3} gives back directly a binomial state, namely
\begin{equation}
    \label{adagger_gives_binomial}
    (a_R^\dagger - a_L^\dagger)|\chi_n\rangle = |\chi_{n+1}\rangle~,
\end{equation}
and both of the two next terms in \eqref{Lanczos_Induction_Anplus1_line3} belong to the sector $\mathcal{H}_{1p}^{(n-1)}$ but, as we have already emphasized, they need not exactly cancel because such a sector has a dimension greater than one. Thus, for the binomial Ansatz \eqref{Binomial_Ansatz} to be correct, the following statement is a necessary and sufficient condition:
\begin{equation}
\label{Binom_Ansatz_necsuf_condition}
     (a_R-a_L)\ket{\chi_n} \overset{!}{=} \frac{b_n^2}{J^2} \ket{\chi_{n-1}}~.
\end{equation}

Let us now analyze the left-hand side of the condition \eqref{Binom_Ansatz_necsuf_condition}, which can be expressed as
\begin{equation}
    \label{Binom_Ansatz_action_of_aRminusaL}
    (a_R-a_L)|\chi_n\rangle = \sum_{k=0}^{n-1} c_k(n)~ (-1)^k\binom{n-1}{k}|k,n-1-k\rangle~,
\end{equation}
where $c_k(n)$ is the prefactor of the coefficients in the linear combination that captures the potential deviation of \eqref{Binom_Ansatz_action_of_aRminusaL} from the binomial Ansatz $\ket{\chi_{n-1}}$. For the condition \eqref{Binom_Ansatz_necsuf_condition} to be fulfilled, it is sufficient (and necessary) that, for a fixed $n$, $c_k(n)$ is constant in $k$ for the corresponding $k$-domain, namely $k=0,\dots,n-1$, in which case we can assign it to be $b_n^2$, solving the induction. After some algebra, we find that $c_k(n)$ admits the following simplified expression:
\begin{eqnarray}
     \label{Ckn_simplified}
     c_k(n) = n \frac{[n-k]_q}{n-k}(1-\tilde{q}q^k) + n\frac{[k+1]_q}{k+1}(1-\tilde{q}q^{n-1-k})~,
 \end{eqnarray}
which enjoys the following symmetry:
 \begin{eqnarray}
     \label{Ckn_symmetry}
     c_k(n)=c_{n-1-k}(n)~.
 \end{eqnarray}
Appendix \ref{appx:Details_Lanczos} provides an explanation of why this symmetry property allows for the binomial Ansatz \eqref{Binomial_Ansatz} for the Krylov elemens $|\psi_n\rangle$ to be the correct solution of the Lanczos algorithm for $n=0,1,2,3$. However, generically for $n\geq 3$, $c_k(n)$ is a non-constant (yet symmetric) discrete function of $k=0,\dots,n-1$, thus failing to allow for the cancellation \eqref{Binom_Ansatz_necsuf_condition}, as a consequence of which it can be shown inductively that the generic Krylov elements will take the form:
    \begin{eqnarray}
        \label{Generic_Krylov_element}
        |\psi_n\rangle = \frac{1}{(b_1/J)\dots (b_n/J)}|\chi_n\rangle + \sum_{m=1}^{\lfloor n/2 \rfloor}|\xi^{(n)}_{n-2m}\rangle~,
    \end{eqnarray}
    where $|\xi^{(n)}_{n-2m}\rangle\in\mathcal{H}_{1p}^{(n-2m)}$ denotes the tail on smaller total chord number sectors that the $n$-th Krylov element develops, besides the leading binomial term $|\chi_n\rangle\in\mathcal{H}_{1p}^{(n)}$. Note that, given a fixed $n$, states $|\xi^{(n)}_{n-2m}\rangle$ with different values of $m$ are orthogonal to each other (and to $|\chi_n\rangle$) because they belong to different total chord number sectors.

From now on, we shall analyze the properties of the tail in \eqref{Generic_Krylov_element}. Section \ref{subsect:Lanczos_Limit_lambda0_qtfixed} shows that in a limit in which $\lambda\to 0$ and $\widetilde{q}$ is kept fixed the coefficients $c_k(n)$ in \eqref{Ckn_simplified} become constant as a function of $k$ for every $n$, thus allowing to obtain a simple closed-form solution to the Lanczos algorithm. Next, section \ref{subsect:numerics} will provide numerical evidence for the fact that the parameter $\lambda$ controls the suppression of such a tail, which motivates the semiclassical analysis in section \ref{subsect:bn_asymptotic_limit}: The semiclassical limit is defined as the limit where $\lambda\to 0$ but $\lambda n$ is held fixed, hence becoming a continuous version of chord number. In this regime, we find an analytical expression for the Lanczos coefficients, allowing for the analytical evaluation of Krylov complexity subsequently presented in section \ref{subsect:analytical_K_Complexity}. Building up on this result, a canonical analysis of this semiclassical K-complexity connecting scrambling dynamics to an instability in the effective Krylov space potential will be presented in section \ref{sect:Hamiltonian_analysis_operator}. Finally, in section \ref{sect:Triple_scaling_operator} we will derive the \textit{triple-scaled} \cite{Lin:2022rbf} Hamiltonian relevant to the gravitational regime out of the operator Lanczos coefficients.

\subsection{Exact solution for heavy operator and \texorpdfstring{$\lambda\to0$}{TEXT} }\label{subsect:Lanczos_Limit_lambda0_qtfixed}
As announced, there exists a  limit in which $c_k(n)$ becomes constant in $k=0,\dots,n-1$ for fixed and arbitrary $n$, in which case it can always (i.e. for every $n$) be pulled out of the $k$-sum in \eqref{Binom_Ansatz_action_of_aRminusaL} and identified with the $b_n^2$ coefficient, assuring condition \eqref{Binom_Ansatz_necsuf_condition}, as we shall prove by induction below. Such a limit is:
\begin{eqnarray}
    \label{Limit_bn_closed_form}
    q\to 1~, \quad \tilde{q}\;\,\, \text{fixed.}
\end{eqnarray}
Recalling that $q=e^{-\lambda}$ and $\tilde{q}=e^{-\Delta \lambda}$, this limit can be equivalently expressed as:
\begin{eqnarray}
    \label{limit_bn_closed_form_v2}
    \lambda \to 0~, \quad \Delta \to +\infty~, \quad \tilde{q}=e^{-\Delta \lambda}\;\,\, \text{fixed.}
\end{eqnarray}
Noting that the $q$-numbers \eqref{qnumber_definition} have the property of reducing to \textit{regular} numbers when $q\to 1$, namely $\lim_{q\to 1} [n]_q=n$, expression \eqref{Ckn_simplified} can be seen to simplify drastically in the proposed limit \eqref{Limit_bn_closed_form}:
\begin{eqnarray}
    \label{Ckn_limit_simplification}
    c_k(n)\overset{q\to 1~,~ \tilde{q}\;\text{fixed}}{\relbar\joinrel\relbar\joinrel\relbar\joinrel\relbar\joinrel\relbar\joinrel\relbar\joinrel\relbar\joinrel\longrightarrow} 2n(1-\tilde{q})\equiv c(n)~. 
\end{eqnarray}
Where, in the last step, the result of the limit has been assigned to $c(n)$ for any value of $k$ because $c_k(n)$ became $k$-independent. In other words, in this limit we have:
\begin{equation}
    \label{annihilation_on_binom_limit}
    \left(a_R-a_L\right)|\chi_n\rangle = c(n)|\chi_{n-1}\rangle \quad \forall n \geq 1~, 
\end{equation}
which implies that \eqref{Binom_Ansatz_necsuf_condition} will be satisfied with $b_n^2=c(n)$. We therefore propose this expression as the induction hypothesis for the Lanczos coefficients (recall we had an Ansatz \eqref{Binomial_Ansatz} for the Krylov elements, but not one for the Lanczos coefficients). Thanks to this cancellation, as promised, the Lanczos step \eqref{Lanczos_Induction_Anplus1_line1} simplifies greatly and yields:
\begin{equation}
    \label{Krylov-non_norm_limit}
    |A_{n+1}\rangle = \frac{J}{(b_1/J)\dots (b_n/J)}\left(a_R^\dagger - a_L^\dagger\right)|\chi_n\rangle = \frac{J}{(b_1/J)\dots (b_n/J)}|\chi_{n+1}\rangle~.
\end{equation}
This is almost the induction step for the Krylov elements binomial Ansatz, but first the state needs to be normalized:
\begin{eqnarray}
    \label{Lanczos_coeff_Ansatz_proof_limit}
    & \frac{b_{n+1}^2}{J^2} = \frac{1}{J^2} \langle A_{n+1} | A_{n+1}\rangle = \frac{\langle \chi_{n+1} | \chi_{n+1}\rangle}{(b_1/J)^2\dots (b_n/J)^2} =\frac{ \langle \chi_{n} | \left( a_R^\dagger - a_L^\dagger \right)^\dagger |\chi_{n+1}\rangle}{(b_1/J)^2\dots (b_n/J)^2}~. \\ \label{Lanczos_coeff_Ansatz_proof_limit_line2}
    & = \frac{ \langle  \chi_{n} | \left( a_R - a_L \right) |\chi_{n+1}\rangle}{(b_1/J)^2\dots (b_n/J)^2} = \frac{ c(n+1)\langle  \chi_{n} |\chi_{n}\rangle}{(b_1/J)^2\dots (b_n/J)^2}~,
\end{eqnarray}
where in the last step we have made use of \eqref{annihilation_on_binom_limit}.
Now, recalling our induction hypothesis for the Krylov elements \eqref{Binomial_Ansatz},
we note that normalization of the Krylov elements $|\psi_n\rangle$ up to step $n$, which is assumed for the sake of the inductive argument, implies the following norm of the binomial states of total chord number (up to) $n$:
\begin{equation}
    \label{Norm_of_binom_limit}
    \langle \chi_n | \chi_n \rangle = \frac{b_1^2\dots b_n^2}{J^{2n}}~. 
\end{equation}
Combining \eqref{Norm_of_binom_limit} with \eqref{Lanczos_coeff_Ansatz_proof_limit_line2}, we reach:
\begin{eqnarray}
    \label{Lanczos_coeff_Ansatz_proof_limit_endofproof}
    b_{n+1}^2 =J^2 c(n+1)~,
\end{eqnarray}
in agreement with our Lanczos coefficients hypothesis.
Now, normalizing \eqref{Krylov-non_norm_limit} with the obtained Lanczos coefficient yields the next normalized Krylov element,
\begin{equation}
    \label{Krylov_element_induction_step_proof_in_q1_limit}
    |\psi_{n+1}\rangle = \frac{1}{(b_1/J)\dots (b_n/J) (b_{n+1}/J)} |\chi_{n+1}\rangle~,
\end{equation}
also in agreement with the induction hypothesis for the Krylov elements \eqref{Binomial_Ansatz}. At this point, \eqref{Lanczos_coeff_Ansatz_proof_limit_endofproof} together with \eqref{Krylov_element_induction_step_proof_in_q1_limit} conclude the induction step, and note that the seeds for both the Krylov elements and the Lanczos coefficients Ansätze have also been explicitly checked in the previous section \ref{subsect:KrylovBasis}, as one can verify by taking the limit of expressions \eqref{b1_Exact}, \eqref{b2_Exact} and \eqref{b3_Exact} for $b_1$, $b_2$ and $b_3$ (respectively) when $q\to 1$ with $\tilde{q}$ fixed. 

Summarizing, we have proved inductively that in the limit \eqref{Limit_bn_closed_form}, which is equivalently the limit \eqref{limit_bn_closed_form_v2}, the Lanczos algorithm can be solved in closed form for arbitrary $n$, with Lanczos coefficients $b_n$ and Krylov basis elements $|\psi_n\rangle$ given by:
\begin{eqnarray}
    \label{Lanczos_closed_form_solution_in_limit}
    b_n = J\sqrt{2n(1-\tilde{q})}~, \qquad |\psi_n\rangle = \frac{J^n}{b_1\dots b_n} \sum_{k=0}^n (-1)^k \binom{n}{k} |k,n-k\rangle~,\qquad \forall n\geq 0~.
\end{eqnarray}

The solution \eqref{Lanczos_closed_form_solution_in_limit} of the Lanczos algorithm in the limit \eqref{limit_bn_closed_form_v2} is, however, apparently too simple: We only get a square-root growth in $n$ for the Lanczos coefficients, similar to what we obtained in the $q\to 1$ limit of the Lanczos coefficients of the TFD state in \cite{Rabinovici:2023yex}, which implies \cite{Caputa:2021sib} a quadratic growth of K-complexity, $C_K(t)~\sim 2(1-\widetilde{q})(Jt)^2$. In particular, this does not feature any imprints of chaotic operator dynamics, namely a linear $b_n$ sequence that would give rise to an exponential profile of K-complexity as a function of time \cite{Parker:2018yvk}. We may understand this behavior by noting that the limit \eqref{limit_bn_closed_form_v2} is the special case where we make the operator much bigger in size than the interaction terms in the Hamiltonian: it is a non-typical operator for which scrambling dynamics are not generically expected. 
Nevertheless, we shall eventually see in section \ref{subsect:bn_asymptotic_limit} that the limit studied in the present section effectively isolates the small-$n$ behavior of the Lanczos coefficients $b_n$ computed in the semiclassical limit, where operators with fixed $\widetilde{q}$ are seen to feature scrambling depending on how close $\widetilde{q}$ is to $1$. In that sense, the Lanczos coefficients computed in the current section may be understood as 
describing the early-time regimes of the semiclassical K-complexity. Systematic corrections to the solution \eqref{Lanczos_closed_form_solution_in_limit} in a small-$\lambda$ expansion with $\widetilde{q}$ (and $n$) fixed that deviate from the limit \eqref{limit_bn_closed_form_v2} are presented in appendix \ref{appx:small_lambda}, where it is shown that $c_k(n)$ remains $k$-independent up to order $\lambda^2$.

\subsection{Numerical experiments} \label{subsect:numerics}
In order to build some intuition, and to back up some of the analytical results in later sections, we shall now present a numerical analysis of the solution of the Lanczos recursion \eqref{Lanczos_recursion_1p}. Our aim is to analyze the tails \eqref{Generic_Krylov_element} of the Krylov elements over the various total chord number sectors which, as argued in section \ref{subsect:KrylovBasis}, are the main obstacle to achieving an analytic expression for the Lanczos coefficients and, more importantly, for identifying the Krylov basis elements $|\psi_n\rangle$ with total chord number eigenstates.

The numerical calculations were obtained using two complementary approaches:
\begin{itemize}
    \item Efficient implementation of the Lanczos algorithm based on the partial re-orthogonalization routine (PRO) described in \cite{Rabinovici:2020ryf} (see \cite{Sanchez-Garrido:2024pcy} for a discussion on the implementation of PRO for generic Hilbert spaces, not just operator space).
    \item High-precision direct implementation of the Lanczos algorithm \eqref{Lanczos_recursion_1p} in the one-particle irrep of the chord algebra, equipped with the inner product defined through \eqref{inner_rec_start}.
\end{itemize}
Both approaches are found to be complementary. We apply the PRO algorithm to truncations of the one-particle Hilbert space consisting of the direct sum of all total chord number sectors up to some value $N$, i.e.:
\begin{equation}
    \label{Hilbert_space_truncation}
    \mathcal{H}_{1p;N}:=\bigoplus_{n=0}^N\mathcal{H}_{1p}^{(n)}~,
\end{equation}
whose dimension is therefore:
\begin{eqnarray}
    \label{truncation_dimension}
    d_N:=\dim\mathcal{H}_{1p;N}=\sum_{n=0}^N (n+1) = \frac{1}{2}(N+1)(N+2)~,
\end{eqnarray}
which behaves as $d_N\sim \frac{N^2}{2}$. This illustrates the problem introduced at the beginning of section \ref{Sec:Operator_KC} and in section \ref{subsect:KrylovBasis}: we have that the number of distinct fixed chord number states in the Hilbert space truncation, $N+1$ (corresponding to total chord number $n=0,\dots,N$), is smaller than the  Krylov space dimension $K$, i.e. the cardinality of the Krylov basis, which is generically bounded by the inequality $K\leq d_N$ \cite{Lanczos:1950zz,Rabinovici:2020ryf,Sanchez-Garrido:2024pcy}. Generically $K$ can be (much) larger than $N$, implying that the Krylov elements $|\psi_n\rangle$ cannot be eigenstates of total chord number with eigenvalue $n$ for all $n=0,\dots,K-1$. Even without a truncation ($N\to +\infty$), the dimensionality of the sectors $\mathcal{H}_{1p}^{(n)}$ could allow for more than one orthogonal Krylov elements to develop a projection over the same sectors, and in the case of a finite truncation ($N<+\infty$) this is in fact compulsory, as we shall see, because $K\sim \frac{N^2}{2}>N$. We will thus focus on building the Krylov elements $|\psi_n\rangle$ for $n=0,\dots,N$ as the ones with $n$ in the range $N< n < K$ will suffer from finite-truncation effects, exploring linear re-combinations of already probed sectors. This unveils the numerical cost of this truncation-based approach: In order to access up to $N$ Krylov elements and Lanczos coefficients free from finite-truncation effects, one needs, at least, a Hilbert space truncation $\mathcal{H}_{1p;N}$, whose dimension scales as $\sim \frac{N^2}{2}$, therefore requiring the numerical construction of operators as $\frac{N^2}{2}\times \frac{N^2}{2}$ matrices\footnote{Note that, due to the \textit{locality} of the Hamiltonian \eqref{Total_Hamiltonian_minus_line2} with respect to the total chord number, the Lanczos coefficients $b_n$ and Krylov elements $|\psi_n\rangle$ will be completely blind to finite-truncation effects for $n\leq N$, and may thus be compared to analytical results where $N$ is formally infinite.}. The PRO algorithm is able to handle these constructions efficiently at machine precision; however, for instances in which the numerical instability of the Lanczos algorithm \cite{Simon1984TheLA,parlett1998symmetric} requires control of a precision finer than the regular double floating point machine precision, we resort to high-precision implementations of the pure Lanczos algorithm, based on symbolic manipulation of the inner product recursion \eqref{inner_rec_start} and of the Lanczos recursion \eqref{Lanczos_recursion_1p}.

\begin{figure}
    \centering
     \includegraphics[scale=0.5]{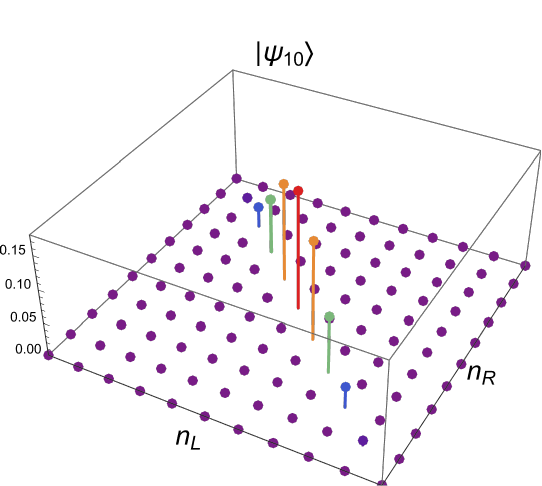}\includegraphics[scale=0.5]{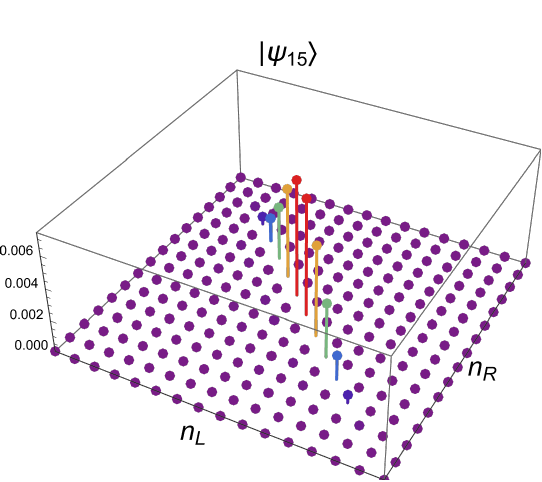}\includegraphics[scale=0.5]{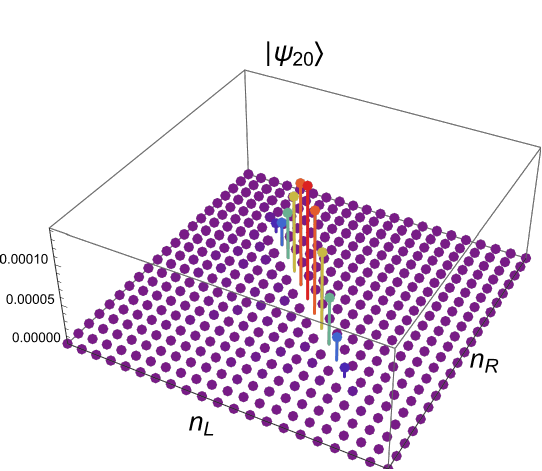} \\ 
     \includegraphics[scale=0.5]{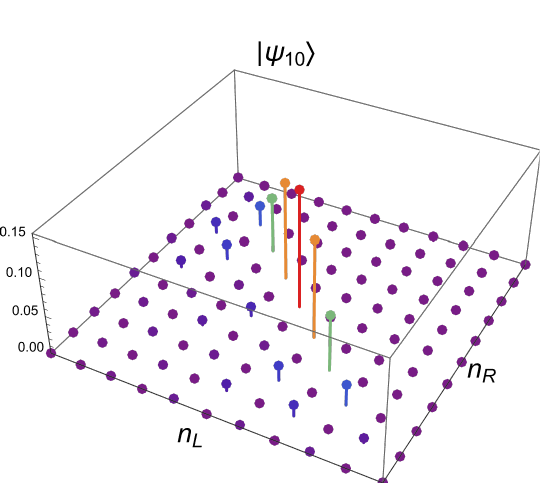}\includegraphics[scale=0.5]{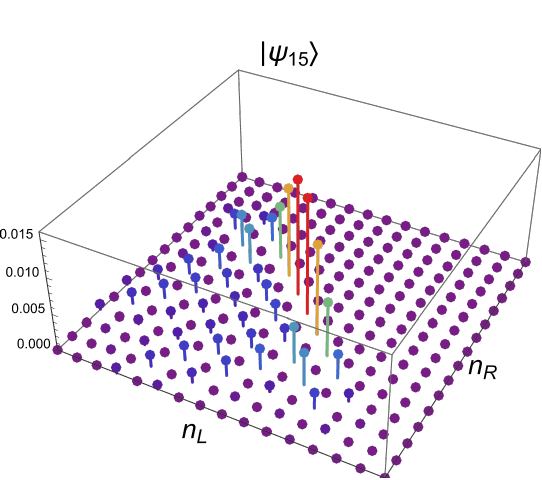}\includegraphics[scale=0.5]{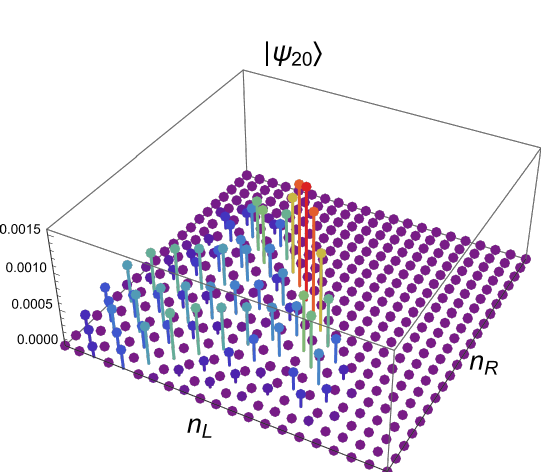}
    \caption{Profile of three Krylov elements $|\psi_n\rangle$, for $n=10,15,20$ (respectively, \textbf{left}, \textbf{center} and \textbf{right}) over the basis $|k,m-k\rangle$; entries are shown in absolute value (from red to purple, in decreasing absolute value). In all cases we fixed $\widetilde{q}=0.56$, and considered two values of $\lambda = 0.05,0.5$ (resp. \textbf{top} and \textbf{bottom}). We can observe that in the case where $\lambda=0.5$ the Krylov elements $|\psi_n\rangle$ develop significant projection over sectors with total chord number $m<n$. }
    \label{fig:3dplots_main}
\end{figure}

As a starter, figure \ref{fig:3dplots_main} presents three-dimensional plots of a few Krylov basis elements $|\psi_n\rangle$ in coordinates over the chord basis elements $\left\{|k,m-k\rangle,~k=0,\dots,m,~m\geq 0\right\}$ for two different values of $\lambda$. More explicitly, the plots show the coefficients $\psi^{(n)}_{mk}$ of the linear combination
\begin{equation}
    \label{Krylov_element_coordinates}
    |\psi_n\rangle = \sum_{m\geq 0} \sum_{k=0}^m \psi^{(n)}_{mk}|k,m-k\rangle~,
\end{equation}
where the states of the tail $|\xi^{(n)}_m\rangle$, introduced in \eqref{Generic_Krylov_element}, may be identified as:
\begin{equation}
    \label{Krylov_tail_coordinates}
    |\xi_m^{(n)}\rangle = \sum_{k=0}^m \psi^{(n)}_{mk}|k,m-k\rangle~.
\end{equation}

For a given Krylov element $|\psi_n\rangle$, the plots in figure \ref{fig:3dplots_main} depict the following features:
\begin{enumerate}
    \item The main contribution to $\ket{\psi_n}$ are the chord states with total chord number equal to $n$. Reducing $\lambda$ suppresses the tail states $|\xi^{(n)}_m\rangle$ with $m<n$ in the tail of the Krylov vector $|\psi_n\rangle$, which is seen to be very localized in the $n$-th total chord number sector for the smallest value of $\lambda$, but which develops a tail on other sectors in the case of the largest $\lambda$ value.
    \item Furthermore, the profile of $|\psi_n\rangle$ over the sector $\mathcal{H}_{1p}^{(n)}$ is very symmetric (note that the plots show $\lvert\psi^{(n)}_{mk}\rvert$) and the largest contribution comes from the states $|k,n-k\rangle$ with $k\approx \frac{n}{2}$. 
    \item As anticipated by \eqref{Generic_Krylov_element}, we observe that the non-zero tail states $|\xi^{(n)}_m\rangle$ of a given Krylov vector $|\psi_n\rangle$ are those such that $m=n-2k$ for $k=1,\dots,\lfloor n/2 \rfloor$. This is a consequence of the fact that the total Hamiltonian \eqref{Total_Hamiltonian_minus_line2} only features hoppings between total chord number sectors differing by one unit. This implies that the diagonal Lanczos coefficients are $a_n=0$ as discussed in section \ref{subsect:KrylovBasis} which, when plugged in the Lanczos recursion \eqref{Lanczos_recursion_1p}, results in the fact that the Krylov elements ``jump'' through total chord number sectors in steps of two units. More formally, it can be shown inductively that even Krylov elements $|\psi_{2n}\rangle$ only have projections over the sectors $\mathcal{H}_{1p}^{(2n)}$ of even total chord number, and similarly for the odd ones, which only have support over sectors of odd total chord number.
\end{enumerate}

Despite the fact that the plots in figure \ref{fig:3dplots_main} are very insightful illustrations of the wave functions $\psi^{(n)}_{mn}$ of the Krylov elements $|\psi_n\rangle$ in the one-particle Hilbert space, we shall note that
these are coordinates
with respect to a non-orthonormal basis $|k,m-k\rangle$. The elements of this basis are in general not normalized with respect to the inner product \eqref{inner_rec_start}, and are in general not orthogonal if they belong to the same total chord number sector. Nevertheless, the fact that states belonging to different sectors are orthogonal will allow us to perform a basis-independent analysis of the localization of Krylov elements on total chord number sectors, by analyzing the norm contributions coming from each one of these. Borrowing expression \eqref{Generic_Krylov_element} and being agnostic about the specific profile of the Krylov element $|\psi_n\rangle$ over the leading Hilbert space sector, we may write
\begin{equation}
    \label{Krylov_vector_generic_for_numerics}
    |\psi_n\rangle = \sum_{m=0}^{\lfloor n/2 \rfloor}|\xi^{(n)}_{n-2m}\rangle~, \qquad |\xi^{(n)}_{n-2m}\rangle \in \mathcal{H}_{1p}^{(n-2m)}~.
\end{equation}
We now recall what the Krylov elements are orthonormal, $\langle \psi_m | \psi_n \rangle = \delta_{mn}$. Denoting the norm contribution of each sector as
\begin{equation}
    \label{Norm_contributions_numerics}
    \psi^{(n)}_m := \sqrt{\langle \xi^{(n)}_m | \xi^{(n)}_m\rangle }
\end{equation}
and recalling that sectors of different total chord number are orthogonal, we have that: 
\begin{equation}
    \label{Norms_Pythagoras}
    \sum_{m=0}^n \big( \psi^{(n)}_m\big)^2 = 1~,
\end{equation}
but note that $\big( \psi^{(n)}_m\big)^2 \neq \sum_{k=0}^m \big( \psi^{(n)}_{mk} \big)^2$ because the set $\left\{ |k,m-k\rangle \right\}_{k=0}^m$ with fixed $m$ is not orthogonal with respect to the chord inner product \eqref{inner_rec_start}. Nevertheless, \eqref{Norms_Pythagoras} is just enough to analyze systematically the norm contributions of the different sectors to a given Krylov element and the localization of the latter within the sector with the largest possible total chord number. In order to study this numerically, we find it useful to consider the following indicators:
\begin{itemize}
    \item Chord expectation value of the Krylov vector, given by:
    \begin{equation}
        \label{chord_eval_Krylov}
        \langle \psi_n | \widehat{n} | \psi_n\rangle = \sum_{m=0}^N ~m~\big( \psi^{(n)}_m\big)^2~.
    \end{equation}
    This indicator provides an average of the chord sectors over which the state $|\psi_n\rangle$ has a projection, weighted by the norm contributions $\big( \psi^{(n)}_m\big)^2$.
    \item Similarly to other many-body physics contexts \cite{evers2008anderson}, we may estimate the number of chord sectors effectively contributing to $|\psi_n\rangle$ through the \textit{participation ratio} ($PR_m$), given by the inverse of the \textit{inverse participation ratio} ($IPR_n$):
    \begin{eqnarray}
        IPR_n = \sum_{m=0}^N \big( \psi^{(n)}_m\big)^4~, \label{chord_inverse_participation_ratio}\\
        PR_n = \frac{1}{IPR_n} ~.\label{chord_participation_ratio}
    \end{eqnarray}
    We note that \eqref{Norms_Pythagoras} implies $1\leq PR_n\leq N$.
\end{itemize}
Numerical simulations assess whether $|\psi_n\rangle \in \mathcal{H}_{1p}^{(n)}$ for every $n=0,\dots N$. If such is the case, one should find $\langle \psi_n | \widehat{n} | \psi_n\rangle=n$ and $PR_n=1$ for all $n=0,\dots,N$.

Figure \ref{fig:PRO_results_lambda0pt05} shows the numerical results for a Hilbert space truncation of $N=50$ chord sectors (whose dimension is $d_N=1326$). For a fixed numerical value of $\widetilde{q}=0.577$ we set a small value of $\lambda = 0.05$ and solve the Lanczos algorithm supplemented by PRO. As already shown in figure \ref{fig:3dplots_main}, numerical observation suggests that, even when the binomial Ansatz does not apply, the projection of the $n$-th Krylov element over the $n^{\rm th}$ sector is very symmetric and peaked near the state $|n/2,n/2\rangle$ (assuming $n$ is even), a good guess for an approximation to the Lanczos coefficients (on which next section will elaborate further) is the evaluation of the $c_k(n)$ coefficient at the center $k=\frac{n}{2}$. Such an estimate is included in figure \ref{fig:PRO_results_lambda0pt05}, showing impressive agreement with the numerically obtained Lanczos coefficients, even in the regime where the limiting form \eqref{Lanczos_closed_form_solution_in_limit}, which is also included in the plot, fails to accurately capture the $b$-sequence.

\begin{figure}
    \centering
    \includegraphics[width=0.4\linewidth]{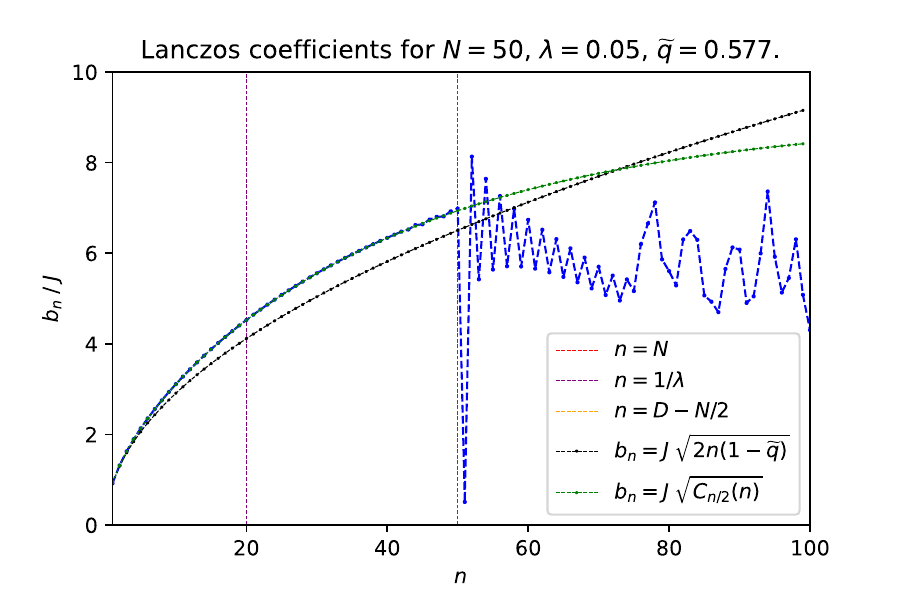} \includegraphics[width=0.4\linewidth]{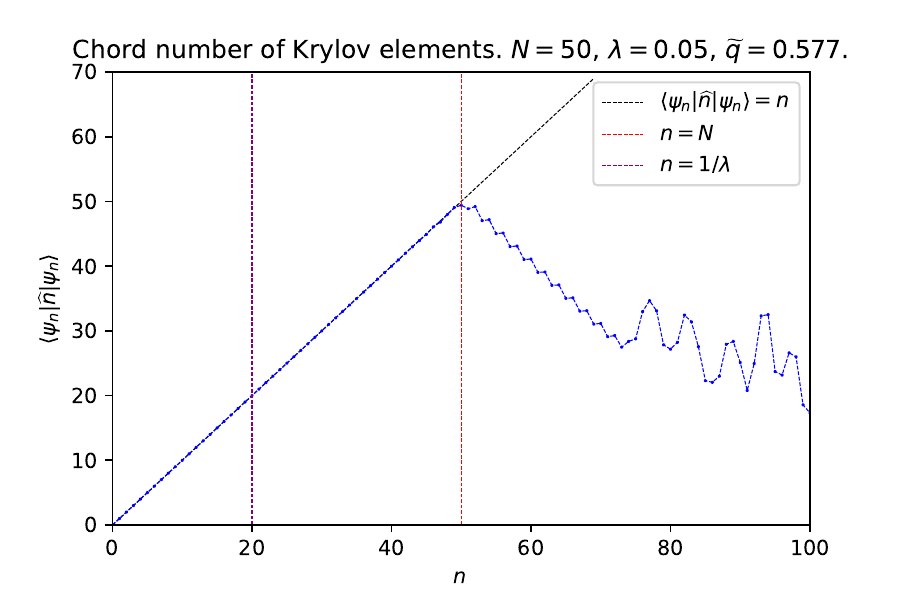} \\ 
    \includegraphics[width=0.4\linewidth]{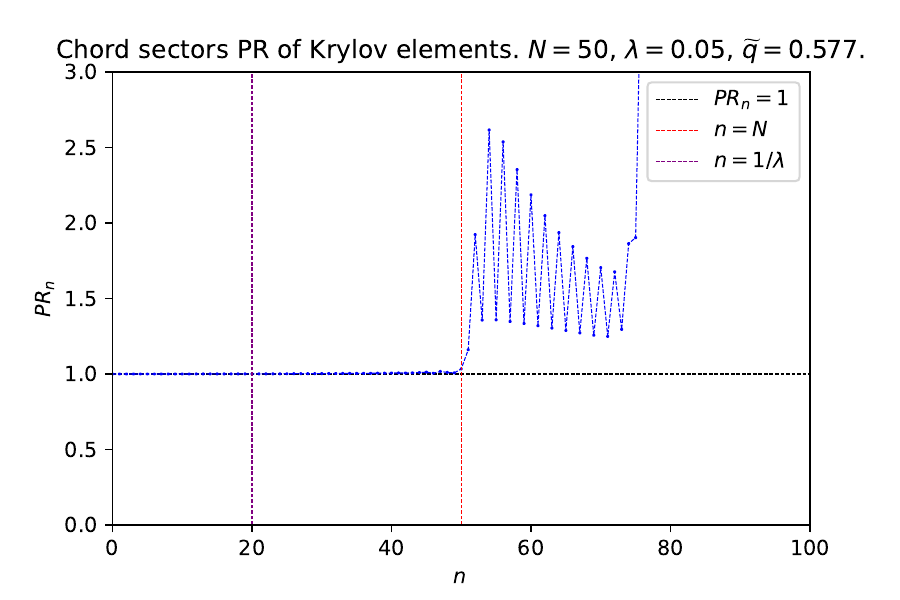} \includegraphics[width=0.4\linewidth]{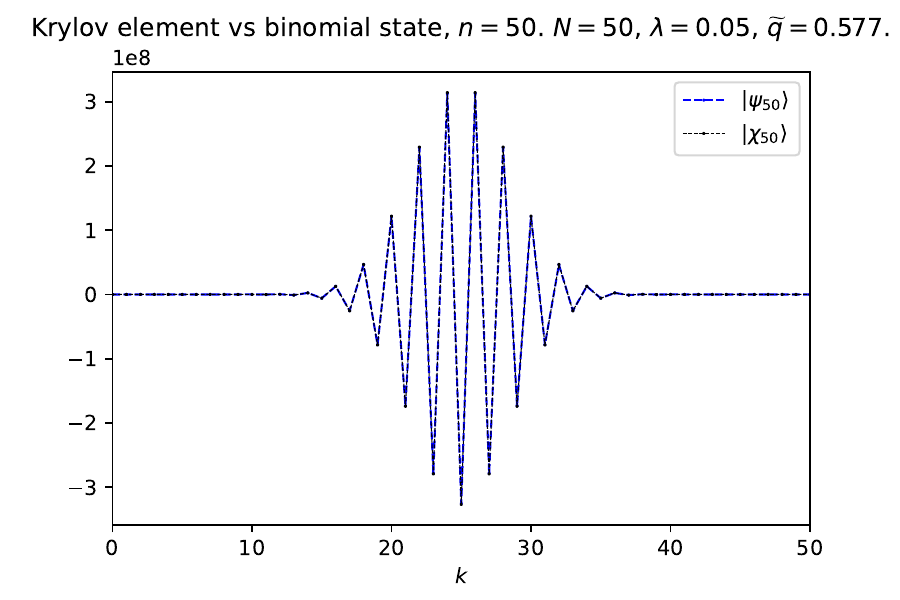}
    \caption{Results of the numerical implementation of the Lanczos algorithm for a system with $\lambda = 0.05$, $\widetilde{q}=0.577$, using a finite Hilbert space truncation that reaches up to the $N=50$ chord sector. \textbf{Top left:} Lanczos coefficients $b_n$ as a function of $n$. Vertical lines indicate, for reference, the values $n=\frac{1}{\lambda}$ and $n=N$. The coefficients $b_n$ for $n>N$ are subject to finite-truncation effects (in fact, the $b_n$ sequence was found to continue up to termination at $K\sim d_N\sim N^2/2$, even though it is not plotted for the sake of this discussion). The analytical estimate $b_n^2=J^2\, c_{n/2}(n)$, given in section \ref{subsect:bn_asymptotic_limit}, is found to agree excellently with the Lanczos coefficients for all $n\leq N$. \textbf{Top right:} Expectation value of total chord number of the Krylov elements $|\psi_n\rangle$. For $n\leq N$ (i.e. before truncation effects kick in) we observe that $\langle \psi_n |\widehat{n}|\psi_n\rangle\approx n$, suggesting that those Krylov elements are peaked on the corresponding chord sector. For $n>N$ the chord number expectation value starts to decrease as a function of $n$, reflecting the fact that the wave function of the subsequent Krylov vectors probes orthogonal directions within the chord sectors that are not included in the span of the Krylov elements with $n\leq N$. \textbf{Bottom left:} Participation ratio of the chord sectors in each Krylov vector. For $n\leq N$ we find that it is equal to one, confirming that each Krylov vector is confined within one chord sector. Putting this together with the plot on the top right, one can conclude that $|\psi_n\rangle \in \mathcal{H}_{1p}^{(n)}$ for $n\leq N$, i.e. the Krylov vectors, which are Krylov complexity eigenstates, are simultaneously total chord number eigenstates. \textbf{Bottom right:} For illustration purposes, this plot compares the profile of the $n=N$ Krylov vector over the sector $\mathcal{H}_{1p}^{(N)}$, i,e, the wave function $\psi^{(N)}_{Nk}$ in \eqref{Krylov_element_coordinates}, and compares it to the (normalized) binomial Ansatz \eqref{Binomial_Ansatz}, showing good qualitative agreement. The horizontal axis in this case is the left chord number $k$, labeling the basis elements $|k,N-k\rangle$ in the aforementioned sector. Such basis elements are not orthogonal, and therefore this plot may be taken as a qualitative demonstration (similar to figure \ref{fig:3dplots_main}), rather than as a quantitative analysis of norm contributions. Nevertheless, the binomial Ansatz is seen to match very closely the numerically obtained wave function. The overall conclusion emerging from this figure is the following: For $n\leq N$, the Krylov elements $|\psi_n\rangle$ are localized in sectors of total chord number $n$, within which they are satisfactorily described by the binomial Ansatz \eqref{Binomial_Ansatz}; furthermore, they solve the Lanczos recursion \eqref{Lanczos_recursion_1p} with Lanczos coefficients that are accurately described by $b_n=J\sqrt{c_{n/2}(n)}$, where $c_k(n)$ is given in \eqref{Ckn_simplified}, an approximation that will be argued for in section \ref{subsect:bn_asymptotic_limit} (always in the regime that is not affected by the truncation protocol).}
    \label{fig:PRO_results_lambda0pt05}
\end{figure}

By contrast, figure \ref{fig:PRO_results_lambda0pt5} presents the corresponding results for an instance of the system where $\lambda$ takes a larger value, $\lambda = 0.5$, as also studied for figure \ref{fig:3dplots_main}. 

\begin{figure}[ht!]
    \centering
    \includegraphics[width=0.45\linewidth]{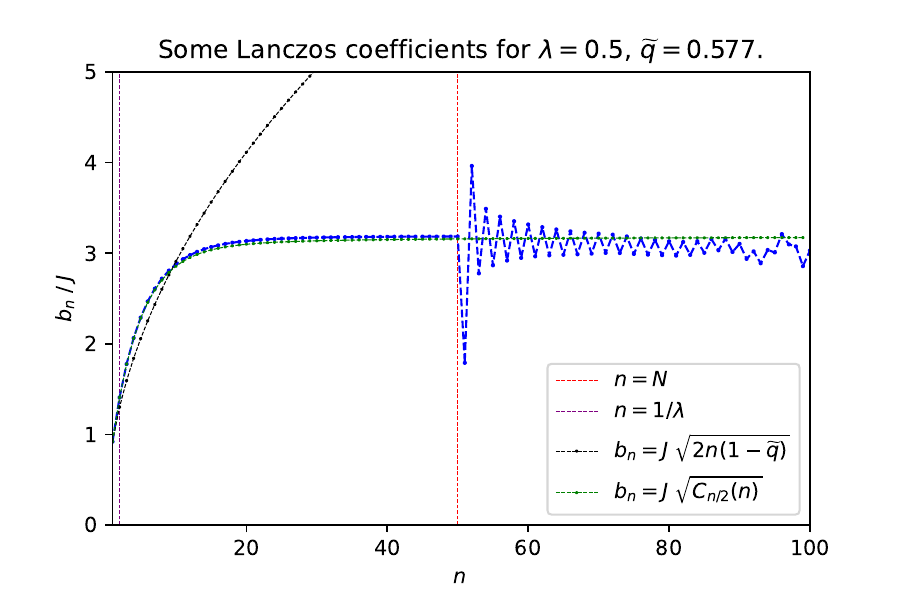} \includegraphics[width=0.45\linewidth]{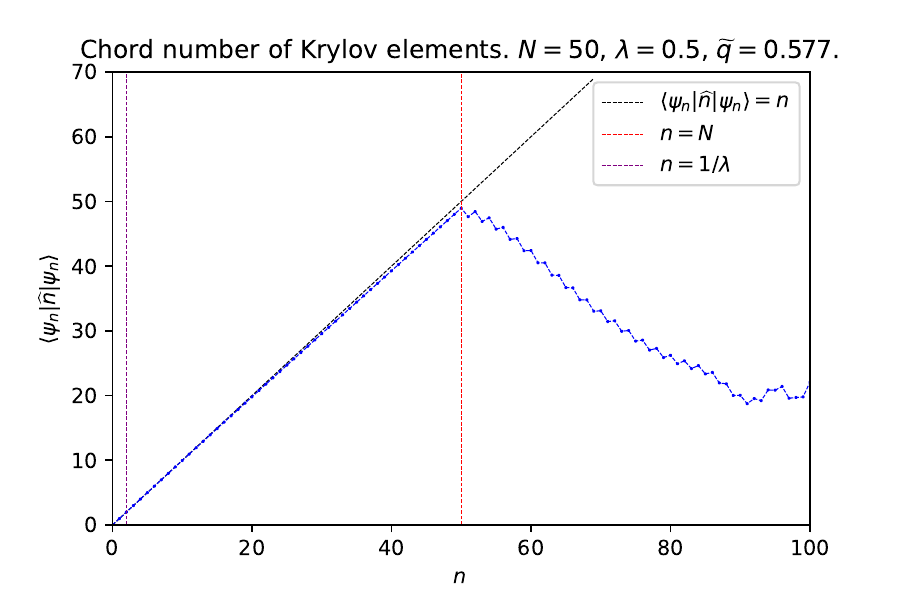} \\ 
    \includegraphics[width=0.45\linewidth]{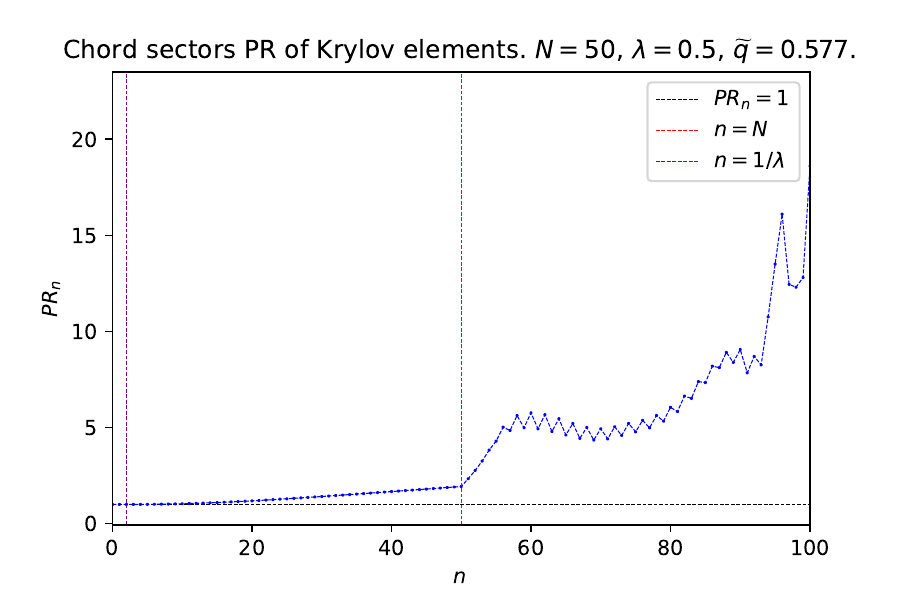} \includegraphics[width=0.45\linewidth]{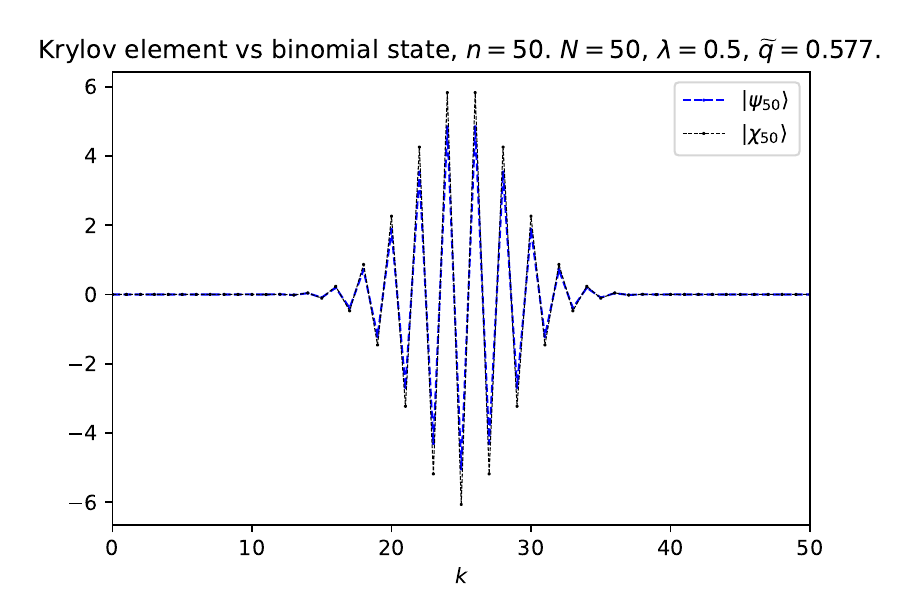}
    \caption{Numerical solution of the Lanczos algorithm with PRO for a finite truncation of the one-particle Hilbert space with chord sectors up to $N=50$, for the parameters $\lambda=0.5$ and $\widetilde{q}=0.577$ (i.e. same parameters as in \ref{fig:PRO_results_lambda0pt05} but with a bigger value of $\lambda$). In plots whose horizontal axis is $n$, the vertical lines $n=\frac{1}{\lambda},N$ have been included for reference. \textbf{Top left:} Lanczos coefficients. We start to observe small deviations between the numerics and the semiclassical approximation $b_n=J\sqrt{c_{n/2}(n)}$ (cf. section \ref{subsect:bn_asymptotic_limit}). \textbf{Top right:} Total chord number expectation value of the Krylov vectors $|\psi_n\rangle$. Even before the truncation, it deviates from a line with unit slope, reflecting the fact that Krylov vectors start to develope a non-negligible tail over sectors of smaller total chord number (cf. figure \ref{fig:3dplots_main}). \textbf{Bottom left:} Consistently, the chord sector participation ratios deviate from $1$. \textbf{Bottom right:} As an illustration, the profile of the Krylov element $|\psi_N\rangle$ over the sector $\mathcal{H}_{1p}^{(N)}$, captured by the wave function $\psi^{(N)}_{Nk}$, is seen to deviate slightly from the (normalized) binomial Ansatz \eqref{Binomial_Ansatz}.}
    \label{fig:PRO_results_lambda0pt5}
\end{figure}

\subsubsection*{Summary of numerics}

We have provided numerical evidence that, keeping $\widetilde{q}$ fixed, reducing $\lambda$ has the effect of suppressing the tails $|\xi^{(n)}_m\rangle$ of the Krylov basis vectors $|\psi_n\rangle$ over the sectors $m<n$, allowing us to describe them accurately with the binomial Ansatz \eqref{Binomial_Ansatz} and featuring Lanczos coefficients $b_n$ that follow closely the form $b_n=J\sqrt{c_{n/2}(n)}$, where the coefficient $c_k(n)$ is given in \eqref{Ckn_simplified}. These numerical observations will be explained in the next section from the perspective of an asymptotic, or semiclassical, limiting procedure. Consequently, in such a limit it is possible to identify the Krylov complexity eigenstates (i.e. the Krylov basis elements) with total chord number eigenstates. We therefore obtain an equality between the Krylov complexity operator \eqref{KC_operator_def} and the restriction of the total chord number operator to the Krylov space of the one-particle state \eqref{Time_evolving_state_HRminusHL_Schr} whose time evolution we are studying. Some further details of the numerical computations presented in this section, together with complementary results, are given in appendix \ref{appx:numerical_details}.

\subsection{Analytical Lanczos coefficients}\label{subsect:bn_asymptotic_limit}
In the previous sections we showed how one can come up with an Ansatz for the Lanczos coefficients $\{b_n\}_{n\geq 1}$ and the Krylov basis $\left\{|\psi_n\rangle\right\}_{n\geq 0}$, which is given in terms of what we called the binomial states $\ket{\chi_n}$, cf. \eqref{Binomial_Ansatz}. This basis solves the Lanczos algorithm exactly until $n=3$ (cf. \cref{subsect:KrylovBasis}), while discrepancies occurring for higher $n>3$, are of higher order in $\lambda$ (cf. \cref{subsect:Lanczos_Limit_lambda0_qtfixed} and \cref{appx:small_lambda}), thus being under control for small $\lambda$, given a fixed $n$. Section \ref{subsect:numerics} presented numerical results for the Lanczos coefficients and the Krylov states, which follow closely the small-$\lambda$ predictions in appendix \ref{appx:small_lambda} up to some $\lambda$-dependent value of $n$ after which the deviations become significant. This is a result of the fact that such an analysis was performed taking $\lambda$ to zero for fixed $n$.

In this section we want to study analytically what happens at higher $n$, and in particular we will propose that in the semiclassical limit the binomial Ansatz is controlled analytically, yielding an accurate expression for the Lanczos coefficients. Such a semiclassical limit is defined as follows \cite{Lin:2022rbf,Lin:2023trc}:
\begin{equation}
    \label{Semiclassical_limit}
    \lambda \to 0~,\quad n\to\infty~,\quad \lambda n\equiv l~ \text{fixed}~,
\end{equation}
where $l$ may be thought of as a dimensionless length introduced by the semiclassical scaling of the chord number variable. As argued in \cref{subsect:KrylovBasis}, operator complexity can be obtained as the Krylov complexity of the $\ket{0,0}$ seed state evolving under the two-sided Hamiltonian $H_R-H_L$, given explicitly in \eqref{Total_Hamiltonian_minus_line2}. 

The states $\{\ket{\psi_n}\}_{n\geq 0}$ are built out of the Lanczos recursion in the one particle sector \eqref{Lanczos_recursion_1p}. To begin, we shall argue that there is a sense in which the binomial Ansatz \eqref{Binomial_Ansatz} is still satisfied by the Krylov basis vectors in the semiclassical limit \eqref{Semiclassical_limit}. The sense in which this is true is the following: each Krylov vector $|\psi_n\rangle$ will have the generic form predicted in \eqref{Generic_Krylov_element}, where the norm of the tail vectors belonging to sectors of chord number smaller than $n$ will be suppressed in $\lambda$ in the semiclassical limit, such that the leading norm contribution comes from the binomial state in the $n$-th sector appearing in \eqref{Generic_Krylov_element}. In order to show inductively the self-consistency of the suppression of these tails, we may consider as an induction hypothesis that both the Krylov vectors $|\psi_n\rangle$ and $|\psi_{n-1}\rangle$ are given by the binomial Ansatz \eqref{Binomial_Ansatz}, and we shall argue that the Lanczos step \eqref{Lanczos_recursion_1p} produces the $(n+1)^{\rm th}$ binomial state, plus a tail whose norm is suppressed in the semiclassical limit. Let us now present the explicit inductive argument. We propose the following induction hypothesis for the Krylov vectors and the Lanczos coefficients in the semiclassical limit:
\begin{equation}
    \label{Induction_hypothesis_semiclassical}
    |\psi_n\rangle = \frac{1}{(b_1/J)\dots (b_n/J)}|\chi_n\rangle + (\text{subleading in }\lambda)~,\qquad b_n^2\underset{\lambda\sim 0}{\sim}J^2 \,c_{n/2}(n)~,
\end{equation}
where, in the Krylov vector Ansatz, ``subleading in $\lambda$'' stands for a vector orthogonal to $|\chi_n\rangle$ whose norm contribution to $|\psi_n\rangle$ is suppressed with respect to the one due to $|\chi_n\rangle$. In order to proceed to the inductive proof of this hypothesis, we should start discussing the induction seed, which is in fact a subtle point given that we have posed an Ansatz for the Krylov basis and the Lanczos coefficients that we expect to be satisfied in the semiclassical limit. It shall nevertheless suffice to say, as explained at the beginning of this section, that the hypothesis \eqref{Induction_hypothesis_semiclassical} is exact up to $n=3$ where the $\lambda$-corrections are exactly zero, and remains valid thereafter with corrections suppressed in an expansion in powers of $\lambda$ given fixed $n$; therefore, the hypothesis \eqref{Induction_hypothesis_semiclassical} proposes to extend this behavior to the large-$n$ regime\footnote{As a check, later in this section equation \eqref{eq:op_lanczos_regimes} will show that the Lanczos coefficients Ansatz in \eqref{Induction_hypothesis_semiclassical} recovers the small-$\lambda$ analysis of appendix \eqref{appx:small_lambda} if we set $\lambda n\ll 1$.}.

We now turn to proving that the induction step is satisfied by \eqref{Induction_hypothesis_semiclassical}. The induction step is precisely a step of the Lanczos algorithm, cf. \eqref{Lanczos_recursion_1p}, which is a two-term recursion. Thus, we assume that \eqref{Induction_hypothesis_semiclassical} is satisfied for $|\psi_n\rangle$, $|\psi_{n-1}\rangle$, $b_n$ and $b_{n-1}$ and we proceed to construct $b_{n+1}$ and $|\psi_{n+1}\rangle$. As for the Krylov vector, proving the self consistency of \eqref{Induction_hypothesis_semiclassical} in the semiclassical limit, amounts to plugging the leading term of the Krylov state, i.e. the binomial state $|\chi_n\rangle$, in the Lanczos step and verifying that it produces the next binomial state $|\chi_{n+1}\rangle$ plus tails that are indeed subleading.
Explicitly, the Lanczos step that constructs the non-normalized ($n+1$)-Krylov vector, $|A_{n+1}\rangle = b_{n+1}|\psi_{n+1}\rangle$, reads:
\begin{align}
    |A_{n+1}\rangle = (H_R-H_L)|\psi_n\rangle - b_n|\psi_{n-1}\rangle \label{Induction_step_semicl_line1} \\ 
    =\frac{J}{(b_1/J)\dots (b_n/J)}\Big( |\chi_{n+1}\rangle - |\zeta_{n-1}\rangle \Big)~,
\end{align}
where, after some algebra, the tail vector $|\zeta_{n-1}\rangle$ reads:
\begin{eqnarray}
    \label{Zeta_tail_vector}
    |\zeta_{n-1}\rangle = \sum_{k=0}^{n-1} \big( c_k(n) - b_n^2/J^2 \big) (-1)^k \binom{n-1}{k} |k,n-1-k\rangle~,
\end{eqnarray}
with the coefficient $c_k(n)$ given in \eqref{Ckn_simplified}. In order to check whether the norm of the tail is suppressed compared to that of $|\chi_{n+1}\rangle$, we compute explicitly the norm of $|A_{n+1}\rangle$, which is nothing but the Lanczos coefficient $b_{n+1}$. This coefficient picks up norm contributions from both the two orthogonal states $|\chi_{n+1}\rangle\in\mathcal{H}_{1p}^{(n+1)}$ and $|\zeta_{n-1}\rangle\in\mathcal{H}_{1p}^{(n-1)}$:
\begin{equation}
    \label{bnplus1_Semicl_induction_step}
    \frac{b_{n+1}^2}{J^2} = \frac{J^{2n}}{b_1^2\dots b_n^2}\Big( \langle\chi_{n+1}|\chi_{n+1}\rangle + \langle \zeta_{n-1}|\zeta_{n-1}\rangle \Big)~.
\end{equation}
Explicitly, the two norm contributions are:
\begin{align}
    &\langle \chi_{n+1}|\chi_{n+1}\rangle = \sum_{k^\prime,k=0}^{n+1}(-1)^{k^\prime + k}\binom{n+1}{k^\prime}\binom{n+1}{k}\langle k^\prime,n+1-k^\prime | k,n+1-k\rangle~, \label{Semicl_limit_norm_contribs_binom} \\
    &\langle \zeta_{n-1}|\zeta_{n-1}\rangle = \sum_{k^\prime,k=0}^{n-1}\!\left[c_{k^\prime}(n)-\frac{b_n^2}{J^2}\right]\!\! \left[c_{k}(n)-\frac{b_n^2}{J^2}\right] \! \! (-1)^{k^\prime + k} \binom{n-1}{k^\prime}\binom{n-1}{k}\langle k^\prime,n-1-k^\prime|k,n-1-k\rangle~.\label{Semicl_limit_norm_contribs_tail}
\end{align}
As promised, we may now analyze both contributions \eqref{Semicl_limit_norm_contribs_binom} and \eqref{Semicl_limit_norm_contribs_tail} in the semiclassical limit \eqref{Semiclassical_limit} and argue that the latter is suppressed in powers of $\lambda$ with respect to the former. Let us start by noting that the overlaps between states belonging to the same sector $\mathcal{H}_{1p}^{(n)}$ are computed via the semiclassical limit of \eqref{inner_product_recursion_solution}, which was worked out in \cite{Lin:2023trc}:
\begin{equation}\label{eq:scal_prod_semiclass}
   \braket{x'|x}= \left\langle n_L^{\prime}, n_R^{\prime} \mid n_L, n_R\right\rangle=[n]_q !\left(\frac{\left(1-c^2\right) / 2}{\cosh \frac{x-x^{\prime}}{2}-c \cosh \frac{x+x^{\prime}}{2}}\right)^{2 \Delta},
\end{equation}
where $c^2=q^n$ and $x=\lambda\frac{n_L-n_R}{2}$. 
Indeed, in this limit, the $n$-chord sector will be weighted with an extra $\frac{1-q^n}{1-q}\sim 1/\lambda$ factor (note that $\lambda n$ is fixed in the semiclassical limit) with respect to the ($n-1$)-sector, due to the $[n]_q!$ term in \eqref{eq:scal_prod_semiclass},
implying that inner products in the sector $\mathcal{H}_{1p}^{(n)}$ scale as $\lambda^{-n}$ in the semiclassical limit.
Hence, the overlap in \eqref{Semicl_limit_norm_contribs_binom} gives the corresponding norm contribution a total scaling of $~\lambda^{-(n+1)}$. On the other hand, the overlap in \eqref{Semicl_limit_norm_contribs_tail} contributes to the tail norm with a factor $~\lambda^{-(n-1)}$, and, in order to assess whether $\langle\zeta_{n-1}|\zeta_{n-1}\rangle$ is suppressed with respect to $\langle \chi_{n+1}|\chi_{n+1}\rangle$, it is only left to show that the contribution due to each of the $c_k(n)-b_n^2/J^2$ terms is at least more suppressed than $1/\lambda$. The coefficients $c_k(n)$, whose definition is given in \eqref{Ckn_simplified}, admit a smooth form in the semiclassical limit\footnote{We also consider $\lambda k\equiv l_L$ to be fixed in the semiclassical limit, in which strictly speaking the $k$-sums will become infinite Riemann sums whose integration measure is given by $\lambda\to 0$. For details, see appendix \ref{app:analytics_details}.}:
\begin{equation}
    \label{ckn_semicl}
    c_k(n)\underset{\lambda\to 0}{\sim} c(l,l_L;\lambda) \equiv \frac{l}{l-l_L} \frac{1-e^{-(l-l_L)}}{\lambda}(1-\widetilde{q}e^{-l_L}) + \frac{l}{l_L}\frac{1-e^{-l_L}}{\lambda}(1-\widetilde{q}e^{-(l-l_L)})~.
\end{equation}
Expression \eqref{ckn_semicl} is in fact of order $1/\lambda$. However, as discussed in appendix \ref{app:analytics_details}, the binomials entering the sums in \eqref{Semicl_limit_norm_contribs_binom} and \eqref{Semicl_limit_norm_contribs_tail} get \textit{squeezed} in the semiclassical limit, effectively becoming Dirac delta functions centered in the middle of the domain of $l_L\equiv \lambda k$.
We refer to appendices \ref{appx:Binom_delta_EasyCase} and \ref{appx:binomials_semicl_generic} for the details of the proof of this fact, but we can summarize it here by noting that, in the semiclassical limit, the asymptotic expansion of the binomial $\binom{n}{k}$ near $\lambda\sim 0$ is:
\begin{align}
    &\qquad \qquad \qquad \qquad \qquad \qquad \qquad\qquad\binom{n}{k}\overset{\text{semicl.}}{=} \binom{l/\lambda}{l_L/\lambda} \label{Binom_asymptotic_main_text_line1} \\
     &\underset{\lambda\sim 0}{\sim} \sqrt{\frac{2l}{\pi\lambda(l^2-4x^2)}} ~ \text{exp}\left\{ \frac{1}{\lambda} \Big( l\log(2l)-(l-2x)\log(l-2x)-(l+2x)\log(l+2x) \Big) \right\}~, \label{Binom_asymptotic_main_text_line2}
\end{align}
where $x=l_L-l/2$. In expression \eqref{Binom_asymptotic_main_text_line2} we can see that $\frac{1}{\lambda}$ enters as a large parameter in the prefactor of the exponent, which is therefore amenable to a saddle-point approximation. Direct computation yields that the saddle point is indeed at $x=0\Leftrightarrow l_L=\frac{l}{2}$. 
Therefore, the sums in \eqref{Semicl_limit_norm_contribs_tail} are asymptotically dominated by the configuration $k= n/2+\delta k$ where $\frac{\delta k}{n}\to 0$ (and similarly for $k^\prime$), for which the induction hypothesis on the Lanczos coefficients \eqref{Induction_hypothesis_semiclassical} assures that $c_{n/2}(n)-b_n^2/J^2\underset{\lambda\sim 0}{\sim}\lambda^0$, implying that the tail norm \eqref{Semicl_limit_norm_contribs_tail} is of order $\lambda^{-(n-1)}$, thus suppressed with respect to \eqref{Semicl_limit_norm_contribs_binom}. 
This shows that the non-normalized Krylov vector \eqref{Induction_step_semicl_line1} is dominated by the binomial state in the $n+1$ sector. Schematically:
\begin{equation}
    \label{Anplusone_induction_conclusion}
    |A_{n+1}\rangle = \frac{J^{n+1}}{b_1\dots b_n}|\chi_{n+1}\rangle + \text{subleading in }\lambda~,
\end{equation}
where the orthogonal tail has a subleading norm in the sense that we have just discussed. 

In order to continue with the inductive proof, we need to finish the computation of the Lanczos coefficient \eqref{bnplus1_Semicl_induction_step}, which we have just shown to be dominated in the semiclassical limit by the contribution coming from the binomial state. We have:
\begin{align}
\label{eq:lower_diag_L}
    \frac{b_{n+1}^2}{J^2}= J^{2n}\frac{\langle\chi_{n+1}|\chi_{n+1}\rangle}{b_1^2\dots b_n^2} = \frac{J^{2n}}{b_1^2\dots b_n^2}
    \sum_{k^\prime,k=0}^{n} (-1)^{k^\prime+k} c_{ \scriptscriptstyle k}{ (n+1)}\,\binom{n}{k^\prime}\binom{n}{k}\braket{k^\prime,n-k^\prime|k,n-k}~,
\end{align}
where in the second equality we have used that $\langle \chi_{n+1}|=\langle\chi_n|(a_R-a_L)$. The sums in \eqref{eq:lower_diag_L} localize, in the semiclassical limit, to the contribution coming from the center of the summation range, due to the same argument in appendix \eqref{app:analytics_details} to which we already alluded earlier. We thus have:\footnote{Even though we are using discrete labels, strictly speaking, in the semiclassical limit the coefficient $c_k(n)$ becomes asymptotically a smooth function $c(l,l_L;\lambda)$, given in \eqref{ckn_semicl}, where $l\equiv \lambda n$. From this perspective, $n$ and $n+1$ are undistinguishable. It is nevertheless harmless and preferred for this analysis to use the same index for the Lanczos coefficient and for the vector whose norm produces it in equation \ref{b_nplus1_induction_proof}.}
\begin{equation}
\label{b_nplus1_induction_proof}
b^2_{n+1}\underset{\lambda\sim 0}{\sim} J^{2n+2}\frac{c_{\scriptscriptstyle\frac{n+1}{2}}{\scriptstyle (n+1)}~\langle \chi_n|\chi_n\rangle}{b_1^2\dots b_n^2}  \underset{\lambda\sim 0}{\sim} J^2 c_{\frac{n+1}{2}}(n+1)~,
\end{equation}
where in the last step the $\langle \chi_n|\chi_n\rangle$ has been canceled out against the denominator using the induction hypothesis \eqref{Induction_hypothesis_semiclassical}. Finally, the ($n+1$)-Krylov vector is obtained normalizing $|A_{n+1}\rangle$ with the Lanczos coefficient \eqref{b_nplus1_induction_proof}:
\begin{equation}
    \label{psi_nplus1_induction_proof}
    |\psi_{n+1}\rangle = \frac{1}{b_{n+1}}|A_{n+1}\rangle = \frac{1}{(b_1/J)\dots (b_{n+1}/J)}|\chi_{n+1}\rangle + \text{subleading in }\lambda~,
\end{equation}
where in the last step we have used \eqref{Anplusone_induction_conclusion}. All in all, we have obtained \eqref{b_nplus1_induction_proof} and \eqref{psi_nplus1_induction_proof}, which satisfy the induction hypothesis \eqref{Induction_hypothesis_semiclassical}. This concludes the proof by induction.

We have proved inductively that in the semiclassical limit the Krylov vectors $|\psi_n\rangle$ are given by binomial states, and therefore $|\psi_n\rangle\in\mathcal{H}_{1p}(n)$, that is, they are eigenstates of the total chord number operator. Putting this together with the fact that the states $|\psi_n\rangle$ are, by definition, eigenstates of the Krylov complexity operator \eqref{KC_operator_def}, it follows that the latter is equal to the chord number operator\footnote{In order to be fully correct, we should say that the Krylov complexity operator $\widehat{C_K}$ is equal to the restriction of the total chord number operator $\widehat{n}$ over the subspace spanned by the Krylov basis associated with the state~$|0,0\rangle$.} and, in particular, Krylov complexity is equal to the chord number expectation value.

Let us now move on to analyzing the obtained Lanczos sequence. The proof presented above holds in the semiclassical limit given any fixed $\Delta>0$, as discussed in appendix \eqref{app:analytics_details}, since the operator dimension is not scaled parametrically with $\lambda$ in the limiting procedure. This provides a limiting form for the Lanczos sequence, expression \eqref{Induction_hypothesis_semiclassical}, which is seen to describe very accurately the Lanczos coefficients of in the the numerical simulations in section \eqref{subsect:numerics} and appendix \eqref{appx:Numerics_further_results} provided sufficiently small values of $\lambda$. However, the $\Delta$-dependence only enters through $\widetilde{q}=e^{-\lambda \Delta}$ in the $c$-coefficients \eqref{Ckn_simplified}, and gets sent to $\widetilde{q}=1$ in the strict semiclassical limit as we can observe from \eqref{ckn_semicl}. Hence, in order to obtain a Lanczos sequence with a non-trivial operator dependence, we can supplement the semiclassical limit \eqref{Semiclassical_limit} with the additional scaling
\begin{equation}
    \label{Large_Delta}
    \Delta\to+\infty~,\qquad \widetilde{q}=e^{-\lambda \Delta}~\quad\text{fixed}~.
\end{equation}
In appendix \ref{app:analytics_details} we argue that this $\Delta$-scaling in the semiclassical limit still provides (and in fact controls parametrically) the localization of the sums that were instrumental in the inductive proof of the semiclassical solution of the Lanczos algorithm, which therefore carries through. With this, the Lanczos coefficients take the form:
\begin{equation}\label{semiclassical_operator_bn}
    b_n=J\sqrt{c_{k=n/2}(n)}=2J\sqrt{\frac{1-q^{n/2}}{1-q}\bigr(1-\Tilde{q}q^{n/2}\bigr)}
\end{equation}
We now wish to study the regimes of this Lanczos sequence and contrast them with the Krylov complexity regimes of the analysis in next section. For this, we may reinstate the Hamiltonian normalization used in \cite{Lin:2022rbf} and reviewed in section \ref{sec.ToolsTrade}, which makes DSSYK reduce to large-$p$ SYK when $\lambda\to 0$ \cite{Lin:2022rbf,Mukhametzhanov:2023tcg} and which is convenient for the continuum approximation of Krylov complexity (cf. \cite{Rabinovici:2023yex} and next section). This normalization is implemented by the change $J\mapsto \frac{J}{\sqrt{\lambda}}$, implying:
\begin{equation}\label{Renormalization_a_la_Lin}
    b_n\longmapsto b_n /\sqrt{\lambda}=\frac{J}{\sqrt{\lambda}}\sqrt{c_{k=n/2}(n)}~,
\end{equation}
where $J$ is a dimensionful parameter that sets the units of energy.
With this, the Lanczos coefficients take the form:
\begin{equation}\label{eq:lanczos_coeff}
    b_n=\frac{2J}{\sqrt{\lambda}}\sqrt{\frac{1-q^{n/2}}{1-q}\bigr(1-\Tilde{q}q^{n/2}\bigr)},
\end{equation}

Next, we will study the behavior of \eqref{eq:lanczos_coeff} in various regimes, in particular for small and large $n$. We will also see that the value of $\Tilde{q}$ plays an important role.

\subsubsection{Regimes of Lanczos coefficients}
Here, we analyze the regimes of the Lanczos coefficients as a function of $n$, while keeping $\Tilde{q}$, and $\lambda$ fixed. The analysis is analogous to the one performed for the Lanczos coefficients for the state K-complexity of the infinite temperature TFD in \cite{Rabinovici:2023yex}. However, in this case, we gain a new regime from the presence of the additional $\Tilde{q}$ parameter. Recalling that when $\Tilde{q}\to 1$ the operator is small, we would expect to see some kind of \textit{operator growth} whose signature is a \textit{linear} growth of Lanczos coefficients, see \cite{Parker:2018yvk}, in this limit.

At small $n$, we perform an expansion keeping first order terms in $q^{n/2}=e^{-\lambda n/2}\approx 1-\frac{\lambda n}{2}$, to find a square-root behavior of the Lanczos coefficients:
\begin{eqnarray}
    b_n \approx J\sqrt{\frac{2(1-\Tilde{q})}{1-q}}\sqrt{n}, \quad n\ll \frac{2}{\lambda}.
\end{eqnarray}
At large $n$, $q^{n/2}=e^{-\lambda n/2}$ is small, and we can approximate $\sqrt{(1-q^{n/2})(1-\Tilde{q}q^{n/2})}\approx (1-q^{n/2}/2)(1-\Tilde{q}q^{n/2}/2)$ such that:
\begin{eqnarray}
    b_n \approx \frac{2J}{\sqrt{\lambda(1-q)}}\left(1-\frac{q^{n/2}}{2}\right)\left(1-\frac{\Tilde{q} \,q^{n/2}}{2}\right), \quad n\gg \frac{2}{\lambda}.
\end{eqnarray}
This result tends to a $\Tilde{q}$-independent constant when $n\to \infty$:
\begin{eqnarray}
    b_n \approx \frac{2J}{\sqrt{\lambda(1-q)}}, \quad n\to\infty .
\end{eqnarray}
Indeed, as can be observed in \cref{fig:lanczos_op_compl}, at large $n\gg2/\lambda$ the Lanczos coefficients reach the same horizontal asymptote, independently of $\Tilde{q}$. This is evidence that the operator Krylov complexity we will build out of them in the next section will present the expected late-time linear growth, regardless of the value of $\Tilde{q}$. The only difference between operators of different $\Tilde{q}$ will then be the onset of this linear behavior. 

We will now take care of the $\Tilde{q}\to1$ limit, where, as we have noted above, we expect to see a signature of operator growth in the behavior of the Lanczos coefficients. Going to the next term in the small $n$ expansion gives rise to the result:
\begin{eqnarray} \label{eq:bn_second_order}
    b_n \approx \frac{2J}{\sqrt{\lambda(1-q)}}\sqrt{(1-\Tilde{q}) \frac{\lambda n}{2} + \frac{(3\Tilde{q}-1)}{2} \left(\frac{\lambda n}{2}\right)^2+O\left(\left(\lambda n/2\right)^3\right)},
\end{eqnarray}
interestingly, in the 
small $n$ regime we see a competition between the $\propto n$ and the $\propto n^2$ term inside the square root. When the $\propto n^2$ term is dominant, this can define a novel linear regime for the Lanczos coefficients, not found in the analysis of the TFD state complexity of \cite{Rabinovici:2023yex}. When $\Tilde{q}$ is big enough this regime is parametrically distinct from the square root behavior, and in particular the transition happens around $n\sim\frac{2}{\lambda}\frac{2(1-\Tilde{q})}{3\Tilde{q}-1}$. We will confirm in the next section that the linear regime observed here, which lasts longer the closer $\Tilde{q}$ is to $1$, is linked to the characteristic exponential chaotic regime \cite{Parker:2018yvk} of operator complexity. More explicitly, for $n\ll 2/\lambda$ around $\Tilde{q}=1$, the result \eqref{eq:bn_second_order} can be written as
\begin{eqnarray} \label{eq:bn_linear_approx}
    b_n \approx \frac{2J}{\sqrt{\lambda(1-q)}}\left(\frac{\lambda n}{2}\right)\sqrt{ \frac{(3\Tilde{q}-1)}{2} }\left(1+\frac{(1-\Tilde{q})}{3\Tilde{q}-1}\frac{2}{\lambda n} \right), \quad \frac{2(1-\Tilde{q})}{3\Tilde{q}-1} \frac{2}{\lambda} \ll n \ll \frac{2}{\lambda}~,
\end{eqnarray}
which is linear in $n$. Note that the linear behavior will not be seen if $2(1-\Tilde{q})/(3\Tilde{q}-1)>1$ or $\Tilde{q}<0.6$.
This result shows that as $\Tilde{q} \to 1$, the Lanczos coefficients tend to 
\begin{eqnarray}
    b_n \approx J\frac{\lambda}{\sqrt{\lambda(1-q)}}\, n, \quad n\ll \frac{2}{\lambda} \text{ and } \Tilde{q}\to 1~.
\end{eqnarray}
For small $\lambda$ this becomes $b_n\approx J\,n$ which implies an exponential growth of Krylov complexity of the form $C_K(t)\propto e^{2Jt}$. In the next section we will show that such a behavior is indeed found in the result for K-complexity. Finally, we mention that for $\Tilde{q}\to1$, the large $n$ behavior of the Lanczos coefficients is 
\begin{eqnarray} \label{eq:bn_qtilde1_largen}
    b_n \approx \frac{2J  }{\sqrt{\lambda(1-q) }}\left(1-q^{n/2}\right), \quad n \gg \frac{2}{\lambda} ~.
\end{eqnarray}

To summarize, we found that  the Lanczos coefficients show the following regimes:
\begin{equation}\label{eq:op_lanczos_regimes}
    b_n\approx
    \begin{cases}
      \frac{2J}{\sqrt{\lambda(1-q)}}\sqrt{(1-\Tilde{q}) \frac{\lambda n}{2} + \frac{(3\Tilde{q}-1)}{2} \left(\frac{\lambda n}{2}\right)^2+O\left(\left(\lambda n/2\right)^3\right)}, & n\ll 2/\lambda \\
        \frac{2J}{\sqrt{\lambda(1-q)}}\left(1-\frac{q^{n/2}}{2}\right)\left(1-\frac{\Tilde{q}q^{n/2}}{2}\right), &  n\gg 2/\lambda
    \end{cases}.
\end{equation}
For $\Tilde{q}\approx 1$, the first regime becomes approximately linear between $\frac{2(1-\Tilde{q})}{3\Tilde{q}-1} \frac{2}{\lambda} \ll n \ll \frac{2}{\lambda}$, with $b_n \approx J\frac{\lambda}{\sqrt{\lambda(1-q)}}\, n$. And, the second regime behaves as $b_n \approx \frac{2J  }{\sqrt{\lambda(1-q) }}\left(1-q^{n/2}\right)$. We attempt to demonstrate all of these statements in \cref{fig:lanczos_op_compl}.

\begin{figure}
    \centering
    \includegraphics[width=0.75\linewidth]{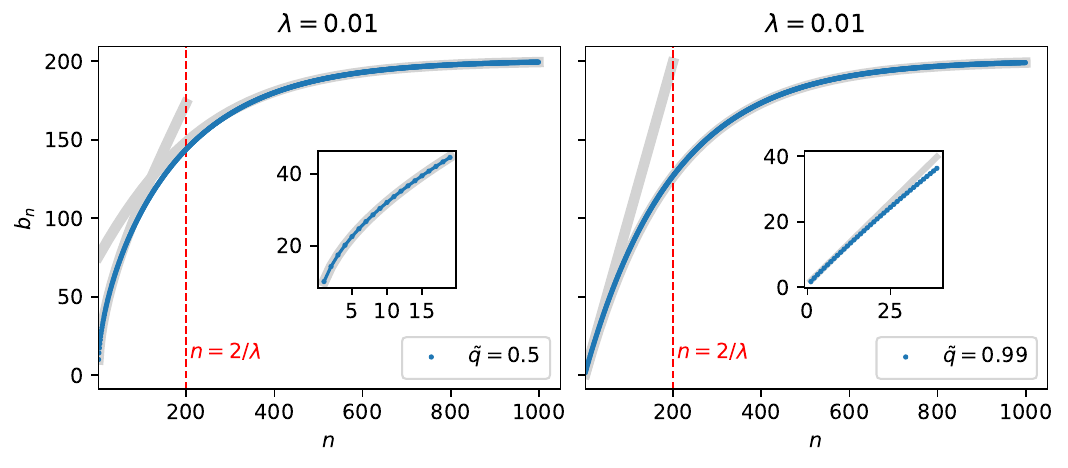}
    \caption{The Lanczos coefficients of \eqref{eq:lanczos_coeff} with $J=1$ and $\lambda=0.01$, for $\Tilde{q}=0.5$ and $\Tilde{q}=0.99$ (in the inset: a zoomed-in version for better visualization of the small $n$ regimes). For $\Tilde{q}=0.5$, the thick gray lines show the small and large $n$ approximations in \eqref{eq:op_lanczos_regimes}. For $\Tilde{q}=0.99$ the gray lines show \eqref{eq:bn_second_order} for small $n$ and \eqref{eq:bn_qtilde1_largen} for large $n$; in the inset we show in gray the linear approximation \eqref{eq:bn_linear_approx}.}
    \label{fig:lanczos_op_compl}
\end{figure}

\subsection{Analytical form of operator K-complexity in the semiclassical limit}\label{subsect:analytical_K_Complexity}
In this section we will find an analytical expression for Krylov complexity in the small $\lambda$ limit. As we mentioned before, and as shown in \cite{Rabinovici:2023yex}, $\lambda$ can serve as a parameter which makes Krylov space continuous. Defining $x\equiv \lambda n$, where it is understood that $\lambda\to 0$ and $n\to \infty$, the wavefunction is identified with a continuous function, $\varphi_n(t)\to f(t,x)$; in the same limit, the Lanczos coefficients become continuous as well, $b_n \to b(x)$. 
With these identifications, the wavefunction equation, $\dot{\varphi}_n(t)=b_n\varphi_{n-1}(t)-b_{n+1}\varphi_{n+1}(t)$, becomes
\begin{eqnarray}
    \dot{f}(t,x)&=& b(x)f(t,x-\lambda)-b(x+\lambda)f(t,x+\lambda) \nonumber\\
    &=& -2\lambda b(x) f'(x,t)-\lambda b'(x)f(t,x) -\frac{1}{2}\lambda^2 b''(x)f(t,x)-\lambda^2 b'(x)f'(t,x)+O(\lambda^3)
\end{eqnarray}
If we define $v(x) \equiv 2 \lambda b(x)$, this takes the form
\begin{eqnarray}\label{Cont_eq}
    \dot{f}(t,x) = -v(x) f'(t,x) - \frac{v'(x)}{2}f(t,x) -\frac{\lambda}{2}\left(v'(x)f'(t,x)+\frac{v''(x)}{2}f(t,x) \right)+O(\lambda^2)
\end{eqnarray}

In this case, the continuous version of the Lanczos coefficients takes the form
\begin{eqnarray}
    b(x) = \frac{2J}{\sqrt{\lambda(1-q)}} \sqrt{(1-e^{-x/2})(1-\tilde{q}e^{-x/2})}
\end{eqnarray}
and, in a similar manner to the matterless case, the velocity field, $v(x) = 4J\sqrt{(1-e^{-x/2})(1-\tilde{q}e^{-x/2})}$, is of $O(\lambda^0)$. Keeping only the $O(\lambda^0)$ in \eqref{Cont_eq}, the equation is equivalent to a chiral wave equation $(\partial_t+\partial_y)g(t,y)=0$ through the change of variables $dy=dx/v(x)$ and $g(t,y)=\sqrt{v(x(y))}f(t,x(y))$. Such an equation moves any initial condition, $g(0,y)$, rigidly in time, allowing us to use a point-particle approximation for the wave-packet.

We note, however, that at small $x$, the first order in $\lambda$ velocity field is $v(x) = 2 \sqrt{2} \sqrt{1 - \tilde{q}} \sqrt{x} +O(x^{3/2})$, while its first derivative is given by $v'(x)=\frac{\sqrt{2} \sqrt{1-\tilde{q}}}{\sqrt{x}}+\frac{3 (3 \tilde{q}-1) \sqrt{x}}{4 \sqrt{2} \sqrt{1-\tilde{q}}}+O\left(x^{3/2}\right)$ whose second term can be large when $\tilde{q}\to1$. When inserted into \eqref{Cont_eq}, the third coefficient, $(\lambda/2)v'(x)$, can compete with the first coefficient, $v(x)$, when $\tilde{q}\to1$, unless $\lambda$ is scaled down accordingly. This can spoil the chiral wave equation and therefore the point particle approximation at early times.
We note however that, if we are taking the $\lambda\to 0$ limit, the approximation will hold at all times for any fixed $\Tilde{q}$, so that the motion of the wave-packet on the Krylov chain can be well described as the motion of a point particle with velocity proportional to the Lanczos coefficients. Analogously to \cite{Rabinovici:2023yex} then, the Krylov complexity is the position of this particle, which can be found by inverting the integral:
\begin{equation}
    \int_0^{n(t)}\frac{dn}{2b_n}=t~.
\end{equation}
If we perform the integration we get:
\begin{equation}
    \frac{\sqrt{\lambda(1-q) } }{J\log q}\tanh ^{-1}\left(\frac{\sqrt{1- q^{n/2}}}{\sqrt{1-\tilde{q}q^{n/2}}}\right)\bigg|_{n=0}^{n(t)}= \frac{\sqrt{\lambda(1-q) } }{J\log q}\tanh ^{-1}\left(\frac{\sqrt{1- q^{n(t)/2}}}{\sqrt{1-\tilde{q} q^{n(t)/2}}}\right).
\end{equation}
By inverting this result and reinstating $q=e^{-\lambda}$, we obtain the operator complexity in the point particle approximation, $n(t)$:
\begin{equation}
\label{eq:op_compl_approx_small_lambda}
    n(t)=\frac{2}{\lambda} \log \left[1+(1-\Tilde{q}) \sinh ^2\left(\sqrt{\frac{\lambda}{1-q}}J\,t\right)\right],
\end{equation}
which in the $\lambda\to0$ limit simplifies to:
\begin{equation}\label{eq:op_compl}
\boxed{\begin{aligned}
    C_K(t)= n(t)&\approx\frac{2}{\lambda}\log \left[1+(1-\Tilde{q})\sinh^2 (J\, t) \right].
\end{aligned}}
\end{equation}
Let us study the behavior of this complexity as a function of time. At early times, $Jt\ll 1$, we can approximate $\sinh^2(Jt)\approx (Jt)^2$:
\begin{equation} \label{KC_quadratic}
    \lambda C_K(t)\approx 2\log[1+(1-\Tilde{q})(J t)^2]\approx 2(1-\Tilde{q})(J t)^2 , \quad t\ll \frac{1}{J}.
\end{equation}
At later times we approximate $\sinh^2(Jt)\approx \frac{1}{4}e^{2Jt}-\frac{1}{2}$, such that $2\log[1+(1-\tilde{q})\sinh^2(Jt)] \approx 2\log\left(\frac{1+\Tilde{q}}{2}\right)+2\log\left(1+\frac{1-\Tilde{q}}{1+\Tilde{q}}\frac{e^{2Jt}}{2}\right)$.  Further assuming that $\frac{1-\Tilde{q}}{1+\Tilde{q}}\frac{e^{2Jt}}{2}\ll 1$ we can approximate
\begin{equation} \label{KC_exp}
        \lambda C_K(t) \approx 2\log\left(\frac{1+\Tilde{q}}{2}\right)+\frac{1-\Tilde{q}}{1+\Tilde{q}}e^{2Jt},\quad \frac{1}{J}\ll t\ll t_{\rm scr}\equiv\frac{1}{2J}\log\frac{2(1+\Tilde{q})}{1-\Tilde{q}},
\end{equation}
where we defined the scrambling time as the time of onset of the linear regime, analogously to what we discussed in \cref{fig:DSSYK_noMatter} for the matterless TFD state complexity.
At even later times, we preserve only the exponential part of $\sinh^2(Jt)\approx \frac{1}{4}e^{2Jt}$ which gives:
\begin{equation} \label{KC_linear}
    \lambda C_K(t)\approx 2\log\left(\frac{1-\Tilde{q}}{4}\right)+4Jt\quad\mathrm{when}\quad t\gg t_{\rm scr}
\end{equation}
To summarize, depending on the time scale and on the parameter $\tilde{q}$, K-complexity has the following behavior:
\begin{equation}\label{eq:op_lanczos_behav}
    \lambda C_K(t)\sim
    \begin{cases}
      2(1-\Tilde{q})(J t)^2 &  t\ll \frac{1}{J} \\
        2\log\left(\frac{1+\Tilde{q}}{2}\right)+\frac{1-\Tilde{q}}{1+\Tilde{q}}e^{2Jt}       &\frac{1}{J} \ll t\ll t_{\rm scr}\\
      2\log\left(\frac{1-\Tilde{q}}{4}\right)+4Jt & t\gg t_{\rm scr}
    \end{cases}
\end{equation}
Let us briefly comment on the regimes we have identified. The operator complexity has the same quadratic early time behavior that was found for the infinite temperature TFD state complexity in \cite{Rabinovici:2023yex}. However, before arriving at the asymptotic linear behavior, there is a new intermediate exponential regime whose length is controlled by $\Tilde{q}$. This regime lasts for for a time proportional to $-\log({1-\Tilde{q}})$, which we can think as the characteristic time needed for a matter insertion perturbation to grow on the system. So we can consider this intermediate exponential regime as long and parametrically different to the $\propto t^2$ behavior when $t_{\rm scr}\gg 1/(2J)$, that is when $\Tilde{q}\to 1$. In these cases, the complexity will stay small following this exponential regime for a very long time before starting the linear growth, which is a manifestation that an energy perturbations that is initially confined to a smaller number of the fermionic modes takes more time to grow. 

\begin{figure}
    \centering
    \includegraphics[width= 0.4\textwidth]{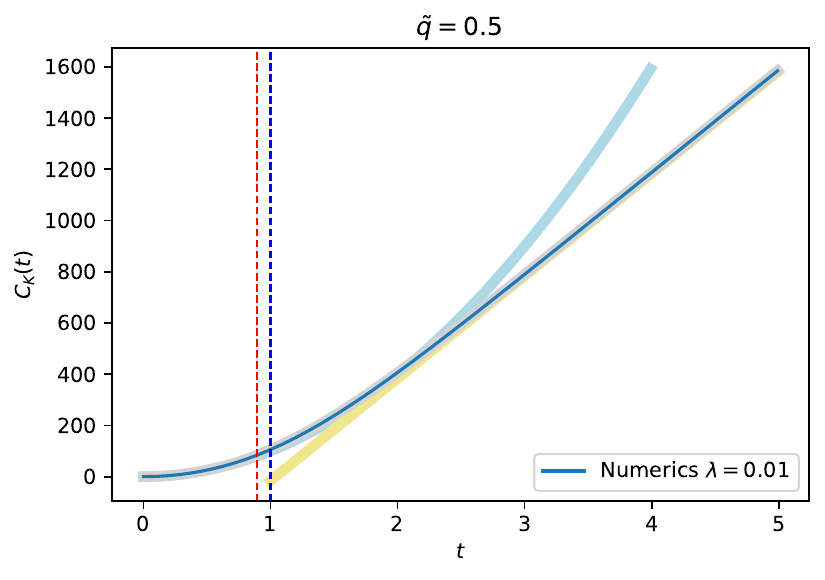}\includegraphics[width= 0.4\textwidth]{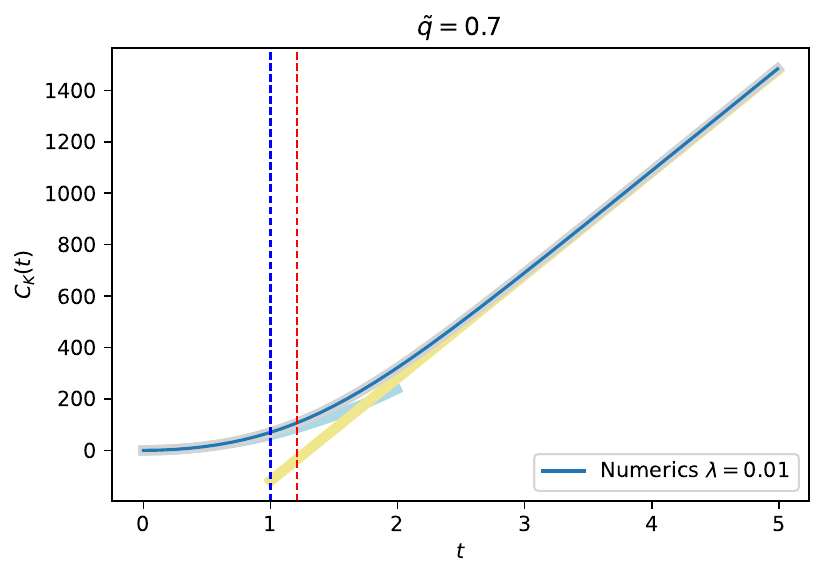} \includegraphics[width= 0.4\textwidth]{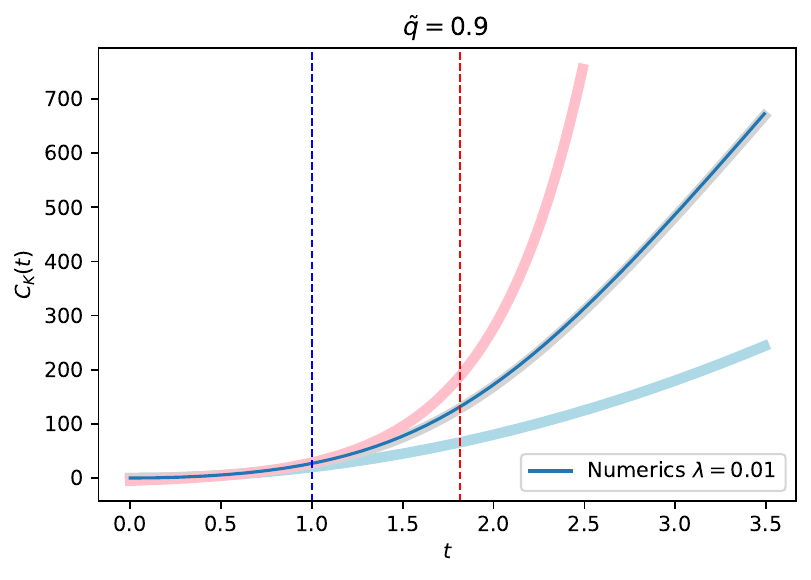}\includegraphics[width= 0.4\textwidth]{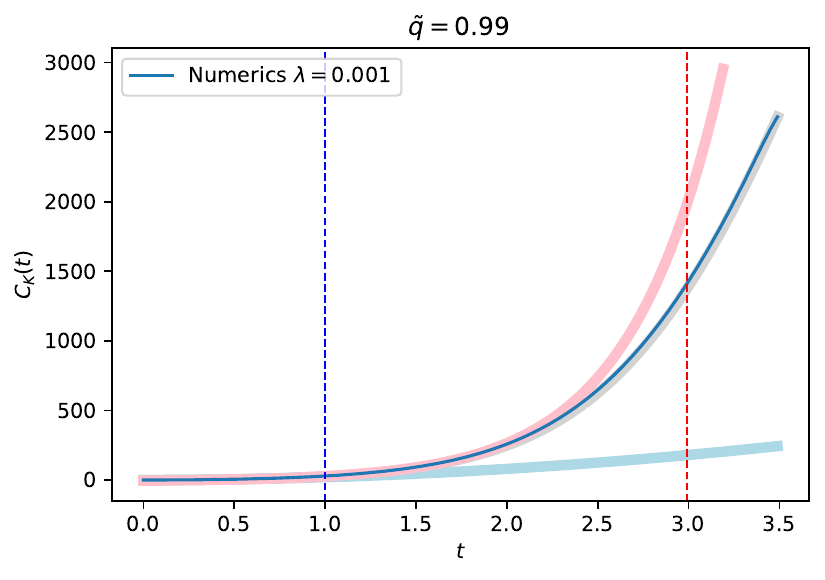}
    \caption{K-complexity in the small $\lambda$ limit with various values of $\Tilde{q}$. The thick grey line shows the analytical result \eqref{eq:op_compl}, while the thin blue line shows numerical results. The blue dashed vertical line shows the time scale below which the behavior is quadratic and given by \eqref{KC_quadratic}. The red dashed vertical line is the `scrambling' time, below which the behavior is exponential and given by \eqref{KC_exp}. We superimpose the quadratic approximation \eqref{KC_quadratic},  the scrambling regime \eqref{KC_exp} and the linear regime \eqref{KC_linear} as thick lightblue, pink and yellow lines respectively.
    Note that for $\Tilde{q}=0.5$ the scrambling time is below the quadratic time, and thus there is no scrambling regime.}
    \label{fig:op_compl}
\end{figure}

Next we want to compare the result we found for operator complexity with those obtained from models of quantum circuits \cite{susskind2014switchbacksbridge}, as they are usually good benchmarks to understand quantum information theoretic quantities.
\subsubsection*{Connection with circuit complexity}
 In particular we will recognize that DSSYK and circuit complexities follow similar behaviors, and that they share the same dependence on the dimension of the system and the size parameters of the operator and Hamiltonian.
The operator circuit complexity of $\mathcal{O}$ can be defined as the minimum number of gates needed to approximate it up to a certain tolerance \cite{Chapman_2022}. Notice that this notion, in contrast to the Krylov complexity we defined, does not depend only on the operator choice and system details, but also on some arbitrary allowed tolerance parameter and set of gates considered.
Now let us consider a system of $S$ qubits evolving with a Hamiltonian $H$ acting at each time step on a random set of $k\ll S$ qubits. We are interested in the circuit operator complexity of the operator $\mathcal{O}=W(t)=e^{iHt} W e^{-iHt}$, which is defined as the precursor of a perturbation $W$ acting on a small number $s_0$ of qubits. 
It has been shown \cite{susskind2014switchbacksbridge}, that this operator circuit complexity has an early exponential time dependence followed by a linear regime, ensuing at a ‘scrambling time' $t_*$ such as:
\begin{equation}\label{eq:tscr_circuit}
   t_*\propto \log\frac{S}{s_0 (k-1)} .
\end{equation}
We note that this circuit complexity behavior is very similar to that of \eqref{eq:op_compl} when $\Tilde{q}\sim 1$. Indeed they differ only at very early times (when the DSSYK complexity $\propto t^2$), and share the long exponential regime and the linear behavior after scrambling. Moreover, the scrambling times related to these two notions of complexity, present a similar dependence on the Hamiltonian and operator size parameters. When $\Tilde{q}\sim 1$, the scrambling time for the DSSYK operator complexity can be written as:
\begin{equation}
    t_{\rm scr}\underset{\Tilde{q}\sim 1}{\sim}\frac{1}{2J}\log\left(\frac{2}{1-\Tilde{q}}\right)\sim\frac{1}{2J}\log\left(\frac{2}{\lambda \Delta}\right)\sim \frac{1}{2J}\log\left(\frac{N}{p \Tilde{p}}\right)
\end{equation}
where we have reinstated in the expression the dimension of the Hamiltonian $p$, the operator $\Tilde{p}$, and the size of the system $N$. These size parameters appear inside the $\log$ in a similar manner as in \eqref{eq:tscr_circuit}, with the natural identifications $k\leftrightarrow p$, $s_0\leftrightarrow\Tilde{p}$ and $S\leftrightarrow N$. We consider this similarity a good sanity check for our result of DSSYK operator K-complexity. 

\subsection{Canonical analysis} \label{sect:Hamiltonian_analysis_operator}
Section \ref{subsect:analytical_K_Complexity} noted that, in the semiclassical limit, controlled by $\lambda\to 0$ and where the dimensionless variable is $x\equiv \lambda n$, the propagation of the wave packet in Krylov space becomes ballistic, admitting an equivalent description in terms of a point particle that follows a trajectory determined by the velocity profile
\begin{equation}
    \label{veloc_from_Lanczos_operator}
    v(x) = \lim_{\lambda\to 0}\Big( 2 \lambda b(x)\Big) = 4J\sqrt{\left( 1-e^{-x/2} \right)\left(1-\widetilde{q}e^{-x/2}\right)}~.
\end{equation}
The velocity field $v(x)$ may be regarded as the parametric equation for the phase space trajectory of the point particle. We may now seek the Hamiltonian that produces this trajectory as a solution of its equation of motion. Given any constant $\mathcal{E}>0$, which may be thought of as setting the energy units, it is possible to show that such a trajectory follows from the following Hamiltonian:
\begin{equation}
    \label{Semiclassical-operator_Hamiltonian}
    H = \mathcal{E}\left\{ \frac{k^2}{2} + 8\left( \frac{J}{\mathcal{E}} \right)^2 V(x) \right\}~,
\end{equation}
where $V(x)$ is given by a Morse potential:
\begin{equation}
    \label{Morse_potential_operator}
    V(x) = (1+\widetilde{q})~ e^{-x/2} - \widetilde{q} e^{-x}~.
\end{equation}
Note that below we present \eqref{Hamiltonian_semicl_operator_from_bx_line2} as a classically equivalent Hamiltonian derived  
from the semiclassical Lanczos coefficients. For $\widetilde{q}\in [0,1]$, the potential \eqref{Morse_potential_operator} takes the form of the reverse of the potential for a diatomic molecule, see figure \ref{fig:Morse_Potential_operator}. In particular, it features an unstable maximum located at the position
\begin{eqnarray}
    \label{Unstable_max_location}
    x_m(\widetilde{q}) = -2 \log\frac{1+\widetilde{q}}{2\widetilde{q}}~.
\end{eqnarray}
We note that for $\widetilde{q}\in [0,1]$ the unstable maximum location satisfies $x_m(\widetilde{q})\leq 0$.

In order to reproduce the semiclassical Krylov complexity result \eqref{eq:op_compl} out of Hamiltonian \eqref{Semiclassical-operator_Hamiltonian}, we need to solve its associated equation of motion given the initial condition $x(0)=0$, as the starting value of K-complexity is always zero. The second initial condition is already dictated by the Lanczos coefficients and encoded in $v(x)$: $\dot{x}(0)=v\big( x(0) \big)=0$. This yields (see appendix \ref{appx:Morse_generic_solution}):
\begin{eqnarray}
    \label{operator_KC_solution_EOM}
    \lim_{\lambda\to 0} \left[\lambda C_K(t)\right] = x(t) = 2\log\left\{ 1 + (1-\widetilde{q}) \sinh^2\left(tJ\right) \right\}~,
\end{eqnarray}
in perfect agreement with \eqref{eq:op_compl}, as expected by construction of \eqref{Semiclassical-operator_Hamiltonian}.

When $\widetilde{q}=0$ (i.e. very heavy operators), equation \eqref{operator_KC_solution_EOM} reduces to the matterless solution studied in \cite{Rabinovici:2023yex}, namely
\begin{equation}
    \label{Matterless_when_qtilde_0}
    \frac{x(t)}{2} = 2\log\cosh(tJ)~,
\end{equation}
where the factor of $\frac{1}{2}$ on the left hand-side appears because the heavy operator insertion effectively splits the system into two identical copies, the left and right half-lengths evolving independently and following \eqref{Matterless_when_qtilde_0}. Mathematically, this is the result of the fact that the one-particle algebra becomes the tensor product of two copies of the zero-particle algebra when $\widetilde{q}=0$ \cite{Lin:2023trc}. On the other hand, when $\widetilde{q}=1$, the solution \eqref{operator_KC_solution_EOM} becomes stationary, as expected for the identity operator.

We are now in position to understand the scrambling dynamics that solution \eqref{operator_KC_solution_EOM} features when $\widetilde{q}$ is close to $1$, as discussed in section \ref{subsect:analytical_K_Complexity}, from the perspective of the Morse potential \eqref{Morse_potential_operator}. The computation of Krylov complexity amounts to solving the equation of motion of the Hamiltonian \eqref{Semiclassical-operator_Hamiltonian} with the initial conditions $\Big( x(0),\dot{x}(0)\Big)=(0,0)$. For any value of $\widetilde{q}\in [0,1[$, the unstable maximum $x_m(\widetilde{q})<0$ appears to the left of the starting position of the point particle (see figure \ref{fig:Morse_Potential_operator}), which will therefore start rolling down the potential towards $x>0$. The closer $\widetilde{q}$ is to $1$, the closer will $x_m(\widetilde{q})$ be to the starting position, and hence the roll-off of the particle gets slightly delayed. In the limiting case $\widetilde{q}=1$, the particle at $t=0$ happens to be sitting exactly on top of the unstable equilibrium point, hence remaining stationary. In short, this analysis connects scrambling dynamics of operator complexity to the dynamics of a point particle evolving in a potential that features an unstable equilibrium point.

\begin{figure}
    \centering
    \includegraphics[width=0.45\linewidth]{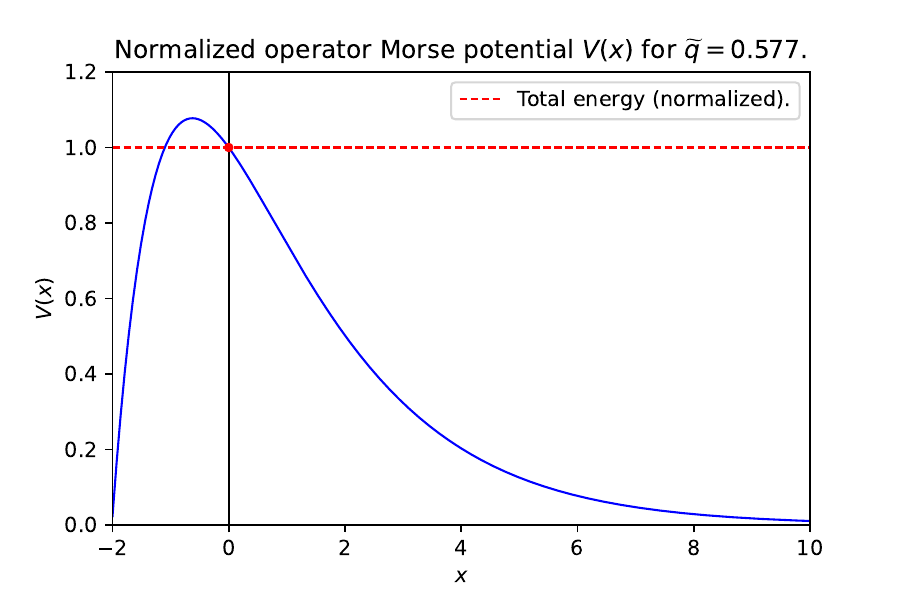} \includegraphics[width=0.45\linewidth]{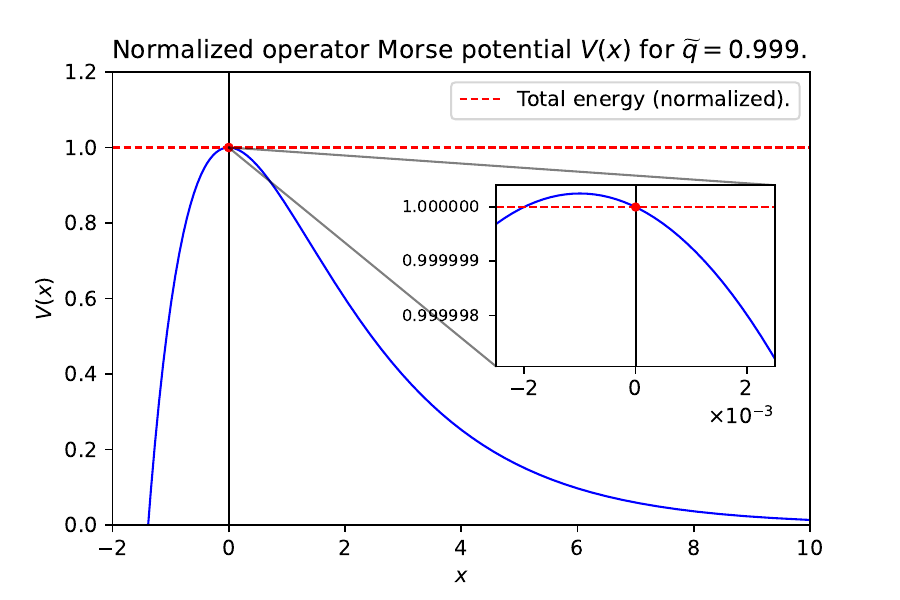} \\
    \includegraphics[width=0.45\linewidth]{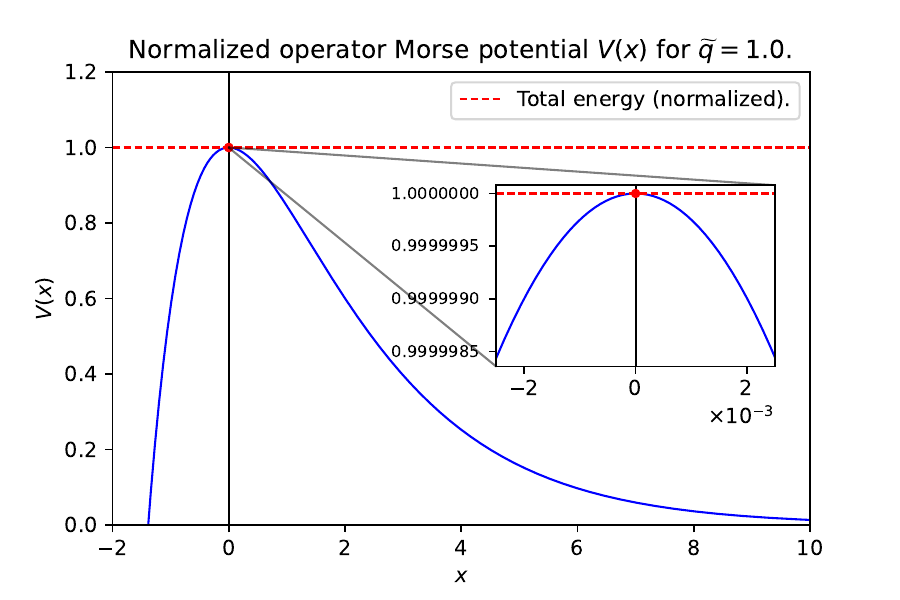}  \includegraphics[width=0.45\linewidth]{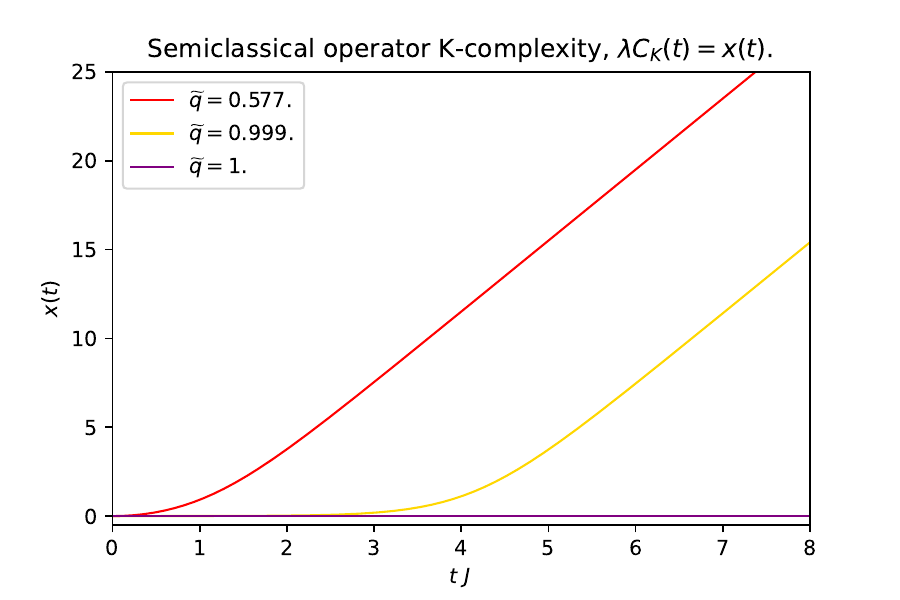}
    \caption{This figure depicts the Krylov space potential that governs the evolution of the operator Krylov complexity in the semiclassical ($\lambda\to 0$) limit, where the K-complexity is given by the position $x(t)$ of a point particle that propagates through Krylov space driven by such a potential, starting at rest at $x(0)=0$. The \textbf{top left}, \textbf{top right} and \textbf{bottom left} graphs correspond to the potential for values $\widetilde{q}=0.577,0.999,1$, respectively. We observe that this potential has an unstable maximum that lies closer to the initial position of the particle the closer $\widetilde{q}$ is to $1$, and hence the asymptotically linear Krylov complexity growth gets delayed by some amount of time, as seen in the \textbf{bottom right} plot. In short, the instability in the potential provides a mechanism for scrambling-like dynamics. More precisely, this behavior is the imprint of the switchback effect caused by the insertion of the operator in the state $\mathcal{O}(t)|TFD\rangle$, whose Krylov complexity we have computed using the fact that in DSSYK such a state gets mapped to the one-particle state $e^{-it(H_R-H_L)}|0,0\rangle$, cf. \eqref{operator_evolution}.}
    \label{fig:Morse_Potential_operator}
\end{figure}

For completeness, we may conclude this section by noting that the Hamiltonian \eqref{Semiclassical-operator_Hamiltonian} is classically equivalent to the Hamiltonian that can be obtained in the usual manner from the combination of the Lanczos coefficients and the displacement operator in Krylov space \cite{Rabinovici:2023yex}. In the semiclassical limit, we have:
\begin{align}
    &H_R-H_L = e^{i\lambda k } b(x) + b(x) e^{-i \lambda k}\label{Hamiltonian_semicl_operator_from_bx_line1} \\
    &= \frac{2J}{\sqrt{\lambda(1-q)}} \left\{ e^{i \lambda k}\sqrt{1-V(x)}+  e^{-i\lambda k} \sqrt{1-V(x)}\right\}~, \label{Hamiltonian_semicl_operator_from_bx_line2}
\end{align} 
where $V(x)$ is the same function defined in \eqref{Morse_potential_operator}.
The Hamiltonian \eqref{Hamiltonian_semicl_operator_from_bx_line2} is different as an operator from \eqref{Semiclassical-operator_Hamiltonian}; however, it can be checked that when $\lambda\to 0$ the equation of motion for $x(t)$ is the same in both cases\footnote{In fact, this is true for any (hermitian) potential function $V(x)$.}, hence the classical equivalence of the Hamiltonians. Furthermore, the semiclassical analysis in section \ref{subsect:analytical_K_Complexity} showed that when $\lambda\to 0$ the evolution generated by \eqref{Hamiltonian_semicl_operator_from_bx_line1} in Krylov space becomes classical in the sense that the Krylov wave function evolves ballistically, and therefore in this limit the Hamiltonians \eqref{Hamiltonian_semicl_operator_from_bx_line1} and \eqref{Semiclassical-operator_Hamiltonian} may be freely interchanged for the purpose of the canonical description of Krylov complexity and, in turn, of total chord number.

\subsection{Triple-scaled Hamiltonian}\label{sect:Triple_scaling_operator}
Let us now derive the form of the Hamiltonian $H_R-H_L$ in the low-energy regime. This is achieved by the so-called \textit{triple-scaling limit} \cite{Lin:2022rbf}, which zooms near the ground state via a large-length limit\footnote{This limit may be understood as a choice of bulk length regularization scheme, as explained in appendix B of \cite{Rabinovici:2023yex}}, namely
\begin{equation}
    \label{Triple_scaling_limit} 
    \lambda\to 0~,\quad x\to\infty~,\quad \frac{e^{-x}}{(2\lambda)^2}\equiv e^{-\widetilde{x}}~~~\text{fixed.}
\end{equation}
This limit captures the low-energy dynamics of DSSYK (and, in turn, of large-$p$ SYK, since $\lambda\to 0)$, where the system features a gravitational dual governed by JT gravity, as reviewed in section \ref{sec:recap_dssyk_nomatter}. In previous work \cite{Rabinovici:2023yex} we showed that in this limit the matterless Hamiltonian reduces to the Liouville Hamiltonian relevant to such a gravitational theory \cite{Lin:2022rbf,Harlow:2018tqv}, where the rescaled Krylov complexity variable $\widetilde{x}$ played the role of the regularized bulk length in AdS units. 
The total chord number in the one-particle algebra has also been argued in \cite{Lin:2022rbf,Lin:2023trc} to become the gravitational length variable when $\lambda\to0$ and, since in section \ref{subsect:bn_asymptotic_limit} we proved that operator Krylov complexity is equal to total chord number, the form of $H_R-H_L$ in terms of the bulk length operator in the JT gravity regime can be obtained out of triple-scaling the operator Lanczos coefficients \eqref{semiclassical_operator_bn}. However, an important subtlety needs to be taken into account: as we make the variable $x$ large to zoom in on the vicinity of the ground state, the operator should be made light in a coordinated manner, in order to prevent its influence from being too disruptive. More technically, for the low-energy limit to be self-consistent, the effect of the operator dependence should be contained within the low-energy Hamiltonian. In order to achieve this, we propose to supplement the triple-scaling protocol \eqref{Triple_scaling_limit} with the prescription of keeping $\Delta$ fixed in $\widetilde{q}=e^{\lambda \Delta}$ as $\lambda$ goes to zero. All in all, the triple-scaled Lanczos sequence, $b^{TS}(\tilde{x})$, in terms of the semiclassical variables is:
\begin{align}
    &b(x)=\frac{2J}{\sqrt{\lambda (1-q)}}\sqrt{(1-e^{-x/2})(1-e^{-\lambda \Delta - x/2})} \\
    &\longmapsto b^{TS}(\widetilde{x}) =  \frac{2J}{\sqrt{\lambda (1-q)}}\sqrt{(1-2 \lambda e^{-\widetilde{x}/2})(1-2 \lambda e^{-\lambda \Delta - \widetilde{x}/2})} \\
    &=b_0 (\lambda) - 4J e^{-\widetilde{x}/2} - \lambda J (1-2\Delta) e^{-\widetilde{x}/2} + \mathit{O}(\lambda^2)~,
\end{align}
where $b_0(\lambda) = \frac{2J}{\lambda}+\mathit{O}(\lambda^0)$ is a constant related to the system's ground-state energy. The triple-scaled Hamiltonian $H^{(-)}$ may be determined by posing
\begin{equation}
    \label{Triple_scaled_Ham_minus_def}
    -H^{(-)}\equiv H_R-H_L = e^{i\lambda \widetilde{k}}b^{TS}(\widetilde{x}) + b^{TS}(\widetilde{x})e^{-i\lambda \widetilde{k}}
\end{equation}
and expanding around $\lambda=0$. Given that the Lanczos coefficients tend to a constant when $\widetilde{x}\to+\infty$, the expression after the second equality in \eqref{Triple_scaled_Ham_minus_def} can be understood as a scattering Hamiltonian whose spectrum is symmetric around zero. Hence, in order to assure that the lowest energy corresponds to the mode of lowest momentum, we introduce a sign flip in the definition of $H^{(-)}$ which is harmless at this point due to the symmetry of the spectrum but which will be crucial after triple-scaling in order to obtain a Hamiltonian correctly bounded from below. As promised, expanding \eqref{Triple_scaled_Ham_minus_def} around $\lambda = 0$ yields:
\begin{equation}
    \label{Triple_scaled_Ham_minus_result}
    H^{(-)} - E_0(\lambda) = 2\lambda J~ \widetilde{k}^2 + 8J \left( 1-\frac{\lambda \Delta}{2} \right) e^{-\widetilde{x}/2} +\mathit{O}(\lambda^2)~,
\end{equation}
where the ground-state energy is $E_0(\lambda)=-2b_0(\lambda)$. In light of \eqref{Triple_scaled_Ham_minus_result}, the following comments are in order:

\begin{itemize}
    \item The first occurrence of the operator scaling dimension in \eqref{Triple_scaled_Ham_minus_result} is of first order in $\lambda$, that is, at the same level as the kinetic term. As announced earlier, this signals that the low-energy approximation is self-consistent; the operator-dependence doesn't disruptively affect any terms leading with respect to the kinetic term in the small-$\lambda$ expansion.
    \item When $\lambda\to 0$, the Hamiltonian takes the form of a rigid potential with no kinetic term, and with no operator dependence, namely:
    \begin{equation}
        \label{Triple_scaled_Ham_minus_order_zero}
        \lim_{\lambda\to 0}\Big( H^{(-)}-E_0(\lambda) \Big) = 8J e^{-\widetilde{x}/2}~.
    \end{equation}
    This admits the following physical interpretation: In our limiting protocol, the triple-scaling of the length has been accompanied by keeping $\Delta$ fixed in $\widetilde{q}=e^{-\lambda \Delta}$ as $\lambda$ was sent to zero. Therefore, setting $\lambda=0$ effectively turns off the operator insertion, in which case we do not expect the Hamiltonian $H_R-H_L$ to generate any time evolution at all\footnote{As a complementary remark, we note that setting $\Delta=0$ in \eqref{Triple_scaled_Ham_minus_result} yields a potential term at order $\lambda^0$ plus a kinetic term at order $\lambda$. It is well-known \cite{GAMMEL1967103,GAMMEL1967229} that treating the kinetic term as a perturbation of the potential yields an ill-defined perturbative expansion with convergence radius equal to zero. In fact, in order to address the case when $\Delta=0$ independently from $\lambda$ we should just go all the way back to the definition of the Lanczos algorithm in section \ref{subsect:KrylovBasis}: When $\Delta = 0$ the Hamiltonian $H_R-H_L$ simply becomes identically zero, not yielding any time evolution at all.
    }, 
    which is consistently the case of \eqref{Triple_scaled_Ham_minus_order_zero}.
    \item More importantly, we have found that the triple-scaled Hamiltonian \eqref{Triple_scaled_Ham_minus_result}  no longer features a Morse potential, but instead it is, to first order in $\lambda$, a Liouville Hamiltonian with a subleading operator-dependent correction. In particular, since its potential does not feature any unstable equilibrium points, it will show no traces of scrambling dynamics. We may understand this as being consistent with our limiting protocol: The operator weighting factor $\widetilde{q}= e^{-\lambda \Delta}$ tends to $1$ when $\lambda$ is sent to zero while $\Delta$ is fixed, implying that the position of the unstable maximum \eqref{Unstable_max_location} of the Morse potential \eqref{Morse_potential_operator} tends to $x_m\to 0$. On top of this limiting form of the Morse potential, the length triple-scaling \eqref{Triple_scaling_limit} amounts to observing this potential at large distances, i.e. far away from the unstable maximum, where the Morse potential effectively looks monotonously decreasing. We may therefore understand this as a late-time regime that focuses on the linear growth of K-complexity (bulk length), which is indeed featured by the solution of the Liouville Hamiltonian \eqref{Triple_scaled_Ham_minus_result}, cf. \cite{Rabinovici:2023yex}.
\end{itemize}

\section{The K-complexity of the \texorpdfstring{$\mathcal{O}$}{TEXT}TFD state}\label{sect:OTFD_KC}
In this section, we want to discuss a new object, namely the state K-complexity that we obtain by solving a Krylov problem with Hamiltonian $H_L+H_R$ and initial state $\ket{0,0}$. This defines the complexity of the state that we obtain by inserting an operator $\mathcal{O}$ of the form \eqref{random_operator} on the infinite-temperature TFD state $\ket{0}$. Notice that this complexity is not the operator complexity, which was obtained from the study of the evolution of $|0,0\rangle$ generated by $H_R-H_L$, because of the crucial relative sign difference in the Hamiltonian evolution chosen here. However, it is an equally interesting object to study because it extends the discussion of \cite{Rabinovici:2023yex}, allowing to compute the complexity of a new class of states obtained by the insertion of operators $\mathcal{O}$ on the TFD that here we call $\mathcal{O}$TFD states. The expected dual geometry of states of this kind is a two-sided black hole perturbed by a shock wave launched from one side at boundary time $t=0$. We shall show that the Krylov complexity operator constructed in this setup is equal to the total chord number operator restricted to the relevant Krylov space, hence becoming the bulk length operator in the triple-scaling limit that zooms in on the JT regime, in full consistency with \cite{Lin:2023trc}. We leave the details of this matching to a future publication.

\subsection{Semiclassical Lanczos coefficients}

In order to compute the Krylov complexity of the $\mathcal{O}$TFD state, let us start by analyzing the corresponding Lanczos algorithm. We shall jump directly to the asymptotic (cf. semiclassical) limit relevant for the eventual construction of the gravitational Hamiltonian. Using an inductive argument completely analogous to that in section \ref{subsect:bn_asymptotic_limit}, we can prove that the Krylov basis elements are total chord number eigenstates, which will in turn imply equality between Krylov complexity and total chord number of the time-evolving state.

The action of $(H_L+H_R)^m\ket{0,0}$, will again result in combination of states $\ket{n_L,n_R}$, that will be used by the Lanczos algorithm to build the orthonormal Krylov basis.
It is possible to write the evolution operator $H_L+H_R$ in terms of chord creation/annhilation operators, analogously to \eqref{Total_Hamiltonian_minus_line2}, as:
\begin{align}
    H_R+H_L &= J\left( a_R^\dagger + a_L^\dagger + a_R + a_L\right)= \\ &= J \left[a_R^\dagger + a_L^\dagger + \alpha_R [n_R]_q \left(1 + \widetilde{q}q^{~n_L}\right) + \alpha_L [n_L]_q \left(1 + \widetilde{q}q^{~n_R}\right)\right] ~. 
\end{align}
If we act with $H_R+H_L$ on some state in the $n$-th total chord number sector $\mathcal{H}_{1p}^{(n)}$, we will obtain a linear combination of states in $\mathcal{H}_{1p}^{(n+1)}$ and $\mathcal{H}_{1p}^{(n-1)}$. In particular we will see, analogously to what we observed in \cref{subsect:bn_asymptotic_limit}, that inserting the $n$-th state of the Krylov ansatz in the Lanczos recursion won't exactly result in a total chord number eigenstate. Indeed we will obtain a state in $\mathcal{H}_{1p}^{(n+1)}$, but we will generally also develop a tail living in $\mathcal{H}_{1p}^{(n-1)}$, which, similarly to \cref{subsect:bn_asymptotic_limit}, will be suppressed in norm with respect to the projection of the Krylov element over the subspace $\mathcal{H}_{1p}^{(n+1)}$.
Just like in section \ref{subsect:bn_asymptotic_limit}, we shall show that the following Ansatz for the Krylov basis satisfies the Lanczos algorithm in the semiclassical limit where $\lambda\to 0$ and $\lambda n$ is fixed:
\begin{equation}
    \label{eq:binomialstates_opstatecompl}
    \ket{\psi_n^+} = \frac{1}{\prod_{k=0}^n (b_n^+/J)}\ket{\chi_n^+},\qquad \ket{\chi_n^+} := \sum_{k=0}^n \binom{n}{k}\ket{k,n-k}~. 
\end{equation}
The binomial states $\ket{\chi_n^+}$ have been defined analogously to $\ket{\chi_n}$ in \eqref{Binomial_Ansatz}, modulo the alternating sign that is now absent in \eqref{eq:binomialstates_opstatecompl}, and similarly the $b_n^+$ are normalization coefficients.

In order to proceed inductively\footnote{The discussion on the induction seed is completely analogous to sections \ref{subsect:KrylovBasis} and \ref{subsect:Lanczos_Limit_lambda0_qtfixed}.}, let us insert $\ket{\psi_n^+}$ in the Lanczos recursion that computes the next unnormalized Krylov vector $\ket{A_{n+1}^+}=b_{n+1}^+\ket{\psi_{n+1}^+}$:
\begin{align}
    \ket{A_{n+1}^+} = (H_R+H_L)|\psi_n^+\rangle - b_n^+|\psi_{n-1}^+\rangle =\\ 
    =\frac{J}{(b_1^+/J)\dots (b_n^+/J)}\Big( |\chi_{n+1}^+\rangle - |\zeta_{n-1}^+\rangle \Big)~.
\end{align}
Here $|\zeta^+_{n-1}\rangle$ is the aforementioned tail in $\mathcal{H}_{1p}^{(n-1)}$, potentially spoiling our Ansatz \eqref{eq:binomialstates_opstatecompl}, which can be rewritten as:
\begin{eqnarray}
    |\zeta^+_{n-1}\rangle = \sum_{k=0}^{n-1} \left[ {c^+_k}(n) - \left(\frac{b^+_n}{J}\right)^{2} \right]\binom{n-1}{k} |k,n-1-k\rangle~,
\end{eqnarray}
where $c_k^+(n)$ are objects analogous to \eqref{Ckn_simplified}, defined as:
\begin{equation}\label{ckntilde}
    c_k^+(n)\equiv n \frac{[n-k]_q}{n-k}(1+\tilde{q}q^k) + n\frac{[k+1]_q}{k+1}(1+\tilde{q}q^{n-1-k})
\end{equation}
The next Lanczos coefficient is given by the norm of $|A_{n+1}^+\rangle$:
\begin{equation}
    \left(\frac{b_{n+1}^{+}}{J}\right)^2 = \frac{1}{(b_1^+/J)^2\dots (b_n^+/J)^2}\Big( \langle\chi_{n+1}^+|\chi_{n+1}^+\rangle + \langle \zeta_{n-1}^+|\zeta^+_{n-1}\rangle \Big)~,
\end{equation}
where these norm contributions are:
\begin{equation}\label{eq:tail_norm_opstate}
    \begin{aligned}
         &\langle \chi_{n+1}^+|\chi_{n+1}^+\rangle = \sum_{k^\prime,k=0}^{n+1}\binom{n+1}{k^\prime}\binom{n+1}{k}\langle k^\prime,n+1-k^\prime | k,n+1-k\rangle~, \\
    &\langle \zeta_{n-1}^+|\zeta^+_{n-1}\rangle = \sum_{k^\prime,k=0}^{n-1}\!\left[c_{k^\prime}^+(n)-\left(\frac{b_n^+}{J}\right)^2\right]\!\! \left[c_{k}^+(n)-\left(\frac{b_n^+}{J}\right)^2\right] \! \!\binom{n-1}{k^\prime}\!\binom{n-1}{k}\!\langle k^\prime,n-1-k^\prime|k,n-1-k\rangle~.
    \end{aligned}
\end{equation}

At this point, the argument to prove that $\langle \zeta_{n-1}^+|\zeta^+_{n-1}\rangle$ gives a subleading contribution to the Lanczos coefficients, is analogous to the induction performed in \cref{subsect:bn_asymptotic_limit}. In sums similar to those above, the binomials in the semiclassical limit become effectively Dirac delta functions around the symmetric configuration $k,k'\sim n/2$ (as explained in \cref{appx:Binom_delta_EasyCase}). In the region contributing to \eqref{eq:tail_norm_opstate}, the expression $c_k^+(n)$ is constant and equal to $c_{n/2}^+(n)$, so that, analogously to \cref{subsect:Lanczos_Limit_lambda0_qtfixed} and \ref{subsect:bn_asymptotic_limit}, we obtain $b_n^+=J\sqrt{c_{n/2}^+(n)}$ from the Lanczos algorithm. Consequently the tails in $\mathcal{H}_{1p}^{(n-1)}$ are suppressed, and the Lanczos recursion builds the following Krylov basis and Lanczos coefficients: 
\begin{equation}\label{eq:krylsol_opstate}
\begin{aligned}
      |\psi_{n+1}^+\rangle &= \frac{1}{b_{n+1}^+}|A_{n+1}^+\rangle = \frac{1}{(b_1^+/J)\dots (b_{n+1}^+/J)}|\chi_{n+1}^+\rangle + \text{subleading in }\lambda~,\\&b_n^+\mapsto b_n^+=\frac{J}{{\sqrt{\lambda}}}\sqrt{c_{n/2}^+(n)}=2J\sqrt{\frac{1-q^{n/2}}{\lambda(1-q)}(1+\Tilde{q}q^{n/2})}~,
\end{aligned}
\end{equation}
where in the last line we have directly switched to the Hamiltonian normalization such that $J\mapsto \frac{J}{\sqrt{\lambda}}$, similarly to \eqref{Renormalization_a_la_Lin}, which shall be more convenient in subsequent sections.
In conclusion, we find that the $n$-th Krylov element $|\psi_n\rangle$ belongs to the sector $ \mathcal{H}_{1p}^{(n)}$ in the $\lambda\to0$ limit. Therefore, the Krylov basis will be made of simultaneous eigenstates of the total chord number operator and of the $\mathcal{O}$TFD state Krylov complexity operator.

\subsection{Semiclassical Krylov complexity}\label{sect:Semiclassical_KC_OTFD}

In an analysis which is analogous to the one in section \ref{subsect:analytical_K_Complexity}, it is possible to show that the Krylov complexity of the evolution of the $\mathcal{O}TFD$ state in the small-$\lambda$ limit is captured by the position of a classical point particle exploring Krylov space following a velocity profile dictated by the twice the value of the Lanczos coefficients. In this semiclassical limit, K-complexity can be obtained from the direct integration of such a velocity profile:

\begin{equation}
    \int_0^{n^+(t)}\frac{dn}{2b_n^+}=t~,
\end{equation}
which yields:
\begin{equation}\label{eq:opstate_contcompl}
    \lambda C_K^+(t)\underset{\lambda\sim 0}{\sim} 2\log(1+(1+\Tilde{q})\sinh^2 (Jt))
\end{equation}
Notice that $\Tilde{q}\to 1$ corresponds to the insertion of the identity on the TFD. Consistently, we have $\lambda C_K^+(t)\overset{\widetilde{q}\to 1}{\rightarrow} 2\log\cosh{ (2 Jt)}$, correctly reproducing the continuum limit of the matterless state complexity of the TFD, up to a $t\mapsto 2t$ rescaling due to the two-sided evolution $H_R+H_L$.
The regimes of this complexity are the following:
\begin{equation}\label{eq:op_lanczos_behav_OTFD}
    \lambda C_K^+(t)\approx
    \begin{cases}
      2(1+\widetilde{q})(J t)^2 & \text{if $t\lesssim \frac{1}{J}$} \\
      2\log\left(\frac{1+\Tilde{q}}{4}\right)+4Jt & \text{if $t\gg \frac{1}{J}$}
    \end{cases}
\end{equation}
We note that in this case there is no early exponential behavior because, for any $\widetilde{q}\in [0,1]$, the factor $1+\widetilde{q}$ is always greater than $1$ and, in particular, it is never small, which is what would yield a scrambling behavior at early times.
Figure \ref{fig:Morse_OTFD_KC} depicts the $\mathcal{O}$TFD state Krylov complexity \eqref{eq:opstate_contcompl} for different values of $\widetilde{q}$. We observe that it features a much milder dependence on the details of the operator insertion, as compared to operator K-complexity \eqref{eq:op_compl}.

\subsection{Canonical analysis}\label{sect:Hamiltonian_Analysis_OTFD}
The semiclassical analysis in the previous section used the fact that in the limit $\lambda\to 0$ the Krylov wave packet of the time-evolving state $e^{-it(H_R+H_L)}|0,0\rangle$  behaves as a delta function that propagates ballistically through Krylov space. Thanks to this, Krylov complexity, which we have also been shown to be equal to the state's total chord number in this limit, can be computed as the position of a point particle that travels through Krylov space following a velocity profile dictated by the semiclassical limit of the Lanczos coefficients in \eqref{eq:krylsol_opstate}:
\begin{equation}
    \label{velocity_profile_OTFD}
    v(x) = \lim_{\lambda\to 0} \Big(2 \lambda b(x)\Big) = 4J \sqrt{(1-e^{-x/2})(1+\widetilde{q}e^{-x/2})}~.
\end{equation}

We may now proceed to a Hamiltonian analysis in the same spirit as that in section \ref{sect:Hamiltonian_analysis_operator}. It can be shown that the classical trajectory $v(x)$ in \eqref{velocity_profile_OTFD} can be obtained from the equation of motion of a Hamiltonian of the same form as \eqref{Semiclassical-operator_Hamiltonian}, where this time the potential is given by:
\begin{equation}
    \label{Potential_OTFD_interpolates_Liouville}
    V(x) = (1-\widetilde{q})e^{-x/2} + \widetilde{q}e^{-x}~.
\end{equation}
Self-consistently, the Krylov complexity computed in section \ref{sect:Semiclassical_KC_OTFD} can be derived by solving the equation of motion of the Hamiltonian \eqref{Semiclassical-operator_Hamiltonian} equipped with the potential \eqref{Potential_OTFD_interpolates_Liouville} given the initial conditions $\Big(x(0),\dot{x}(0)\Big)=\Big(0,v(0)\Big)=(0,0)$, yielding (cf. appendix \ref{appx:Morse_generic_solution}):
\begin{equation}
    \label{KC_from_EOM_OTFD}
    \lim_{\lambda\to 0}\Big( \lambda C_K^+(t) \Big) = x(t) = 2\log\left\{ 1+(1+\widetilde{q})\sinh^2(tJ) \right\}
\end{equation}
We note that this solution exhibits no scrambling behavior because for any $\widetilde{q}\in[0,1]$ the prefactor of the hyperbolic sine is never small. The solution transitions from an initial polynomial behavior to late-time linear growth, as studied in the previous section. Physically, we understand this in the following terms: The thermofield double state already features non trivial evolution in the absence of matter, as studied in \cite{Rabinovici:2023yex}, and the addition of the operator insertion on the initial state only modifies the complexity profile in time quantitatively without qualitatively disrupting it.

In connection to the previous considerations, we can give a physical interpretation to the potential \eqref{Potential_OTFD_interpolates_Liouville}. Since $\widetilde{q}\in [0,1]$, $V(x)$ takes the form of a Morse potential that doesn't feature a relative sign between the $e^{-x/2}$ and the $e^{-x}$ terms, and instead it is a monotonously decaying potential (asymptoting to zero) which interpolates between the Liouville potential $e^{-x}$ when $\widetilde{q}=1$ and another Liouville-like potential $e^{-x/2}$ when $\widetilde{q}=0$, see figure \ref{fig:Morse_OTFD_KC}. Indeed, when $\widetilde{q}=1$ the operator becomes the identity, and the potential $V(x)$ takes the limiting form of double the Liouville potential of the matterless case studied in \cite{Rabinovici:2023yex}, since in this limit $H_R=H_L$. On the other hand, an extremely heavy operator with $\widetilde{q}=0$ splits the system into two identical DSSYK copies (cf. discussion in section \ref{sect:Hamiltonian_analysis_operator}), and therefore $V(x)$ becomes the Liouville potential for the half-length $x/2$. The corresponding limiting forms of the solution \eqref{KC_from_EOM_OTFD} are consistent with this discussion. 

\begin{figure}
    \centering
    \includegraphics[width=0.45\linewidth]{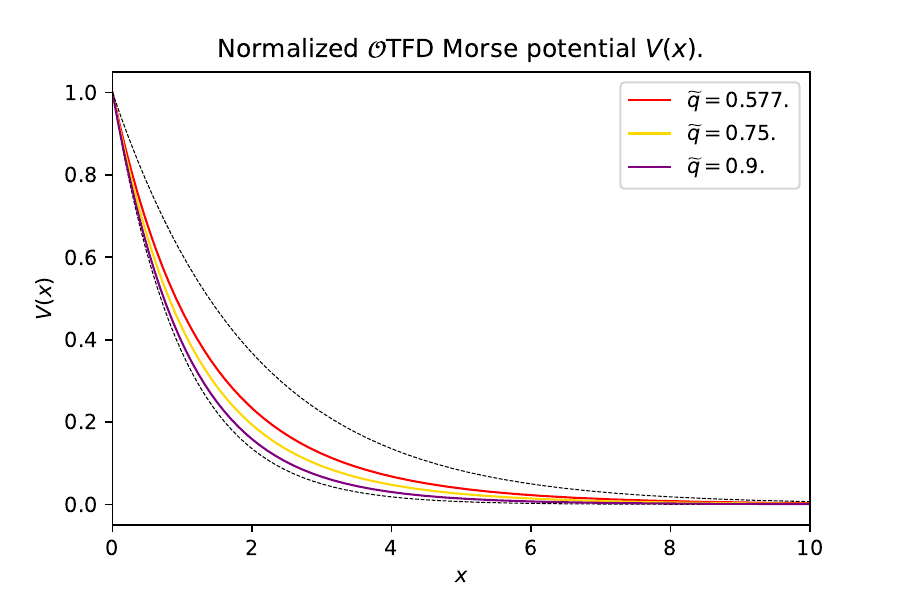}   \includegraphics[width= 0.45\textwidth]{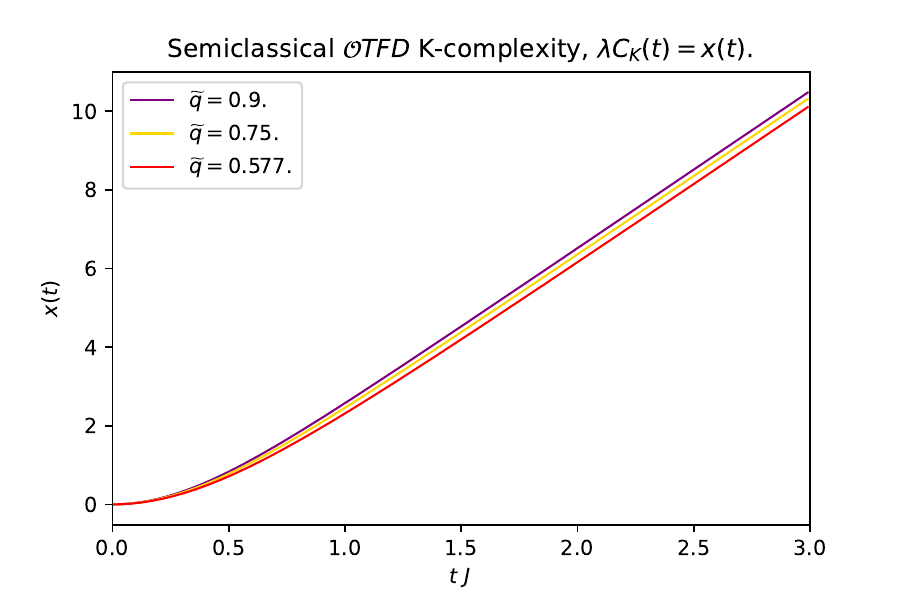} 
    \caption{\textbf{Left:} Krylov space potential for the $\mathcal{O}TFD$ state in the semiclassical limit ($\lambda\to 0$), where Krylov complexity (shown to be equal to the total chord number) is given by the position of a point particle propagating through Krylov space as it rolls down the potential, starting from $x(0)=0$ at rest. The function $V(x)$ is a Morse-like potential that interpolates between the matterless Liouville potential $e^{-x}$ when $\widetilde{q}=1$ (lower dashed line) and another matterless Liouville potential $e^{-x/2}$ when $\widetilde{q}=0$ (upper dashed line). The specific cases of the potential when $\widetilde{q}=0.577,0.75,0.9$ have been plotted. \textbf{Right:} Evolution of Krylov complexity as a function of time for the same values $\widetilde{q}=0.577,0.75,0.9$. As discussed in the text, operator-dependence is rather mild and, in particular, there are no signs of scrambling, as expected from \eqref{KC_from_EOM_OTFD}.}
    \label{fig:Morse_OTFD_KC}
\end{figure}

Finally, we stress that the discussion around \eqref{Hamiltonian_semicl_operator_from_bx_line1} also applies to the present case where we study the evolution of $|0,0\rangle$ under $H_R+H_L$: The Hamiltonian \eqref{Semiclassical-operator_Hamiltonian} equipped with the potential \eqref{Potential_OTFD_interpolates_Liouville} is classically equivalent to the actual Hamiltonian that we may obtain out of the semiclassical limit of the Lanczos coefficients of the $\mathcal{O}TFD$ state and, since in the semiclassical limit the dynamics generated by the latter Hamiltonian become classical anyway, we can equivalently describe evolution in Krylov space in this limit with the canonical Hamiltonian specified by the potential $V(x)$.

\subsection{Triple-scaled Hamiltonian out of the Lanczos coefficients}\label{sect:Triple_scaling_OTFD}
In this section we will use the triple-scaling limit to derive the low-energy regime of the Hamiltonian $H_R+H_L$ that generates the evolution of the TFD state perturbed by an operator insertion. We will express it in terms of the Krylov complexity operator, which we have proved to be equal to chord number, therefore becoming the bulk length operator in this low-energy limit that takes us to the regime where DSSYK is dual to JT gravity.

Similarly to how we proceeded in section \ref{sect:Triple_scaling_operator}, we complement the triple-scaling of the Krylov space position variable \eqref{Triple_scaling_limit} with keeping the operator scaling dimension $\Delta$ fixed in $\widetilde{q}=e^{-\lambda \Delta}$ as $\lambda$ is sent to zero, so as to achieve a consistent low-energy Hamiltonian. With this prescription, the triple-scaled Lanczos coefficients $b^{TS}$ take the form:
\begin{align}
&b(x)=\frac{2J}{\sqrt{\lambda (1-q)}}\sqrt{(1-e^{-x/2})(1+\widetilde{q}e^{- x/2})}  \label{Triple_scaled_Lanczos_OTFD_line1}\\
    &\longmapsto b^{TS}(\widetilde{x}) =  \frac{2J}{\sqrt{\lambda (1-q)}}\sqrt{(1-2 \lambda e^{-\widetilde{x}/2})(1+2 \lambda e^{-\lambda \Delta - \widetilde{x}/2})} \label{Triple_scaled_Lanczos_OTFD_line2}\\
    &=b_0 (\lambda) - 2 \lambda J \Big( \Delta e^{-\widetilde{x}/2} + 2 e^{-\widetilde{x}} \Big) + \mathit{O}(\lambda^2)~, \label{Triple_scaled_Lanczos_OTFD_line3}
\end{align}
where $b_0(\lambda) = \frac{2J}{\lambda}+\mathit{O}(\lambda^0)$ will be related to the ground state energy just like in section \ref{sect:Triple_scaling_operator}, as we will shortly see. The triple-scaled Hamiltonian, which we denote $H^{(+)}$, can be derived out of the triple-scaled Lanczos coefficients in the usual manner:
\begin{equation}
    \label{Triple_scaled_OTFD_Ham_def}
    -H^{(+)}\equiv H_R+H_L = e^{i\lambda \widetilde{k}} b^{TS}(\widetilde{x}) + b^{TS}(\widetilde{x}) e^{-i\lambda \widetilde{k}}~.
\end{equation}
Where, just like in section \ref{sect:Triple_scaling_operator}, the minus sign in the definition of $H^{(+)}$ is necessary in order to achieve a low-energy Hamiltonian bounded from below. Plugging \eqref{Triple_scaled_Lanczos_OTFD_line3} in \eqref{Triple_scaled_OTFD_Ham_def} we obtain:
\begin{equation}
\label{Triple_scaled_Hamiltonian_OTFD_result}
H^{(+)} - E_0(\lambda) = 4\lambda J \left( \frac{\widetilde{k}^2}{2} + \Delta e^{-\widetilde{x}/2} + 2 e^{-\widetilde{x}} \right) + {\cal O}\left( \lambda^2  \right)~, 
\end{equation}
where the ground-state energy is $E_0(\lambda) = -2 b_0(\lambda)$. This is the same gravitational Hamiltonian that was derived in \cite{Lin:2022rbf} out of the triple-scaling of the total chord number operator in the case of symmetric configurations. In our analysis, we proved that the Krylov complexity operator for the state $|\psi(t)\rangle=e^{-it(H_R+H_L)}|0,0\rangle$ is equal to the total chord number operator restricted to the Krylov space explored by the latter state and, consistently, the low-energy Hamiltonian expressed in terms of the triple-scaled K-complexity operator $\widetilde{x}$, derived out of the triple-scaled Lanczos sequence, has been found to be given by the expected form. This is a strong consistency check of the equivalence between total chord number and Krylov complexity in the setup at hand. The triple-scaled Hamiltonian \eqref{Triple_scaled_Hamiltonian_OTFD_result} is the gravitational Hamiltonian describing the bulk dual of the state $|\psi(t)\rangle$, where the Krylov complexity (or total chord number) variable $\widetilde{x}$ plays the role of bulk length normalized by the AdS length $l_{AdS}$.

For reference, let us denote the triple-scaled potential appearing in the Hamiltonian \eqref{Triple_scaled_Hamiltonian_OTFD_result} as $\mathcal{V}(\widetilde{x})$, where:
\begin{equation}
    \label{Triple_scaled_potential_OTFD}
    \mathcal{V}(\widetilde{x}) = 8\lambda J e^{-\widetilde{x}} + 4\lambda J \Delta e^{-\widetilde{x}/2}~.
\end{equation}
We can see that, when $\Delta=0$, the potential \eqref{Triple_scaled_potential_OTFD} reduces to the Liouville potential of \eqref{H_triple_scaled} describing bulk length in JT without matter, times an overall factor of two since $H_R+H_L\equiv 2H_{L,R}$ when the operator is the identity. More interestingly, we note that when $\Delta>0$ the operator dependence enters consistently in the low-energy Hamiltonian, namely through a term of order $\lambda$ in \eqref{Triple_scaled_Hamiltonian_OTFD_result}, where the net effect of the operator insertion is to lift the matterless Liouville potential by adding to it a term (of order $\lambda$) controlled by $\Delta$, see figure \ref{fig:Triple_scaled_OTFD_potential}. From \eqref{Triple_scaled_potential_OTFD}, we estimate the energy inserted by the operator (i.e. the shockwave energy from the bulk perspective) as the value of the shift in the potential energy at $\widetilde{x}=0$:
\begin{equation}
    \label{shock_energy}
    E_{op}\equiv E_{shock} = 4J\lambda \Delta~.
\end{equation}
That is, we find that the bulk dual of the operator insertion is a shock of energy $E_{shock}=\mathit{O}(\lambda)$. In \cite{Rabinovici:2023yex} we found that, in the absence of matter, the bulk dual of the triple-scaled Hamiltonian is a two-dimensional black hole with temperature $T=\frac{\lambda J}{\pi}$, which is also of order $\lambda$ (consistently with the low-energy limiting procedure). Combining both results, we have that $E_{shock}\propto T$ which
means that the matter generating the shockwave in the bulk has an energy within the typical range set by the black hole temperature (cf. \cite{Shenker_2014}).

\begin{figure}
    \centering
    \includegraphics[width=0.5\linewidth]{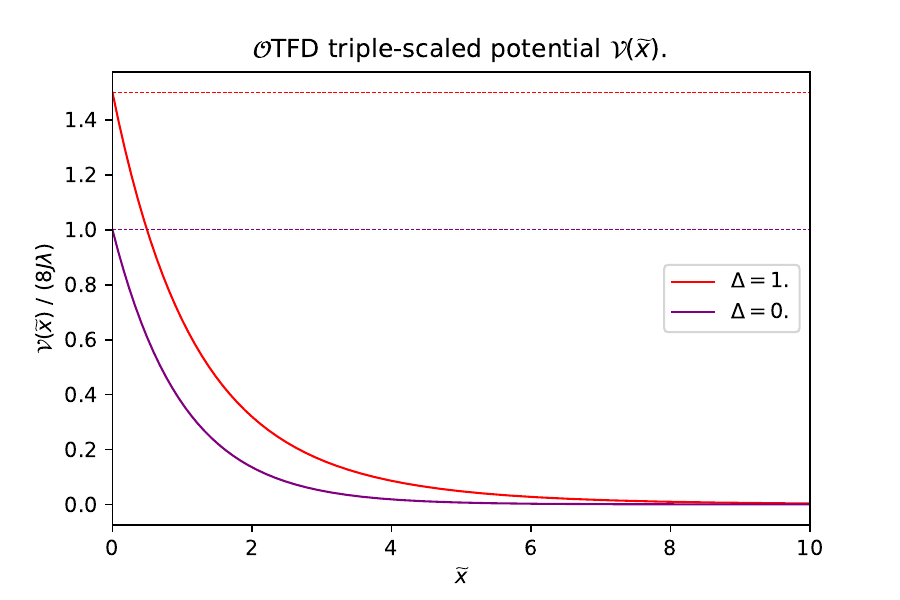}
    \caption{Triple-scaled potential \eqref{Triple_scaled_potential_OTFD} for the $\mathcal{O}TFD$ state as a function of the regularized Krylov complexity variable (shown to be equal to chord number and, in turn, to bulk length in the JT regime). The plot depicts two instances of the potential, one where $\Delta=0$ and it is given by (twice) the matterless Liouville potential studied in \cite{Rabinovici:2023yex}, and one where $\Delta=1$. Dashed horizontal lines mark the maximum value of the potential in each case. We interpret their difference as due to the energy injected in the system by the operator insertion, to be understood as the bulk shockwave energy as stated in \eqref{shock_energy}.}
    \label{fig:Triple_scaled_OTFD_potential}
\end{figure}

The specific bulk computation matching the solutions of the equation of motion dictated by the Hamiltonian \eqref{Triple_scaled_Hamiltonian_OTFD_result} needs nevertheless a careful analysis. Such a Hamiltonian, describing the evolution of the TFD state perturbed by an operator inserted exactly at $t=0$, has been derived in a small-$\lambda$ where $\Delta$ is fixed (i.e. not parametrically scaled with $\lambda$). We return to the issue of bulk matching in the discussion section which we now turn to.

\section{Discussion}
Let us begin by recapping what has been achieved in this paper. Making use of chord-diagram technology and by exploiting a suitable semi-classical limit, where $\lambda \rightarrow 0$, we were able to obtain the Lanczos coefficients and associated complexity profiles of two different complexity measures in DSSYK. These, together with the main results pertaining to each are summarized in the table below:

\begin{center}
\begin{tabular}{|p{7cm}|p{7cm}|}
\hline
    \textbf{Operator K-complexity} & ${\cal O}|$TFD$\rangle$ \textbf{K-complexity}  \\ \hline
    $\mathcal{O}\ket{0}=\ket{0,0}$ seed & $\mathcal{O}\ket{0}=\ket{0,0}$ seed  \\
    $H = H_R - H_L$ evolution & $H = H_R + H_L$ evolution \\ \hline
    Lanczos coefficients: & Lanczos coefficients:\\
    $b_n = \frac{2J}{\sqrt{\lambda}} \sqrt{\frac{1-q^{n/2}}{1-q} \left( 1-\tilde q q^{n/2}  \right)}$ & 
    $b_n^+ = \frac{2J}{\sqrt{\lambda}} \sqrt{\frac{1-q^{n/2}}{1-q} \left( 1 + \tilde q q^{n/2}  \right)}$ \\ \hline
    Krylov basis states are eigenstates of total chord number: &    Krylov basis states are eigenstates of total chord number:\\
    $\ket{\psi_n}=\frac{J^n}{b_1\dots b_n}\sum_{k=0}^n (-1)^k\binom{n}{k}\ket{k,n-k}$& $\ket{\psi_n^+}=\frac{J^n}{b_1^+\dots b_n^+}\sum_{k=0}^n \binom{n}{k}\ket{k,n-k}$
    \\ \hline
    Operator complexity in the semiclassical limit: &  ${\cal O}|$TFD$\rangle$-complexity in the semiclassical limit: \\ \parbox{5cm}{\begin{align*}
    \lambda C_K(t) =&  2 \log \Big[ 1 + (1-\tilde q) \sinh^2 Jt\Big],
    \end{align*}} & \parbox{5cm}{\begin{align*}
    \lambda C_K^+(t) =&  2 \log \Big[ 1 + (1+\tilde q) \sinh^2 \left(Jt\right)\Big],
    \end{align*}}\\
    when $\Tilde{q}\to 1$ has long exponential behavior & No exponential behavior\\
\hline
\end{tabular}
\end{center}

The semiclassical, $\lambda \rightarrow 0$, results in the table above are understood to be valid for operator insertions with fixed $\widetilde{q}$. Let us now move on to discuss the matching of these results to bulk geometric quantities, in the relevant limit, that is the triple-scaled limit of the quantities in the table.  

\subsubsection*{A bulk dual of operator complexity}
The results above identify in both cases the Krylov basis as a certain linear combination of \textit{states of fixed chord number}, which strongly suggests a direct relation with bulk length, similar to what was shown for the matterless case in \cite{Rabinovici:2023yex}, reducing to the length of a two-boundary wormhole in the triple-scaling limit. More concretely one would expect that the light $\mathcal{O}-\mathcal{O}$ matter chord in DSSYK translates to a shockwave insertion in JT gravity, which deforms the matterless wormhole length to account for the presence of the matter insertion. With this in mind, we can try to identify the geodesic length which geometrizes Krylov operator complexity in JT gravity with matter, leaving the detailed calculation for future work \cite{Rabinovici:2025ToAppear}. The ${\cal O}$TFD state is the result of
the operator ${\cal O}$ 
acting on the TFD, which is represented geometrically by the two-dimensional black hole in JT gravity \cite{Mertens:2022irh}, which in turn can be obtained by a dimensional reduction of the BTZ black hole. As has proven useful in similar calculations of the OTOC \cite{Shenker_2014}, the infalling matter can be approximated as a shockwave deformation of the black hole itself, which again can be constructed as a dimensional reduction of the BTZ-shockwave shell geometry \cite{Shenker_2014,Marco2023}. The corresponding geometry is obtained by matching the 2D black hole of the matterless case \cite{Sanchez-Garrido:2024pcy} to an analogous solution with added energy, $4\lambda J \Delta$, where we identify bulk parameters as

\begin{equation}
      2\lambda J =\frac{r_s}{l_{AdS}^2}\qquad\qquad  \Delta\sim\frac{E}{M} \,,
\end{equation}
where $M$ is the mass of the BTZ black before dimensional reduction. Generally, the above procedure can only be expected to provide a bulk geometric representation of the late-time behavior of operator Krylov complexity, for light insertions, owing to the approximations inherent in the shockwave geometry. It would be interesting to study the full complexity profile from a dual geometric perspective, by considering a solution that corrects the crude shockwave approximation at early times. In particular, it would be very interesting to describe early time scrambling dynamics in JT gravity out of Krylov complexity in the adequate low-energy limit, which might imply refining the triple-scaling limit studied in section \ref{sect:Triple_scaling_operator}. Scrambling dynamics in JT gravity have previously been discussed in works such as \cite{Jensen:2016pah}, by coupling the Schwarzian action to suitable conformal matter and evaluating the OTOC.

\subsubsection*{Outlook}
The detailed results and complexity profiles obtained add to the set of explicit examples of Krylov complexities in non-trivial interacting models with holographic duals. However, since the very definition of the model via a double scaling $N,p\rightarrow \infty$ involves a thermodynamic type limit, we have not been able to capture finite Hilbert-space size type effects in the present study, that is to say we have not been able to probe the saturation regime \cite{Rabinovici:2020ryf, Rabinovici:2022beu} of either type of complexity. It would be extremely interesting to try and extend the study of Krylov complexity, and especially its bulk dual into the saturation regime, possibly along the lines of the work of \cite{Iliesiu:2021ari,Iliesiu:2024cnh}, in order to complete the dictionary for finite systems.

Another interesting direction to consider is the relation of our results to the chord algebra of \cite{Lin:2022rbf, Kolchmeyer:2023gwa,lin2023symmetry, Xu:2024hoc}. It would be interesting to see if the fixed-chord number binomial states we have introduced in this paper play a special algebraic role, and whether such an algebraic understanding would enable us to push further into the multi-particle sector of the Hilbert space. The recent paper \cite{Xu:2024gfm} contains a number of relevant results that link the binomial states identified in our work as the Krylov basis to the algebra of DSSYK. 

Finally, it would be rewarding to elaborate further on the behavior of Krylov complexity if time-folds are introduced into the evolution contour, and in particular to address the switchback effect for operator Krylov complexity.

\section*{Acknowledgements}
We would like to thank Martí Berenguer, Felix Haehl, Javier Mas, Alexey Milekhin, Radu Moga, Pietro Pelliconi, Alfonso Ramallo and Juan Santos-Suárez for helpful conversations and comments. 

MA would like to thank Scuola Normale Superiore and University of Pisa for support, where part of this work was presented as a Master's thesis defense in October 2023. The research of MA, ASG and JS is supported in part by the Fonds National Suisse de la Recherche Scientifique (Schweizerischer Nationalfonds zur Förderung der wissenschaftlichen Forschung) through Project Grant 200021\_215300, and and the National Center for Competence in Research, NCCR51NF40-141869 ``The Mathematics of Physics (SwissMAP)". ASG is supported by the UKRI Frontier Research Guarantee [EP/X030334/1]. 

\newpage
\appendix

\section{Details on the Lanczos algorithm for arbitrary \texorpdfstring{$q$}{TEXT} and \texorpdfstring{$\widetilde{q}$}{TEXT}}\label{appx:Details_Lanczos}

This Appendix collects some finer details on the analytical study of the Lanczos algorithm performed in section \ref{subsect:KrylovBasis} for fixed system and operator parameters $q$ and $\tilde{q}$, given the initial state $|\psi_0\rangle = |0,0\rangle$ and the total Hamiltonian $H_R-H_L$. Here we set $J=1$ to simplify our notation, notice that, in order to reinstate it and obtain expressions analogous to those of the main text of the paper, it is sufficient to send $b_n\to b_n/J$.

Let us begin by providing the explicit derivation of the first few Krylov elements $|\psi_n\rangle$ for $n=1,2,3$ which, as announced in section \ref{subsect:KrylovBasis}, fulfill the binomial Ansatz \eqref{Binomial_Ansatz}. The iterative implementation of the corresponding Lanczos steps is the following:

\begin{enumerate}
\setcounter{enumi}{-1}
    \item $|\psi_0\rangle = |0,0\rangle$ with $\langle 0,0| 0,0\rangle=1$
    \item $|\psi_1\rangle$:
    \begin{itemize}
        \item$|A_1\rangle = (H_R-H_L)|0,0\rangle -\langle 0,0|H_R-H_L|0,0\rangle |0,0\rangle = |0,1\rangle - |1,0\rangle$
    where in the final step we used $H_R-H_L= a_R^\dagger - a_R - a_L^\dagger-a_L$ and that $a_{L/R}|0,0\rangle=0$, as well as the property of the inner product that it is zero if $n_L'+n_R'\neq n_L+n_R$ which annihilates $\langle 0,0|H_R-H_L|0,0\rangle$, since $H_R-H_L$ changes either $n_L$ or $n_R$. 
    \item The first Lanczos coefficient is then given by
    \begin{align}
    \label{b1}
        b_1^2 &= \langle A_1|A_1\rangle = (\langle 0,1| - \langle 1,0|)(|0,1\rangle - |1,0\rangle) \nonumber\\
        &= \langle 0,1|0,1\rangle - \langle 0,1|1,0\rangle - \langle 1,0|0,1\rangle + \langle 1,0|1,0\rangle = 2(1-\tilde{q}) = \mu_2
    \end{align}
    where we used the inner products computed in the examples below the recurrence relations. Note that we found the same result as $\mu_2$ in \eqref{mu2}.
    \item Thus, the first Krylov element is given by
    \begin{equation}
        |\psi_1\rangle = \frac{1}{\sqrt{2(1-\tilde{q})}}(|0,1\rangle - |1,0\rangle)
    \end{equation}
    \end{itemize}
    
    \item $|\psi_2\rangle$:

    \begin{itemize}
        \item We start by constructing the non-normalized Krylov vector:
    \begin{align}
        |A_2 \rangle &= (H_R-H_L)|\psi_1\rangle - \langle \psi_1|(H_R-H_L)|\psi_1\rangle\, |\psi_1\rangle - \langle \psi_0|(H_R-H_L)|\psi_1\rangle\, |\psi_0\rangle \\
        &= (H_R-H_L)|\psi_1\rangle -b_1 |\psi_0\rangle
    \end{align}
    where in the second line we used the fact that $\langle \psi_1|(H_R-H_L)|\psi_1\rangle=0$ (because $H_{L/R}$ change the total $n_L+n_R$) and the definition of $b_1$. We now compute: 
    \begin{align}
        (H_R-H_L)|\psi_1\rangle &= \frac{1}{b_1}(a_R^\dagger +a_R-a_L^\dagger -a_L)(|0,1\rangle - |1,0\rangle)\\
        &=\frac{1}{b_1}[|0,2\rangle -2|1,1\rangle+|2,0\rangle+2(1-\tilde{q})|0,0\rangle]\\
        &=\frac{1}{b_1}[|0,2\rangle -2|1,1\rangle+|2,0\rangle]+b_1|0,0\rangle
    \end{align}
    where it is important to remember that $a_{L/R}$ are \textit{not} simply annihilation operators, but are rather given by \eqref{aL} and \eqref{aR}.  Thus we find
    \begin{equation}
        |A_2\rangle = \frac{1}{b_1}(|0,2\rangle -2|1,1\rangle+|2,0\rangle)
    \end{equation}
    and 
    \begin{equation}
        b_2^2 = \langle A_2|A_2\rangle = \frac{1}{b_1^2}[2 \langle 0,2|0,2\rangle-8\langle 0,2|1,1\rangle +2\langle 0,2|2,0\rangle +4\langle 1,1|1,1\rangle]
    \end{equation}
    where we used the right-left symmetry to gather terms. Using the recursion relations one can compute that
    $\langle 0,2|0,2\rangle=1+q$, $\langle 0,2|1,1\rangle= \tilde{q}(1+q)$, $\langle 0,2|2,0\rangle=\tilde{q}^2(1+q)$ and $\langle 1,1|1,1\rangle=1+\tilde{q}^2q$, and we find that:
    \begin{equation}
    \label{b2}
        b_2^2= 3 + q - \tilde{q} (1 + 3 q)~.
    \end{equation}
    It can be checked that this is precisely the result gotten from the moments \eqref{mu2} and \eqref{mu4} according to the relationship $b_2^2 = \mu_4/\mu_2-\mu_2$.

    The 2nd Krylov element is thus
    \begin{equation}
        |\psi_2\rangle = \frac{1}{b_1 b_2}\left(|0,2\rangle -2|1,1\rangle+|2,0\rangle\right)
    \end{equation}
    \end{itemize}

    \item $|\psi_3\rangle$:
    \begin{itemize}
        \item The non-normalized Krylov vector is
    \begin{align}
        |A_3\rangle &= (H_R-H_L)|\psi_2\rangle - b_2|\psi_1\rangle \\
        &= \frac{1}{b_1 b_2}(|0, 3\rangle - 3 |1, 2\rangle + 3 |2, 1\rangle - |3, 0\rangle)
    \end{align}
    \item And the next Lanczos coefficient is:
    \begin{equation}
    \label{b3}
        b_3^2 = \frac{(1 + q) (10 + (-5 + \tilde{q}) \tilde{q} + 
   q + (-14 + \tilde{q}) \tilde{q} q + (1 + 5 \tilde{q} (-1 + 2 \tilde{q})) q^2)}{b_2^2}~.
    \end{equation}
    This is the same result one gets using the moments from the operator chord diagrams, see Section \ref{Sec:Moments_from_chord_diagrams}.
    \item The next Krylov element is then given by
    \begin{equation}
        |\psi_3\rangle = \frac{1}{b_1 b_2 b_3}\left(|0,3\rangle -3|1,2\rangle+3|2,1\rangle -|3,0\rangle\right)~.
    \end{equation}
    \end{itemize}
\end{enumerate}

As announced in section \ref{subsect:KrylovBasis}, one can understand why the binomial Ansatz $|\psi_n\rangle$ \eqref{Binomial_Ansatz} is correct for $n=0,\dots,3$ in terms of the symmetry \eqref{Ckn_symmetry} of the coefficients $c_k(n)$ introduced in \eqref{Ckn_simplified}:
 \begin{itemize}
     \item It correctly recovers the seed of the Lanczos algorithm: $|\psi_0\rangle = |0,0\rangle$.
     \item In order to study the case of $|\psi_1\rangle$, we note that due to the boundary condition of the Lanczos recursion, $|\psi_{-1}\rangle = \mathbf{0}$, setting $n=0$ in \eqref{Lanczos_Induction_Anplus1_line1} we find directly that $|A_1\rangle=(H_R-H_L)|\chi_0\rangle = (a_R^\dagger - a_L^\dagger)|\chi_0\rangle = |\chi_1\rangle$. This is formally compatible with the fact that the domain $k=0,\dots,n-1$ for the sum in \eqref{Binom_Ansatz_action_of_aRminusaL} is empty.
     \item More interestingly, $|\psi_2\rangle$ is addressed by setting $n=1$ in the condition \eqref{Binom_Ansatz_necsuf_condition}. For $n=1$, there is only one value of $k$ summed over in the $k$-sum of \eqref{Binom_Ansatz_action_of_aRminusaL}, namely $k=0$, so $c_0(1)$ can trivially be seen as an overall prefactor, which does coincide with the Lanczos coefficient $b_1^2$ given in \eqref{b1_Exact}, namely:
     \begin{eqnarray}
    c_0(1) = 2(1-\tilde{q}) = b_1^2, \quad n=1 \label{C_0(1)_b1_squared}
\end{eqnarray}
    \item For $n=2$, the $k$-sum in \eqref{Binom_Ansatz_action_of_aRminusaL} only ranges in $k=0,1$, and precisely the symmetry \eqref{Ckn_symmetry} ensures $c_0(2)=c_1(2)$, so that $c_k(2)$ is constant within the $k$-sum and can indeed be pulled out as a prefactor, which does coincide with the coefficient $b_2^2$ given in \eqref{b2_Exact}:
    \begin{eqnarray}
    c_0(2) = c_1(2) = 3 + q - \tilde{q} (1 + 3 q) = b_2^2~, \quad n=2 \label{C_k(2)_b2_squared}
\end{eqnarray}    
    This implies the cancellation \eqref{Binom_Ansatz_necsuf_condition} for $n=2$, thus ensuring that $|\psi_3\rangle$ is still of the form \eqref{Binomial_Ansatz}.
    \item In order to assess $|\psi_4\rangle$ we need to consider $c_{k}(n)$ for $n=3$. In this case we still have \eqref{Ckn_symmetry} ensuring $c_0(3)=c_2(3)$, but these need not be equal to $c_1(3)$, and in fact they are not:
    \begin{eqnarray}
    c_0(3) = 4 + q + q^2 - \tilde{q} (1 + q + 4 q^2) = c_2(3) \label{ckn_n3_k_0_2}\\
    c_1(3) = -3 (1 + q) (-1 + \tilde{q} q) \neq c_0(3), c_2(3) \label{ckn_n3_k1} ~.
    \end{eqnarray}
    Since, for fixed $n=3$, the sum \eqref{Binom_Ansatz_action_of_aRminusaL} ranges in $k=0,1,2$, we conclude that this time $c_k(n=3)$ can no longer be pulled out of the sum as a prefactor, invalidating \eqref{Binom_Ansatz_necsuf_condition} for $n=3$. This implies that $|A_4\rangle$ in \eqref{Lanczos_Induction_Anplus1_line1}, and therefore $|\psi_4\rangle$, will not satisfy the binomial Anstaz\footnote{A comment on the indexing might be helpful: $c_k(n)$ is the coefficient that probes the ``deviation'' of $(a_R-a_L)|\chi_n\rangle$ from $|\chi_{n-1}\rangle$, and the $k$-domain ranges for $k=0,\dots,n-1$. If given a fixed $n$ we have that $c_k(n)$ is constant in $k$, $c_k(n)\equiv c(n)$ $\forall k=0,\dots n-1$, then the condition \eqref{Binom_Ansatz_necsuf_condition} is fulfilled, and hence $|\psi_{n+1}\rangle$ is granted to fulfill the binomial Ansatz \eqref{Binomial_Ansatz}. In this case, however, $c(n)$ gives the Lanczos coefficient $b_n^2$, rather than $b_{n+1}^2$. For instance, the fact that $c_k(2)$ is constant in $k=0,1$ implied that $|\psi_3\rangle$ also follows the binomial Ansatz, but the Lanczos coefficient $b_3^2$ actually has nothing to do with $c_k(3)$, because for $n=3$ $c_k(n)$ is no longer constant in $k$!} \eqref{Binomial_Ansatz}, but will instead be given by a linear combination of $|\chi_{4}\rangle$ and a non-zero element of $\mathcal{H}_{1p}^{(2)}$.
 \end{itemize}

\section{Small-\texorpdfstring{$\lambda$}{TEXT} analysis for heavy operators}\label{appx:small_lambda} 

Section \ref{subsect:Lanczos_Limit_lambda0_qtfixed} showed that in the limit where $\lambda\to 0$ keeping $\widetilde{q}=e^{-\Delta \lambda}$ fixed (which implies taking a large operator dimension $\Delta\to +\infty$ such that the product $\Delta\lambda$ is fixed) the binomial Ansatz for the Krylov basis elements $|\psi_n\rangle$ is correct for every $n\in\mathbb{N}_0$, admitting a simple analytic expression for the Lanczos coefficients, cf. \eqref{Lanczos_closed_form_solution_in_limit}. In this appendix we discuss the leading corrections to this result given by a systematic expansion around $\lambda=0$ where $\widetilde{q}$ is still kept fixed. In the two-dimensional parameter space given by the tuple $(\lambda,\Delta)$ this procedure corresponds to moving along one of the hyperbolas in figure \ref{fig:ParameterSpace}. To begin, it is interesting to note that  $c_k(n)$ remains $k$-independent to first order in $\lambda$:
\begin{eqnarray}
\label{ckn_expansion_lambda_fixed_qt}
    c_k(n) = 2n(1-\tilde{q}) + \frac{1}{2}n (n-1)(3\tilde{q}-1) \lambda + O(\lambda^2)~,
\end{eqnarray}
where the $k$-dependence starts at second order in $\lambda$. Notice that also here we will set $J=1$.

In order to perform a small-$\lambda$ analysis of the deviations from the binomial Ansatz \eqref{Binomial_Ansatz} at fixed $\widetilde{q}$ we shall take as a starting point the exact Krylov elements $|\psi_n\rangle$ for $n=0,1,2,3$, which do agree with such an Ansatz as argued in section \ref{subsect:KrylovBasis}, together with their exact Lanczos coefficients \eqref{b1_Exact}-\eqref{b3_Exact}. From there, we may go further in the Lanczos algorithm:
\begin{eqnarray}
\label{A4_Lanczos_step_small_lambda_analysis}
    \ket{A_4} &=& (H_R-H_L)\ket{\psi_3} -b_3 \ket{\psi_2} \nonumber\\
    &=& \frac{1}{b_1 b_2 b_3}\big\{\, \ket{\chi_4} + \sum_{k=0}^2[c_k(3)-b_3^2] (-1)^k \, \binom{2}{k} |k,2-k\rangle \big\}~.
\end{eqnarray}
It turns out that the difference in the square brackets is second order in $\lambda$:
\begin{eqnarray}
\label{ck3_minus_brsquared_small_lambda}
    c_k(3)-b_3^2 =\frac{1}{2}\,  (-1)^k \, (1 - \tilde{q}) \lambda^2 + O(\lambda^3), \quad k=0,1,2~,
\end{eqnarray}
where the $k$-dependence enters only at the third order in $\lambda$. This implies that:
\begin{eqnarray}
\label{A4_minus_binom_lambda_expans}
    \ket{A_4} - \frac{1}{b_1 b_2 b_3 } \ket{\chi_4}  = O(\lambda^2)~,
\end{eqnarray}
where, more formally speaking, the right-hand side stands for a vector whose norm is of order $\lambda^2$. Indeed, by inspection of the chord inner-product recursion \eqref{inner_rec_start} one can conclude that the $\lambda$-dependence of the overlaps between basis elements of a given total chord number sector, $\langle k^\prime, n-k^\prime|k,n-k\rangle$, always starts at order $\lambda^0$ for fixed $\widetilde{q}$, hence allowing to read off the leading $\lambda$-dependence of the norm of the difference \eqref{A4_minus_binom_lambda_expans} from the $c_k(3)-b_3^2$ factor in \eqref{A4_Lanczos_step_small_lambda_analysis}. Let us now check how this affects the next Lanczos coefficient:
\begin{eqnarray}
    b_4 ^2 &=& \langle A_{4} | A_{4}\rangle = \frac{1}{b_1^2 b_2^2 b_3^2}\Big\{\langle \chi_4 | \chi_4\rangle \nonumber \\
    &&+ \sum_{k,m=0}^2 [c_k(3)-b_3^2][c_m(3)-b_3^2](-1)^{k+m} \binom{2}{k} \binom{2}{m} \langle m,2-m |k,2-k\rangle\Big\}
\end{eqnarray}
where we used the fact that the inner product is zero unless $n_L'+n_R'=n_L+n_R$, which makes the cross terms vanish. Again noting that the overlaps $\langle m,2-m |k,2-k\rangle$ start at order $\lambda^0$, we find that the difference between the Lanczos coefficient and the norm of the binomial state is controlled by $\lambda$ as dictated by \eqref{ck3_minus_brsquared_small_lambda} as follows:
\begin{eqnarray}
    b_4^2 - \frac{1}{b_1^2 \, b_2^2 \, b_3^2} \langle \chi_4 | \chi_4\rangle  = O(\lambda^4) 
\end{eqnarray}

Since the expansion \eqref{ckn_expansion_lambda_fixed_qt} is valid for arbitrary $n$, we may iterate this argument and conclude that generically for fixed $\tilde{q}$ and given some $n\in\mathbb{N}_0$, the following statements hold:
\begin{eqnarray}
    |A_n\rangle - \frac{1}{b_1\dots b_{n-1}}|\chi_n\rangle = \mathit{O}(\lambda^2)~, \label{An_elements_small_lambda}\\ b_n^2 - \frac{1}{b_1^2\dots b_{n-1}^2}\langle\chi_n|\chi_n\rangle = \mathit{O}(\lambda^4)~, \label{bn_coeffs_small_lambda}
\end{eqnarray}
where the line \eqref{An_elements_small_lambda} is telling that the $n$-th non-normalized Krylov vector fulfills the binomial Ansatz up to a vector whose norm is of order $\lambda^2$, for every $n\in\mathbb{N}_0$, and where the second line \eqref{bn_coeffs_small_lambda} relates the square of the $n$-th Lanczos coefficient to the norm of the binomial states. Both expressions may be recast more compactly as follows:

\begin{eqnarray}
    |A_n\rangle - \frac{1}{\sqrt{\langle \chi_{n-1}|\chi_{n-1}\rangle}}|\chi_n\rangle = \mathit{O}(\lambda^2)~, \label{An_elements_small_lambda_compact}\\ b_n^2 - \frac{\langle\chi_n|\chi_n\rangle}{\langle \chi_{n-1}|\chi_{n-1}\rangle} = \mathit{O}(\lambda^4)~, \label{bn_coeffs_small_lambda_compact}
\end{eqnarray}

Now, recalling \eqref{adagger_gives_binomial} and \eqref{Binom_Ansatz_action_of_aRminusaL}, together with the expansion \eqref{ckn_expansion_lambda_fixed_qt} we can see that:
\begin{equation}
    \label{binomial_norm_small_lambda}
    \langle \chi_n|\chi_n\rangle = \langle \chi_{n-1}|(a_R-a_L)|\chi_n\rangle = \Big( 2n(1-\tilde{q}) + \frac{1}{2}n (n-1)(3\tilde{q}-1) \lambda \Big)\langle \chi_{n-1}|\chi_{n-1}\rangle + \mathit{O}(\lambda^2)~,  
\end{equation}
which, together with \eqref{bn_coeffs_small_lambda_compact} leads to the sought subleading correction to the $b$-sequence:
\begin{equation}
    \label{Subleading_correction_to_bn}
    b_n = \sqrt{2n(1-\tilde{q}) + \frac{1}{2}n (n-1)(3\tilde{q}-1) \lambda + \mathit{O}(\lambda^2)} = \sqrt{2n(1-\tilde{q})} + \frac{n (n-1)(3\tilde{q}-1)}{4\sqrt{2n(1-\tilde{q})}}\lambda  + \mathit{O}(\lambda^2)~.
\end{equation}

A word of caution regarding the order of limits might be insightful at this point: This appendix (together with section \ref{subsect:Lanczos_Limit_lambda0_qtfixed}) has studied a limit where, while keeping $\widetilde{q}$ fixed, $\lambda$ is sent to zero for arbitrary $n\in\mathbb{N}_0$, which is also kept fixed. This means that, for every $n$, making $\lambda$ sufficiently small will make the Krylov elements $|\psi_n\rangle$ and Lanczos coefficients $b_n$ get arbitrarily close to the binomial Ansatz \eqref{Binomial_Ansatz} and to the limiting coefficients \eqref{Lanczos_closed_form_solution_in_limit}, respectively. However, the Taylor series coefficients that control these deviations in \eqref{An_elements_small_lambda_compact} and in \eqref{Subleading_correction_to_bn} at higher orders in $\lambda$ are $n$-dependent: This implies that, even if $\lambda$ is fixed to some numerically small value $\lambda\ll 1$, there will always exist a $\lambda$-dependent value of $n$ for which the corrections to the binomial Ansatz and the corresponding Lanczos coefficients will become of order one, invalidating the limiting form. 
Observing \eqref{Subleading_correction_to_bn} we can see that the coefficient of $\lambda$ goes like $\sim n^{3/2}$ (up to a numerical constant), and will become of $\mathit{O}(1)$ for $n\sim \lambda^{-3/2}$.

\begin{figure}
    \centering
    \includegraphics[width=0.45\textwidth]{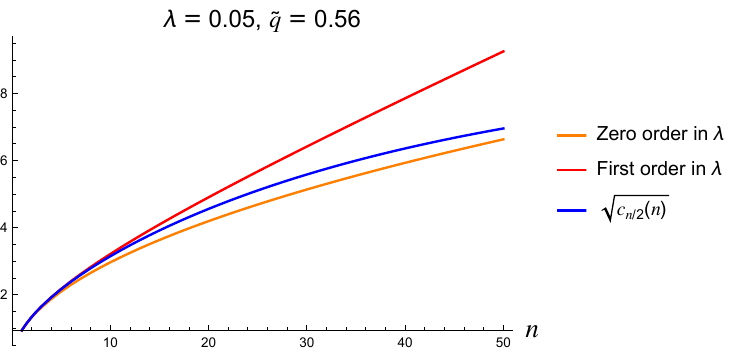}           ~~~~~~~~~~~~\includegraphics[width=0.45\textwidth]{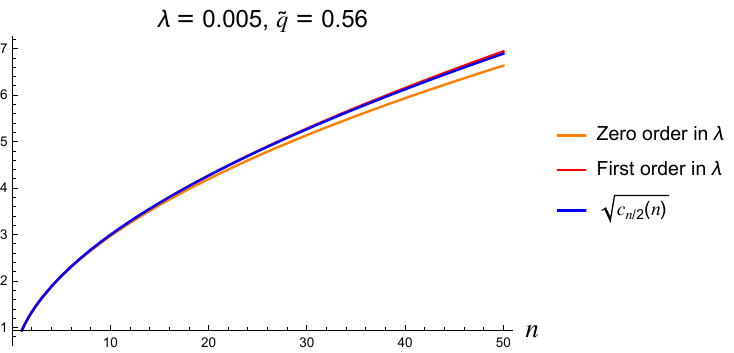}
    \caption{The coefficient $\sqrt{c_k(n)}$ is independent of $k$ for arbitrary $n$ up to (but not including) second order in a small-$\lambda$ expansion. These plots show a comparison between the zeroth and first order truncations of the Taylor series in powers of $\lambda$ of $\sqrt{c_k(n)}$, and the function $\sqrt{c_{n/2}(n)}$, plotted as a function of $n$, for $\lambda = 0.05$ (\textbf{left}) and for $\lambda=0.05$ (\textbf{right}). In both cases $\widetilde{q}$ is fixed to the numerical value $\widetilde{q}= 0.56$.}
    \label{fig:ckn_zeroth_and_first_orders_in_lambda_qtilde_fixed}
\end{figure}

In order to give an analytical expression for $b_n$ and $|\psi_n\rangle$ that remains robust upon increasing $n$ to arbitrary values, it will be useful to resort to the asymptotic (or semiclassical) limit presented in section \ref{subsect:bn_asymptotic_limit}. Such an analysis reveals that the limiting form of the Lanczos coefficients is in fact $b_n\sim \sqrt{c_{n/2}(n)}$. In figure \ref{fig:ckn_zeroth_and_first_orders_in_lambda_qtilde_fixed} we plot as a function of $n$ the value of this estimate, together with the zeroth and first order corrections in $\lambda$ of $\sqrt{c_k(n)}$, both of which are $k$-independent.

\section{Details on numerics}\label{appx:numerical_details}

This appendix will give further details on the numerical implementation of the model for the application of the Lanczos algorithm, presented in section \ref{appx:Numerics_setup}. A collection of further numerical results for various system and operator parameters is also gathered in section \ref{appx:Numerics_further_results}. For a specific discussion of the Lanczos algorithm and the partial re-orthogonalization (PRO) routine implemented in this project, the reader is referred to \cite{Rabinovici:2020ryf} (for an application of this routine to generic Hilbert spaces besides operator space, see \cite{Lanczos:1950zz,Sanchez-Garrido:2024pcy, parlett1998symmetric}).

\subsection{Setup}\label{appx:Numerics_setup}

This section of the appendix contains the details on how the objects relevant to the one-particle sector of the chord Hilbert space were built for the numerical computations presented in this paper.

\subsubsection{Truncation scheme and computational basis}\label{appx:numerics_subsect_truncationAndBasis}

As announced in section \ref{subsect:numerics}, for the instances in which the Lanczos algorithm was not solved using symbolic manipulations and high precision arithmetics, the strategy adopted consisted on building a finite-truncation of the one-particle Hilbert space, given in equation \ref{Hilbert_space_truncation}, which we restate here:
\begin{equation}
    \label{Hilbert_space_truncation_appx}
    \mathcal{H}_{1p;N} = \bigoplus_{n=0}^N \mathcal{H}_{1p}^{(n)}~.
\end{equation}
I.e., the truncated Hilbert space is the span of sectors of total chord number up to and including $N$. We may recall as well that each sector is spanned by a chord eigenbasis $|n_L,n_R\rangle$ with constant $n_L+n_R$, namely
\begin{equation}
    \label{Sector_span_basis_appx}
    \mathcal{H}_{1p}^{(n)}=\text{span}\left\{ |k,n-k\rangle~:~k=0,\dots,n \right\}~,
\end{equation}
its dimension being given by $\dim \mathcal{H}_{1p}^{(n)}$. The dimension of the truncation is $d_N = (N+1)(N+2)/2$, as computed in \eqref{truncation_dimension}, and we can generalize this to the dimension of any subspace given by the span of all sectors up to (and including) the $n$-th sector, given analogously by
\begin{equation}
    \label{Dimension_up_to_n_appx}
    d_n = \sum_{m=0}^n \dim \mathcal{H}_{1p}^{(m)}=\frac{1}{2}(n+1)(n+2)~,
\end{equation}
which will be a useful quantity to keep track of.

Since computationally objects such as states and operators are represented, respectively, by vectors and matrices whose items represent coordinates with respect to some specific basis, let us now devote some words to the algorithmic prescription for encoding the two-label chord basis,
\begin{equation}
    \label{chord_basis_appx}
    \mathcal{B}_{\text{chord}}=\left\{|k,n-k\rangle~k=0,\dots,n\right\}_{n=0}^N
\end{equation}
in the computational basis,
\begin{equation}
    \label{comp_basis_appx}
    \mathcal{B}_{\text{comp}}=\left\{ \mathbf{e_i} \right\}_{i=0}^{d_N-1}~,
\end{equation}
where each vector $\mathbf{e_i}$ is computationally represented by a column with all zeros but a one at the $i$-th position (indexing starts at zero). The algorithm for encoding \eqref{chord_basis_appx} into \eqref{comp_basis_appx} consists on a bijective correspondence between the tuple $(n,k)$ labeling a state in the chord basis and the index $i$ that identifies computational basis elements. Such a bijection is possible if one thinks of a ``lexicographic'' arrangement of the tuples $(n,k)$, i.e. sorting them such that tuples with smaller $n$ are first and, for same $n$, those with smaller $k$ are first. More specifically, one side of the bijection takes a tuple $(n,k)$ and gives the corresponding index $i(n,k)$ of the vector $\mathbf{e_{i(n,k)}}$ that represents the state $|k,n-k\rangle$:
\begin{equation}
    \label{ifromnk}
    i(n,k)=d_{n-1} + k~,
\end{equation}
where we recall that indexing of the computational basis starts at zero. The reciprocal of \eqref{ifromnk} is a map that takes an index $i$ labeling a state $\mathbf{e_i}$ of $\mathcal{B}_{\text{comp}}$ and produces the tuple $\big(n(i),k(i)\big)$ that denotes the corresponding state $\big|k(i),n(i)-k(i)\big\rangle$ on $\mathcal{B}_{\text{chord}}$:
\begin{eqnarray}
    n(i) = \max\left\{ n~:~ i - d_{n-1}\geq 0 \right\}~, \label{nkfromi_n}\\
    k(i) = i - d_{n(i)-1}~. \label{nkfromi_k}
\end{eqnarray}
Note that composing \eqref{ifromnk} with \eqref{nkfromi_n}-\eqref{nkfromi_k}, or viceversa, yields correctly the identity map. Numerically, it is nevertheless preferable to build iteratively two dictionaries, $(n,k)\to i$ and $i \to (n,k)$, by looping through the tuples $(n,k)$ in the convenient order (i.e. nesting a $k$-loop inside an $n$-loop), updating a counter $i$ at every step, which is then used to define the corresponding entry in both dictionaries simultaneously. This spares performing the maximization required in \eqref{nkfromi_n} every time an item of the dictionary $i \to (n,k)$ is built.  

Once the interpretation of the computational basis $\mathcal{B}_{\text{comp}}$ in terms of the chord eigenstates of $\mathcal{B}_{\text{chord}}$ is clear, one can proceed to build operators $\mathcal{O}$ as matrices $(O_{ij})$ in coordinates over the computational basis, where such coordinates may be obtained as follows:
\begin{itemize}
    \item For every state in the chord basis, $|k,n-k\rangle$, compute the action of $\mathcal{O}$, and express it as a linear combination of states in $\mathcal{B}_{\text{chord}}$:
    \begin{equation}
        \label{O_action_chordbasis_appx}
        \mathcal{O}|k,n-k\rangle = \sum_{n^\prime = 0}^N \sum_{k^\prime = 0}^{n^\prime} \mathcal{O}^{nk}_{n^\prime k^\prime} |k^\prime,n^\prime-k^\prime\rangle~.
    \end{equation}
    In particular, note that $\mathcal{O}^{nk}_{n^\prime k^\prime} \neq \langle k^\prime,n^\prime-k^\prime | \mathcal{O} | k,n-k\rangle $ because the basis $\mathcal{B}_{\text{chord}}$ is not orthonormal with respect to the chord inner product (we will elaborate further on the inner product in section \ref{appx:numerics_inner_product}), so one really needs to construct the coordinates of the image of $|k,n-k\rangle$ under the map $\mathcal{O}$, with respect to the chord basis.
    \item For every pair of tuples $(n,k)$ and $(n^\prime,k^\prime)$ entering in \eqref{O_action_chordbasis_appx}, find the corresponding computational basis indices $i(n,k)$ and $i^\prime (n^\prime,k^\prime)$. 
    \item Assign the matrix elements $O_{i^\prime (n^\prime,k^\prime),i(n,k)}=\mathcal{O}^{nk}_{n^\prime k^\prime}$.
\end{itemize}
Note that the instructions may be seen as a generic prescription, rather than as a structured description of an algorithm. In practice, the construction of the matrix $(O_{ij})$ according the the rules given may be achieved by looping through the states $|k,n-k\rangle$ in the chord basis and assessing to what state(s) $|k^\prime,n^\prime - k^\prime\rangle$ they are mapped (and with what weight) by the operation $\mathcal{O}$, something that can only be done if one originally has access to the definition of $\mathcal{O}$ in terms of its action on the chord eigenstates, which is the case for all the operators of interest in this work.

With this prescription for encoding basis elements in the computational basis and representing operators as matrices in such a basis, we were able to construct numerically all the relevant operators of the chord algebra reviewed in section \ref{sec.ToolsTrade}, namely the number operators $n_L,n_R,n$, the ladder operators $a_L^\dagger,a_R^\dagger,\alpha_L,\alpha_R$, and the Hamiltonians $H_L,H_R$, with which the total Hamiltonians $H_R\pm H_L$, relevant for the different holographic prescriptions, may be built.

We were specifically interested in the time evolution of the state $|0,0\rangle$ generated, in the Schrödinger picture, by $H_R-H_L$, whose K-complexity describes operator growth. In order to solve the Lanczos algorithm in this context, one needs to construct the matrix representing $H_R-H_L$ in the computational basis, noting that the seed state $|0,0\rangle$ is encoded in the initial element of the computational basis, $\mathbf{e_0}$. But the last ingredient that remains to be implemented numerically is the inner product that defines the notion of orthogonality, which takes the form of a non-trivial tensor when expressed in coordinates over $\mathcal{B}_{\text{comp}}$, as we shall discuss in the next subsection.

\subsubsection{Inner product and normalized basis}\label{appx:numerics_inner_product}

The chord basis \eqref{chord_basis_appx} is not orthogonal with respect to the inner product \eqref{inner_rec_start}. Therefore, encoding this basis in the computational basis \eqref{comp_basis_appx} comes at the cost of numerically dealing with a non-canonical inner-product matrix. Even if the vectors $\mathbf{e_i}$ are numerically given by columns of ones and zeros, they are not orthonormal and we need to construct an inner-product matrix $g\equiv(g_{ij})$, where $g_{ij}=g(\mathbf{e_i},\mathbf{e_j})$, where $g$ represents the inner-product tensor. The computational prescription for building the matrix elements $g_{ij}$ is analogous to the one described in the previous section regarding the construction of operators, i.e. it boils down to performing the assignment
\begin{equation}
    \label{inner_product_matrix_assignment_appx}
    g_{ij} = \langle k(i),n(i)-k(i)|k(j),n(j)-k(j)\rangle~.
\end{equation}
The above assignment can indeed be performed directly using the closed-form solution of the inner product recursion given in \eqref{inner_product_recursion_solution}. However, this may not be the most efficient approach due to the need of performing a sum of $n(j)-k(j)$ terms; a more time-wise efficient construction (which comes at the cost of a higher propagation of numerical error) consists on the implementation of a \textit{memoized} recursive function that explicitly solves the recursion \eqref{inner_rec_start}. Memoizing a recursive function consists on supplementing it with a caché-like variable that gets progressively updated every time the function is called, storing already-computed values of the inner product so that they don't need to be recomputed every time a recursion step or a function call stumbles on them. In programming languages like Python, where function arguments are passed by assignment, such a caché may be chosen to be a variable of the dictionary type, whose items may be progressively updated inside the function, remaining accessible in the global memory scope. In other languages, like C++, the caché variable may need to be passed by reference to the recursive function.

In order to get rid of the numerically disturbing (q-)factorial growth of the overlaps between chord states, which boils down to the boundary condition \eqref{Inner_product_recursion_boundarycondition}, we chose to work, for the purpose of the numerical analyses only, on a \textit{renormalized} chord basis,
\begin{equation}
    \label{Basis_renormalization_for_numerics}
    |k,n-k\rangle \longmapsto \sqrt{[n]_q} |k,n-k\rangle~.
\end{equation}
This is analogous to the implicit renormalization of the matterless chord eigenbasis assumed throughout \cite{Rabinovici:2023yex}. Here, however, this procedure doesn't quite make the basis orthonormal, but it gets rid of a numerically annoying q-factorial growth of the overlaps with the total chord number. Without changing the notation (for the sake of clarity), let us rewrite the form of the inner-product recursion \eqref{inner_rec_start} and its boundary condition \eqref{Inner_product_recursion_boundarycondition} after subjecting them to the redefinition \eqref{Basis_renormalization_for_numerics}. The expressions below\footnote{Every state $|n_L,n_R\rangle$ may be denoted by a tuple $(n,k)$, where $k=n_L$ and $n=n_L+n_R$. For instance, the state $|k,n-k-1\rangle = |k,(n-1)-k\rangle$ corresponds to the tuple $(n-1,k)$, and so forth for all the states appearing in \eqref{Inner_prod_rec_appx_normalized_recstep}.} therefore only apply to the discussion in this appendix:
\begin{align}
&\langle k^\prime, n^\prime-k^\prime | k,n-l\rangle = 0 \quad\text{if}\quad n^\prime \neq n~, \label{Inner_prod_rec_appx_normalized_orthog}\\
&\langle k^\prime, n-k^\prime | k,n-l\rangle =  \nonumber\\&\frac{[k]_q}{[n]_q}\langle k^\prime-1, n-k^\prime|k-1,n-k\rangle + \tilde{q} q^k \frac{[n-k]_q}{[n]_q}\langle k^\prime-1,n-k^\prime | k,n-k-1\rangle~, \label{Inner_prod_rec_appx_normalized_recstep}\\
&\langle 0,n|k,n-k\rangle = \tilde{q}^k~. \label{Inner_prod_rec_appx_normalized_bdrycondit}
\end{align}
This new basis normalization does not affect the map between the indices characterizing states in $\mathcal{B}_{\text{chord}}$ and $\mathcal{B}_{\text{comp}}$ presented in section \ref{appx:numerics_subsect_truncationAndBasis}, as it only amounts to rescaling the states. It will however affect the expressions of the ladder operators $a_{L,R}^\dagger$ and $\alpha_{L,R}$, given in \eqref{aLdagger}-\eqref{aRdagger} and in \eqref{alphaL}-\eqref{alphaR}, respectively, because they are not diagonal in the chord basis; such expressions therefore need to be modified accounting for the basis renormalization \eqref{Basis_renormalization_for_numerics} in order to build the matrices representing those operators in coordinates over the renormalized basis. Nevertheless, operator identities such as the relation between $H_{L,R}$ and the ladder operators are basis-independent and therefore remain unchanged and applicable to any matrix representation of the operators as long as they are all representations with respect to the same basis.

The fact that the map encoding $\mathcal{B}_{\text{chord}}$ to $\mathcal{B}_{\text{comp}}$ preserves a notion of ordering of the successive total chord number sectors results in the fact that the matrix $(g_{ij})$ features a very neat block-diagonal structure, reflecting orthogonality of the different sectors. Hence, we may find it convenient to denote the blocks of the matrix as $g^{(n)}_{k^\prime k}$, containing the overlaps within a given total chord number sector:
\begin{equation}
    \label{inner_product_matrix_in_sector}
    g^{(n)}_{k^\prime k} := \langle k^\prime,n-k^\prime | k,n-k\rangle~.
\end{equation}
Now, recalling the construction of the chord inner product \cite{Lin:2022rbf}, it is possible to show that the $(n+1)\times (n+1)$ matrix $g^{(n)}$ at a given sector $n$ enjoys the two following symmetries:
\begin{itemize}
    \item The usual inner product symmetry (recall that the chord Hilbert space is a real Hilbert space):
    \begin{equation}
        \label{IPS_appx}
        \langle k^\prime,n-k^\prime | k,n-k\rangle = \langle k,n-k | k^\prime,n-k^\prime\rangle \quad \Longrightarrow\quad g^{(n)}_{k^\prime k} = g^{(n)}_{k k^\prime}~.
    \end{equation}
    \item Left-right symmetry:
    \begin{equation}
        \label{LRS_appx}
        \langle k^\prime,n-k^\prime | k,n-k\rangle = \langle n - k^\prime,k^\prime | n-k,k\rangle\quad\Longrightarrow\quad g^{(n)}_{k^\prime k} = g^{(n)}_{n-k^\prime,n- k}~.
    \end{equation}
\end{itemize}
In words, \eqref{IPS_appx} and \eqref{LRS_appx} imply that the matrix $g^{(n)}$ is symmetric with respect to both of its diagonals. The number of matrix elements that are independent in the matrix is therefore equal to the number of items in one of the four triangular sectors (upper, lower, right or left) in which the matrix is divided by the main and the secondary diagonals. The number of matrix elements belonging to a triangle (including its borders), which we shall denote $s(n)$, is obtained from a counting problem whose solution depends on whether $n$ is even or odd, since in the former case the $(g^{(n)}_{k^\prime k})_{k^\prime,k=0}^n$ matrix has a central item, whereas in the latter it doesn't. The generic solution is:
\begin{equation}
    \label{inner_number_indep_items}
    s(n) = \left(1 +\Big\lfloor \frac{n}{2} \Big\rfloor\right)\left(1 +\Big\lceil \frac{n}{2} \Big\rceil\right)= \left\{\begin{aligned} &\left(1+\frac{n}{2}\right)^2\qquad\;~ n\text{ even}~, \\ &\left(1+\frac{n}{2}\right)^2-\frac{1}{4}\quad n\text{ odd}~.
    \end{aligned}\right.
\end{equation}
Hence, correctly accounting for all the symmetries of the matrix $g^{(n)}$ will imply that the number of independent overlaps that need to be computed in each chord sector when numerically building the inner product is $s(n)$ given in \eqref{inner_number_indep_items}, rather than the full size of the block $g^{(n)}$, which is $(n+1)^2>s(n)$. It is a reduction of roughly a factor of four, i.e. $\lim_{n\to+\infty}\frac{s(n)}{(n+1)^2}=\frac{1}{4}$, as expected since only one triangular sector of the matrix is independent.

For illustration, figure \eqref{fig:inner_product_plots} depicts the inner-product matrix $g_{ij}$ for a Hilbert space truncation $N=50$. Both the block-diagonal structure due to orthogonality of chord sectors and the symmetries within a given chord sector are manifest.

\begin{figure}
    \centering
    \includegraphics[width=0.45\linewidth]{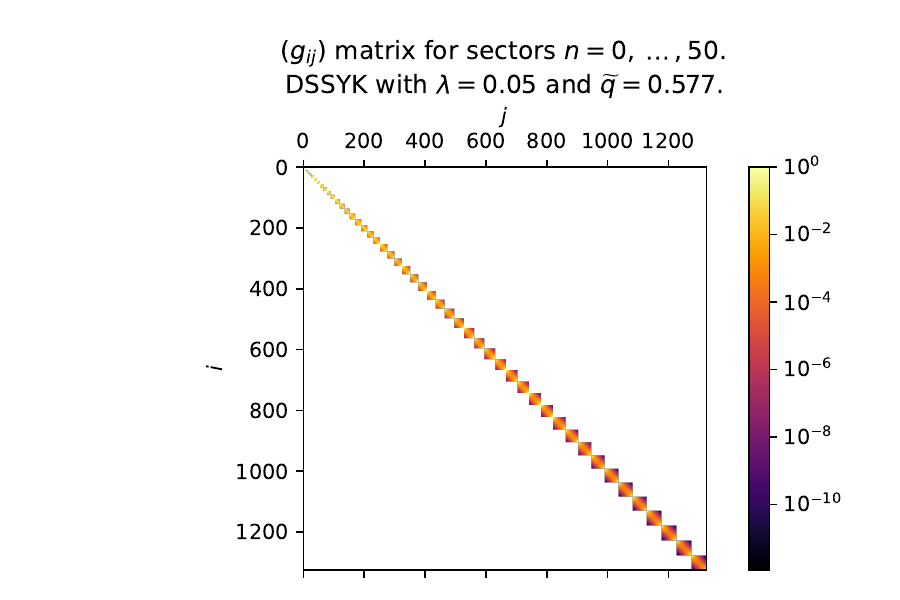} \includegraphics[width=0.45\linewidth]{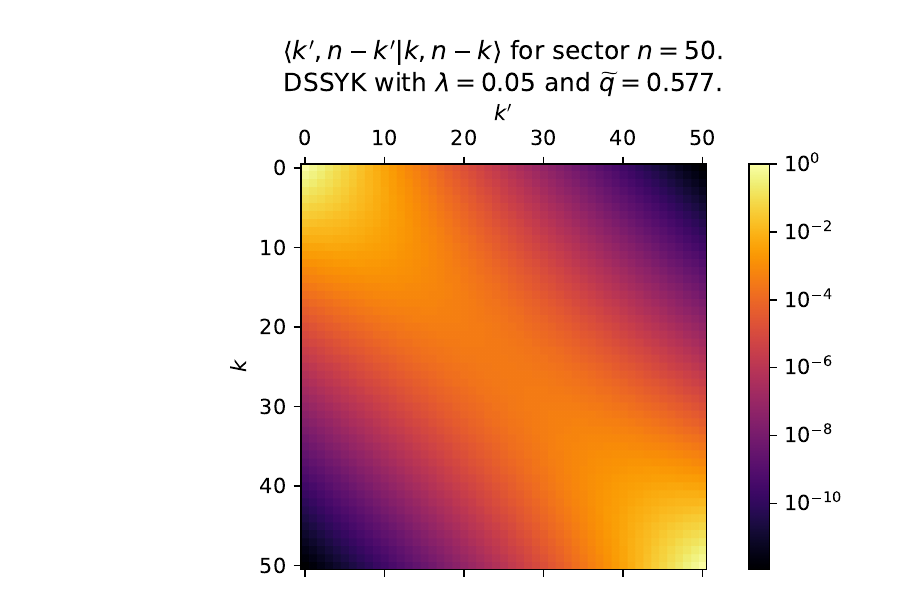}  \\
    \includegraphics[width=0.45\linewidth]{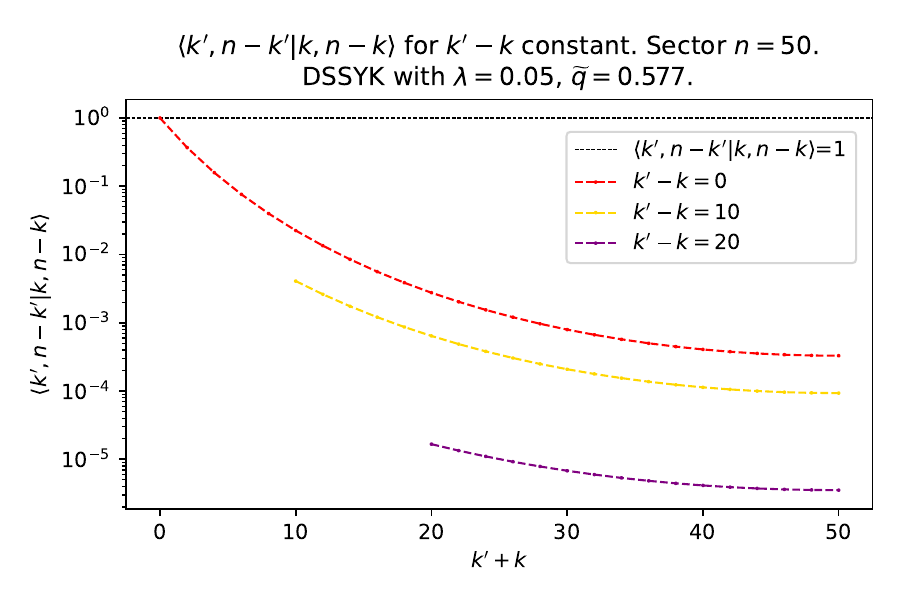}  \includegraphics[width=0.45\linewidth]{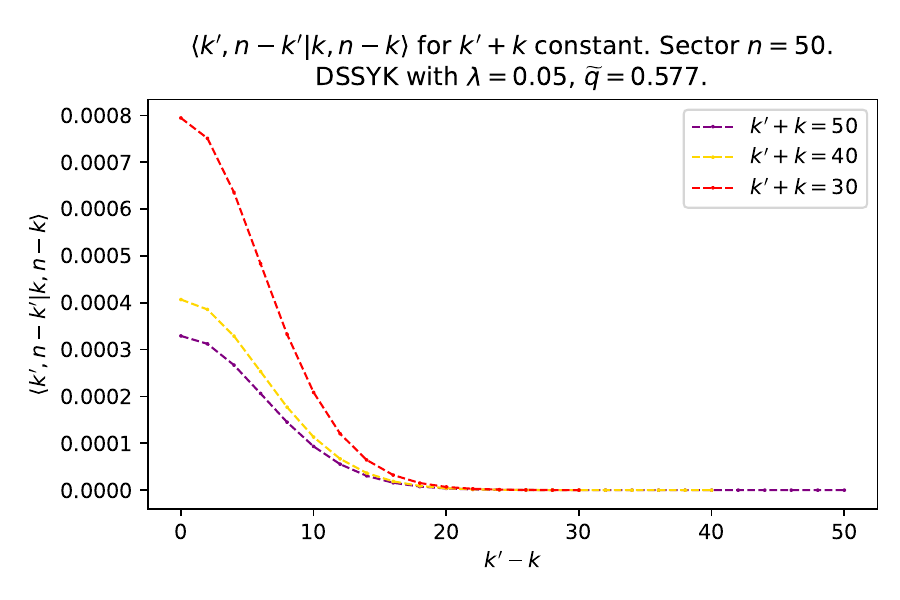} 
    \caption{Inner product matrix \eqref{inner_product_matrix_assignment_appx}. \textbf{Top left:} Matrix plot, including all total chord number sectors up to $n=50$. The encoding \eqref{ifromnk} of the chord basis in the computational basis allows to observe a clear block-diagonal structure reflecting the fact that sectors of different total chord number are orthogonal. \textbf{Top right:} Matrix plot of the overlaps in a fixed sector with total chord number $n=50$, i.e. the matrix $\Big(g^{(n)}_{k k^\prime}\Big)$ in \eqref{inner_product_matrix_in_sector}. In agreement with \eqref{IPS_appx} and \eqref{LRS_appx}, we observe that this matrix is symmetric around both its main and secondary diagonal, taking the form of a saddle. \textbf{Bottom left:} Plots of the overlaps $\Big(g^{(n)}_{k k^\prime}\Big)$ for constant difference $k^\prime - k$, as a function of the sum $k^\prime + k$, which amounts to moving in the top-right figure along lines that are parallel to the main diagonal. The red line corresponds to $k=k^\prime$, and it therefore gives the norms of the states $|k,n-k\rangle$; we note that the only states with unit norm in the basis $|k,n-k\rangle$, renormalized as stated in \eqref{Basis_renormalization_for_numerics}, are $|0,n\rangle$ and $|n,0\rangle$, as the plot confirms. \textbf{Bottom right:} $\Big(g^{(n)}_{k k^\prime}\Big)$ for constant sum $k^\prime + k$, as a function of the difference $k^\prime - k$, which sweeps through lines of the top-right plot that are parallel to the secondary diagonal. We observe that, the larger the difference between $k$ and $k^\prime$, the smaller the overlap.}
    \label{fig:inner_product_plots}
\end{figure}

A comment on numerical reliability of results is in order here. The limit $\widetilde{q}\to 1$ (corresponding to $\Delta\to 0$ for $\lambda$ fixed) may be thought of as an İnönü-Wigner contraction of the one-particle algebra to the zero-particle one\footnote{Intuitively, from the perspective of chord diagrams, crossings with the operator chord will have weight $\widetilde{q}=1$, so effectively they may be regarded as a matterless diagram with no distinction between left and right}. In this case, two-sided states $|n_L,n_R\rangle$ effectively behave as their one-sided counterpart $|n_L+n_R\rangle$, in the sense that
\begin{equation}
    \label{inner_prod_qt_1}
    \langle n_L^\prime,n_R^\prime | n_L,n_R\rangle \overset{\widetilde{q}\to 1}{\longrightarrow} \langle n_L^\prime+n_R^\prime | n_L+n_R\rangle =\delta_{n_L^\prime + n_R^\prime,n_L+n_R}~,
\end{equation}
as can be shown with \eqref{Basis_renormalization_for_numerics}, or just recalling the derivation of the chord inner product recursion from chord diagram techniques in \cite{Lin:2022rbf} (but conveniently renormalizing the states as in \eqref{Basis_renormalization_for_numerics}), where the fact that $\widetilde{q}=1$ would make distinction between left and right irrelevant.
Consequently, appropriate linear combinations of $|n_L,n_R\rangle$ states belonging to the same total chord number sector can yield states that are effectively null. For example, in the sector of total chord number $n=1$, the state
\begin{equation}
    \label{null_state_example}
    |\sigma_{-}\rangle = |0,1\rangle - |1,0\rangle
\end{equation}
becomes a null state when $\widetilde{q}\to 1$, reflecting the fact that it becomes effectively equivalent to the matterless state
\begin{equation}
    \label{null_state_matterless_equivalent}
    |\sigma_{-}\rangle \longmapsto  |1\rangle - |1\rangle =0~,
\end{equation}
We can see this explicitly by considering the inner-product matrix in the $n=1$ sector:
\begin{equation}
    \label{Inner_prod_one_particle_sector}
   \Big( g^{(n=1)}_{k^\prime k}\Big) = \begin{pmatrix}
       1 & ~\widetilde{q}~ \\
       ~\widetilde{q}~ & 1
   \end{pmatrix}~.
\end{equation}
This matrix has eigenvalues $\sigma_{\pm}=1\pm\widetilde{q}$, and therefore when $\widetilde{q}\to 1$ we have that $\sigma_{-}\to 0$, implying that the corresponding eigenvector becomes a null state. Such an eigenvector is indeed \eqref{null_state_example}. In full generality, it is possible to prove\footnote{The proof goes as follows: Using \eqref{inner_prod_qt_1} we see that, when $\widetilde{q}=1$, the overlap matrix $\Big( g_{k^\prime k}^{(n)} \Big)$ for a fixed $n$-sector becomes an $(n+1)\times (n+1)$ matrix of ones (i.e. all its items, not just the diagonal, are exactly $1$), which fortunately is easy to diagonalize in a real space \cite{DonAntonio_StackExchange}. Denoting the eigenvectors of the matrix as $\sum_{k=0}^{n}c_k |k,n-k\rangle$, we can see that the $n$-dimensional zero eigenspace is given by the coefficients locus $\left\{ c_k :~\sum_{k=0}^n c_k = 0 \right\}$, while there is a one-dimensional eigenspace of eigenvalue $n+1$ defined by the locus $\left\{ c_k:~c_0 = c_1=\dots=c_n \right\}$, QED.} that, for arbitrary total chord number $n\geq 0$, the inner-product matrix in that sector, $g^{(n)}$, has exactly $n$ zero eigenvalues and one positive eigenvalue when $\widetilde{q}=1$. This means that, when $\widetilde{q}=1$, all but one linearly-independent directions in a given chord sector become null: The whole sector $\mathcal{H}_{1p}^{(n)}$ \textit{shrinks} onto the matterless state $|n\rangle\in\mathcal{H}_{0p}$.

While the above is a mathematically interesting structure, it is bad news for the numerics: Taking $\widetilde{q}$ closer and closer to $1$, in an attempt to describe lighter operators, brings us closer to the special point $\widetilde{q}=1$ where null states proliferate, and it does so by featuring \textit{almost-zero-norm states} when $\widetilde{q}$ is close to $1$, making numerics unstable. In those situations, it will turn out that the Lanczos-PRO algorithm operating at double floating point precision will not be sufficient for stabilizing the recursion, and high precision implementations may be preferred\footnote{For exactly $\widetilde{q}=1$ we did verify numerically that the Krylov space of the state $|0,0\rangle$, given the Hamiltonian $H_R-H_L$, is one-dimensional as it should, since in the absence of operator insertions, the TFD state is stationary under the evolution generated by this Hamiltonian, verifying $(H_R-H_L)^n|0,0\rangle=\mathbf{0}$ for $n>0$ when $\widetilde{q}=1$. At a technical level, this can be verified by direct application of $H_R-H_L$, given in equation \eqref{Total_Hamiltonian_minus_line2}: We find that $(H_R-H_L)|0,0\rangle =|\sigma_{-}\rangle$, i.e. the state \eqref{null_state_example}, which becomes null when $\widetilde{q}=1$, implying that the first Lanczos coefficient $b_1 = \sqrt{\langle\sigma_{-}|\sigma_{-}\rangle} = \sqrt{2\sigma_{-}} = \sqrt{2(1-\widetilde{q})}$ becomes zero in this limit, cf. \eqref{b1_Exact}, thus terminating the Lanczos algorithm at this $n=1$ step.}.

For completeness, we may say that the opposite limit in $\widetilde{q}$, namely $\widetilde{q}\to 0$, is completely harmless from the point of view of numerics. This limit just amounts to making the operator very heavy ($\Delta\to +\infty$ for fixed $\lambda$), which pinches off all chord diagrams splitting them into two matterless diagrams. The one-particle algebra becomes in this case a tensor product (rather than a co-product \cite{Lin:2023trc}) of two zero-particle algebras, $\mathcal{H}_{1p}\to \mathcal{H}_{0p}\otimes \mathcal{H}_{0p}$, and states $|n_L,n_R\rangle$ become effectively equivalent to $|n_L\rangle\otimes|n_R\rangle$ (modulo prefactors).

\subsubsection{Hermiticity with respect to the chord inner product}\label{appx:numerics_hermiticity_conditions}

The fact that the inner product takes the non-canonical form of the matrix $g_{ij}$ in coordinates over the computational basis $\mathcal{B}_{\text{comp}}$ with which we work implies the need for adjusting the adjoint operation, as this notion is relative to the inner product. Consequently, the hermiticity condition for an operator $\mathcal{O}$ will not be equivalent to the symmetry of the matrix $(O_{ij})$ containing its coordinates over $\mathcal{B}_{\text{comp}}$. In this section we shall elaborate on how hermiticity was accounted for in our numerical implementations.

We shall begin by revising the expression of the bra of a state in coordinates over a non-orthonormal basis. We remind that the bra ``$\langle v|$'' is nothing but the one-form $\omega_v$ obtained from a vector (ket) $|v\rangle$ through the musical isomorphism, according to which such a one-form is defined through its action on any other vector $|u\rangle$ of the Hilbert space as
\begin{equation}
    \label{bra_def}
    \omega_v (|u\rangle) = g(|v\rangle,|u\rangle)~,\quad \forall |u\rangle\in\mathcal{H}_{1p}~,
\end{equation}
which motivates the Dirac notation $\langle v|x\rangle$. In coordinates over the besis $\mathcal{B}_{\text{comp}}$, we have that $|v\rangle$ is represented by a column vector, $|v\rangle \overset{*}{=}\mathbf{v}$, and similarly for $|u\rangle$, such that their overlap is given by
\begin{equation}
    \label{Overlap_in_coords}
    \langle v|u\rangle = \mathbf{v^T~g~u}~, 
\end{equation}
where we recall that $\mathbf{g}$ denotes the inner product matrix. From \eqref{Overlap_in_coords} we can directly read off the representation of the bra in coordinates over $\mathcal{B}_{\text{comp}}$:
\begin{equation}
    \label{bra_in_coords}
    \bra{v}\overset{*}{=} \mathbf{v^T~g}~.
\end{equation}
That is, numerically, in order to build the bra of a given vector we needed to perform the operation \eqref{bra_in_coords}. Each time a vector is constructed numerically, it is  generally recommendable, if memory permits, to store its bra in a separate array. In this way, eventual computations of inner products can directly be called as the canonical (computational) inner product between the bra array and the ket array, instead of performing the operation \eqref{Overlap_in_coords}. This is an efficient solution whenever the same bra is going to be used more than once for the computation of overlaps. Therefore, in the numerical implementation of the Lanczos-PRO algorithm, each time a Krylov element was built, its bra was accordingly constructed and stored, for the purpose of making use of it when convenient in the re-orthogonalization steps \cite{Rabinovici:2020ryf}.

Similarly, the hermitian adjoint of an operator $\mathcal{O}$ \textit{with respect to the inner product $g$}, i.e. $\mathcal{O}^\dagger$, is represented in coordinates over the computational basis by a matrix $(\mathcal{O}^\dagger)_{ij}$ which is in general different from just the transpose of $O_{ij}$. The matrix expression of the adjoint $\mathcal{O}^\dagger$ can just be read of from the definition of the adjoint operator, which is an operator $\mathcal{O}^\dagger$ satisfying
\begin{equation}
    \label{Hermitian_adjoint_def}
    g(\mathcal{O}|v\rangle,|u\rangle)= g(|v\rangle,\mathcal{O}^\dagger|u\rangle),\quad \forall~ |v\rangle,|u\rangle\in\mathcal{H}_{1p}~.
\end{equation}
In matrix notation, and making use of \eqref{Overlap_in_coords}, this implies:
\begin{equation}
    \label{adjoint_arbitrary_inner}
    \mathbf{O}^\dagger = \mathbf{g^{-1}~O^T~g}~.
\end{equation}
Note that \eqref{adjoint_arbitrary_inner} is consistent with \eqref{bra_def}, in the sense that both expressions combine to verify $\big(\mathcal{O}|v\rangle\big)^\dagger = \langle v|\mathcal{O}^\dagger$.

In light of \eqref{adjoint_arbitrary_inner}, we see that an operator $\mathcal{O}$ is hermitian with respect to the chord inner product if and only if its matrix representation over the basis $\mathcal{B}_{\text{comp}}$ verifies
\begin{equation}
    \label{hermiticity_condition_v1}
    \mathbf{O} = \mathbf{g^{-1}~O^T~g} \qquad\Longleftrightarrow\qquad (\mathbf{g~O}) = (\mathbf{g~O})^\mathbf{T}~.
\end{equation}
That is, if $\mathcal{O}$ is hermitian with respect to the chord inner product, the matrix $\mathbf{gO}$ should be explicitly symmetric, but $\mathbf{O}$ need not be so. An equivalent and insightful form of \eqref{hermiticity_condition_v1} is the following:
\begin{equation}
    \label{hermiticity_condition_v2}
    \mathbf{O} = \mathbf{g^{-1}~O^T~g} \qquad\Longleftrightarrow\qquad \mathbf{g^{1/2}~O~g^{-1/2}} = (\mathbf{g^{1/2}~O~g^{-1/2}})^{\mathbf{T}}~,
\end{equation}
where $\mathbf{O^\prime}\equiv \mathbf{g^{1/2}~O~g^{-1/2}}$ may be seen as the expression of the operator $\mathcal{O}$ in coordinates over a new basis for which the passage matrix is $\mathbf{P=g^{-1/2}}$. Without being specific about such a basis, we can anticipate that it is orthonormal given the corresponding transformation of the inner product\footnote{Note that the multiplicity of the square roots of a $\mathbf{g}$ is in correspondence with the fact that there exist multiple orthonormal bases. Additionally, $\mathbf{g}$ is correctly invertible (for $\widetilde{q}\in [0,1[~$) because, as a well-defined inner product, it has no zero eigenvalues (it is in fact a positive-definite and symmetric matrix).}:
\begin{equation}
    \label{inner_change_basis_orthonormal}
    \mathbf{g^\prime} = \mathbf{P^T~g~P} = \mathbf{g^{-1/2}~g~g^{-1/2}}=\mathbf{\mathds{1}}~.
\end{equation}
In retrospect, \eqref{hermiticity_condition_v2} comes to say that hermitian operators take the form of symmetric matrices when expressed in coordinates over an orthonormal basis, consistently. In \cite{Lin:2022rbf}, operations of the form $\mathbf{H}\to \mathbf{g^{1/2}~H~g^{-1/2}}$ are performed with both the zero-particle and one-particle Hamiltonians in order to obtain their expression in terms of the length operator and its canonical conjugate momentum but, in the case of the one-particle sector, only in the semiclassical limit, where the states of the same chord sector are argued to become orthogonal, hence making $\mathbf{g}$ become a diagonal (i.e. not just block-diagonal) matrix, whose inverse may thus be easily computed analytically.

As an illustration, figure \ref{fig:Hamiltonian} shows the Hamiltonian $H_{\text{total}}\equiv H_R-H_L$ given in \eqref{Total_Hamiltonian_minus_line2} in coordinates over $\mathcal{B}_{\textbf{comp}}$, together with the transformation $\mathbf{g H_{\text{total}}}$, which makes it take the form of an explicitly symmetric matrix, confirming that the operator constructed is hermitian with respect to the chord inner product (within the working machine precision).

\begin{figure}
    \centering
    \includegraphics[width=0.45\linewidth]{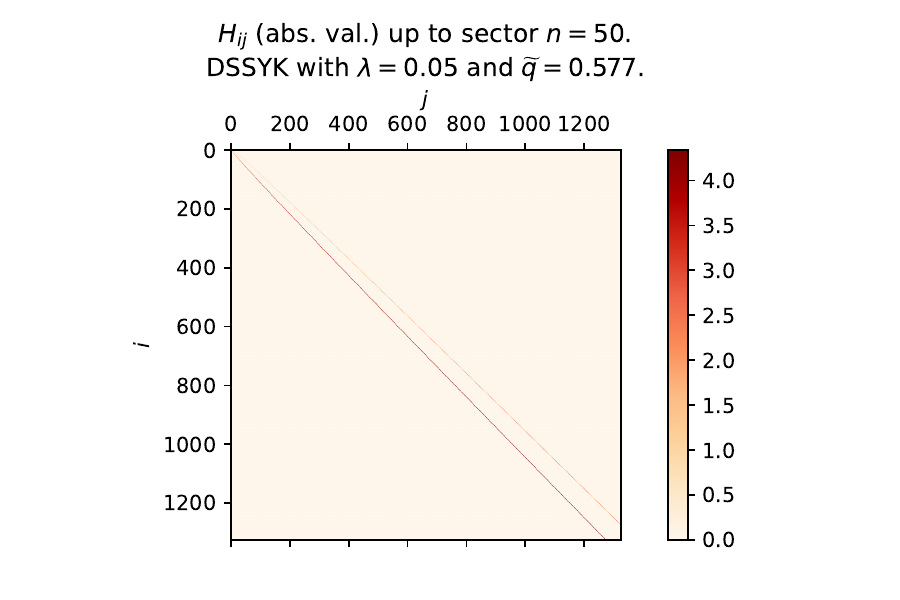} \includegraphics[width=0.45\linewidth]{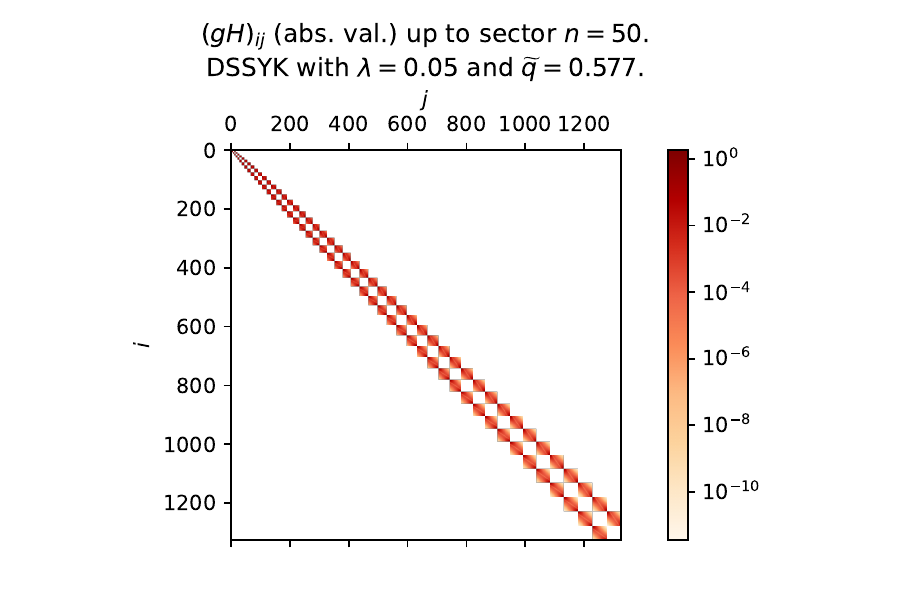} \\
    \includegraphics[width=0.45\linewidth]{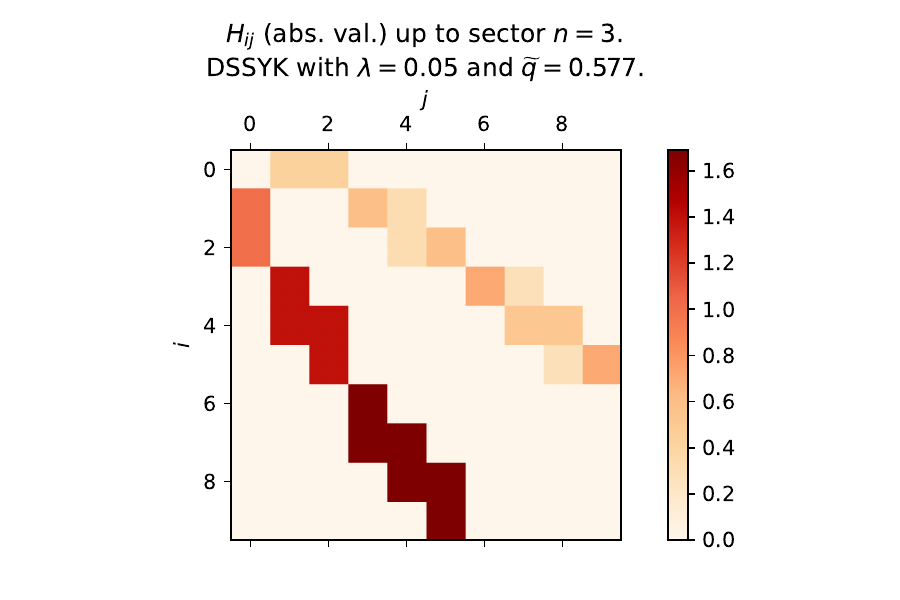} \includegraphics[width=0.45\linewidth]{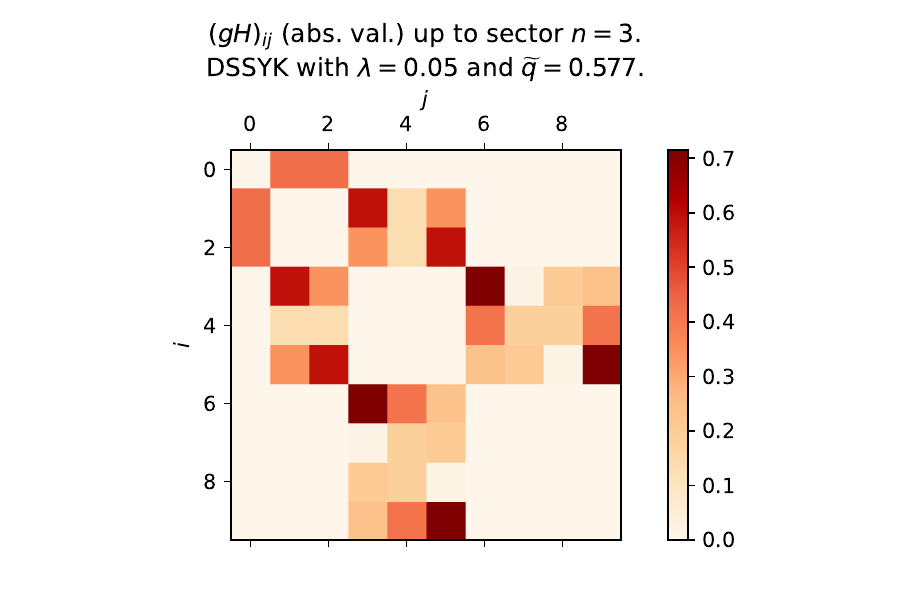} \\
    \includegraphics[width=0.45\linewidth]{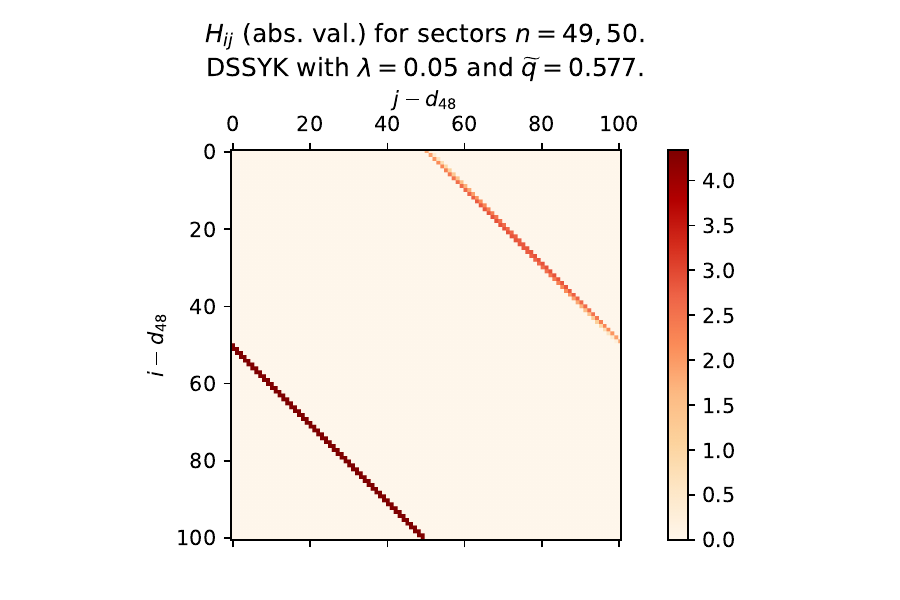} \includegraphics[width=0.45\linewidth]{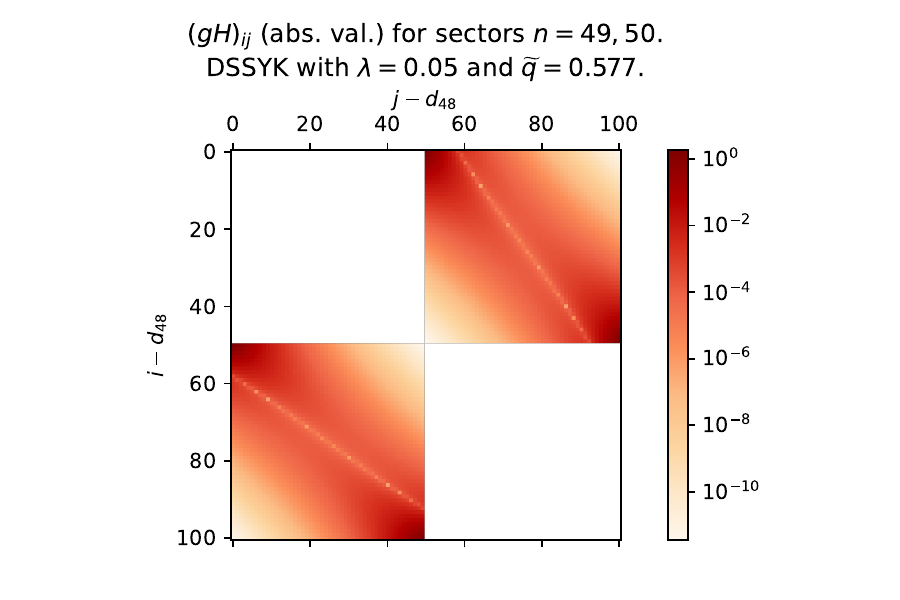} \\
    \includegraphics[width=0.45\linewidth]{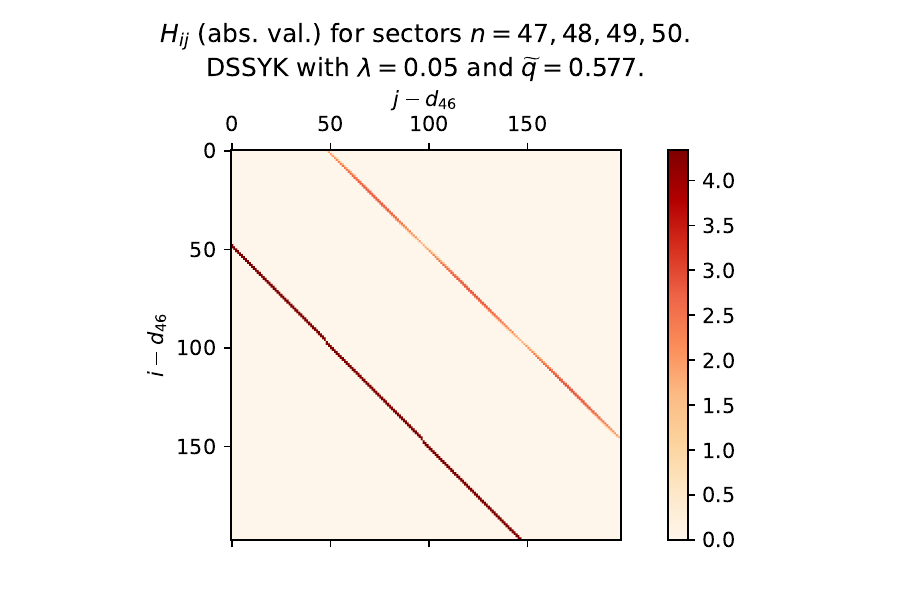} \includegraphics[width=0.45\linewidth]{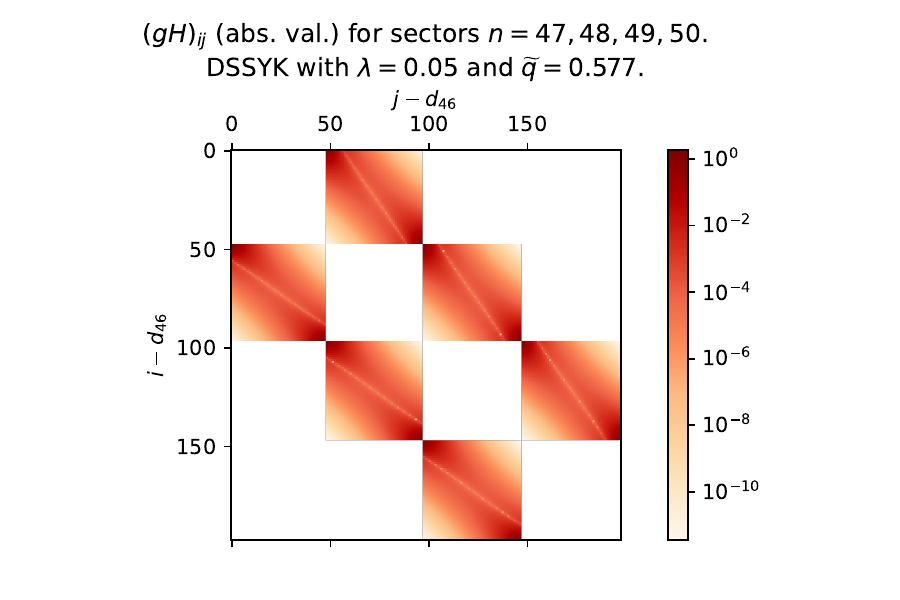}
    \caption{Plots of various sectors of the matrices $\mathbf{H}$ (\textbf{left}) and $\mathbf{gH}$ (\textbf{right}), where $H\equiv H_{\text{total}}=H_R-H_L$, in coordinates over the computational basis that encodes the chord basis. In all cases we can appreciate the block-tridiagonal structure of the Hamiltonian, and additionally the right plots are manifestly symmetric, in agreement with the hermiticity condition \eqref{hermiticity_condition_v1} applied to $H$. Note that, in plots where the color code involves a logarithmic scale, numerically very small values are assigned the color white.}
    \label{fig:Hamiltonian}
\end{figure}

Most of the discussions in sections \ref{appx:numerics_inner_product} and \ref{appx:numerics_hermiticity_conditions} review well-established facts of linear algebra adapted to the problem at hand, where we needed to do numerics using a non-orthonormal basis, which happens to be the physically natural basis in DSSYK with matter (operator). For this reason, it was found convenient to spell out explicitly the relevant technical details in this appendix.

\begin{figure}
    \centering
    \includegraphics[width=0.7\linewidth]{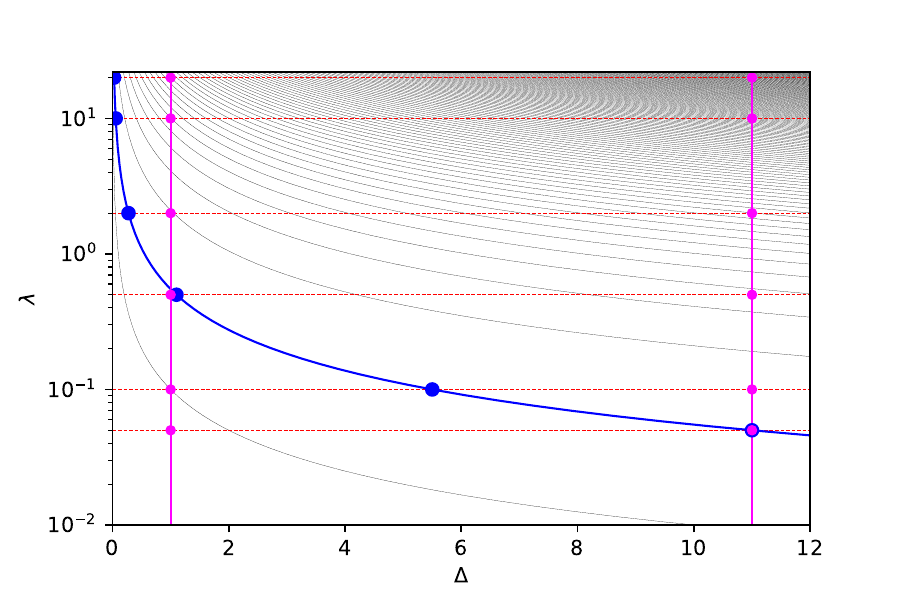}
    \caption{Sketch of the parameter space of the system, DSSYK with 't Hooft coupling $\lambda$ plus double-scaled operator of dimension $\Delta$, where the one-parameter families considered in our numerical analyses are highlighted. Dashed black lines are families of constant $\widetilde{q}=e^{-\lambda \Delta}$, which increases towards the origin of the graph, and in particular the blue line is the family in which $\widetilde{q}=0.577$. Red dashed lines correspond to $\lambda=0.05,0.1,0.5,2,10,20$. Magenta lines are families of constant $\Delta=1,11.002$. The dots represent the actual tuples $(\Delta,\lambda)$ for which numerical simulations were carried out. Note that $\lambda = 0$ is the large-$p$ SYK limit \cite{Maldacena:2016hyu,Mukhametzhanov:2023tcg}, while $\lambda\to+\infty$ is the RMT limit \cite{Berkooz:2018jqr,Berkooz:2024lgq}.}
    \label{fig:ParameterSpace}
\end{figure}
\subsection{Further numerical results}\label{appx:Numerics_further_results}

This section presents numerical results that complement the main ones shown in section \ref{subsect:numerics}. We implemented either the Lanczos-PRO algorithm \cite{Rabinovici:2020ryf, parlett1998symmetric} at double floating point machine precision or the pure Lanczos algorithm at high precision for various system parameters lying within the one-parameter families depicted in figure \ref{fig:ParameterSpace}, which are families of two kinds: either fixed $\widetilde{q}=e^{-\lambda \Delta}$, or fixed $\Delta$. In all cases we compared the output Lanczos coefficients $b_n$ to the expression $\sqrt{c_{n/2}(n)}$, argued in section \ref{subsect:bn_asymptotic_limit} to be the form of the Lanczos coefficients in the asymptotic limit, and we also studied the localization of Krylov basis elements to fixed chord number sectors. In all the one-parameter families studied, generically we observe that reducing $\lambda$ has the effect of improving the accuracy of the Ansatz, as shown in figures \ref{fig:qt0pt577_lambda0pt1} to \ref{fig:Delta11pt002_lambda20pt0} and as generically analyzed in appendix \ref{appx:small_lambda}. 

For the sake of clarity, let us provide a list grouping all the relevant figures into three categories depending on the one-parameter families in parameter space that they belong to:

\begin{itemize}
    \item Figures \ref{fig:qt0pt577_lambda0pt1} to \ref{fig:qt0pt577_lambda20pt0} complement figures \ref{fig:PRO_results_lambda0pt05} and \ref{fig:PRO_results_lambda0pt5} in the main text and correspond to a $\lambda$-study along the hyperbola of constant $\widetilde{q}=0.577$, i.e. the blue line in figure \ref{fig:ParameterSpace}. As $\lambda$ increases, the deviations from the asymptotic limit become significant, and in particular the chord sector participation ration in the Krylov vectors deviates from one, signaling that they are no longer effectively localized in fixed total chord number sectors, developing significant tails on various sectors, lowering the chord number expectation value and making the Lanczos coefficients differ from the Ansatz $b_n=J\sqrt{c_{n/2}(n)}$. Numerically, note that the deviations of the $b$-sequence are more pronounced in the region with $n\eqsim 1/\lambda$, as for small enough values of $n$ compare to $1/\lambda$ the small-$\lambda$ analysis of appendix \eqref{appx:small_lambda} becomes applicable, while for $n\gg 1/\lambda$ the numerics effectively probe again the asymptotic limit (cf. section \ref{subsect:bn_asymptotic_limit}). This is why, for indicative purposes, a vertical line at $n=1/\lambda$ has been included in the plots of $n$-dependent quantities, besides the vertical line at $n=N$ signaling when $n$ approaches the total chord number truncation value (cf. \ref{appx:Numerics_setup}), after which spurious finite-size effects kick in.
    \item Figures \ref{fig:Delta11pt002_lambda0pt1} to \ref{fig:Delta11pt002_lambda20pt0} complement figure \ref{fig:PRO_results_lambda0pt05} and correspond to the analysis with fixed $\Delta=11.002$, i.e. the rightmost magenta line in \ref{fig:ParameterSpace} (which intersects in a point with the fixed-$\widetilde{q}$ family). The conclusions on the $\lambda$-dependence on results are completely analogous to what has been described in the previous point regarding the analysis of constant $\widetilde{q}$.
    \item For illustration purposes, figures \ref{fig:Delta1pt0_lambda0pt05_lambda0pt1_lambda0pt5} to \ref{fig:Delta1pt0_lambda20pt0} present yet another constant-$\Delta$ analysis, but this time with $\Delta=1$ (cf. the leftmost vertical magenta line in figure \ref{fig:ParameterSpace}). When both $\Delta$ and $\lambda$ are numerically small, implying that $\widetilde{q}$ is similar to $1$, the numerics become unstable for the reasons described towards the end of section \ref{appx:numerics_inner_product}. These plots still allow to conclude that decreasing $\lambda$ for fixed $\Delta$ makes the asymptotic Ansatz $b_n=J\sqrt{c_{n/2}(n)}$ better, but comparison with the analysis at fixed $\Delta=11.002$ also reveals that, for fixed $\lambda$, decreasing $\Delta$ makes results deviate from the Ansatz (cf. figures corresponding to points lying on the same horizontal line in figure \ref{fig:ParameterSpace}).
\end{itemize}


\begin{figure}
    \centering
    \includegraphics[width=0.45\linewidth]{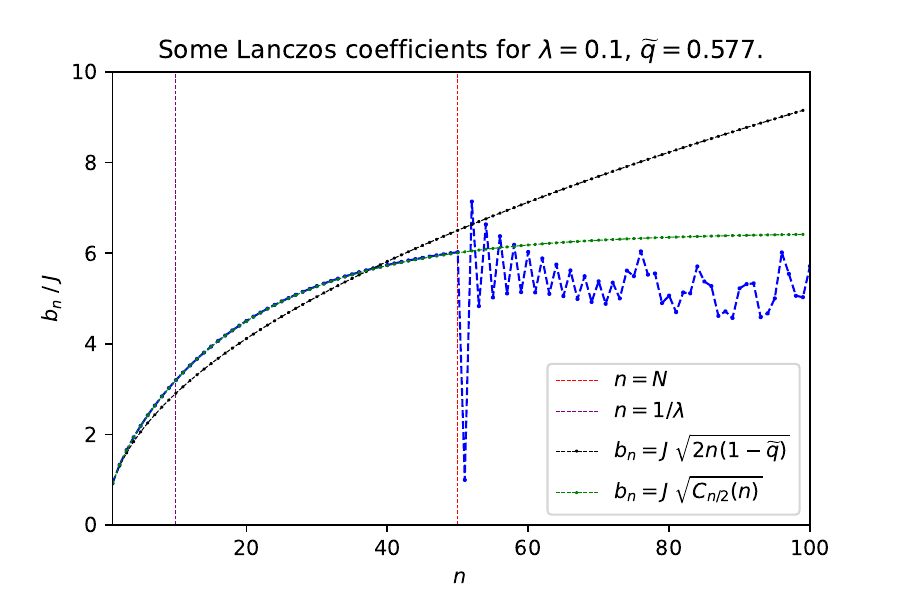}  \includegraphics[width=0.45\linewidth]{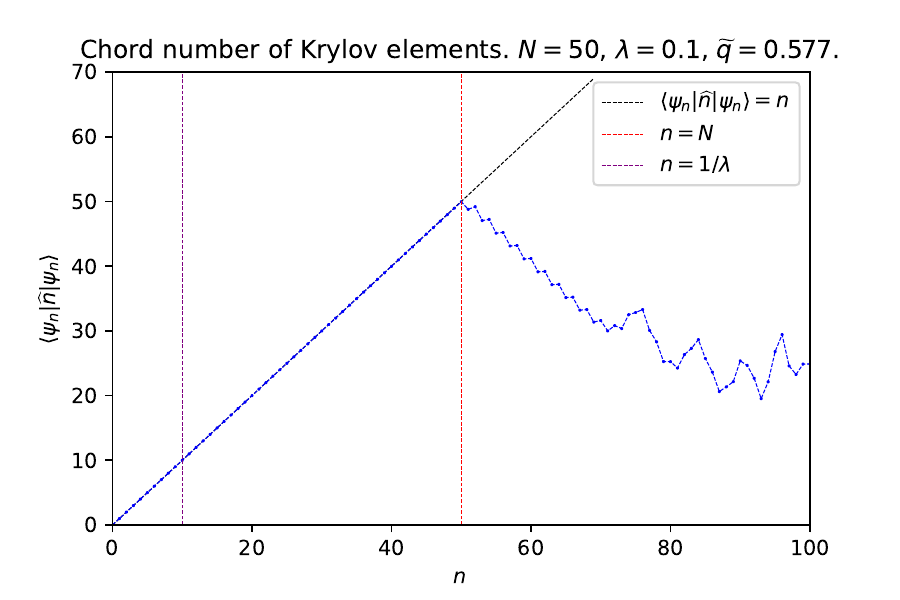} \\
    \includegraphics[width=0.45\linewidth]{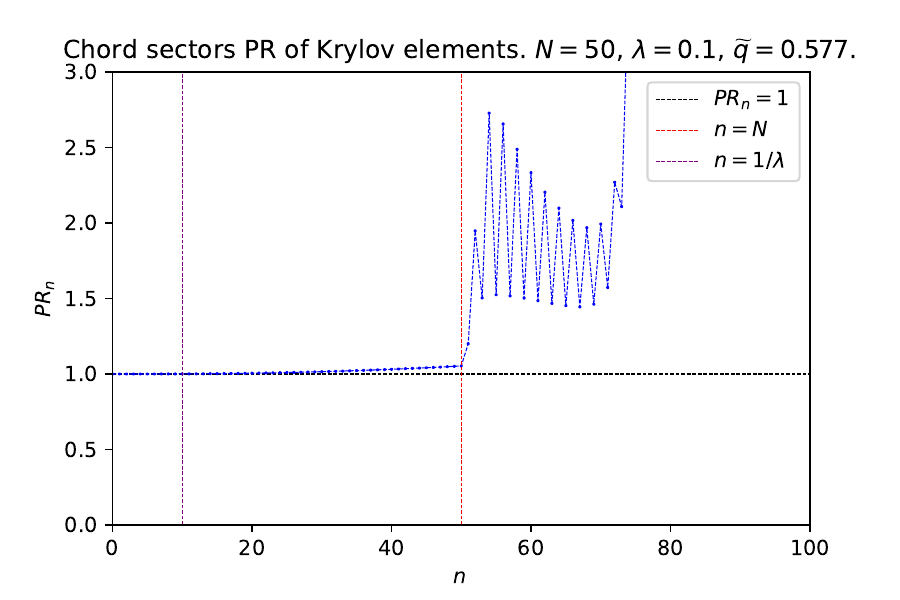}  \includegraphics[width=0.45\linewidth]{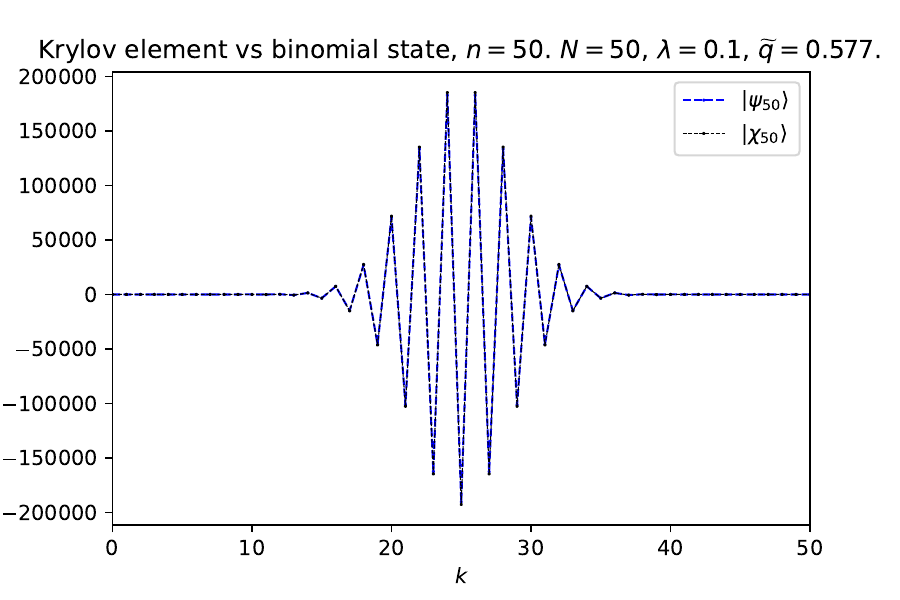}
    \caption{$\widetilde{q}=0.577$, $\lambda = 0.1$.}
    \label{fig:qt0pt577_lambda0pt1}
\end{figure}

\begin{figure}
    \centering
    \includegraphics[width=0.45\linewidth]{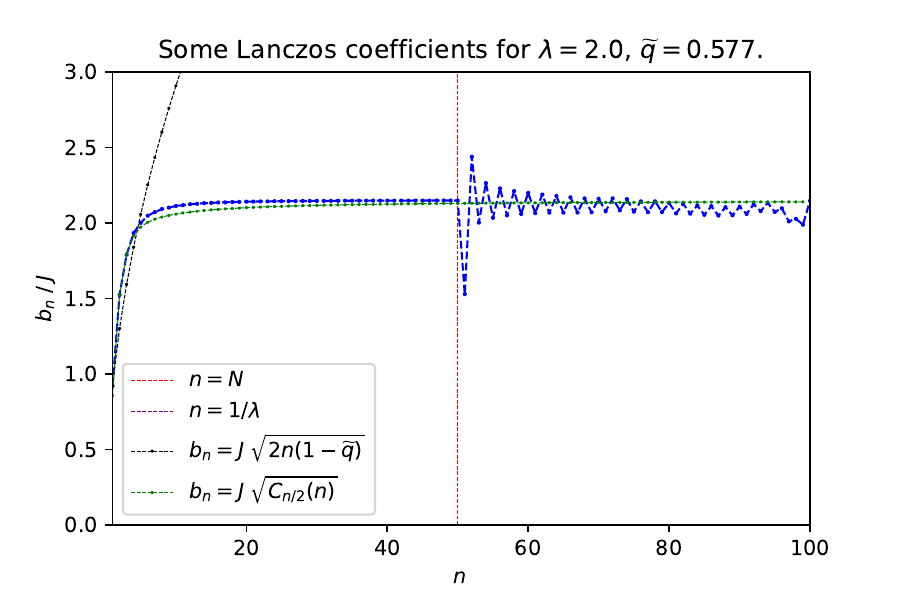}  \includegraphics[width=0.45\linewidth]{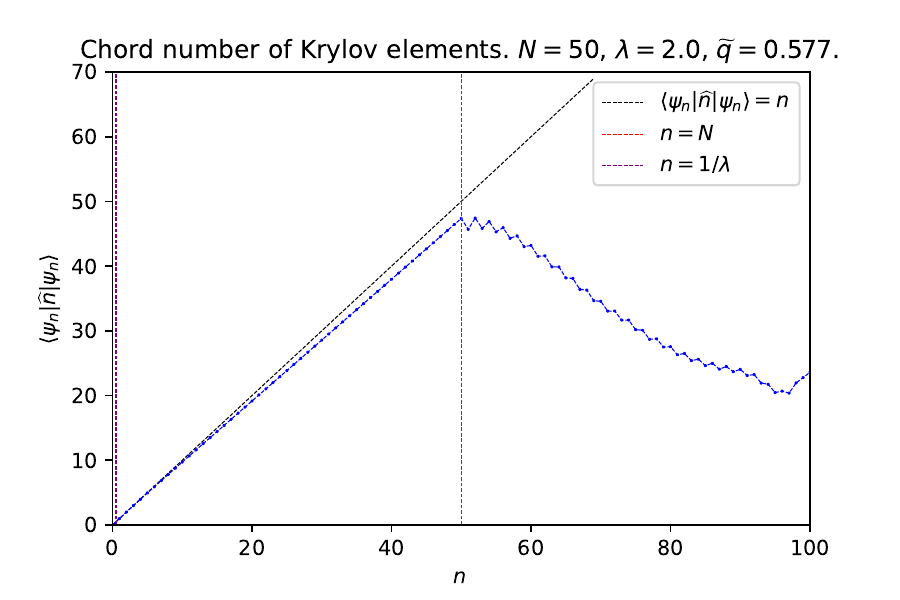} \\
    \includegraphics[width=0.45\linewidth]{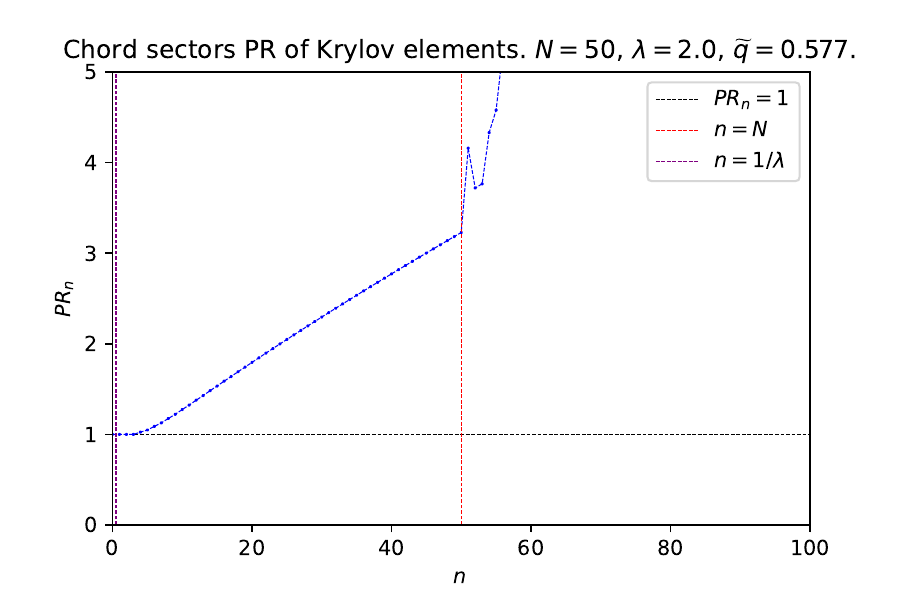}  \includegraphics[width=0.45\linewidth]{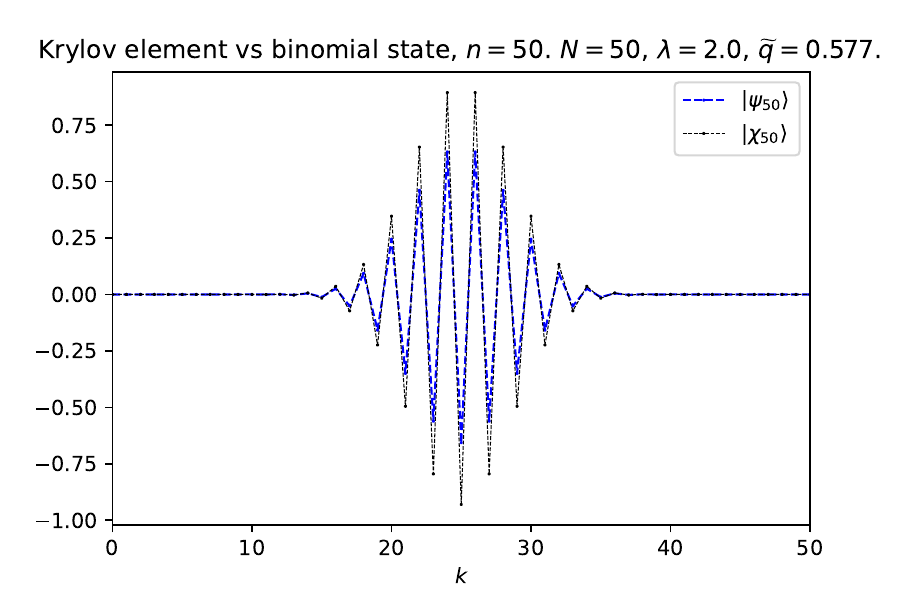}
    \caption{$\widetilde{q}=0.577$, $\lambda = 2$.}
    \label{fig:qt0pt577_lambda2pt0}
\end{figure}

\begin{figure}
    \centering
    \includegraphics[width=0.45\linewidth]{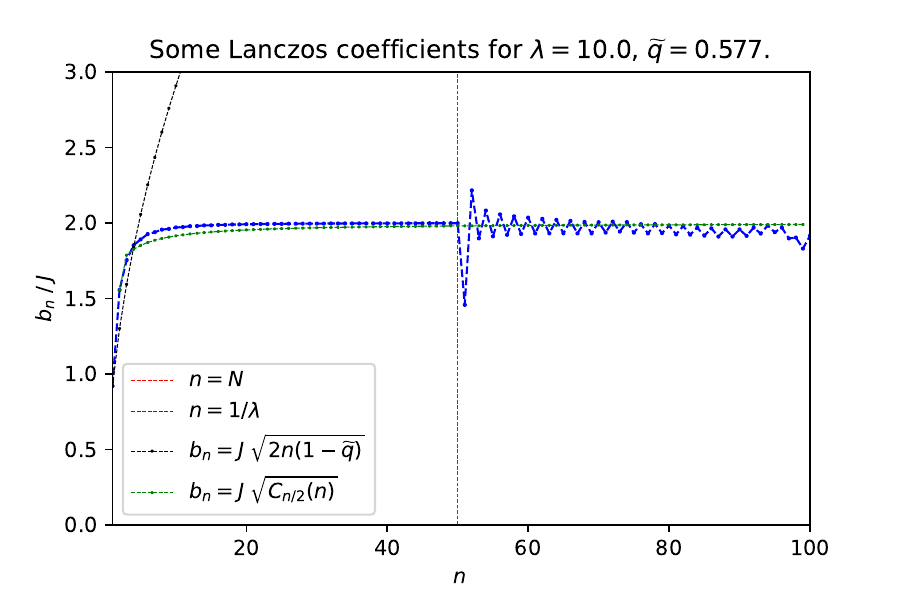}  \includegraphics[width=0.45\linewidth]{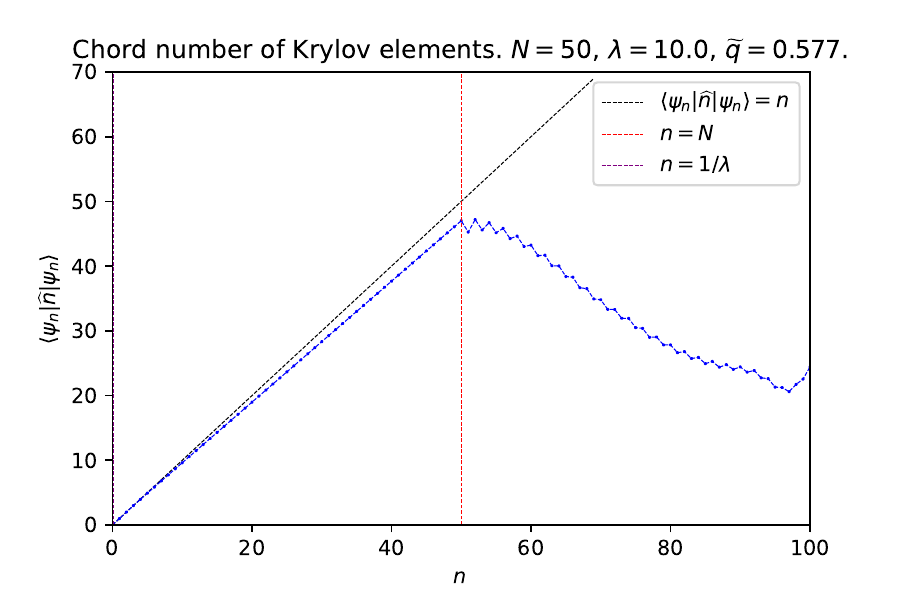} \\
    \includegraphics[width=0.45\linewidth]{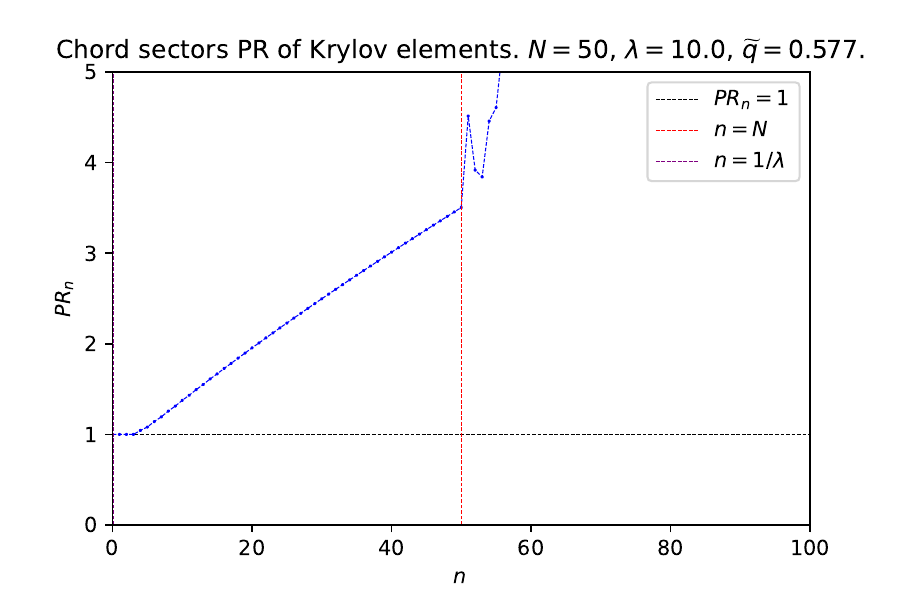}  \includegraphics[width=0.45\linewidth]{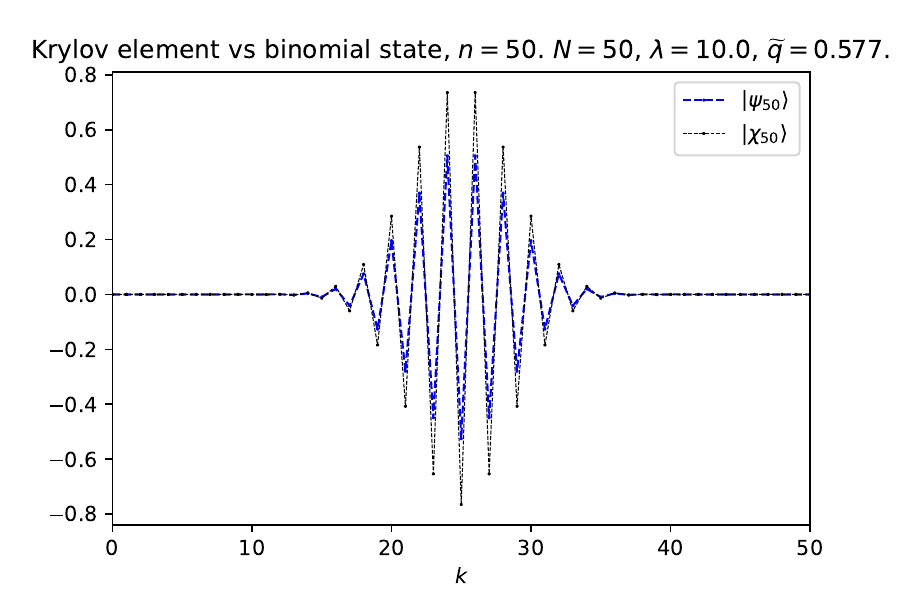}
    \caption{$\widetilde{q}=0.577$, $\lambda = 10$.}
    \label{fig:qt0pt577_lambda10pt0}
\end{figure}

\begin{figure}
    \centering
    \includegraphics[width=0.45\linewidth]{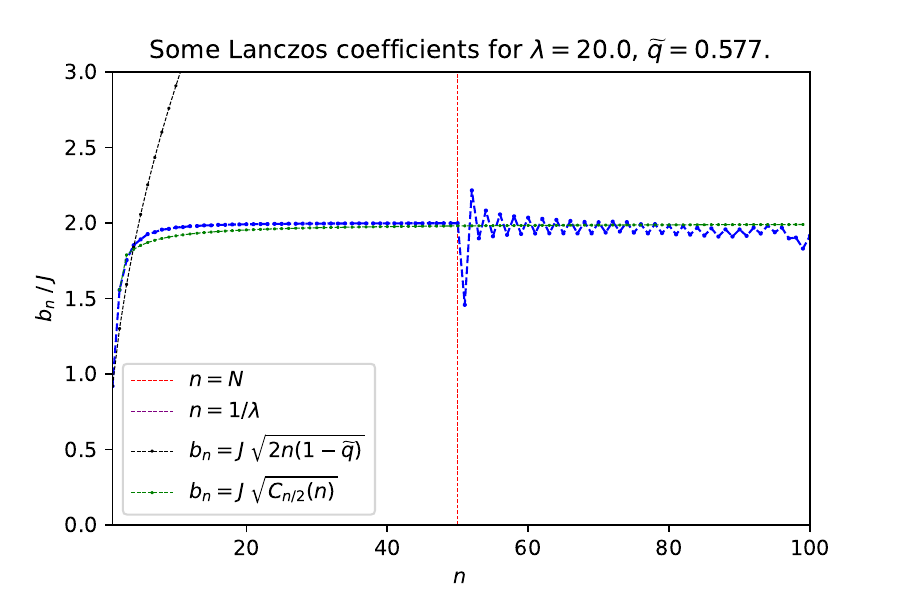}  \includegraphics[width=0.45\linewidth]{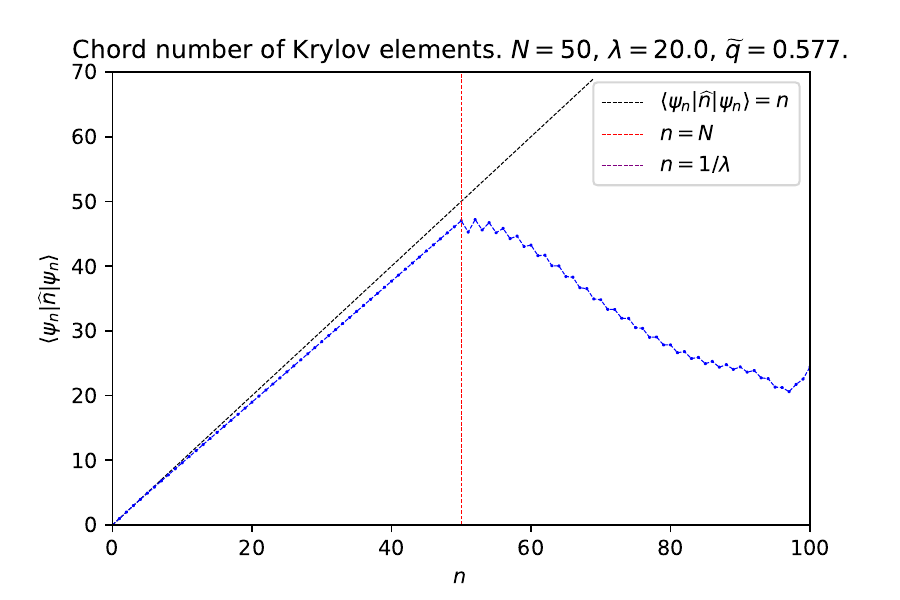} \\
    \includegraphics[width=0.45\linewidth]{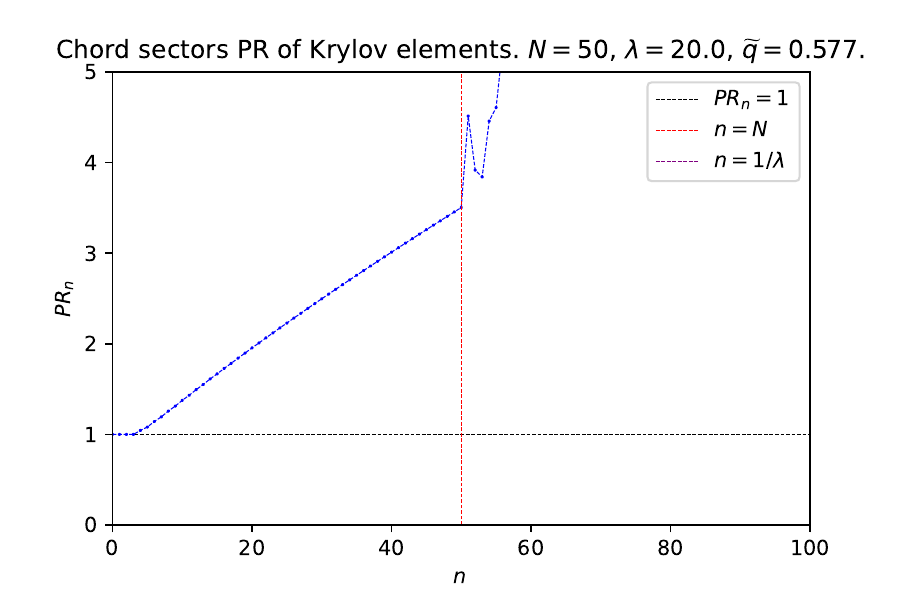}  \includegraphics[width=0.45\linewidth]{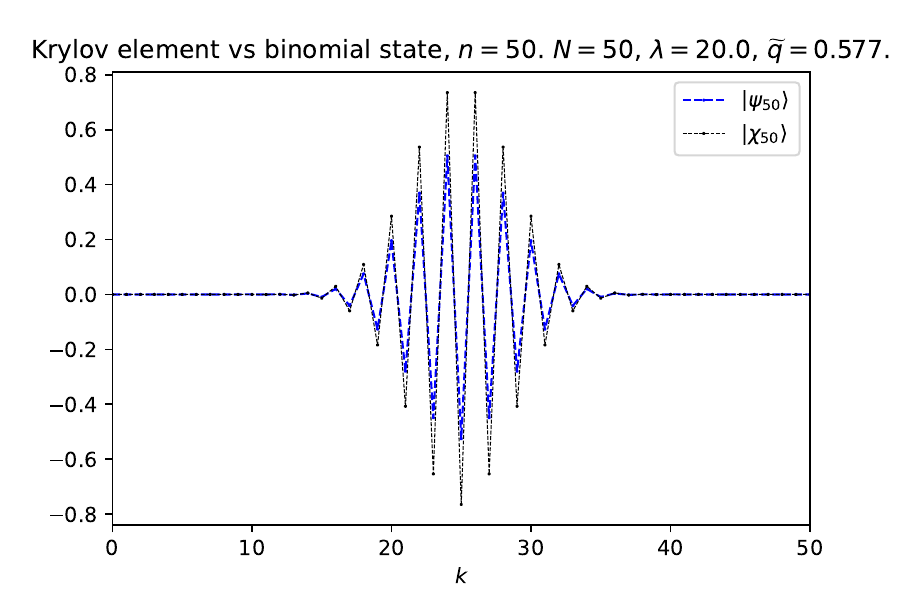}
    \caption{$\widetilde{q}=0.577$, $\lambda = 20$.}
    \label{fig:qt0pt577_lambda20pt0}
\end{figure}


\begin{figure}
    \centering
    \includegraphics[width=0.45\linewidth]{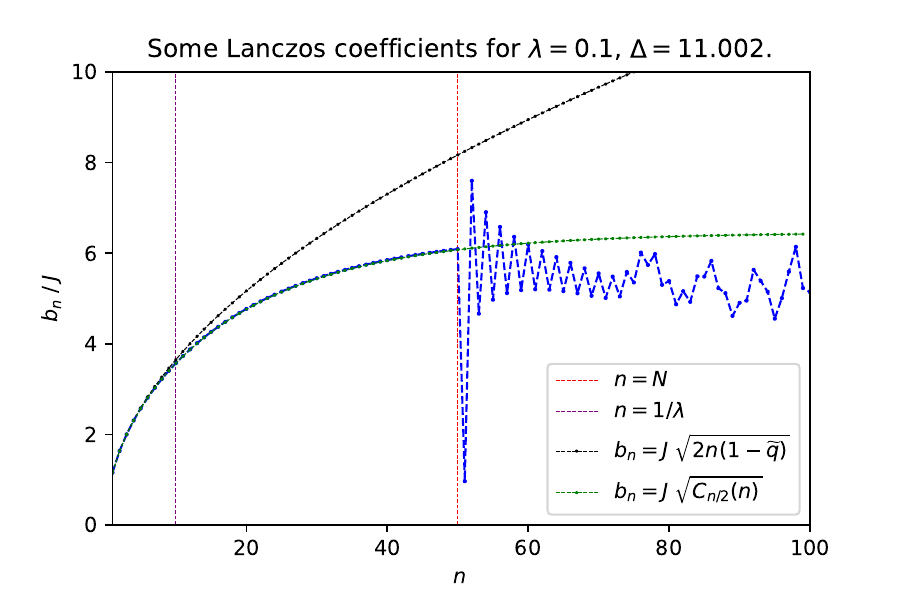}  \includegraphics[width=0.45\linewidth]{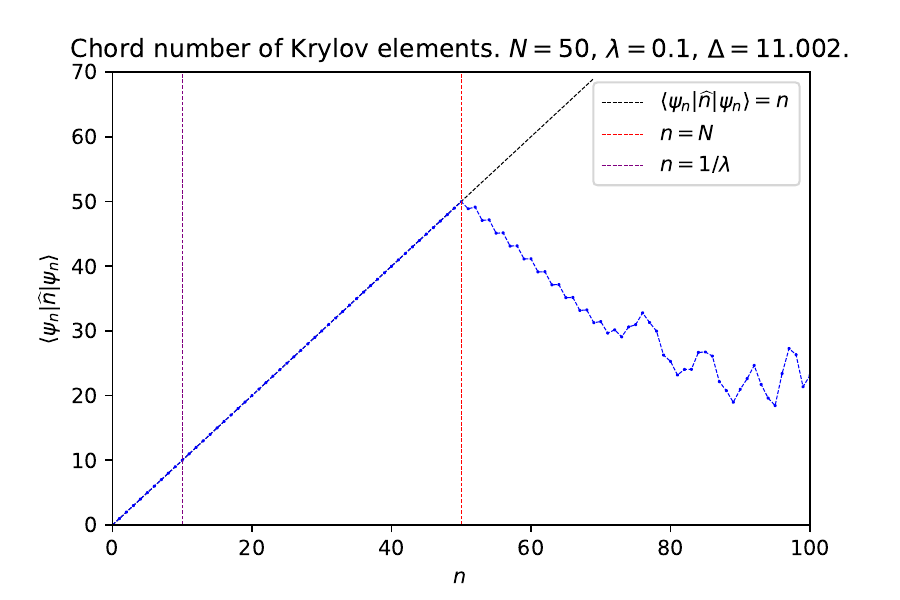} \\
    \includegraphics[width=0.45\linewidth]{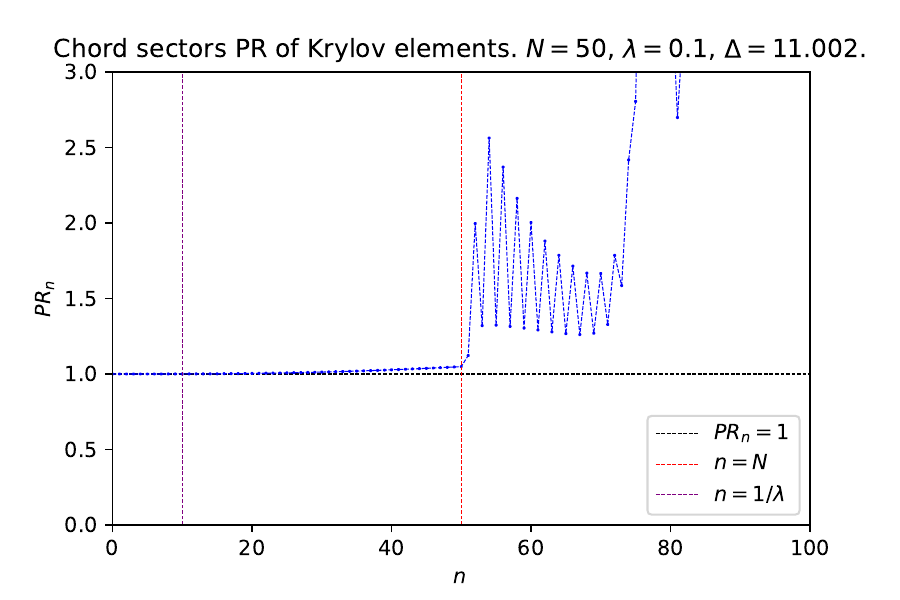}  \includegraphics[width=0.45\linewidth]{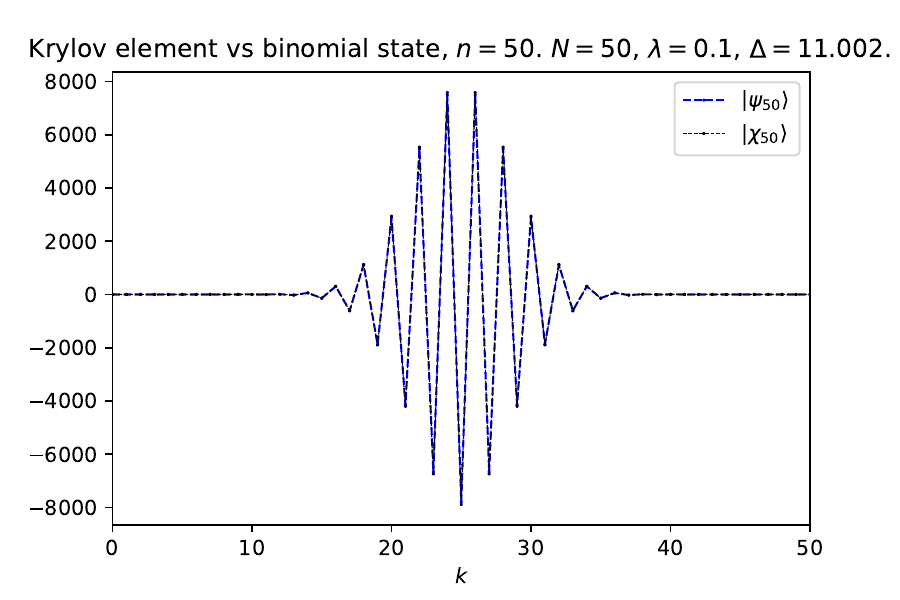}
    \caption{$\Delta=11.002$, $\lambda = 0.1$.}
    \label{fig:Delta11pt002_lambda0pt1}
\end{figure}

\begin{figure}
    \centering
    \includegraphics[width=0.45\linewidth]{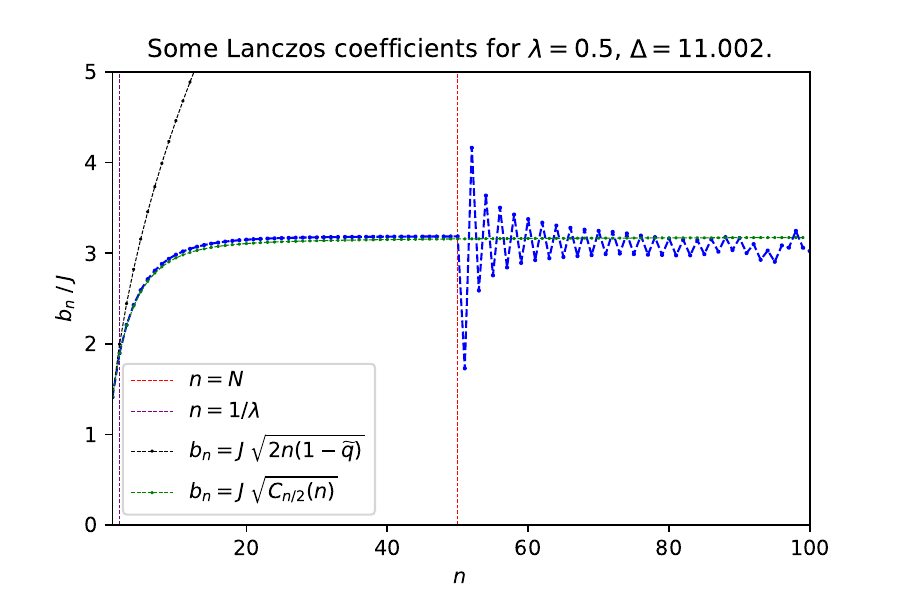}  \includegraphics[width=0.45\linewidth]{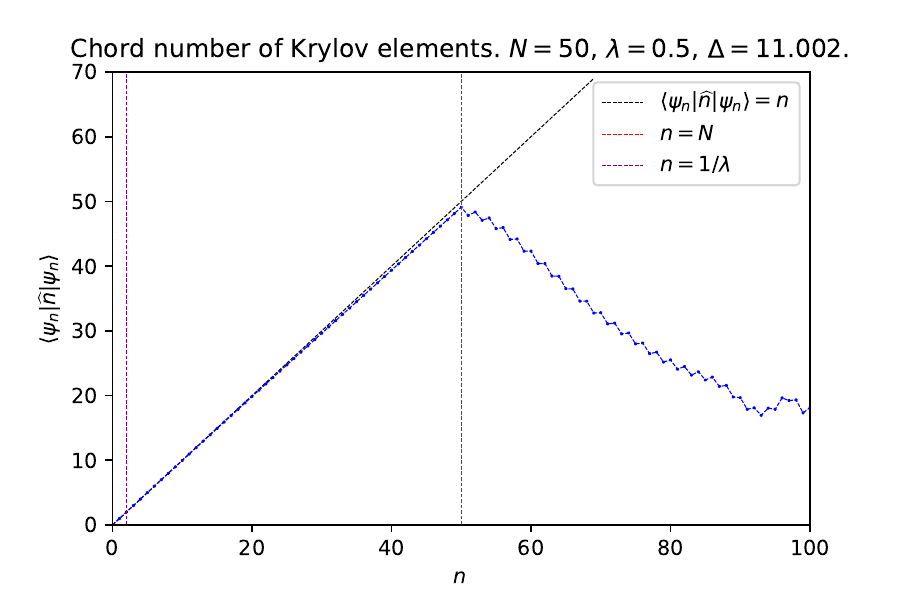} \\ 
    \includegraphics[width=0.45\linewidth]{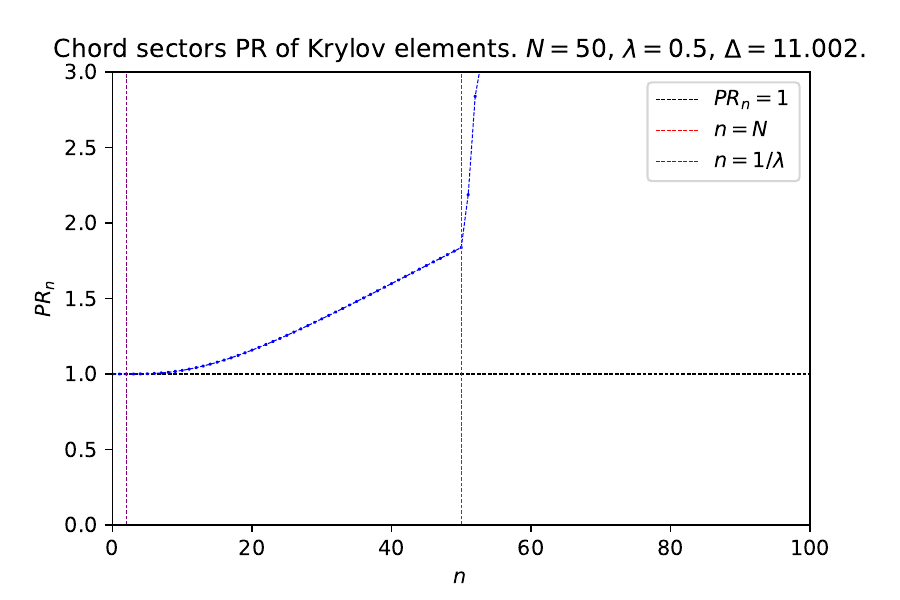}  \includegraphics[width=0.45\linewidth]{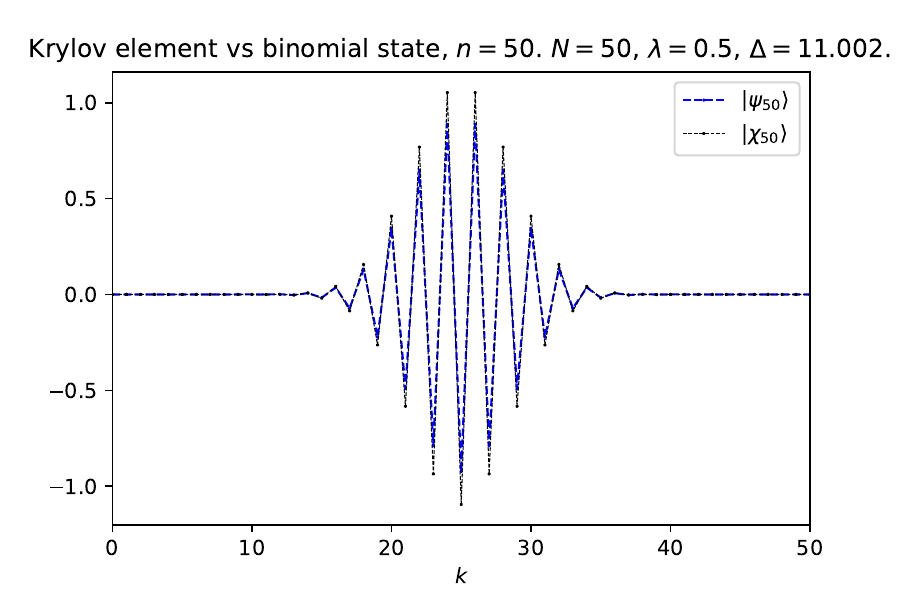}
    \caption{$\Delta=11.002$, $\lambda = 0.5$.}
    \label{fig:Delta11pt002_lambda0pt5}
\end{figure}

\begin{figure}
    \centering
    \includegraphics[width=0.45\linewidth]{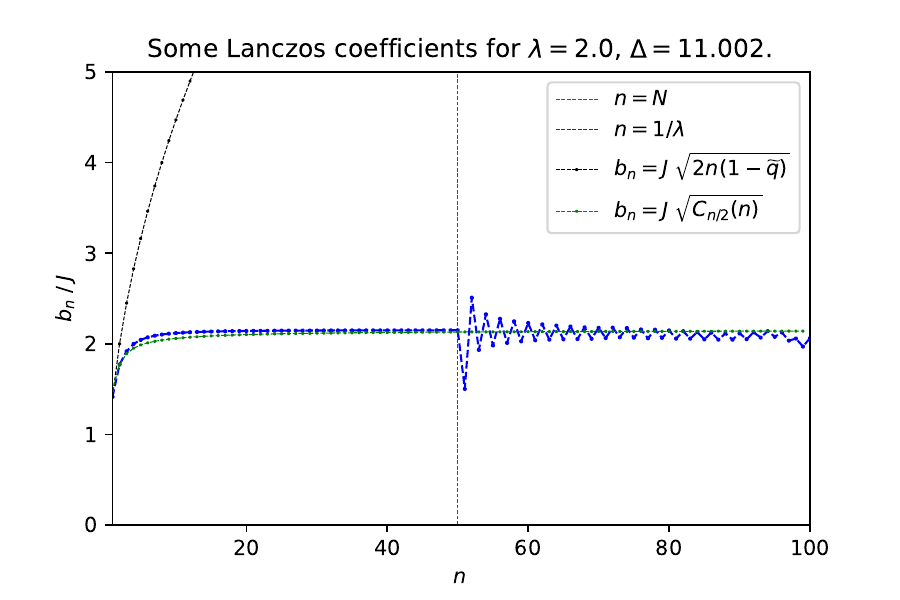}  \includegraphics[width=0.45\linewidth]{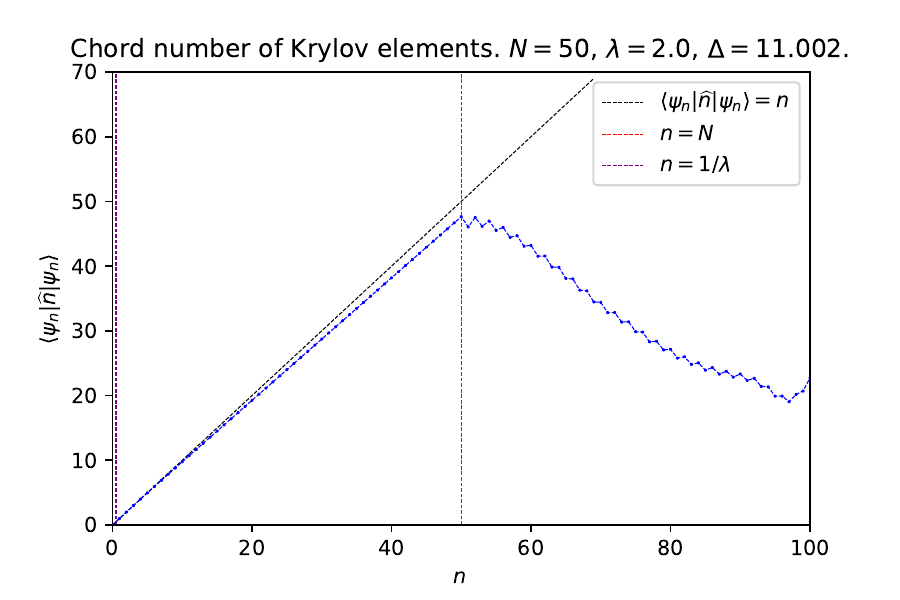} \\
    \includegraphics[width=0.45\linewidth]{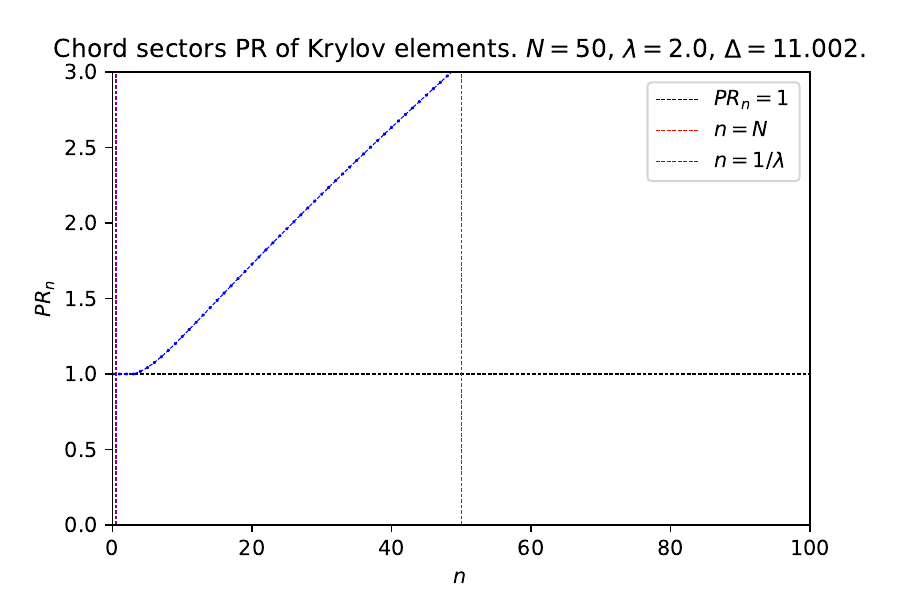}  \includegraphics[width=0.45\linewidth]{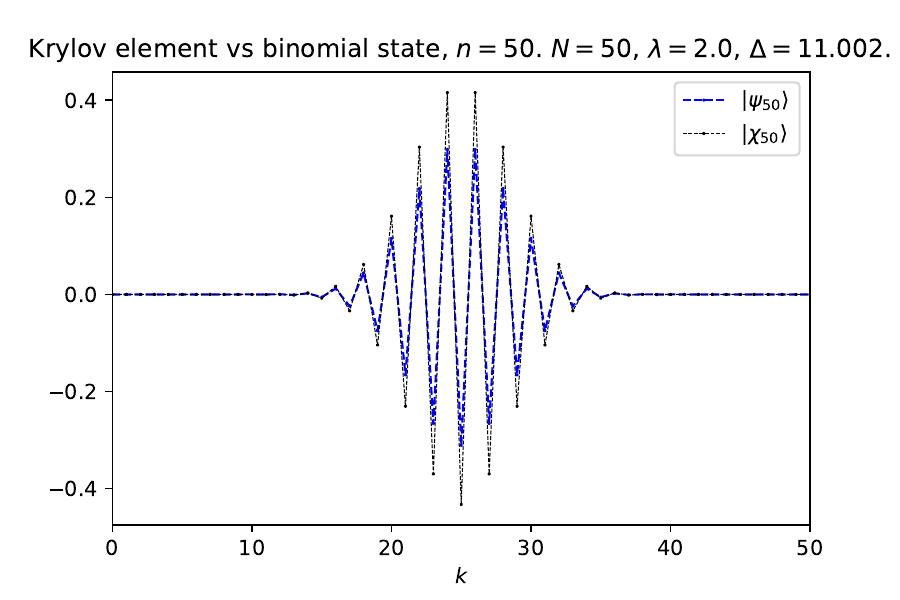}
    \caption{$\Delta=11.002$, $\lambda = 2$.}
    \label{fig:Delta11pt002_lambda2pt0}
\end{figure}

\begin{figure}
    \centering
    \includegraphics[width=0.45\linewidth]{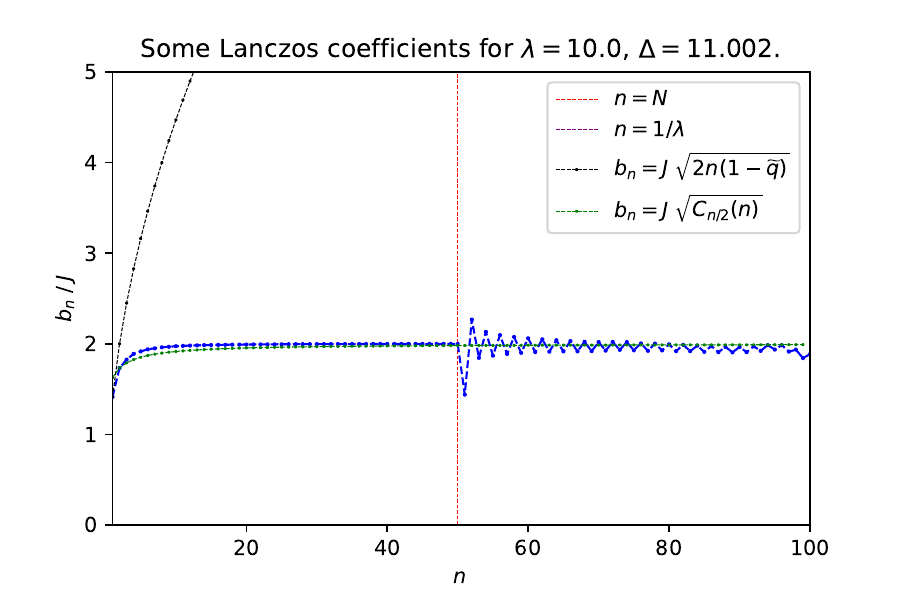}  \includegraphics[width=0.45\linewidth]{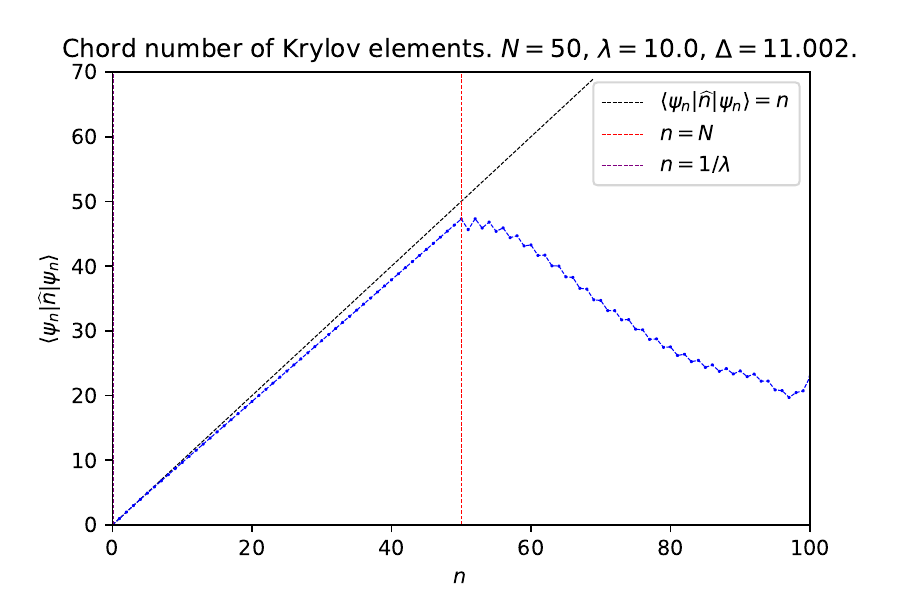} \\
    \includegraphics[width=0.45\linewidth]{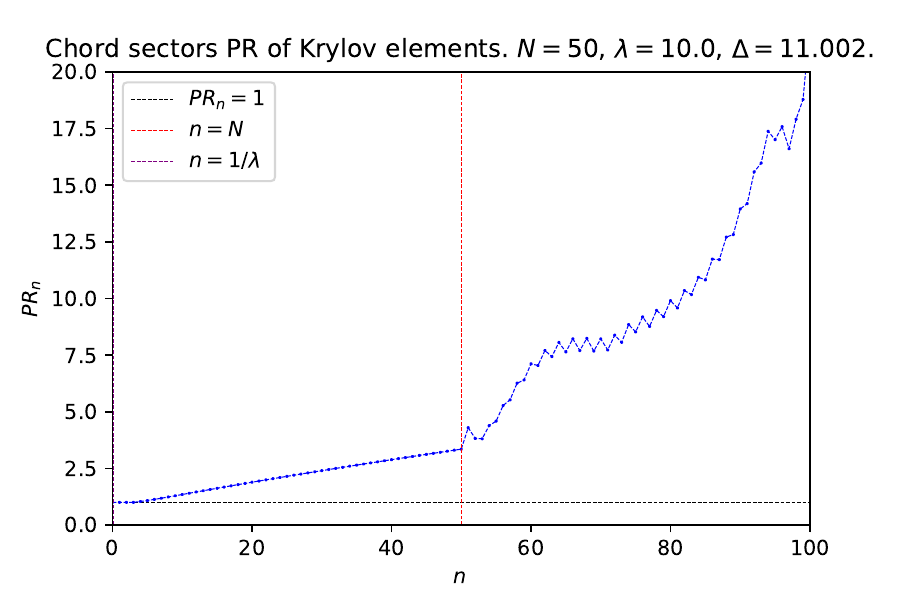}  \includegraphics[width=0.45\linewidth]{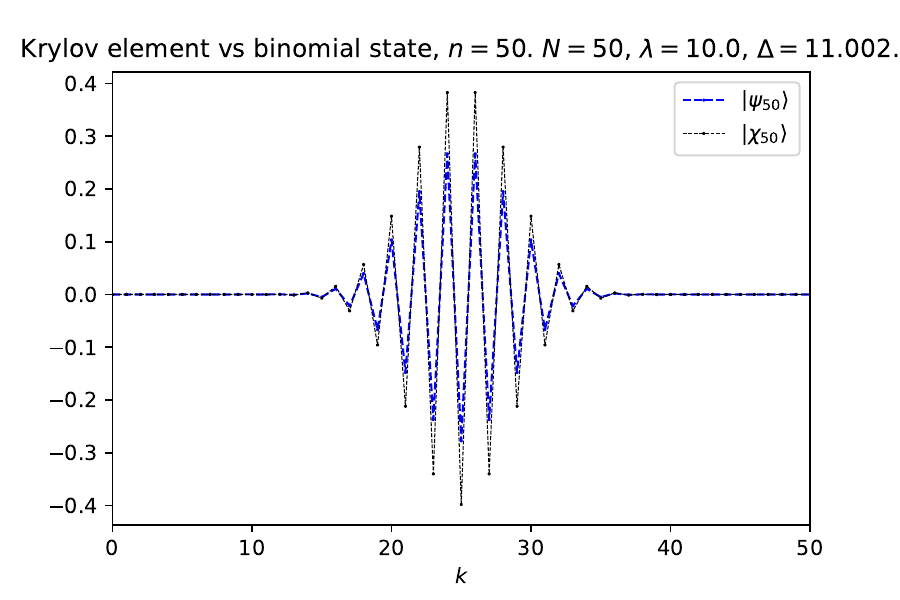}
    \caption{$\Delta=11.002$, $\lambda = 10$.}
    \label{fig:Delta11pt002_lambda10pt0}
\end{figure}

\begin{figure}
    \centering
    \includegraphics[width=0.45\linewidth]{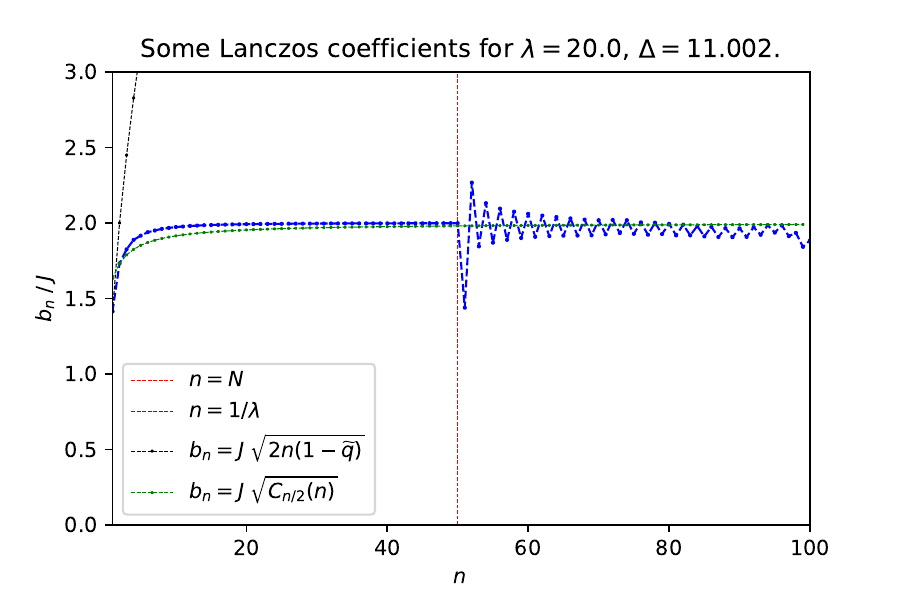}
    \includegraphics[width=0.45\linewidth]{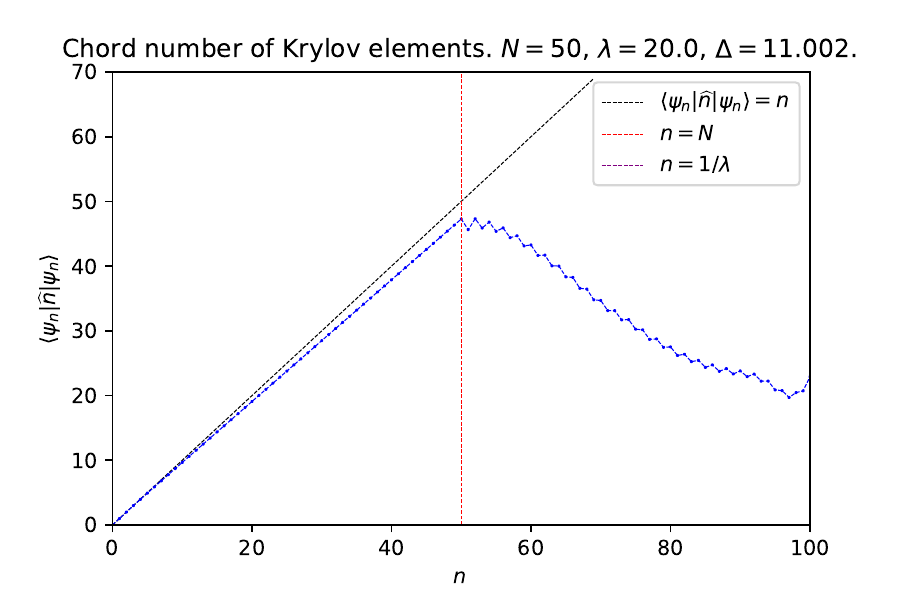} \\
    \includegraphics[width=0.45\linewidth]{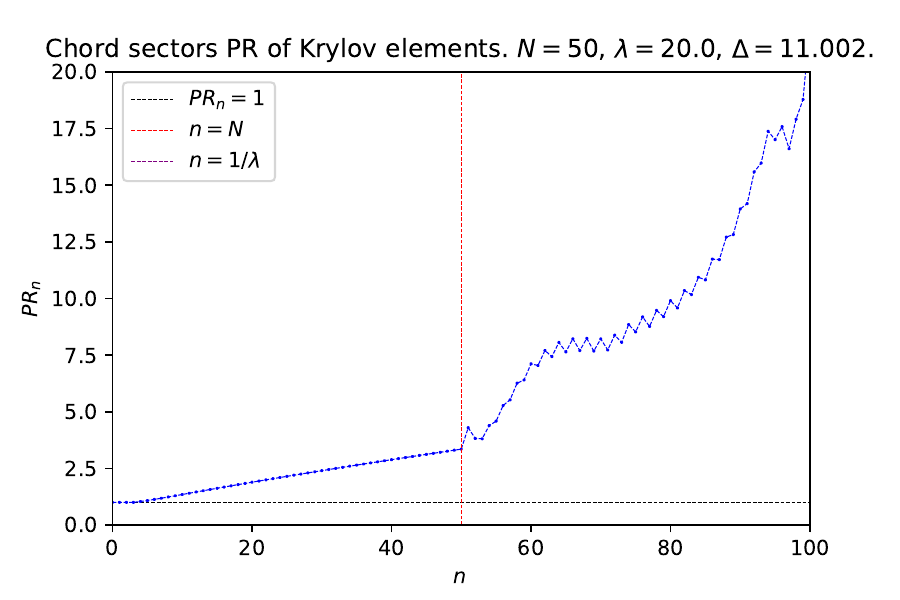}  \includegraphics[width=0.45\linewidth]{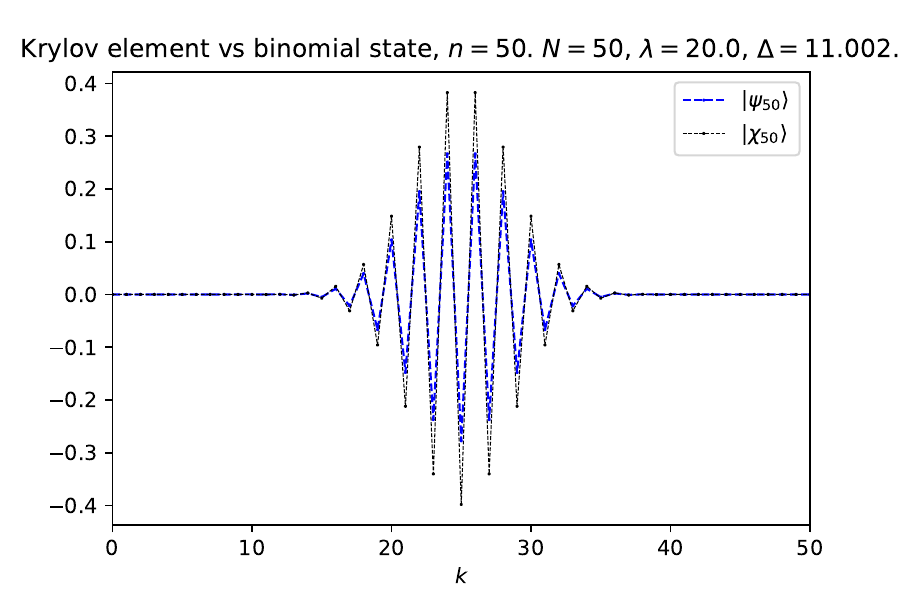}
    \caption{$\Delta=11.002$, $\lambda = 20$.}
    \label{fig:Delta11pt002_lambda20pt0}
\end{figure}


\begin{figure}
    \centering
    \includegraphics[width=0.45\linewidth]{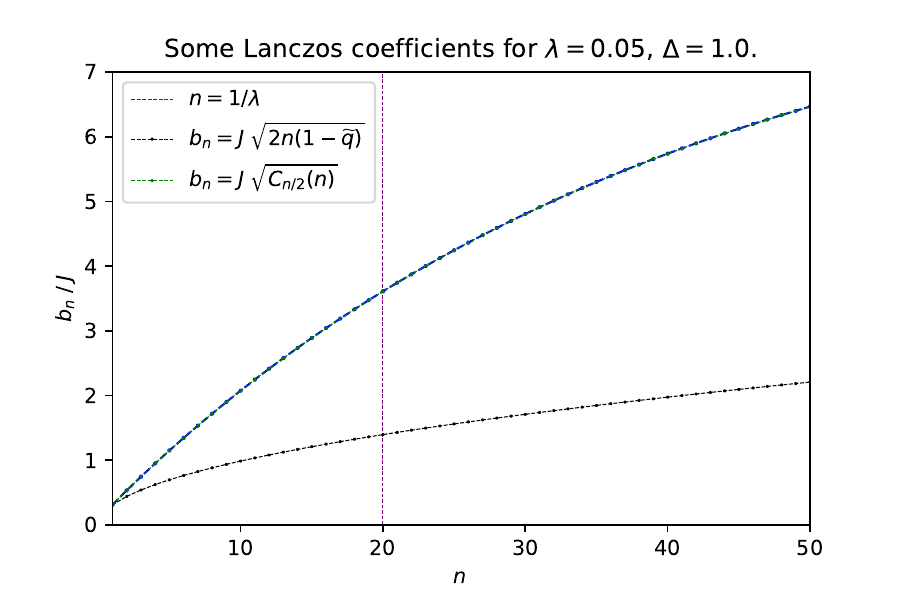} \includegraphics[width=0.45\linewidth]{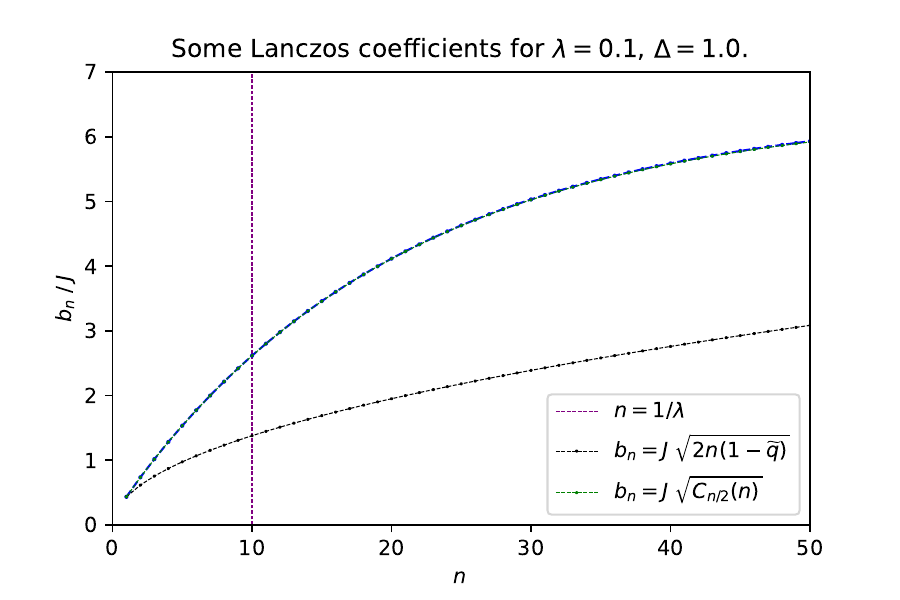} \\
    \includegraphics[width=0.45\linewidth]{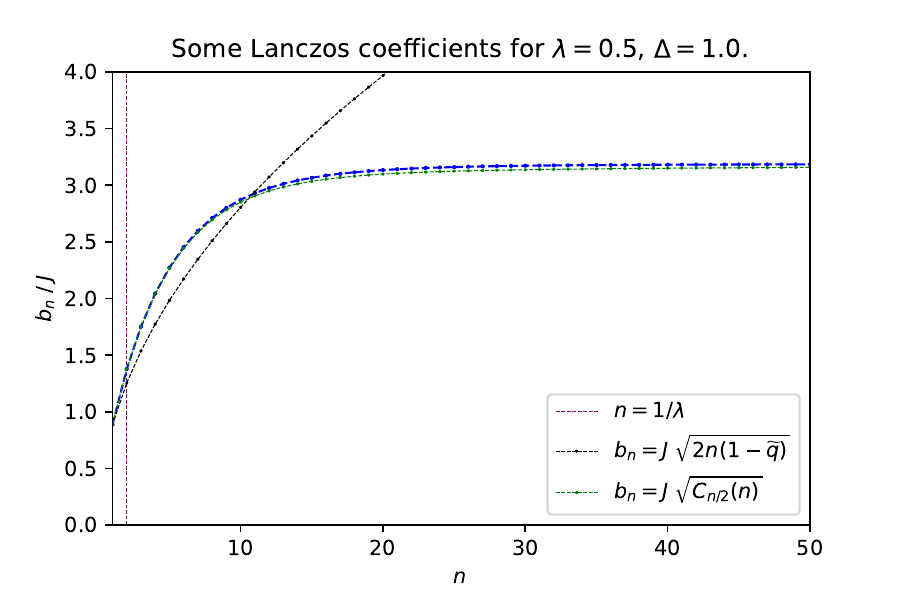}
    \caption{This figure merges together the Lanczos sequences for fixed $\Delta=1$ and $\lambda=0.05,0.1,0.5$. For these parameter values, $\widetilde{q}$ is sufficiently close to $1$ as to make the numerical instabilities of the Lanczos algorithm unbearable at double floating point machine precision (cf. discussion in section \ref{subsect:inner_product}). In these cases, only the first few Lanczos coefficients were computed using high precision arithmetics, and no study of chord expectation values and sector participation ratios was performed.}
    \label{fig:Delta1pt0_lambda0pt05_lambda0pt1_lambda0pt5}
\end{figure}

\begin{figure}
    \centering
    \includegraphics[width=0.45\linewidth]{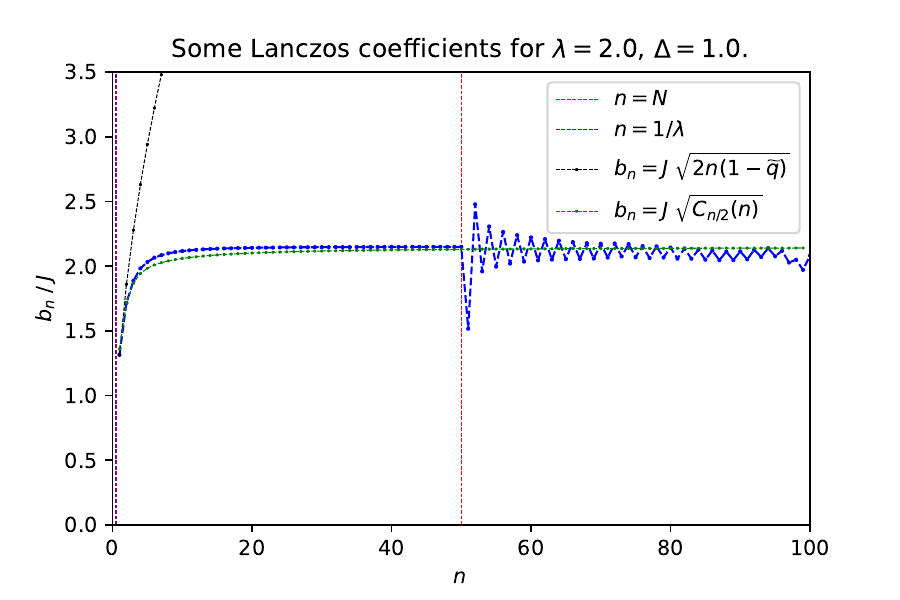}  \includegraphics[width=0.45\linewidth]{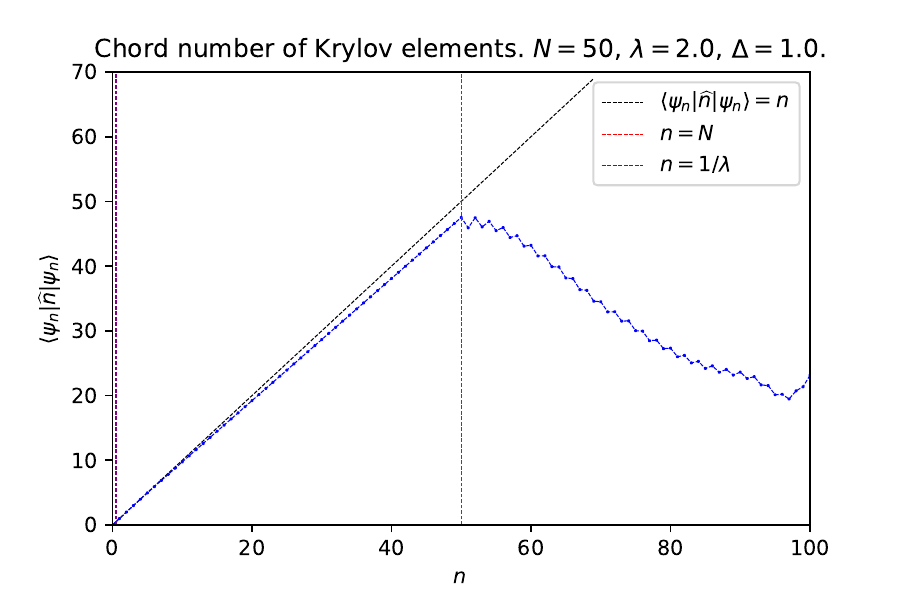}  \\
    \includegraphics[width=0.45\linewidth]{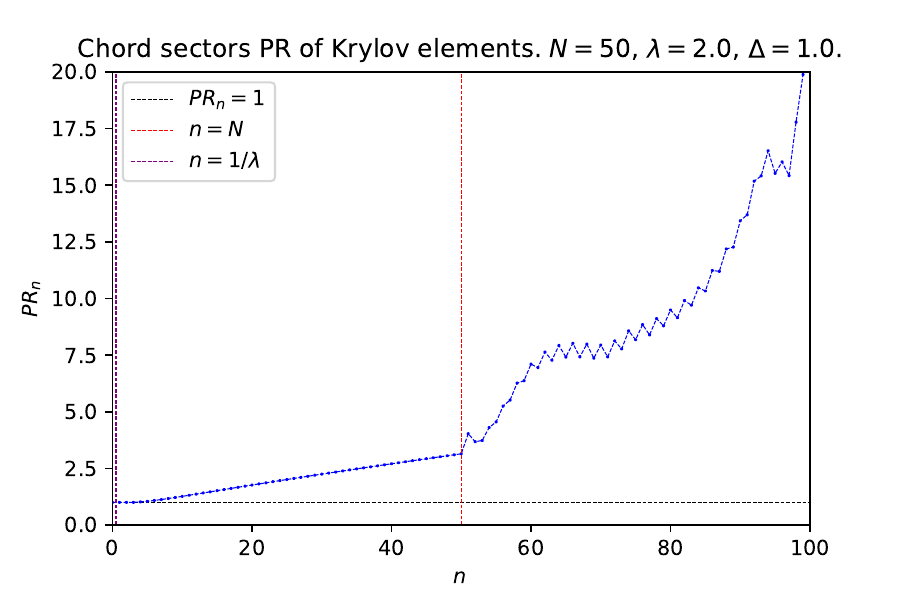}  \includegraphics[width=0.45\linewidth]{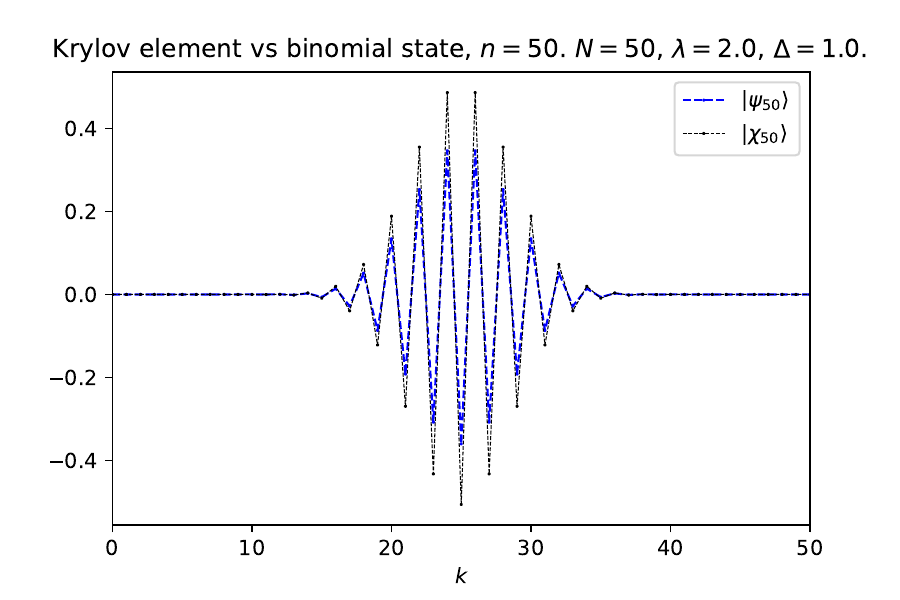}
    \caption{$\Delta=1$, $\lambda = 2$.}
    \label{fig:Delta1pt0_lambda2pt0}
\end{figure}

\begin{figure}
    \centering
    \includegraphics[width=0.45\linewidth]{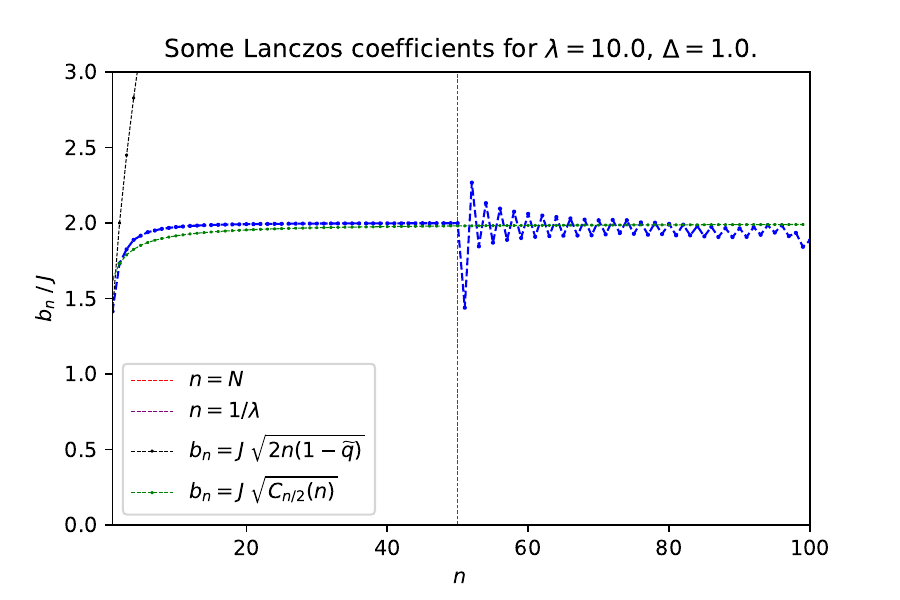}  \includegraphics[width=0.45\linewidth]{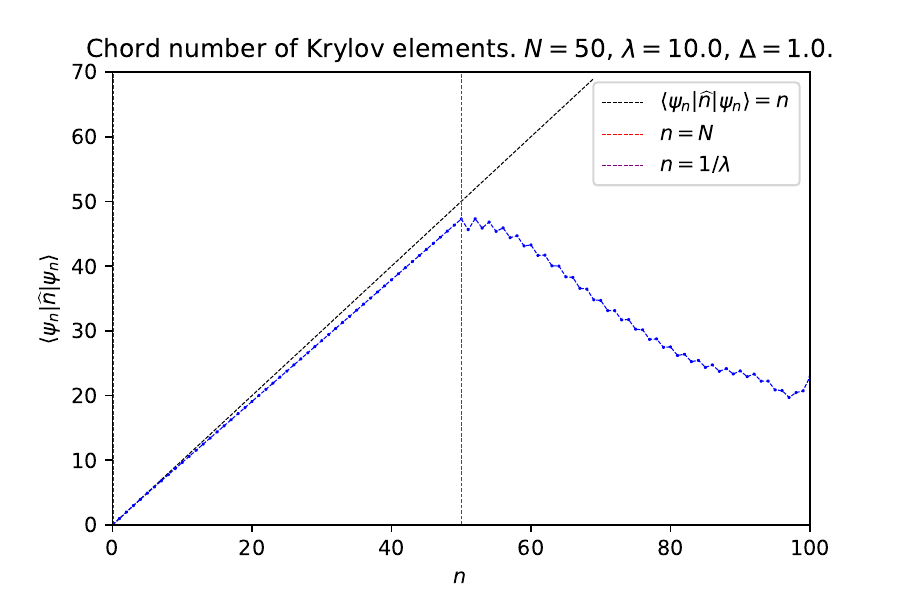}  \\
    \includegraphics[width=0.45\linewidth]{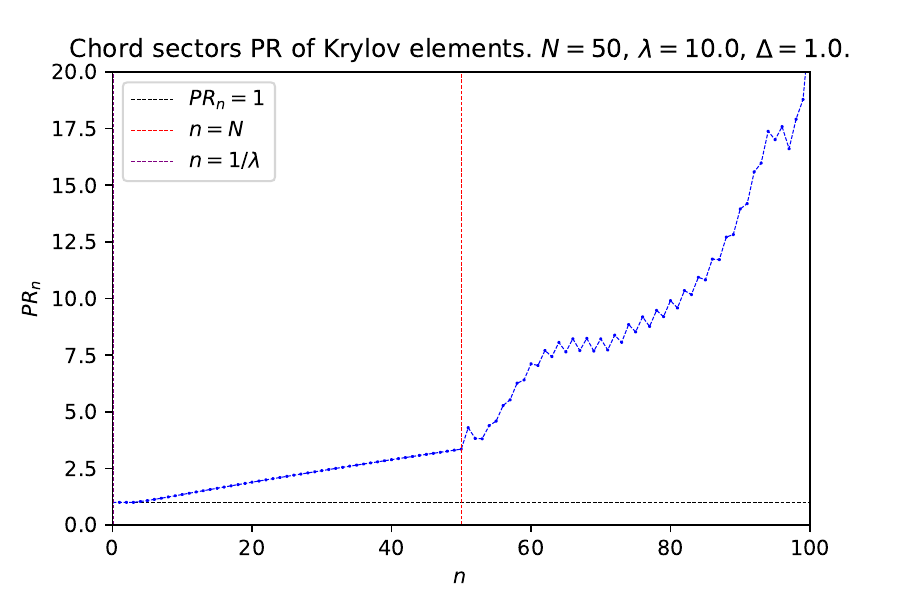}  \includegraphics[width=0.45\linewidth]{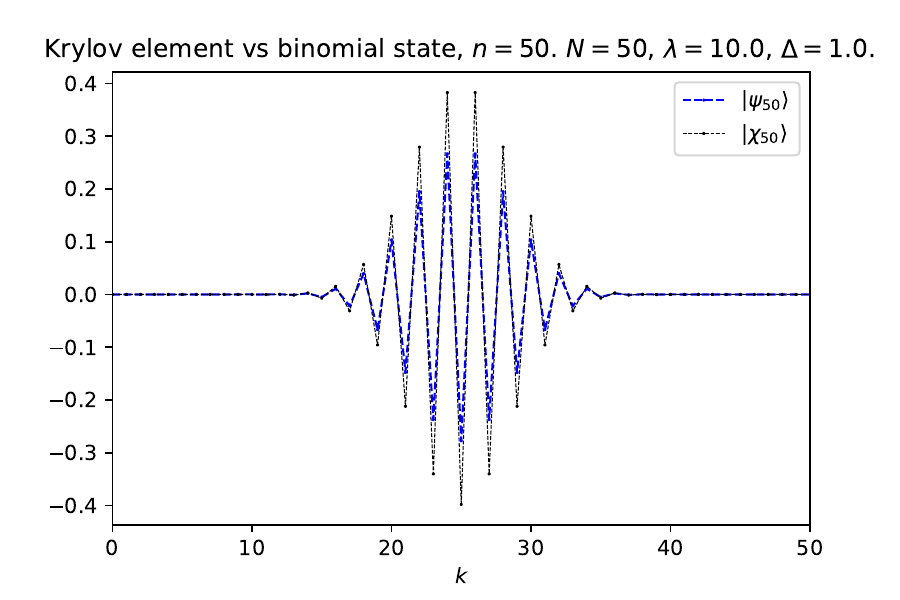}
    \caption{$\Delta=1$, $\lambda = 10$.}
    \label{fig:Delta1pt0_lambda10pt0}
\end{figure}

\begin{figure}
    \centering
    \includegraphics[width=0.45\linewidth]{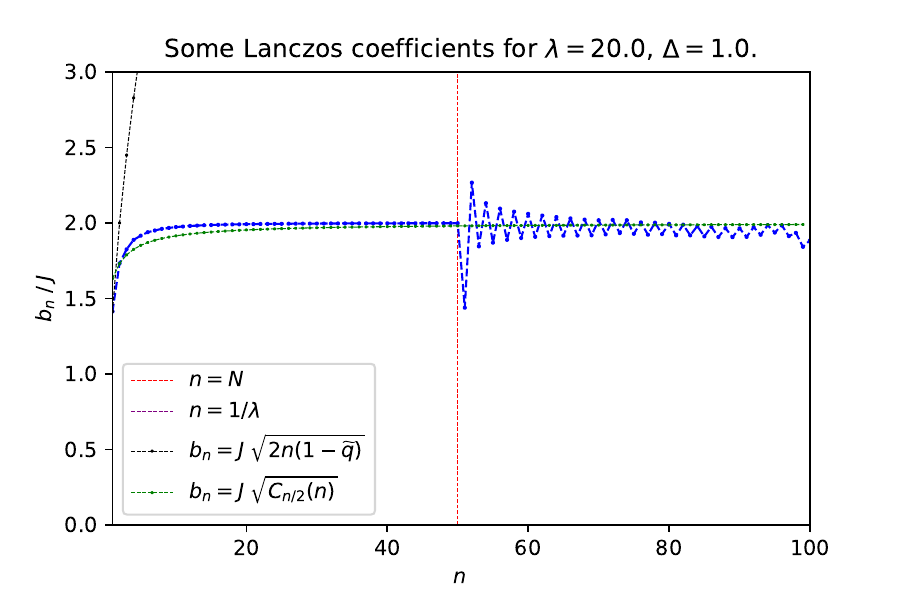}  \includegraphics[width=0.45\linewidth]{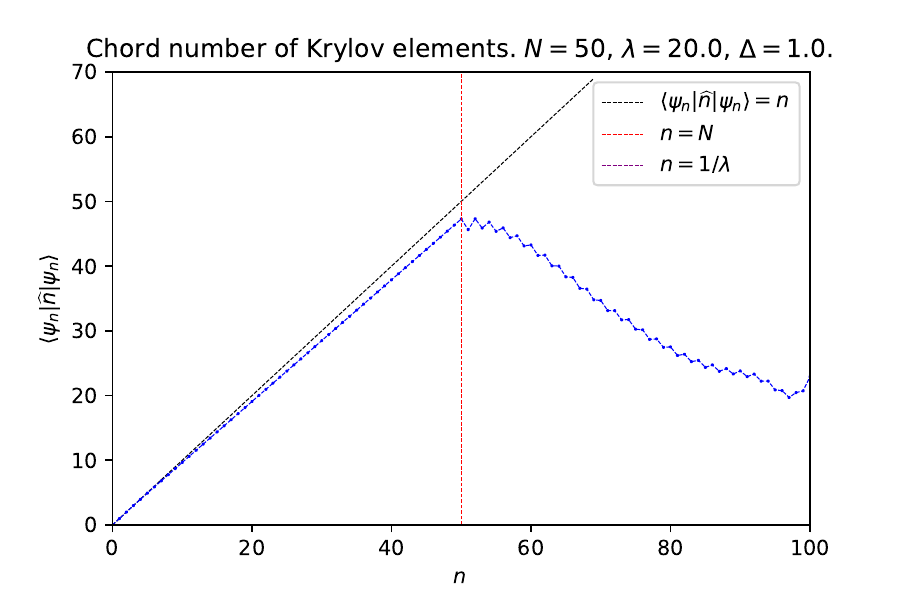} \\
    \includegraphics[width=0.45\linewidth]{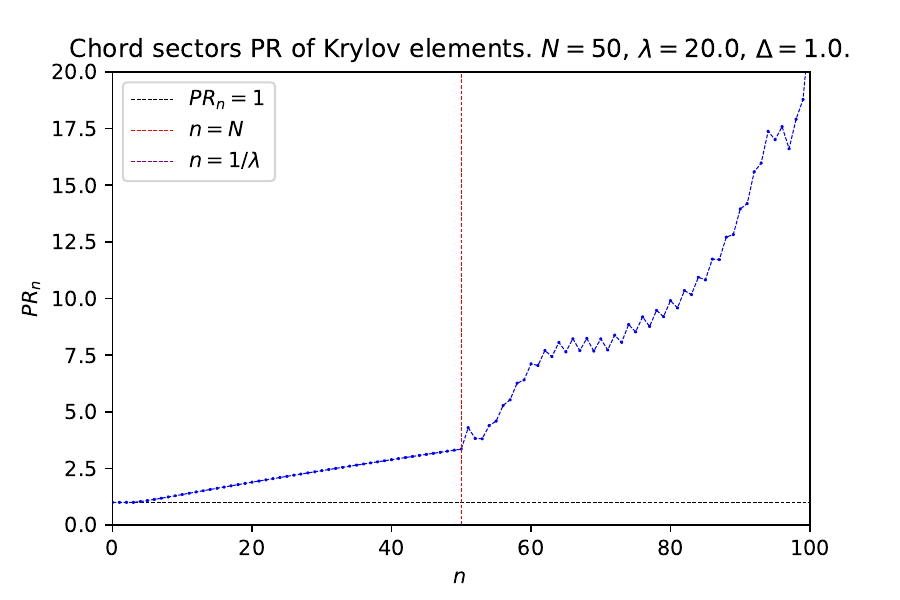}  \includegraphics[width=0.45\linewidth]{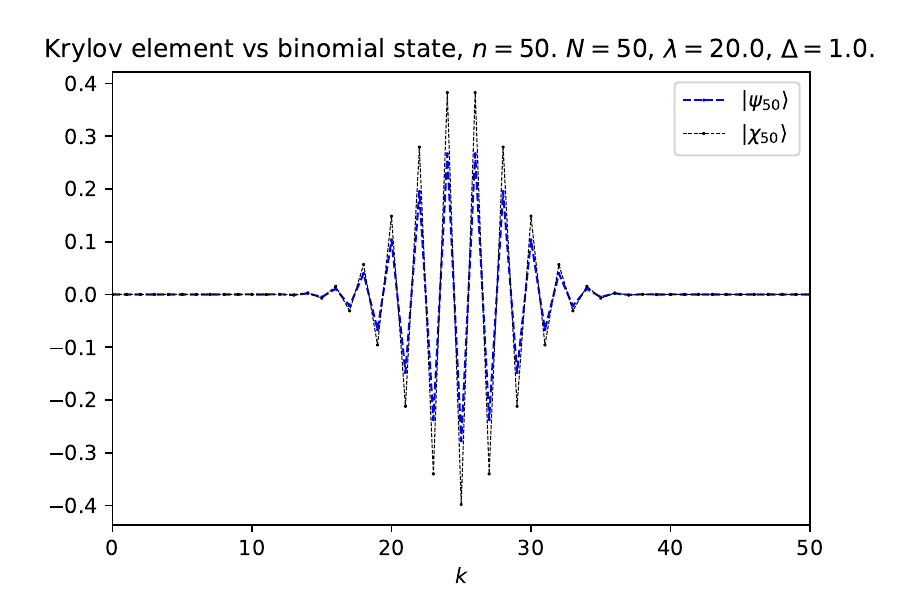}
    \caption{$\Delta=1$, $\lambda = 20$.}
    \label{fig:Delta1pt0_lambda20pt0}
\end{figure}

\clearpage
\section{Details on asymptotic analysis}\label{app:analytics_details}
The objective of this appendix is showing that the leading contribution to sums of the kind in \eqref{eq:lower_diag_L} is coming out of elements inside a particular where $c_k(n)$ is constant. We will show that we can consider this region as a square centered around $k=m\sim n/2$, with side length given essentially by $\frac{1}{\lambda}$, the characteristic variation scale of the semiclassical variables. Notice that in this appendix, we set $J=1$ to simplify the notation during our analysis. In order to reinstate the $J$ dependence, it will be sufficient to perform the change $b_n\to b_n/J$.\\
We will start with a slightly different problem, by examining the sum:
\begin{equation}\label{eq:sum_opstate}
    \frac{1}{b_n^+(b_{n-1}^+)^2...(b_1^+)^2}\sum_{k,m=0}^{n-1} c_k^+(n)\,\binom{n-1}{m}\binom{n-1}{k}\braket{m,n-1-m|k,n-1-k},
\end{equation}
where $c_k^+(n)$ are objects analogous to \eqref{Ckn_simplified}, defined as:
\begin{equation}\label{ckntilde_appx}
    c_k^+(n)\equiv n \frac{[n-k]}{n-k}(1+\tilde{q}q^k) + n\frac{[k+1]}{k+1}(1+\tilde{q}q^{n-1-k})
\end{equation}
This sum is easier to understand because of the absence of the alternating signs accompanying the binomial coefficients. Also we note that \eqref{eq:sum_opstate} is an interesting computation itself. Indeed this is the sum we obtain when we compute $\bra{\psi_{n-1}^+}a_L+a_R\ket{\psi_{n}^+}$, which are lower diagonal entries of the matrix $(H_L+H_R)$ on the basis made of states $\ket{\psi_n^+}$ defined as:
\begin{equation}
       \ket{\psi_n^+}=\frac{1}{b_n^+...b_1^+}\sum_k \binom{n}{k}\ket{k,n-k},
\end{equation}
 where the $b_n^+$ are again normalization coefficients. For analogous reasons as those outlined in \cref{Sec:Operator_KC} for the operator complexity and its basis $\{\ket{\psi_n}\}_n$, the basis $\{\ket{\psi_n^+}\}_n$ will be the Krylov basis, built from the seed state given by an operator $\mathcal{O}$ insertion on the infinite temperature TFD, under the evolution operator $H_L+H_R$. So \eqref{eq:sum_opstate} will give as the result the Lanczos coefficients that compute a new Krylov complexity, the $\mathcal{O}$TFD state complexity, that we will study in \cref{sect:OTFD_KC}. \\

 Let us start by examining the inner product that we use to compute the state overlaps in \eqref{eq:sum_opstate}.
In the semiclassical limit where we send 
 $\lambda\to 0$, $n_{L,\,R}\to\infty$ so that $l_{L,R}=\lambda n_{L,R}$ are fixed, the inner product of \eqref{inner_product_recursion_solution} becomes  \cite{Lin:2023trc}:
\begin{equation}
   \braket{x'|x}= \left\langle n_L^{\prime}, n_R^{\prime} \mid n_L, n_R\right\rangle=[n] !\left(\frac{\left(1-c^2\right) / 2}{\cosh \frac{x-x^{\prime}}{2}-c \cosh \frac{x+x^{\prime}}{2}}\right)^{2 \Delta},
\end{equation}
where $c^2=q^n$ and $x=\lambda\frac{n_L-n_R}{2}$, $x'=\lambda\frac{n_L'-n_R'}{2}$. So we have:
\begin{equation}
    \braket{m,n-m|k,n-k}=[n] !\left(\frac{\left(1-q^n\right) / 2}{\cosh \lambda\frac{k-m}{2}-q^{n/2} \cosh \lambda\frac{n-k-m}{2}}\right)^{2 \Delta}
\end{equation}
Now with this analytical form we want to understand in which regions of the plane $(k,m)$ the scalar product is suppressed. We have an exponential suppression controlled by $\Delta$ when we move of $\delta k$ outside the diagonal $k=m$ :
\begin{equation}
   \braket{k,n-k|k+\delta k,n-\delta k-k} \propto \left(\cosh \lambda\frac{\delta k}{2}-q^{n/2} \cosh \lambda\frac{n-2k-\delta k}{2}\right)^{-2 \Delta}\propto_{\lambda \delta k\gtrsim1} e^{-\lambda \delta k \Delta}
\end{equation}
In particular we notice that in the limit $\Delta\to\infty$, the overlap is null outside of a strip of characteristic width $\sim 1/\lambda$ centered on the diagonal (as first reported in \cite{Lin:2022rbf}). This is the scale of variation of our semiclassical variables (for example $\lambda n$, kept fixed), given by the condition $\delta k \lambda \sim 1$. The scalar product is approximately constant in this region, but notice that if $\Delta$ is held finite the size of this area is controlled by $1/\Delta$.\\
If, instead, we move on the diagonal, we have that $k=m=n/2$ is a minimum and the overlap increases if we go towards the edges of the summation region (\cref{fig:ps_behavior})\footnote{As a safety check, the points where the denominator is null are outside the summation bounds.}:
\begin{equation}
   \braket{n/2+\delta k,n/2-\delta k|n/2+\delta k,n/2-\delta k} \propto \left(1-q^{n/2} \cosh \lambda\delta k\right)^{-2 \Delta}
\end{equation}
Again, in the $\lambda\to0$ limit, this starts to deviate from the constant contribution in the minimum when $\delta k\propto 1/\lambda$. This is again a symptom that in this asymptotic region the scalar product depends on the semiclassical variables, whose unit variation corresponds to a $1/\lambda$ variation in the chord numbers.\\
Notice that these behaviors that we observed from the asymptotic analytical expressions for the inner product are in agreement with those obtained in \cref{fig:inner_product_plots} by solving the defining recursion relation \eqref{RRL} numerically.\\

\begin{figure}
  \centering
  \begin{minipage}[b]{0.49\textwidth}
  \includegraphics[width= \textwidth]{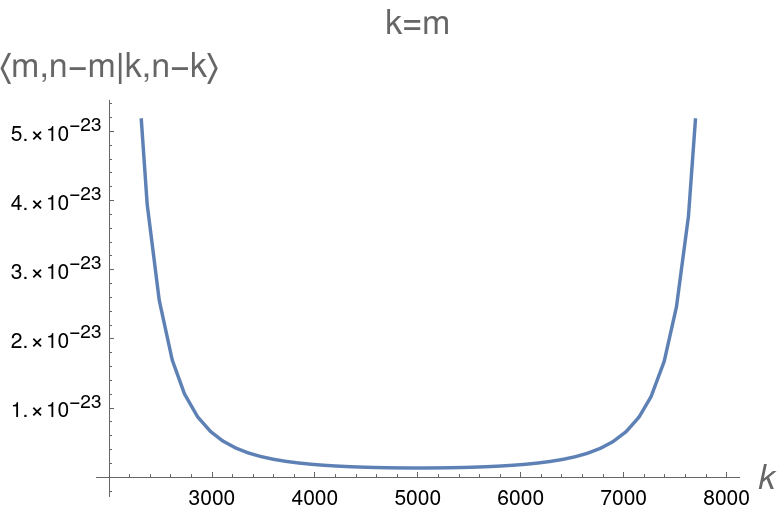} 
  \end{minipage}
  \hfill
  \begin{minipage}[b]{0.49\textwidth}
    \includegraphics[width=\textwidth]{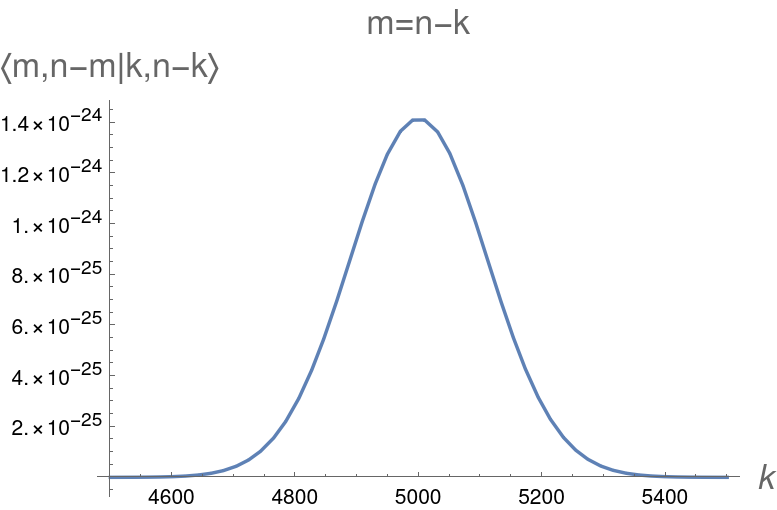}
  \end{minipage}
   \caption{Characteristic behavior of the state overlap $\braket{m,n-m|k,n-k}$ where $\lambda=0.01$, $\Delta=40$, $n=1000$, without the $[n]!$ multiplicative factor. The overlap is minimum in $n/2$ moving along the diagonal (left), while moving off the diagonal results in an exponential suppression (right).  } 
    \label{fig:ps_behavior}
\end{figure}
Let us now study $c_k^+(n)$. We find that the function  $k=n/2$ the function has a minimum, and by expanding around it we find:
\begin{equation}
    c_k^+(n)\sim_{k\sim n/2} c^+_{k=n/2}(n)+\mathrm{const.}\times \frac{(k-n/2)^2}{n^2}\sim c^+_{k=n/2}(n)+\mathrm{const.}\times (\lambda \delta k)^2,
\end{equation}
where $\delta k=k-n/2$,
so we can estimate that the correction becomes non-negligible when $\delta k\sim 1/\lambda$. \\

By putting everything together we have that $c_k^+(n)\braket{m,n-1-m|k,n-1-k}$ has a saddle point around $k=m\sim n/2$, and in a square region of side $1/\lambda$ we can approximate it with a constant.
Now, this region is actually the only one contributing to the leading order of the sum \eqref{eq:sum_opstate} in the semiclassical limit. This happens because the binomial coefficients in the sum make the contribution from these regions parametrically bigger than all others. We can see this for example from the Chernoff bound for the sum of binomial coefficients\footnote{See e.g. \url{https://en.wikipedia.org/wiki/Chernoff_bound}.}:\\
\begin{equation}
    \sum_{k=0}^{n/2-1/\lambda}\binom{n}{k}<2^n e^{-\frac{2}{\lambda n}\frac{1}{\lambda}},
\end{equation}
so the neglected tails of the sum are exponentially suppressed by a parametrically large coefficient, and they will give a subleading contribution to \eqref{eq:sum_opstate} or to norm computations of $\ket{\psi_n^+}$.\\

Now let us go back to our original problem of computing the sum \eqref{eq:lower_diag_L}. In this case the alternating sign attached to the binomials becomes potentially problematic, and we lose the localizing power of the binomial coefficient's sum we leveraged in the previous case. An example comes from considering the case $\Delta\to0$, where summing with alternating signs the binomial coefficients gives a null result. So with the alternating signs, even though binomials are bigger in the central region, we may have non-trivial cancellations so that even terms closer to the edges of the sum parameters can contribute. We can bring this case under analytical control by considering the limit $\Delta\to\infty$ of heavy operators. This limit aids in the localization of \eqref{eq:lower_diag_L}, because it suppresses the scalar product outside of a strip of characteristic width $1/\lambda$ centered on the diagonal $k=m$. Now let us study the sum of binomial coefficients with alternating signs on a rectangular strip centered in $k=m=n/2$, with widths $\sim 1/\lambda$ across the diagonal and $a$ along it:
\begin{equation}
   \sum_{i=-\frac{1}{\lambda},k=0}^{i=\frac{1}{\lambda},k=n/2-a}(-1)^{i}\binom{n}{k}\binom{n}{k+i}\approx\sum_{k=0}^{k=n/2-a}\binom{n}{k}\left(\frac{n+\frac{1}{\lambda}-k}{n}\binom{n}{k-\frac{1}{\lambda}}+\frac{\frac{1}{\lambda}+k}{n}\binom{n}{k+\frac{1}{\lambda}}\right),
\end{equation}
where we parametrized $m=k+i$. In this case we are summing only in a $1/\lambda$-wide interval across the diagonal because of the suppression coming from the scalar product. Now the picture for the sum along the diagonal is more similar to that of the previous case. Indeed, in an analogous manner we can bound the contribution from a tail outside the strip we described, when we take $a\sim 1/\lambda$:
\begin{equation}
\begin{aligned}
     &\sum_{k=0}^{k=n/2-\frac{1}{\lambda}}\binom{n}{k}\left(\frac{n+\frac{1}{\lambda}-k}{n}\binom{n}{k-\frac{1}{\lambda}}+\frac{\frac{1}{\lambda}+k}{n}\binom{n}{k+\frac{1}{\lambda}}\right)\lessapprox\mathrm{const.}\binom{n}{n/2} \left(\sum_k\binom{n}{k}\right)\propto e^{-\frac{1}{\lambda}},
\end{aligned} 
\end{equation}
so tails are parametrically suppressed with respect to the sum on the full strip.\\
Analogous considerations to those we performed for $c_k^+(n)$ show that $c_k(n)$ is approximately constant in a width of $1/\lambda$ from $k=n/2$. So we have that sums of the type of \eqref{eq:lower_diag_L} essentially saturate in the $\frac{1}{\lambda}\times \frac{1}{\lambda}$ square region, where all terms in the addends, other than the binomials, are approximately constant.\\

We have performed the analysis of the sums \eqref{eq:lower_diag_L}, \eqref{eq:sum_opstate} in chord number space, and we have verified that the leading contributions are localized in a region of the sum parameters whose dimension is controlled by the scale of the semiclassical variables $\sim \lambda^{-1}$. In the following sections we will show how to reinstate the previous arguments using instead the semiclassical lengths variables, where we can phrase them as a saddle point analysis controlled by $1/\lambda$.
\subsection{Asymptotic behavior of the binomial coefficients in the semiclassical limit}\label{appx:Binom_semiclassical}
This section intends to give a formal analysis of the behavior of the binomial coefficients, which are ubiquitous in the sums defining Krylov vectors and Lanczos coefficients, in the semiclassical limit \eqref{Semiclassical_limit}.
As we shall see, the effect of this limit is to \textit{squeeze} the binomials turning them into Dirac delta functions of the semiclassical variables, putting on solid footing our arguments in favor of the relevant sums being dominated by the contributions due to the middle of the summation domain. The study can also be phrased in terms of a saddle-point analysis which is controlled by the large parameter $\lambda^{-1}$ in the semiclassical limit.

\subsubsection{Binomials as representations of the Dirac delta function}\label{appx:Binom_delta_EasyCase}

Let us consider in this section the semiclassical limit of a discrete sum involving objects that admit a smooth limiting form, weighted by the binomial coefficients. In this way we can isolate the semiclassical analysis of the binomial coefficient as a separate problem. In particular, we consider the computation of the following norm:
\begin{equation}
    \label{Norm_binomal_plus}
     \langle \chi^{(+)}_{n+1}|\chi^{(+)}_{n+1}\rangle = \sum_{k^\prime,k=0}^{n} c^{(+)}_k(n+1) \binom{n}{k^\prime}\binom{n}{k}\langle k^\prime , n-k^\prime | k,n-k\rangle~,
\end{equation}
where $c_k^{(+)}(n)$ was defined in \eqref{ckntilde}.
Equation \eqref{Norm_binomal_plus} is the norm of the binomial state that contributes to the Krylov basis vectors of the state $|0,0\rangle$ evolving under the total Hamiltonian $H_R+H_L$, namely:
\begin{equation}
    \label{Binom_plus}
    |\chi^{(+)}_n\rangle = \sum_{k=0}^n \binom{n}{k} |k,n-k\rangle~.
\end{equation}

We now consider the exercise of taking the semiclassical limit \eqref{Semiclassical_limit} of \eqref{Norm_binomal_plus}, i.e. we shall take $\lambda\to 0$ while $n\to+\infty$ keeping $\lambda n\equiv l$ fixed as our semiclassical (dimensionless) length variable. Consistently with this prescription, we will take $\lambda k\equiv l_L$, in a way such that the sums in \eqref{Norm_binomal_plus} will become Riemann integrals whose measure will be given by $\lambda\to 0$. Both the $c^{(+)}_k(n)$ coefficient and the overlap $\langle k^\prime , n-k^\prime | k,n-k\rangle$ admit smooth limiting forms in the semiclassical limit as functions of the semiclassical variables $l$ and $l_L$. The latter is given by \eqref{eq:scal_prod_semiclass}, while the former is just analogous to \eqref{ckn_semicl}, namely: 
\begin{equation}
    \label{ckn_plus_semicl}
    c^{(+)}_k(n)\underset{\lambda\to 0}{\sim} c^{(+)}(l,l_L;\lambda) \equiv \frac{l}{l-l_L} \frac{1-e^{-(l-l_L)}}{\lambda}(1+\widetilde{q}e^{-l_L}) + \frac{l}{l_L}\frac{1-e^{-l_L}}{\lambda}(1+\widetilde{q}e^{-(l-l_L)})~.
\end{equation}
It is therefore left to study the semiclassical form of the binomial coefficients in \eqref{Norm_binomal_plus}, together with the explicit mechanism through which the sums become integrals. We note that each or the two binomials in \eqref{Norm_binomal_plus} can be written as:
\begin{equation}
    \label{Binom_semicl_nascent}
    \binom{n}{k} = \frac{\Gamma(n+1)}{\Gamma(k+1)\Gamma(n-k+1)}\overset{\text{semicl.}}{\longrightarrow} \frac{\Gamma(l/\lambda+1)}{\Gamma(l_L/\lambda +1) \Gamma(l/\lambda - l_L/\lambda+1)}=\lambda ~ \eta_{\lambda}\left(l,l_L-\frac{l}{2}\right)~,
\end{equation}
where the leftover $\lambda$ prefactor in the rightmost term of \eqref{Binom_semicl_nascent} will be combined with the corresponding $k$-sum to yield a continuous integral, and we have defined the function $\eta_\lambda (l,x)$ of the length offset $x=l_L-\frac{l}{2}=l_L-l_R$ as:
\begin{equation}
    \label{eta_lambda_def}
    \eta_\lambda (l,x):= \frac{1}{\lambda} \eta\left( \frac{l}{\lambda},\frac{x}{\lambda} \right)~,\quad\text{for}\quad \eta(l,x)=\left\{ \begin{aligned}
        & \frac{\Gamma(l+1)}{\Gamma(l/2+x+1)\Gamma(l/2-x+1)}~,\quad |x|\leq l/2~, \\
        &0~,\qquad \qquad\qquad\qquad\qquad\qquad\quad~ \text{else.}
    \end{aligned} \right.
\end{equation}
Regarded as a function of $x$, a function $\eta_\lambda(l,x)$ defined out of an absolutely integrable and symmetric function $\eta(l,x)$ is a \textit{nascent} Dirac delta function in the sense that, when $\lambda$ goes to zero, it becomes asymptotically equivalent to a distribution that is proportional to a delta function centered at $x=0$, as one may verify integrating it against a test function in the $\lambda\to 0$ limit. Indeed, considering the normalization
\begin{equation}
    \label{Normalization_eta_function}
    f(l):=\int_{\mathbb{R}}dx ~\eta(l,x)~,
\end{equation}
we may take a test function $g(x)$ and integrate it against $\eta_\lambda (l,x)$:
\begin{equation}
    \label{etalambda_integral_test}
    \int_{\mathbb{R}}dx ~\eta_\lambda(l,x) g(x)=\int_{\mathbb{R}}dx~ \eta(l/\lambda,x) g(\lambda x) \underset{\lambda\sim 0}{\sim} g(0)~f\left(\frac{l}{\lambda}\right)~,
\end{equation}
where we have made use of the definition of $\eta_\lambda$ in \eqref{eta_lambda_def} and changed variables conveniently. In other words, we have the asymptotic statement:
\begin{equation}
    \label{binom_asympt_delta}
    \eta_\lambda (l,x)\underset{\lambda\sim 0}{\sim} f\left(\frac{l}{\lambda}\right)~\delta(x)~,
\end{equation}
i.e. the binomials become asymptotically proportional to Dirac delta functions of the semiclassical variables centered at $x=0\Leftrightarrow l_L=\frac{l}{2}$. This proof, however, has been slightly cavalier: In order to be rightfully considered as a nascent delta function in the variable $x$, $\eta_\lambda(l,x)$ should have been obtained out of $\eta(l,x)$ by \textit{only} replacing $x\mapsto \frac{x}{\lambda}$ (besides the overall $1/\lambda$ prefactor in \eqref{eta_lambda_def}), but here we have that also $l$ is rescaled to $l/\lambda$, and this implies that the implicit exchange of the limit $\lambda\to 0$ and the integral sign in \eqref{etalambda_integral_test} is dubious, as one cannot exclude the possibility that the remaining $\lambda$-dependence in $\eta(l/\lambda,x)$ competes with $g(\lambda x)$ making the integral localize at a point potentially distinct from $x=0$. For the specific nascent delta function in \eqref{eta_lambda_def} this is happily not the case, as we may show by considering its explicit asymptotic expansion at $\lambda\sim 0$, given by:
\begin{align}
&\eta_\lambda(l,x)\underset{\lambda\sim 0}{\sim} \sqrt{\frac{2l}{\pi\lambda(l^2-4x^2)}}{(2l)^{l/\lambda}} \Big(l-2x\Big)^{-\frac{l-2x}{\lambda}}\Big(l+2x\Big)^{-\frac{l+2x}{\lambda}} \label{eta_lambda_asymptotic_line1}\\
&=\sqrt{\frac{2l}{\pi\lambda(l^2-4x^2)}} ~ \text{exp}\left\{ \frac{1}{\lambda} \Big( l\log(2l)-(l-2x)\log(l-2x)-(l+2x)\log(l+2x) \Big) \right\}~. \label{eta_lambda_asymptotic_line2}
\end{align}
We may now recognize that the asymptotic behavior at $\lambda\sim 0$ of any integral of $\eta_\lambda (l,x)$ multiplied by smooth functions of the semiclassical length offset $x$ can be computed via a saddle-point analysis of the function in the exponent of \eqref{eta_lambda_asymptotic_line2}. Direct computation shows that its maximum is indeed located at $x=0$, and therefore the integral will localize at this value.

Altogether, the norm of the $|\chi^{(+)}_{n+1}\rangle$ state posed in \eqref{Norm_binomal_plus} takes the following form in the semiclassical limit:
{\footnotesize	 \begin{align}
    \label{norm_binom_plus_semicl}
    &\langle \chi^{(+)}_{n+1}|\chi^{(+)}_{n+1}\rangle \underset{\lambda\sim 0}{\sim}[n]_q!\int_{\mathbb{R}^2}dl_Ldl_L^\prime~ c^{(+)}(l,l_L;\lambda) \eta_\lambda\left(l,l_L-\frac{l}{2}\right) \eta_\lambda\left(l,l_L^\prime-\frac{l}{2}\right) \left( \frac{\left(1-e^{-l}\right)/2}{\cosh\left(\frac{l_L-l_L^\prime}{2}\right)-e^{-l/2}\cosh\left( \frac{l-l_L-l_L^\prime}{2} \right)} \right)^{2\Delta} \\
    &\underset{\lambda \sim 0}{\sim} c^{(+)}\left(l,\frac{l}{2};\lambda\right)~ [n]_q!\int_{\mathbb{R}^2}dl_Ldl_L^\prime~  \eta_\lambda\left(l,l_L-\frac{l}{2}\right) \eta_\lambda\left(l,l_L^\prime-\frac{l}{2}\right) \left( \frac{\left(1-e^{-l}\right)/2}{\cosh\left(\frac{l_L-l_L^\prime}{2}\right)-e^{-l/2}\cosh\left( \frac{l-l_L-l_L^\prime}{2} \right)} \right)^{2\Delta} \\
    &\underset{\lambda \sim 0}{\sim} c^{(+)}\left(l,\frac{l}{2};\lambda\right) \langle \chi^{(+)}_{n}|\chi^{(+)}_n\rangle~,
\end{align}}%
where the ``$\underset{\lambda \sim 0}{\sim}$'' symbols are used to formally denote asymptotic equivalence at $\lambda\sim 0$. In the second line, the $c^{(+)}(l,l_L;\lambda)$ factor has been pulled outside of the integral taking its value at $l_L=l/2$ thanks to the nascent delta functions $\eta_\lambda$ in the integrand or, equivalently, invoking the saddle-point analysis that follows from writing such functions in the form \eqref{eta_lambda_asymptotic_line2}. As announced earlier, we don't intend to perform the leftover integral explicitly, but we just note that it coincides with the semiclassical limit of the norm of the previous binomial state $|\chi^{(+)}_n\rangle$. More formally, we have proved the following identity:
\begin{equation}
    \label{Norm_binom_plus_semicl_formal}
    \lim_{\substack{\lambda\to 0 \\ n\to\infty\\\lambda n \text{ fixed}}} \frac{\langle \chi^{(+)}_{n+1}|\chi^{(+)}_{n+1}\rangle}{c^{(+)}_{(n/2)}(n)\langle \chi^{(+)}_{n}|\chi^{(+)}_{n}\rangle} = 1~.
\end{equation}

In general, any sum involving binomial coefficients times other functions that have a smooth form in the semiclassical limit localizes in such a limit to the contributions coming from the middle of the summation domain thanks to the fact that the binomials get squeezed into Dirac delta functions. This can be used to prove formally the expression of the $b_n$ coefficients of the $\mathcal{O}TFD$ state studied in section \ref{sect:OTFD_KC}, together with the fact that for such a state the Krylov basis vectors are proportional to the binomial states $|\chi^{(+)}_n\rangle$, which are in turn total chord number eigenstates.

\subsubsection{Saddle-point analysis of generic sums in the semiclassical limit}\label{appx:binomials_semicl_generic}

In this section we shall consider the case in which the sums under consideration involve the binomials plus some other coefficients that do \textit{not} have a smooth limiting form in the semiclassical limit. The analysis of the nascent delta functions needs to be performed with more care in these situations. Specifically, let us imagine that we wish to work with the two-sided Hamiltonian $H_R+zH_L$, for some $z\in \mathbb{R}$ (required for hermiticity). Applying the Lanczos algorithm with this Hamiltonian to the state $|0,0\rangle$ would make us stumble with some \textit{generalized} binomial states, defined as\footnote{In this appendix we are slightly changing notation, in order to adapt it to the generalized binomial states of the form \eqref{Generalized_binomial_state}. The states $\ket{\chi_n^{(+)}}$ are recognized as the Krylov basis of the $\mathcal{O}$TFD complexity, while here we denote with $\ket{\chi_n^{-}}$ what in the rest of the text were the Krylov states $\ket{\chi_n}$ of operator complexity.}
\begin{equation}
    \label{Generalized_binomial_state}
    |\chi^{(z)}_n\rangle := (a_R^\dagger + z a_L^\dagger)^n~|0,0\rangle = \sum_{k=0}^{n}z^k\binom{n}{k}|k,n-k\rangle~.
\end{equation}
The norm of these states defines a function of the variable $z$ (and the system parameters $\lambda$ and $\Delta$) given by:
\begin{equation}
    \label{Norm_binom_generalized}
    \langle\chi^{(z)}_{n+1} | \chi^{(z)}_{n+1} \rangle = \sum_{k^\prime,k=0}^{n}c^{(z)}_k(n+1)~z^{k^\prime + k}\binom{n}{k^\prime}\binom{n}{k}\langle k^\prime,n-k^\prime | k,n-k\rangle=:\mathcal{F}_{n+1}(z;\lambda,\Delta)~,
\end{equation}
where we have introduced a generalized $c$-coefficient
\begin{equation}
    \label{ckn_generalized}
    c^{(z)}_k(n) := n\frac{[n-k]_q}{n-k}\left( 1+z\widetilde{q}q^k \right) + z^2 n\frac{[k+1]_q}{k+1}\left( 1+\frac{1}{z}\widetilde{q}q^{n-1-k} \right)~.
\end{equation}
The second equality in \eqref{Norm_binom_generalized} can be proved using the hermitian adjoint of \eqref{Generalized_binomial_state}. 
For $z=1$, the function $\mathcal{F}_{n+1}(z;\lambda,\Delta)$ becomes the norm of the states \eqref{Binom_plus} studied in the previous section, while for $z=-1$ we get sums with alternating signs, which are of relevance throughout section \ref{Sec:Operator_KC} for the study of the actual binomial states \eqref{Binomial_Ansatz} that contribute to the Krylov states of the state $|0,0\rangle$ evolving under $H_R-H_L$.
For any $z>1$ (resp. for $0<z<1$) the function $z^{l_L/\lambda}$ becomes non-perturbatively large (resp. small) in the semiclassical limit when $\lambda\sim 0$ and, even more worryingly, for $z<0$ it oscillates wildly and does not have a smooth limiting form. This implies that the arguments in section \ref{appx:Binom_delta_EasyCase} do not apply immediately to sums of the form of that in \eqref{Norm_binom_generalized}: One needs to take the semiclassical limit of $z^k\binom{n}{k}$ as a whole. The $z$-dependence in $c_k^{(z)}(n)$ is not disruptive because this coefficient still admits a smooth limiting form as a function of $l$ and $l_L$ in the semiclassical limit regardless of $z$, which is:
\begin{equation}
    \label{ckn_generalized_semicl}
    c^{(z)}_k(n)\underset{\lambda\sim 0}{\sim} c^{(z)}(l,l_L;\lambda) \equiv \frac{l}{l-l_L} \frac{1-e^{-(l-l_L)}}{\lambda}(1+z\widetilde{q}e^{-l_L}) +z^2 \frac{l}{l_L}\frac{1-e^{-l_L}}{\lambda}(1+\frac{1}{z}\widetilde{q}e^{-(l-l_L)})~.
\end{equation}

In what follows, we will assume $z>0$, for which $\mathcal{F}(z;\lambda,\Delta)$ is well-defined in the semiclassical limit, admitting an integral representation. We may then extend the relevant properties to the full complex plane of $z\in \mathbb{C}$ (except for singular points) by analytic continuation\footnote{We remind that a sufficient condition for the analytic continuation of a function to be unique is being analytic on an interval. Defining $\mathcal{F}(z;\lambda,\Delta)$ for $z>0$ will therefore be enough. Note that the generalized $c$-coefficient in \eqref{ckn_generalized} is manifestly a meromorphic function of $z$.}$^{\text{,}}$\footnote{Note that if $z$ is not real, the function $\mathcal{F}(z;\lambda,\Delta)$ is no longer equal to the norm of generalized binomial states, because in that case $(a_R^\dagger + z a_L^\dagger)^\dagger=a_R + \overline{z}a_L$, but nothing prevents us from formally studying its analytic continuation to the full complex plane. On a related note, the Hamiltonian $H_R+zH_L$ is not hermitian if $\text{Im}(z)\neq 0$.}, in order to eventually focus on the values of $z$ that are relevant for this paper.

Taking the semiclassical limit of \eqref{Norm_binom_generalized} analogously to how we proceeded in section \eqref{appx:Binom_delta_EasyCase} yields the following integral expression:
\begin{align}
    &\mathcal{F}_{n+1}(z;\lambda,\Delta) \label{Norm_binom_generalized_integral_line1}\\
    &\underset{\lambda\sim 0}{\sim} [n]_q! \int_{\mathbb{R}^2} dl_L dl_L^\prime c^{(z)}(l,l_L;\lambda)\eta_{\lambda}(l,l_L;z)\eta_\lambda(l,l_L^\prime;z) \left( \frac{\left(1-e^{-l}\right)/2}{\cosh\left(\frac{l_L-l_L^\prime}{2}\right)-e^{-l/2}\cosh\left( \frac{l-l_L-l_L^\prime}{2} \right)} \right)^{2\Delta}~, \label{Norm_binom_generalized_integral_line2}
\end{align}
where we have defined
\begin{equation}
    \label{Nascent_generalized}
    \eta_\lambda(l,l_L;z) := \frac{1}{\lambda}\eta\left(\frac{l}{\lambda},\frac{l_L}{\lambda}\right)~,\qquad \text{for} \qquad \eta(l,l_L)=\left\{ \begin{aligned}
        & \frac{z^{l_L} ~\Gamma(l +1)}{\Gamma(l_L +1) \Gamma(l-l_L+1)}~,\quad 0\leq l_L\leq l~, \\
        &0~,\qquad\qquad\qquad \qquad\qquad~\text{else}~.
    \end{aligned} \right.  
\end{equation}
As announced, this time it is not \textit{a priori} obvious where the center of this nascent delta function will be. We have therefore chosen to define it as taking $l_L$ as its second argument, rather than taking $x=l_L-l/2$, which was the preferred choice in section \ref{appx:Binom_delta_EasyCase}. In order to identify the point on which $\eta_\lambda(l,l_L;z)$ localizes we shall directly consider its asymptotic expansion at $\lambda\sim 0$:
\begin{equation}
    \label{nascent_generalized_asymptotic}
    \eta_\lambda (l,l_L;z)\underset{\lambda\sim 0}{\sim} \sqrt{\frac{l}{2\pi\lambda l_L (l-l_L)}} \text{exp}\left\{\frac{1}{\lambda}\Big( l\log(l) -l_L\log(l_L/z)-(l-l_L)\log(l-l_L) \Big)\right\}~.
\end{equation}
The function \eqref{nascent_generalized_asymptotic} will be sharply localized, for $\lambda\sim 0$, at a point $l_L=l_L^{(*)}(z;l)$ given by the extremum (which is in fact a maximum when $z>0$) of the exponent. Direct calculation yields that the $z$-dependent center of $\eta_\lambda(l,l_L;z)$ is:
\begin{equation}
    \label{z_dependent_center}
    l_L^{(*)}(z;l)=\frac{lz}{z+1}~.
\end{equation}
This shows that in the semiclassical limit the expression $\frac{z^k}{\lambda}\binom{n}{k}$ becomes asymptotically proportional to a Dirac delta function $\delta\Big( l_L - l_L^{(*)}(z;l) \Big)$, allowing to write:
\begin{equation}
    \label{Integral_localization_generic}
    \mathcal{F}_{n+1}(z;\lambda,\Delta) \underset{\lambda\sim 0}{\sim} c^{(z)}\Big(l,l_L^{(*)}(z;l);\lambda\Big) \mathcal{F}_n(z;\lambda,\Delta)~,
\end{equation}
or equivalently
\begin{equation}
    \label{Integral_localization_generic_formal}
    \lim_{\lambda\to 0}\frac{ \mathcal{F}_{n+1}(z;\lambda,\Delta) }{c^{(z)}\Big(l,l_L^{(*)}(z;l);\lambda\Big) \mathcal{F}_n(z;\lambda,\Delta)}=1~.
\end{equation}

Strictly speaking, the above has been derived assuming $z>0$, and we may now extend these relations to the whole complex plane of $z\in\mathbb{C}$, although we shall not discuss cases with $\text{Im}(z)\neq 0$ because they don't come from the analysis of a hermitian two-sided Hamiltonian $H_R+zH_L$. For the particular value $z=1$, expression \eqref{z_dependent_center} gives us that the integral \eqref{Norm_binom_generalized_integral_line2} localizes at $l_L=\frac{l}{2}$, as we had already shown in the case-specific analysis of section \ref{appx:Binom_delta_EasyCase}. Generally speaking, for $z\geq 0$ we have that the localization point is $l_L^{(*)}\in[0,l[$, which belongs to the domain of $l_L$. More interestingly, we note that the case of interest for operator complexity, which involves time evolution generated by the Hamiltonian $H_R-H_L$, is addressed by $z=-1$: for this value of $z$, the center $l_L^{(*)}(z;l)$ in \eqref{z_dependent_center} depicts a vertical asymptote, i.e. the localization point is $l_L^{(*)}\equiv \infty$, which formally signals that the integral might pick contributions from the edges of the integration domain. We can understand this qualitatively as follows: the expression $(-1)^{(l_L/\lambda)}\binom{l/\lambda}{l_L/\lambda}$ is, in absolute value, equal to $\binom{l/\lambda}{l_L/\lambda}$ and hence for small $\lambda$, both functions get squeezed and become sharply peaked around $l_L=l/2$; however, while the latter does so in a symmetric fashion, the former is wildly oscillating, with frequency proportional to $\frac{1}{\lambda}$ (since $(-1)^{l_L/\lambda}=e^{i\pi l_L/\lambda}$ in the principal branch). In the semiclassical limiting procedure, there always exists a limiting scheme such that the oscillating function is odd around the center\footnote{This is because both $k$ and $k+1$ are undistinguishable in this limit, where $k\to\infty$ such that $\lambda k = l_L$.}: integrating it against a smooth test function $g(l_L)$ that comes from the semiclassical limit of some discrete coefficient $g_k$ in the original discrete sum will yield exactly zero (up to surface terms) if such a test function is even around $l_L=l/2$, because the contributions to the integral coming from the right of the peak will get exactly canceled out by those to the left. This is why $(-1)^k\binom{n}{k}$ cannot be claimed to converge (in the semiclassical limit) to a Delta function even if its absolute value does: there are various ways to take the semiclassical limit of the sum, which yield different results if the test function comes from terms $g_k$ that are symmetric around the middle of the summation domain. This means that formally the limit does not exist\footnote{We may also note that $\sum_{k=0}^n (-1)^k\binom{n}{k}=0$ for all $n>0$. In the semiclassical limit, $\eta_\lambda(l,l_L;z=-1)$ is therefore not normalizable, and as such it can never tend in the sense of distributions to a well-defined delta function.}. Nevertheless, we can still make the following point: If the test function integrated against $\eta_\lambda(l,l_L;z=-1)$ \textit{is} ever so slightly asymmetric around $l_L=l/2$, the integrand contributions in any neighborhood to the right of $l_L=l/2$ will not exactly cancel out against those coming from the left, and in fact this non-zero difference will get amplified the smaller $\lambda$ gets. For the set of test functions that are not symmetric around $l_L=\frac{l}{2}$, $\eta_\lambda (l,l_L;z=-1)$ is indeed acceptable as a delta function at $l_L=l/2$. 

As a simple illustration, consider the following sum:
\begin{equation}
    \label{appx_alternating_sign_sum_illustration}
    S=\sum_{k=0}^n (-1)^k\binom{n}{k}g_k~,
\end{equation}
where we assume that $g_k$ is such that it takes a smooth form in the semiclassical limit, e.g. $g_k\equiv g(\lambda k)\to g(l_L)$. Additionally assuming, for simplicity, that $n$ is even, we can define $k_*\equiv (n-1)/2$ and, after some algebra, we reach:
\begin{equation}
    \label{appx_alternating_sign_sum_illustration_rearranged}
    S = (-1)^{k_*}\sum_{y=0}^{k_{*}}(-1)^y\binom{n}{k_*-y}\big(g_{k_*-y}-g_{k_*+1+y}\big)~.
\end{equation}
In the above expression, the alternating sign is now harmless because the summation domain is no longer symmetric with respect to the center of symmetry of the binomial: in order to assess whether in the semiclassical limit the sum is parametrically dominated by the contribution with $y=0$ we only need to check whether the differences $g_{k_*-y}-g_{k_*+1+y}$ are not zero, i.e. whether the test function is not symmetric around the symmetry center of the binomial\footnote{One may raise the concerning point that the overall phase of the sum, dominated by the $y=0$ term, is not well-defined in the limit. Fortunately, the sums such as \eqref{Norm_binom_generalized} that are of interest for us feature two binomials, each of them with their corresponding alternating sign, so that the overall phase $(-1)^{2k_*}=1$ cancels out.}. This is a concrete manifestation of the qualitative discussion we made in the paragraph preceding equation \eqref{appx_alternating_sign_sum_illustration}. 

Considering our integral of interest, \eqref{Norm_binom_generalized_integral_line2}, we note that $c^{(-)}(l,l_L;\lambda)$ is even around $l_L=l/2$, but the rightmost term of the integrand, coming from the semiclassical limit of the inner product, is not whenever $\Delta>0$ (in fact it is not factorizable as the product of functions of $l_L$ and $l_L^\prime$ separately), its symmetries being those of the inner product spelled out in \eqref{IPS_appx} and \eqref{LRS_appx}. Consequently, for any value of $\Delta>0$, the integrand of \eqref{Norm_binom_generalized_integral_line2} is always non-symmetric to some extent around $l_L=\frac{l}{2}$, and such lack of symmetry is accentuated the smaller $\lambda$ is, which allows us to treat $\eta_\lambda(l,l_L;z=-1)$ as a delta function at $l_L=l/2$ and to accordingly pull outside of the integral the desired factor of $c^{(-)}(l,l/2;\lambda)$.

In summary, for the $z=-1$ case of interest for operator complexity, the integral \eqref{Norm_binom_generalized_integral_line2} effectively localizes at $l_L=l/2$ given $\Delta>0$ thanks to the asymmetric profile of the inner product factor if regarded as a function of $l_L$, and we may write
\begin{equation}
    \label{Norm_binom_minus_asympt_conclusion}
    \langle \chi_{n+1}^{(-)}|\chi_{n+1}^{(-)}\rangle \sim c^{(-)}_{n/2}(n)\langle \chi_n^{(-)}|\chi^{(-)}_n\rangle~.
\end{equation}

In the numerical implementations of section \ref{subsect:numerics} and appendix \ref{appx:Numerics_further_results}, the accuracy of \eqref{Norm_binom_minus_asympt_conclusion} may be tested by fixing the value of $\Delta$ and scanning through values of $\lambda$ (i.e. the vertical paths on figure \ref{fig:ParameterSpace}). We consistently observed that, given a value of $\Delta>0$, arbitrarily small values of $\lambda$ make expression \eqref{Norm_binom_minus_asympt_conclusion} become arbitrarily accurate. On the other hand, we note that numerically fixing the value of $\lambda$ and decreasing $\Delta$ towards zero will produce a departure from \eqref{Norm_binom_minus_asympt_conclusion}, as the inner product gets progressively flattened (hence becoming more symmetric around the center of the integration domain). We can confirm this by observing the numerical results in appendix \ref{appx:Numerics_further_results} corresponding to points lying on the same horizontal line in parameter space (cf. figure \ref{fig:ParameterSpace}). For exactly $\Delta=0$ the inner product becomes a constant, and therefore even around $l_L=l/2$, in which case we don't have localization of the integral regardless of the value of $\lambda$ and it just vanishes\footnote{This would be an instance of a sum of the form \eqref{appx_alternating_sign_sum_illustration_rearranged} in which all the terms are identically zero due to the vanishing of the differences inside the parenthesis. We can verify via direct computation that the binomial states $|\chi_n^{(-)}\rangle$ are null states for any finite $n$ when $\Delta=0$, since they belong to the kernel of the inner-product matrix $g_{k^\prime,k}^{(n)}$ of the corresponding total chord number sector, as discussed in appendix \ref{appx:numerics_inner_product}.}, signaling that $|\chi_n^{(-)}\rangle$ becomes a null state in this limit: this is in agreement with the discussion in section \ref{appx:numerics_inner_product}, where the one-particle algebra is argued to contract to the zero-particle algebra when $\Delta=0$ (or equivalently $\widetilde{q}=1$ if $\lambda$ is fixed). 

Finally, let us remark that, the bigger $\Delta$ is, the more peaked the inner product term in the integrand \eqref{Norm_binom_generalized_integral_line2} will be, making it more non-symmetric as a function of $l_L$. This suggests that we can parametrically control the accuracy of \eqref{Norm_binom_minus_asympt_conclusion} if we additionally scale $\Delta\to\infty$ while $\lambda\to 0$ in the semiclassical limit, putting on more solid footing the discussion presented at the beginning of this appendix, where the product $\Delta\lambda$ is kept fixed on top of the semiclassical limit.

\section{Generic classical solution of Morse-like Hamiltonians}\label{appx:Morse_generic_solution}

For reference, this appendix gathers the generic classical solution of the Hamiltonians of the Morse type that are ubiquitous throughout this work. These are of the form:
\begin{equation}
    \label{Morse_Hamiltonian_generic}
    H = \alpha p^2 + U(x)~,
\end{equation}
where $\alpha>0$ the potential energy is
\begin{equation}
    \label{Morse_potential_generic}
    U(x)=\beta e^{-x/2} + \gamma e^{-x}~,
\end{equation}
where $\beta,\gamma\in\mathbb{R}$. When $\beta<0$ and $\gamma>0$, the potential $U(x)$ is popularly known to describe a diatomic molecule with bound states\footnote{See \cite{Gao_2022} for a two-dimensional gravity application of this potential.}. In our analysis, however, we will only be interested in the two following situations:
\begin{itemize}
    \item $\beta,\gamma>0$, yielding a monotonously decreasing potential that asymptotes to zero.
    \item $\beta>0$ and $\gamma<0$, in which case $U(x)$ features an unstable maximum (note that this is the reverse of the potential for the diatomic molecule).
\end{itemize}
In any case, given some generic coefficients $\beta,\gamma\in\mathbb{R}$, it is possible to find the most generic form of solutions whose energy $E$ does not exceed the maximum value of $U(x)$, hence describing trajectories that \textit{roll down} the potential. All the solutions of interest for the current paper fall within this class\footnote{Note that a trajectory with energy bigger than $\max_{x\in\mathbb{R}} U(x)$ can go from $x\to -\infty$ at $t\to -\infty$ to $x\to + \infty$ at $t\to +\infty$. Bulk length in a black hole background does not feature this type of behavior, and neither does Krylov complexity, which is known to be an even function of time \cite{Sanchez-Garrido:2024pcy}.}. Given a trajectory with $E\leq \max_{x\in\mathbb{R}} U(x)$, there will always exist a time $t_0$ at which all its energy is potential, hence satisfying $\dot{x}(t_0)=0$. We can therefore find the most generic solution $x(t)$ by solving the equation of motion of \eqref{Morse_Hamiltonian_generic} given the initial conditions:
\begin{equation}
    \label{Morse_generic_initial_conditions}
    x(t_0)=x_0~,\qquad \dot{x}(t_0)=0~.
\end{equation}
Direct integration of the equation of motion, or the use of a sufficiently inspired Ansatz, yields:
\begin{equation}
    \label{Morse_generic_solution}
    x(t)=x_0 + 2\log\left\{ 1 + b(\beta,\gamma;x_0) \sinh^2\Big( (t-t_0)~c(\alpha,\beta,\gamma;x_0) \Big) \right\}~,
\end{equation}
where $(t_0,x_0)$ are the two independent integration constants that parametrize the solution and the $b$ and $c$ coefficients are given by:
\begin{align}
    &b(\beta,\gamma;x_0) = \frac{\beta + 2\gamma e^{-x_0/2}}{\beta + \gamma e^{-x_0/2}}~,\label{Morse_solution_b_coefficient} \\
    &c(\alpha,\beta,\gamma;x_0) = \frac{1}{2}e^{-x_0/4} \sqrt{\alpha\Big( \beta + \gamma e^{-x_0/2} \Big)} ~.\label{Morse_solution_c_coefficient}
\end{align}
This is a solution with total energy:
\begin{equation}
    \label{Morse_generic_total_energy}
    E\equiv E(\beta,\gamma;x_0) = U(x_0)\leq \max_{x\in\mathbb{R}}U(x)~.
\end{equation}

\section{Operators in the effective averaged theory}\label{Appx:Lanczos_W_T}

This appendix suggests an alternative route to operator Krylov complexity in double-scaled SYK based on the use of an effective operator evolving in the Heisenberg picture. Even though this approach was not taken in the present paper, we shall spell out the setup for reference.

\subsection{Effective operator approach}

In \cite{Berkooz:2018jqr} we find the following prescription for computing the disorder average of an operator self-correlation function, which is obtained based on the analysis of chord diagrams with one matter chord:
\begin{equation}
    \label{Averaged_correlation_effective_operator}
    \Big\langle \text{Tr}\left[\mathcal{O}\mathcal{O}(t)\right] \Big\rangle = \langle 0| W e^{itH} W e^{-itH} | 0 \rangle~,
\end{equation}
where, on the right-hand side, $|0\rangle\in\mathcal{H}_{0p}$ is the zero-chord state (representing the unperturbed infinite-temperature TFD state in the matterless sector of the averaged theory \cite{Lin:2022rbf,Rabinovici:2023yex}), $H$ is the effective Hamiltonian in the zero-particle sector of the Hilbert space, given by equation \eqref{H_a_ad}, and the effective operator $W$ is an operator that acts over the zero-particle sector as:
\begin{eqnarray}
    \label{effective_operator}
    W=\widetilde{q}^{~n} = e^{-\Delta \lambda n}~.
\end{eqnarray}
That is, the operator's infinite-temperature auto-correlation function is given, after disorder average, by the self-correlation of the time-evolving effective operator $W(t)=e^{itH}We^{-itH}$ measured in the zero-chord state. Note that a BCH expansion of the right-hand side of \eqref{Averaged_correlation_effective_operator} yields the averaged moments written in section \ref{Sec:Moments_from_chord_diagrams}. To the extent that the scaling dimension $\Delta$ is holographically related to the mass (or energy) of the bulk perturbation, and $\lambda n$ is related to bulk length, we can say that this picture is morally consistent with the geodesic approximation of correlation functions \cite{Berkooz:2018jqr,Berkooz:2024lgq,Lin:2023trc,Corradini,STRASSLER1992145}.

\subsection{Comparison between operator and state approaches}

The relation between the operator and state approaches may be manifested by combining \eqref{averaged_moments_00state} with \eqref{Averaged_correlation_effective_operator}. The disorder-averaged infinite-temperature auto-correlation of the operator $\mathcal{O}$ given in \eqref{random_operator} is:
\begin{align}
    \label{two_pt_equivalent_formulations}
    \Big\langle \text{Tr}\left[ \mathcal{O} \mathcal{O}(t) \right]  \Big\rangle = C(t) = S(t)~, \\
    C(t) \equiv \langle 0 | W W(t)|0\rangle ~, \label{two_pt_fn_of_W}\\
    S(t) \equiv \langle 0,0 |e^{-it(H_R-H_L)}|0,0\rangle~, \label{survival_amplitude_of_TFD}
\end{align}
where $C(t)$ in \eqref{two_pt_fn_of_W} is the self-correlation function of the operator $W(t)$ and $S(t)$ is the fidelity of the Schrödinger evolution $|\psi(t)\rangle = e^{-it(H_R-H_L)}|0,0\rangle$ of the perturbed state $|0,0\rangle$ as introduced in \eqref{Time_evolving_state_HRminusHL_Schr}. The matching \eqref{two_pt_equivalent_formulations} suggests that, at least morally speaking, the Krylov complexity of the Heisenberg-evolving operator $W(t)$ is related to the Krylov complexity of the Schrödinger-evolving state $|\psi(t)\rangle$, which in the averaged theory represents the perturbed infinite-temperature thermofield double state, $\mathcal{O}(t)|TFD\rangle$. However, as we shall discuss in section \ref{appx:Lanczos_alg_in_op_approach}, the K-complexities of both objects need not be mathematically equal despite the fact that $S(t)=C(t)$. This is a specific feature of DSSYK, where the TFD state is captured by the zero-chord state $|0\rangle$, which has the interpretation of a state belonging to a ``two-sided'' Hilbert space, but yet does not belong to the tensor product of two identical Hilbert spaces.

In more general, potentially higher-dimensional setups, in which the thermofield double can be written as state belonging to the tensor product of two identical Hilbert spaces, \cite{Sanchez-Garrido:2024pcy} proves that the (operator) Krylov complexity of an operator $\mathcal{O}(t)$ is \textit{equal} (i.e. not just qualitatively related) to the (state) Krylov complexity of the TFD double perturbed by such an operator insertion, e.g. $\mathcal{O}_L(t)|TFD\rangle$, which after some algebra can be rewritten as $e^{-it(H_R-H_L)}\mathcal{O}_L|TFD\rangle$, i.e. a state evolving unitarily in the Schrödinger picture with the Hamiltonian $H_R-H_L$, to which the Lanczos algorithm is manifestly applicable. The proof in \cite{Sanchez-Garrido:2024pcy} involves the explicit map between the Krylov problems defined on the different Hilbert spaces, and in particular the bijection between the operator and state Krylov basis elements, as well as the Lanczos coefficients in both the operator and the state versions of the Lanczos algorithm. The fact that the operator's auto-correlation function and the state's survival amplitude are equal is a necessary condition for the Lanczos coefficients (and consequently K-complexity) to coincide, but it is not sufficient\footnote{Given two elements of the \textit{same} Hilbert space (either state space or operator space) the fact that their fidelities coincide is both a necessary and sufficient condition for their Lanczos coefficients to be equal because the Lanczos algorithms being mapped have the same formulation (in particular, they are defined through the same inner product). However, here we are discussing the fidelities of time-evolving elements of \textit{different} Hilbert spaces.}. In fact, as already announced, in the current example of DSSYK we have $C(t)=S(t)$ in \eqref{two_pt_equivalent_formulations} but, as we will see in next section, the Lanczos coefficients of the operator $W(t)$ differ from those of $|\psi(t)\rangle$ computed in section \ref{Sec:Operator_KC}.

\subsection{Lanczos algorithm in the operator approach}\label{appx:Lanczos_alg_in_op_approach}

For the sake of illustration, let us compute explicitly the first Lanczos coefficient of the effective operator $W(t)$ evolving in the Heisenberg picture with the Liouvillian $\mathcal{L}\equiv[H,\cdot]$. The expression of $H$ in coordinates over the normalized chord basis of the zero-particle Hilbert space is given in \eqref{H_matterless}, where it is explicitly seen to take a symmetric and tridiagonal form with zero diagonal. Having noted the initial condition $W$ and the time evolution generator $\mathcal{L}$, in order to implement the operator Lanczos algorithm it is left to decide which is the relevant operator inner product in this case \cite{Parker:2018yvk,viswanath1994recursion}. Expression \eqref{two_pt_equivalent_formulations} shows that the averaged operator two-point function is given by $C(t)$, defined in \eqref{two_pt_fn_of_W}; this object can only be seen as the \textit{survival amplitude} of $W(t)$ if we take the operator inner product to be the projection over the zero-chord state, i.e.
\begin{equation}
    \label{Effective_operator_inner}
    \big(\mathcal{A} \big| \mathcal{B} \big) = \langle 0| \mathcal{A}^\dagger \mathcal{B} |0\rangle~,\quad\forall~\mathcal{A},\mathcal{B}\in\widehat{\mathcal{H}_{0p}}~,
\end{equation}
where $\widehat{\mathcal{H}_{0p}}$ denotes the space of (bounded) linear operators acting over $\mathcal{H}_{0p}$. With the product $(\cdot | \cdot)$ we can see that $C(t)=\big(W\big|W(t)\big)$. We may now implement the operator Lanczos algorithm, as we announced, for the objects $W$, $\mathcal{L}$ and $(\cdot|\cdot)$. The seed of the Lanczos algorithm is the initial operator itself,
\begin{equation}
    \label{Operator_Lanczos_alg_seed}
    |W_0) = |W)~,
\end{equation}
which can be checked to be correctly normalized to one according to \eqref{Effective_operator_inner}. Next, we may construct the non-normalized first Krylov basis element,
\begin{equation}
    \label{Operator_Lanczos_alg_A1}
    |A_1) = \Big| [H,W_0] \Big) = \Big| [H,W] \Big)~.
\end{equation}
Due to the Gaussianity of the disordered model, reflected in the fact that $H$ has zero diagonal in \eqref{H_matterless}, it can also be seen that \eqref{Operator_Lanczos_alg_A1} is already orthogonal to \eqref{Operator_Lanczos_alg_seed} according to the product \eqref{Effective_operator_inner}, not requiring any $a$-coefficients \cite{Lanczos:1950zz,viswanath1994recursion,parlett1998symmetric} in the construction of $|A_1)$ written in \eqref{Operator_Lanczos_alg_A1}. The first operator Lanczos coefficient, which we shall denote as $b^{(op)}_1$, may now be computed as the norm of $|A_1)$ according to \eqref{Effective_operator_inner}:
\begin{equation}
    \label{Operator_Lanczos_alg_b1}
    \left(b^{(op)}_1\right)^2 = \Big(A_1\Big|A_1\Big) = \langle 0 | [H,W]^\dagger [H,W]|0\rangle = -\langle 0 | [H,W]^2|0\rangle = 1-\widetilde{q}~.
\end{equation}
This coefficient passes the check of becoming zero when $\widetilde{q}=1$, in which case the operator $W$ becomes the identity and therefore features no time evolution. However, the $\left(b^{(op)}_1\right)^2$ obtained in \eqref{Operator_Lanczos_alg_b1} differs (by a factor of $2$) from the $b_1^2$ obtained from the state Lanczos algorithm applied to $|\psi(t)\rangle=e^{-it(H_R-H_L)}|0,0\rangle$, given in \eqref{b1_Exact}! 

Let us take a moment to discuss this discrepancy. The fidelity of the state $|\psi(t)\rangle$, given in \eqref{survival_amplitude_of_TFD}, admits a Taylor expansion
\begin{equation}
    \label{Taylor_of_fidelity_appx}
    S(t) = \sum_{n\geq 0}\frac{(it)^{2n}}{(2n)!}\mu_{2n}~,
\end{equation}
where the moments $\mu_{2n}$ are the ones studied in section \ref{Sec:Moments_from_chord_diagrams}, i.e.
\begin{eqnarray}
    \label{moments_of_fidelity_appx}
    \mu_{2n} = \langle 0,0| (H_R-H_L)^{2n}|0,0\rangle~.
\end{eqnarray}
On the other hand, the two-point function of the effective operator admits an analogous Taylor expansion,
\begin{equation}
    \label{Taylor_of_twopt_appx}
    C(t)=\sum_{n\geq 0}\frac{(it)^{2n}}{(2n)!}m_{2n}~,
\end{equation}
with moments
\begin{equation}
    \label{moments_of_twopt_appx}
    m_{2n} = \Big( W\Big| \mathcal{L}^{2n} \Big|W \Big)~,
\end{equation}
By construction, as spelled out in \eqref{two_pt_equivalent_formulations}, $C(t)=S(t)$ and therefore $\mu_{2n}=m_{2n}$, as one may also check explicitly unwrapping the expressions \eqref{moments_of_fidelity_appx} and \eqref{moments_of_twopt_appx}. Since the moments of both the operator's two point function and the state's fidelity are equal, one might conclude that their Lanczos coefficients are equal, since such coefficients are in one-to-one correspondence with the aforementioned moments. This reasoning is, nevertheless, flawed: The bijection relating the $\mu_{2n}$ moments to the state Lanczos coefficients is, in this particular case, \textit{not} the same bijection as the one relating $m_{2n}$ to the operator Lanczos coefficients. Therefore, as announced in the previous section, the equality $\mu_{2n}=m_{2n}$, which follows from the relation $S(t)=C(t)$, is in this case not enough to grant equality of the Lanczos coefficients.

The operator inner product \eqref{Effective_operator_inner}, naturally induced from the definition of $C(t)$ in \eqref{two_pt_fn_of_W}, where disorder-averaged correlation functions at infinite temperature are captured by an expectation value in the zero-chord state, is in fact the responsible for the bijection between $m_{2n}$ and the operator Lanczos coefficients being structurally different from the bijection between the $\mu_{2n}$ and the state Lanczos coefficients. Notably, the effective Liouvillian $\mathcal{L}=[H,\cdot]$ governing the Heisenberg evolution of $W(t)$ is \textit{not} hermitian with respect to the operator inner product \eqref{Effective_operator_inner}, as one can verify by noting that generically
\begin{equation}
    \label{non_hermiticity_Liouvillian_operator_approach}
    \langle 0 | [H,A]^\dagger B|0\rangle \neq \langle 0 | A^\dagger [H,B]|0\rangle\quad \Longleftrightarrow\quad \Big( \mathcal{L} A \Big| B\Big)\neq \Big(  A \Big|\mathcal{L}\Big| B\Big)
\end{equation}
for two arbitrary operators $A,B\in\widehat{\mathcal{H}_{0p}}$. The recursion method relating the moments \eqref{moments_of_twopt_appx} to the operator Lanczos coefficients $b_n^{(op)}$ changes depending on whether the time evolution generator (in this case, $\mathcal{L}$) is hermitian or not with respect to the operator inner product \cite{Lanczos:1950zz,viswanath1994recursion}. In the state approach, $H_R-H_L$ is a hermitian operator with respect to the one-particle sector inner product, as shown in \cite{Lin:2022rbf,Lin:2023trc} and extensively discussed in sections \ref{sec.ToolsTrade} and \ref{Sec:Operator_KC}. As an illustration, let us consider the case $n=1$: The (operator) moment $m_2$ is given by
\begin{equation}
    \label{operator_moment_m2}
    m_2 = \Big(W\Big|\mathcal{L}^2\Big|W\Big)=2(1-\widetilde{q})~,
\end{equation}
while the square of the first operator Lanczos coefficient, explicitly computed above in \eqref{Operator_Lanczos_alg_b1}, is given by the square of the norm of $|A_1)$, which we may formally rewrite as:
\begin{equation}
    \label{Operator_Lanczos_alg_b1_formal_expression}
    \left(b^{(op)}_1\right)^2=(A_1|A_1) = \Big( W \Big| \mathcal{L}^\dagger \mathcal{L} \Big|W \Big)=1-\widetilde{q}~.
\end{equation}
We now see explicitly that $\left(b^{(op)}_1\right)^2 \neq m_2$ because $\mathcal{L}^\dagger\neq \mathcal{L}$, as we have argued. In the state approach, hermiticity of the two-sided Hamiltonian $H_R-H_L$ allows to show via an analogous manipulation that $b_1^2=\mu_2$, as one can explicitly check comparing \eqref{b1_Exact} and \eqref{mu2}. Summarizing, the identity $S(t)=C(t)$ in \eqref{two_pt_equivalent_formulations} implies that $\mu_2=m_2$, but $\mu_2$ gets mapped to $b_1^2=2(1-\widetilde{q})$ via the recursion method for hermitian time-evolution generators, while $m_2$ maps to $\left(b^{(op)}_1\right)^2=1-\widetilde{q}$ through the recursion method for non-hermitian time-evolution generators. Not inconsistently, $S(t)=C(t)$ but the Lanczos coefficients that one would obtain from the operator approach need not be mathematically equal to those in the state approach. 

Furthermore, in order to correctly implement the Lanczos algorithm for this case in which $\mathcal{L}$ is non-hermitian with respect to the relevant operator inner product, one would actually need to resort to the extension of the Lanczos algorithm for non-hermitian generators \cite{Lanczos:1950zz}, which essentially consists on two parallel Lanczos algorithms, where roughly speaking there is one recursion for the bras and an analogous recursion for the kets that need to be simultaneously kept track of. Finally, as an additional technical complication, we may note that, as defined, the product \eqref{Effective_operator_inner} is not an entirely well-defined inner product, since it admits zero-norm operators. Given an operator $A\in\widehat{\mathcal{H}_{0p}}$ we have that
\begin{equation}
    \label{zero_norm_operator}
    (A|A)=0\quad\Longleftrightarrow\quad \langle n|A|0\rangle=0~\forall~n\geq 0~,
\end{equation}
i.e. all operators whose kernel contains the state $|0\rangle$ have zero norm according to the inner product \eqref{Effective_operator_inner}. Physically, this is not so important because in order to make contact with disorder-averaged observables we shall eventually only consider expectation values in the zero chord state, so the zero-norm operators may be regarded as \textit{effectively} null operators in the sense that they will not contribute the sought observables. With this motivation, one may attempt to ``fix'' the inner product by repackaging the operators in $\widehat{\mathcal{H}_{0p}}$ into universality classes where operators that differ by a null operator belong to the same class.

We leave the implementation of the (duly modified) Lanczos algorithm in this operator approach as an interesting open problem. It would be enlightening to compare the eventually obtained operator Krylov complexity in this approach to the Krylov complexity of states in the one-particle Hilbert space that has constituted the center of the analysis in this paper. Even if they are morally related concepts, as they both probe the complexity due to an operator insertion perturbing the thermofield double state, they are mathematically different objects that require a separate analysis.

\bibliography{references}
\end{document}